\documentclass[10pt]{article}
\usepackage{epsfig}
\usepackage{axodraw}
\usepackage{graphicx}
\usepackage{rotate}
\usepackage{latexsym}
\usepackage{amssymb}
\def\csuma{Fakult\"at f\"ur Physik, Albert-Ludwigs Universit\"at, Freiburg,
Germany\\
Institut f\"ur Theoretische Teilchenphysik, Universit\"at Karlsruhe,
Karlsruhe, Germany}
\def\csumb{Dipartimento di Fisica, Universit\`a di Padova 
and INFN--Padova, Italy\\
Departament de F\'\i sica Te\`orica and IFIC,
Universitat de Val\`encia -- CSIC, Spain\\
Institut f\"{u}r Theoretische Physik, Universit\"{a}t Bern, Bern, 
Switzerland}
\def\csumc{Dipartimento di Fisica Teorica, Universit\`a di Torino, Italy\\
INFN, Sezione di Torino, Italy}
\def\csumd{Max-Plank-Institut f\"ur Physik (Werner-Heisenberg Institut)\\
F\"ohringer Ring 6, D-80805 Munich, Germany}
\def\support{\footnote{Work supported by the
European Union under contract HPRN-CT-2000-00149 and by MIUR under contract
\\ 2001023713$\_$006.}}
\def\Title#1{\begin{center} {\Large\bf #1 } \end{center}}
\newcommand{\Authors}[2]{\begin{center}{ \sc #1 \hspace{0.1cm} {\rm and}
\hspace{0.1cm} #2} \end{center}}

\def\Author#1{\begin{center}{ \sc #1} \end{center}}
\def\Address#1{\begin{center}{ \it #1} \end{center}}

\newenvironment{Abstract}{\begin{quotation}  }{\end{quotation}}

\def\Acknowledgments{\bigskip  \bigskip \begin{center}
          \large\bf Acknowledgments\end{center}}
\def\email#1{\footnote{#1}}
\makeatletter
\def\section{\@startsection{section}{0}{\z@}{5.5ex plus .5ex minus
 1.5ex}{2.3ex plus .2ex}{\large\bf}}
\def\subsection{\@startsection{subsection}{1}{\z@}{3.5ex plus .5ex minus
 1.5ex}{1.3ex plus .2ex}{\normalsize\bf}}
\def\subsubsection{\@startsection{subsubsection}{2}{\z@}{-3.5ex plus
-1ex minus  -.2ex}{2.3ex plus .2ex}{\normalsize\sl}}

\renewcommand{\@makecaption}[2]{%
   \vskip 10pt
   \setbox\@tempboxa\hbox{\small #1: #2}
   \ifdim \wd\@tempboxa >\hsize     
       \small #1: #2\par          
     \else                        
       \hbox to\hsize{\hfil\box\@tempboxa\hfil}
   \fi}
 \def\citenum#1{{\def\@cite##1##2{##1}\cite{#1}}}
\def\citea#1{\@cite{#1}{}}
\newcount\@tempcntc
\def\@citex[#1]#2{\if@filesw\immediate\write\@auxout{\string\citation{#2}}\fi
  \@tempcnta\z@\@tempcntb\m@ne\def\@citea{}\@cite{\@for\@citeb:=#2\do
    {\@ifundefined
       {b@\@citeb}{\@citeo\@tempcntb\m@ne\@citea\def\@citea{,}{\bf }\@warning
       {Citation `\@citeb' on page \thepage \space undefined}}%
    {\setbox\z@\hbox{\global\@tempcntc0\csname b@\@citeb\endcsname\relax}%
     \ifnum\@tempcntc=\z@ \@citeo\@tempcntb\m@ne
       \@citea\def\@citea{,}\hbox{\csname b@\@citeb\endcsname}%
     \else
      \advance\@tempcntb\@ne
      \ifnum\@tempcntb=\@tempcntc
      \else\advance\@tempcntb\m@ne\@citeo
      \@tempcnta\@tempcntc\@tempcntb\@tempcntc\fi\fi}}\@citeo}{#1}}
\def\@citeo{\ifnum\@tempcnta>\@tempcntb\else\@citea\def\@citea{,}%
  \ifnum\@tempcnta=\@tempcntb\the\@tempcnta\else
  {\advance\@tempcnta\@ne\ifnum\@tempcnta=\@tempcntb \else\def\@citea{--}\fi
    \advance\@tempcnta\m@ne\the\@tempcnta\@citea\the\@tempcntb}\fi\fi}
\makeatother
\newcommand{\nl}{\nonumber\\}
\newcommand{\nn}{\nonumber}

\newcommand{\lpar}{\left(}                            
\newcommand{\rpar}{\right)}

\newcommand{\bq}{\begin{equation}}                    
\newcommand{\eq}{\end{equation}}
\newcommand{\bqa}{\arraycolsep 0.14em\begin{eqnarray}}
\newcommand{\eqa}{\end{eqnarray}}
\newcommand{\ba}[1]{\begin{array}{#1}}
\newcommand{\ea}{\end{array}}
\newcommand{\ben}{\begin{enumerate}}
\newcommand{\een}{\end{enumerate}}
\newcommand{\bei}{\begin{itemize}}
\newcommand{\eei}{\end{itemize}}
\newcommand{\eqn}[1]{Eq.(\ref{#1})}
\newcommand{\eqns}[2]{Eqs.(\ref{#1})--(\ref{#2})}

\newcommand{\tabn}[1]{Tab.~\ref{#1}}

\newcommand{\fig}[1]{Fig.~\ref{#1}}
\newcommand{\figs}[2]{Figs.~\ref{#1}--\ref{#2}}
\newcommand{\sect}[1]{Section~\ref{#1}}

\newcommand{\sectm}[2]{Sects.~\ref{#1} -- \ref{#2}}

\newcommand{\appendx}[1]{Appendix~\ref{#1}}

\def\Re{\mathop{\operator@font Re}\nolimits}
\def\Im{\mathop{\operator@font Im}\nolimits}
\newcommand{\ord}[1]{{\cal O}\lpar#1\rpar}

\newcommand{\ib}{i}

\newcommand{\wb}{W}

\newcommand{\zb}{Z}

\newcommand{\hb}{H}

\newcommand{\barf}{\overline f}

\newcommand{\mw}{M_{_W}}

\newcommand{\mz}{M_{_Z}}

\newcommand{\mh}{M_{_H}}

\newcommand{\mm}{m_{\mu}}

\newcommand{\mws}{M^2_{_W}}

\newcommand{\mzs}{M^2_{_Z}}

\newcommand{\mhs}{M^2_{_H}}

\newcommand{\gb}{g}

\newcommand{\stw}{s_{\theta}}             
\newcommand{\ctw}{c_{\theta}}

\newcommand{\sla}[1]{/\!\!\!#1}

\newcommand{\spro}[2]{{#1}\cdot{#2}}
\newcommand{\gfd}{\gamma_5}

\newcommand{\gapu}[1]{\gamma^{#1}}

\newcommand{\egam}[1]{\Gamma\lpar#1\rpar}               

\newcommand{\intmomii}[3]{\int\,d^{#1}#2\,\int\,d^{#1}#3}
\newcommand{\intfx}[1]{\int_{\scriptstyle 0}^{\scriptstyle 1}\,d#1}

\newcommand{\MSB}{\overline{MS}}

\newcommand{\ep}{\epsilon}

\newcommand{\tHs}{\mu}

\newcommand{\tHss}{\mu^2}
\newcommand{\Reb}{{\rm{Re}}}

\newcommand{\tpfi}{\lpar 2\pi\rpar^4\ib}

\newcommand{\upar}[1]{u}

\newcommand{\ssA}{{\scriptscriptstyle{A}}}
\newcommand{\ssB}{{\scriptscriptstyle{B}}}
\newcommand{\ssC}{{\scriptscriptstyle{C}}}
\newcommand{\ssD}{{\scriptscriptstyle{D}}}
\newcommand{\ssE}{{\scriptscriptstyle{E}}}
\newcommand{\ssF}{{\scriptscriptstyle{F}}}
\newcommand{\ssG}{{\scriptscriptstyle{G}}}
\newcommand{\ssH}{{\scriptscriptstyle{H}}}
\newcommand{\ssI}{{\scriptscriptstyle{I}}}
\newcommand{\ssJ}{{\scriptscriptstyle{J}}}
\newcommand{\ssK}{{\scriptscriptstyle{K}}}
\newcommand{\ssL}{{\scriptscriptstyle{L}}}
\newcommand{\ssM}{{\scriptscriptstyle{M}}}
\newcommand{\ssN}{{\scriptscriptstyle{N}}}

\newcommand{\ssP}{{\scriptscriptstyle{P}}}
\newcommand{\ssQ}{{\scriptscriptstyle{Q}}}
\newcommand{\ssR}{{\scriptscriptstyle{R}}}
\newcommand{\ssS}{{\scriptscriptstyle{S}}}

\newcommand{\ssU}{{\scriptscriptstyle{U}}}
\newcommand{\ssV}{{\scriptscriptstyle{V}}}

\newcommand{\ssX}{{\scriptscriptstyle{X}}}

\newcommand{\ssZ}{{\scriptscriptstyle{Z}}}

\newcommand{\bqas}{\begin{eqnarray*}}
\newcommand{\eqas}{\end{eqnarray*}}

\def\app#1#2 {{\it Acta. Phys. Pol.} {\bf#1},#2}
\def\cpc#1#2 {{\it Computer Phys. Comm.} {\bf#1},#2}
\def\np#1#2 {{\it Nucl. Phys.} {\bf#1},#2}
\def\pl#1#2 {{\it Phys. Lett.} {\bf#1},#2}
\def\prep#1#2 {{\it Phys. Rep.} {\bf#1},#2}
\def\prev#1#2 {{\it Phys. Rev.} {\bf#1},#2}
\def\prl#1#2 {{\it Phys. Rev. Lett.} {\bf#1},#2}
\def\zp#1#2 {{\it Zeit. Phys.} {\bf#1},#2}
\def\sptp#1#2 {{\it Suppl. Prog. Theor. Phys.} {\bf#1},#2}
\def\mpl#1#2 {{\it Modern Phys. Lett.} {\bf#1},#2}
\def\jetp#1#2 {{\it Sov. Phys. JETP} {\bf#1},#2}
\def\fpj#1#2 {{\it Fortschr. Phys.} {\bf#1},#2}
\def\afp#1#2 {{\it Acta.Phys. Polon.} {\bf#1},#2}
\def\err#1#2 {{\it Erratum} {\bf#1},#2}
\def\ijmp#1#2 {{\it Int. J. Mod. Phys} {\bf#1},#2}
\def\nc#1#2 {{\it Nuovo Cimento} {\bf#1},#2}
\def\ap#1#2 {{\it Ann. Phys.} {\bf#1},#2}
\def\cmp#1#2 {{\it Comm. Math. Phys.} {\bf#1},#2}
\def\el#1#2 {{\it Europhys. Lett.} {\bf#1},#2}
\def\hpa#1#2 {{\it Helv. Phys. Acta} {\bf#1},#2}
\def\yf#1#2 {{\it Yad. Fiz.} {\bf#1},#2}
\def\nim#1#2 {{\it Nucl. Instrum. Meth.} {\bf#1},#2}
\def\spz#1#2 {{\it Sov. Pisma Zhetf} {\bf#1},#2}
\def\jetpl#1#2 {{\it JETP Lett.} {\bf#1},#2}
\def\sjnp#1#2 {{\it Sov. J. Nucl. Phys.} {\bf#1},#2}
\def\ptp#1#2 {{\it Progr. Theor. Phys. (Kyoto)} {\bf#1},#2}
\def\rmp#1#2  {{\it Rev. Mod. Phys.} {\bf#1},#2}
\def\zhetf#1#2 {{\it ZhETF} {\bf#1},#2}
\def\prs#1#2 {{\it Proc. Roy. Soc.} {\bf#1},#2}
\def\phys#1#2 {{\it Physica} {\bf#1},#2}

\newcommand{\intfxx}[2]{\int_{\scriptstyle 0}^{\scriptstyle 1}\,d#1\,
                        \int_{\scriptstyle 0}^{\scriptstyle 1}\,d#2}
\def\bfi{\begin{figure}}
\def\efi{\end{figure}}
\newcommand{\intmomsii}[3]{\int\,d^{#1}#2\,d^{#1}#3}

\newcommand{\dsimp}[1]{\int\,dS_{#1}}
\newcommand{\dcub}[1]{\int\,dC_{#1}}

\newcommand{\dcubs}[2]{\int\,dCS\lpar #1\,;\,#2 \rpar}
\newcommand{\aaa}{\ssA}
\newcommand{\aba}{\ssE}
\newcommand{\aban}[1]{#1;\ssE}
\newcommand{\aca}{\ssI}
\newcommand{\acan}[1]{#1;\ssI}
\newcommand{\ada}{\ssM}
\newcommand{\adan}[1]{#1;\ssM}

\newcommand{\bba}{\ssG}
\newcommand{\bban}[1]{#1;\ssG}

\newcommand{\bca}{\ssK}
\newcommand{\bcan}[1]{#1;\ssK}
\newcommand{\bbb}{\ssH}
\newcommand{\bbbn}[1]{#1;\ssH}

\newcommand{\bX}{{\overline X}}
\newcommand{\bY}{{\overline Y}}

\newcommand{\ox}{{\overline x}}
\newcommand{\oy}{{\overline y}}
\newcommand{\chiu}[1]{\chi_{_{#1}}}
\newcommand{\xiu}[1]{\xi_{_{#1}}}

\newcommand{\GS}{{\cal G}raph{\cal S}hot}

\newcommand{\CIM}[1]{\int\,{\cal D} #1}
\newcommand{\cS}{S}
\newcommand{\cV}{V}
\newcommand{\cC}{{\cal C}}

\newcommand{\ono}{\scriptscriptstyle 1 N 1}
\newcommand{\Z}{\scriptscriptstyle 0}
\newcommand{\sa}{\scriptscriptstyle 1}
\newcommand{\msb}{\scriptscriptstyle 2}
\newcommand{\sgena}{\scriptscriptstyle i1}
\newcommand{\sgenb}{\scriptscriptstyle i2}
\newcommand{\sijgen}{\scriptscriptstyle ij}
\newcommand{\sagen}{\scriptscriptstyle 1i}
\newcommand{\sbgen}{\scriptscriptstyle 2i}
\newcommand{\sajgen}{\scriptscriptstyle 1j}

\newcommand{\sasca}{\scriptscriptstyle 1\,;\,1}
\newcommand{\sascb}{\scriptscriptstyle 1\,;\,2}
\newcommand{\sbsca}{\scriptscriptstyle 2\,;\,1}
\newcommand{\sbscb}{\scriptscriptstyle 2\,;\,2}
\newcommand{\saa}{\scriptscriptstyle 11}
\newcommand{\sab}{\scriptscriptstyle 12}
\newcommand{\sbb}{\scriptscriptstyle 22}
\newcommand{\sbc}{\scriptscriptstyle 23}
\newcommand{\sbd}{\scriptscriptstyle 24}
\newcommand{\sba}{\scriptscriptstyle 21}
\newcommand{\szz}{\scriptscriptstyle 00}
\newcommand{\saasca}{\scriptscriptstyle 11\,;\,1}
\newcommand{\saascb}{\scriptscriptstyle 11\,;\,2}
\newcommand{\sabsca}{\scriptscriptstyle 12\,;\,1}
\newcommand{\sabscb}{\scriptscriptstyle 12\,;\,2}
\newcommand{\sbbsca}{\scriptscriptstyle 22\,;\,1}
\newcommand{\sbbscb}{\scriptscriptstyle 22\,;\,2}
\newcommand{\saagen}{\scriptscriptstyle 11i}
\newcommand{\saaa}{\scriptscriptstyle 111}
\newcommand{\saab}{\scriptscriptstyle 112}
\newcommand{\saac}{\scriptscriptstyle 113}
\newcommand{\saad}{\scriptscriptstyle 114}
\newcommand{\sbbgen}{\scriptscriptstyle 22i}
\newcommand{\sbba}{\scriptscriptstyle 221}
\newcommand{\sbbb}{\scriptscriptstyle 222}
\newcommand{\sbbc}{\scriptscriptstyle 223}
\newcommand{\sbbd}{\scriptscriptstyle 224}

\newcommand{\sabgen}{\scriptscriptstyle 12i}
\newcommand{\saba}{\scriptscriptstyle 121}
\newcommand{\sabb}{\scriptscriptstyle 122}
\newcommand{\sabc}{\scriptscriptstyle 123}
\newcommand{\sabd}{\scriptscriptstyle 124}
\newcommand{\sabe}{\scriptscriptstyle 125}
\newcommand{\saca}{\scriptscriptstyle 131}
\newcommand{\saaagen}{\scriptscriptstyle 111i}
\newcommand{\saaaa}{\scriptscriptstyle 1111}
\newcommand{\saaab}{\scriptscriptstyle 1112}
\newcommand{\saaac}{\scriptscriptstyle 1113}
\newcommand{\saaad}{\scriptscriptstyle 1114}
\newcommand{\saaae}{\scriptscriptstyle 1115}
\newcommand{\saaaf}{\scriptscriptstyle 1116}
\newcommand{\sbbbgen}{\scriptscriptstyle 222i}
\newcommand{\sbbba}{\scriptscriptstyle 2221}
\newcommand{\sbbbb}{\scriptscriptstyle 2222}
\newcommand{\sbbbc}{\scriptscriptstyle 2223}
\newcommand{\sbbbd}{\scriptscriptstyle 2224}
\newcommand{\sbbbe}{\scriptscriptstyle 2225}
\newcommand{\sbbbf}{\scriptscriptstyle 2226}
\newcommand{\sbbbg}{\scriptscriptstyle 2227}
\newcommand{\sbbbh}{\scriptscriptstyle 2228}
\newcommand{\sbbbi}{\scriptscriptstyle 2229}
\newcommand{\sbbbj}{\scriptscriptstyle 22210}
\newcommand{\sabbgen}{\scriptscriptstyle 122i}
\newcommand{\sabba}{\scriptscriptstyle 1221}
\newcommand{\sabbb}{\scriptscriptstyle 1222}
\newcommand{\sabbc}{\scriptscriptstyle 1223}
\newcommand{\sabbd}{\scriptscriptstyle 1224}
\newcommand{\sabbe}{\scriptscriptstyle 1225}
\newcommand{\sabbf}{\scriptscriptstyle 1226}
\newcommand{\sabbg}{\scriptscriptstyle 1227}
\newcommand{\sabbh}{\scriptscriptstyle 1228}
\newcommand{\sabbi}{\scriptscriptstyle 1229}
\newcommand{\sabbj}{\scriptscriptstyle 12210}
\newcommand{\sabbl}{\scriptscriptstyle 12211}
\newcommand{\sabbm}{\scriptscriptstyle 12212}
\newcommand{\sabbn}{\scriptscriptstyle 12213}
\newcommand{\sabbo}{\scriptscriptstyle 12214}
\newcommand{\sabbp}{\scriptscriptstyle 12215}
\newcommand{\sabbq}{\scriptscriptstyle 12216}
\newcommand{\sabbr}{\scriptscriptstyle 12217}
\newcommand{\sabbs}{\scriptscriptstyle 12218}
\newcommand{\saabgen}{\scriptscriptstyle 112i}
\newcommand{\saaba}{\scriptscriptstyle 1121}
\newcommand{\saabb}{\scriptscriptstyle 1122}
\newcommand{\saabc}{\scriptscriptstyle 1123}
\newcommand{\saabd}{\scriptscriptstyle 1124}
\newcommand{\saabe}{\scriptscriptstyle 1125}
\newcommand{\saabf}{\scriptscriptstyle 1126}
\newcommand{\saabg}{\scriptscriptstyle 1127}
\newcommand{\saabh}{\scriptscriptstyle 1128}
\newcommand{\saabi}{\scriptscriptstyle 1129}
\newcommand{\saabj}{\scriptscriptstyle 11210}
\newcommand{\saabl}{\scriptscriptstyle 11211}
\newcommand{\saabm}{\scriptscriptstyle 11212}
\newcommand{\saabn}{\scriptscriptstyle 11213}
\newcommand{\saabo}{\scriptscriptstyle 11214}
\newcommand{\saabp}{\scriptscriptstyle 11215}
\newcommand{\saabq}{\scriptscriptstyle 11216}
\newcommand{\saabr}{\scriptscriptstyle 11217}

\newcommand{\cagen}{\scriptscriptstyle Ai}
\newcommand{\caa}{\scriptscriptstyle A1}
\newcommand{\cab}{\scriptscriptstyle A2}
\newcommand{\cac}{\scriptscriptstyle A3}
\newcommand{\cad}{\scriptscriptstyle A4}
\newcommand{\cbgen}{\scriptscriptstyle Bi}
\newcommand{\cba}{\scriptscriptstyle B1}
\newcommand{\cbb}{\scriptscriptstyle B2}
\newcommand{\cbc}{\scriptscriptstyle B3}
\newcommand{\cbd}{\scriptscriptstyle B4}
\newcommand{\ccgen}{\scriptscriptstyle Ci}
\newcommand{\cca}{\scriptscriptstyle C1}
\newcommand{\ccb}{\scriptscriptstyle C2}

\newcommand{\bmid}{\Bigr|}
\newcommand{\oDUV}{{\overline\Delta}_{\ssU\ssV}}
\newcommand{\DUV}{{\Delta}_{\ssU\ssV}}
 
\newcommand{\lstm}[1]{\{m\}_{#1}}

\begin{document}
\begin{titlepage}
%
\vspace{10mm}
\vfill
\def\thefootnote{\fnsymbol{footnote}}
\Title{{\LARGE Two-Loop Tensor Integrals in Quantum Field Theory}\support} 
\vspace{10mm}
\Authors{Stefano Actis\email{actis@to.infn.it}} 
{Giampiero Passarino\email{giampiero@to.infn.it}}
\Address{\csumc}
\Author{Andrea Ferroglia\email{andrea.ferroglia@physik.uni-freiburg.de}}
\Address{\csuma}
\Author{Massimo Passera\email{massimo.passera@pd.infn.it}}
\Address{\csumb}
\Author{Sandro Uccirati\email{uccirati@mppmu.mpg.de}}
\Address{\csumd}
\vfill
\begin{Abstract}
\noindent 
A comprehensive study is performed of general massive, tensor, two-loop
Feynman diagrams with two and three external legs. Reduction to generalized
scalar functions is discussed. Integral representations, supporting the same
class of smoothness algorithms already employed for the numerical evaluation
of ordinary scalar functions, are introduced for each family of diagrams.
\end{Abstract}
\vfill
\begin{center}
\end{center}
\end{titlepage}
\def\thefootnote{\arabic{footnote}}
\setcounter{footnote}{0}
\small
\thispagestyle{empty}
\tableofcontents
\setcounter{page}{1}
\normalsize
\clearpage
\section{Introduction}
This paper is the fifth in a series devoted to the numerical evaluation of
multi-loop, multi-leg Feynman diagrams.  In~\cite{Passarino:2001wv}
(hereafter I) the general strategy was outlined and
in~\cite{Passarino:2001jd} (hereafter II) a complete list of results was
derived for two-loop functions with two external legs, including their
infrared divergent on-shell derivatives. Results for one-loop multi-leg
diagrams were shown in~\cite{Ferroglia:2002mz} and additional material can
be found in~\cite{Ferroglia:2002yr}. Two-loop three-point functions for
infrared convergent configurations were considered in~\cite{Ferroglia:2003yj} 
(hereafter III), where numerical results can be found.

In this article we study the problem of deriving a judicious and efficient
way to deal with tensor Feynman integrals, namely those integrals that occur
in any field theory with spin and non trivial structures for the numerators
of Feynman propagators. Admittedly the topic of this paper is rather
technical, but it is needed as a basis for any realistic calculation of
physical observables at the two-loop level.

The complexity of handling two-loop tensor integrals is reflected in the
following simple consideration: the complete treatment of one-loop tensor
integrals was confined to the appendices of~\cite{Passarino:1979jh}, while
the reduction of general two-loop self-energies to standard scalar integrals
already required a considerable fraction of~\cite{Weiglein:hd}; the
inclusion of two-loop vertices requires the whole content of this paper.
Past experience in the
field has shown the convenience of gathering a complete collection of results
needed for a broad spectrum of applications in one place. We devote the present
article to this task.

While a considerable amount of literature is devoted to the evaluation of
two-loop scalar vertices~\cite{Davydychev:2002hy}, fewer papers deal with
the tensor ones~\cite{Anastasiou:2000mf}; for earlier attempts to reduce and
evaluate two-loop graphs with arbitrary masses we refer the reader to the
work of Ghinculov and Yao~\cite{Ghinculov:2000cz}.

In recent years, the most popular and quite successful tool in dealing with
multi-loop Feynman diagrams in QED/QCD (or in selected problems in different
models, characterized by a very small number of scales), has been the
Integration-By-Parts Identities (IBPI)
method~\cite{'tHooft:1972fi}. However, reduction to a set of Master
Integrals (MI) is poorly known in the enlarged scenario of multi-scale
electroweak physics.

Our experience with one-loop multi-leg diagrams~\cite{Ferroglia:2002mz} shows
that the optimal algorithm to deal with realistic calculations should be able
to treat both scalar and tensor integrals on the same footing.  This algorithm
should not introduce multiplications of the tensor integrals by negative powers
of Gram determinants, as the latter's zeros, although unphysical, may be
dangerously close to the physically allowed region.  The numerical quality of
tensor integrals also worsens if they are expressed in terms of linear
combinations of MI; the coefficients of these combinations have zeros
corresponding to real singularities of the diagram~\cite{Landau:1959fi}, and
the singular behavior is usually badly overestimated leading to numerical
instabilities.

       From the point of view of numerical integration, it really makes
little difference if tensor integrals are expressed in terms of generalized
scalar configurations, or in terms of smooth integral representations which
do not grant any privilege to a particular member of the same class of
integrals. Of course, at the end of the day we are always left with the
problem of numerical cancellations (an issue related to the strategy of
trading one difficult integral for many simpler ones), and the optimal
algorithm should minimize the number of smooth integrals in the final
answer. There is no evidence that employing our approach one encounters more
objectionable features than in reducing everything to MI; rather, in our
opinion, the feasibility of the latter has still to be proved in the complex
environment of the full-fledged Standard Model (SM), even if there are
complete applications in QED~\cite{Mastrolia:2003yz} and in
QCD~\cite{Bern:2003ck}.

We have not included four-point functions in the classification, although
they are certainly needed to compute physical observables for
fermion--anti-fermion annihilations or scattering processes, not yet a top
priority in handling electroweak radiative corrections in the SM at the
two-loop level. Note, however, that there is intense activity in (QED) QCD
scattering processes~\cite{Bern:2003ck} and~\cite{Bern:2002tk}. Addressing 
the full set of corrections is, by necessity, a long term
project which we undertake step-by-step (an attitude which should not be
confused with narrow focusing).

A large fraction of physical processes, in particular gauge bosons decays into
fermion--anti-fermion pairs and accurate predictions for gauge boson complex
poles, only require two- and three-point functions.  Also for the analysis
of the two-loop SM renormalization~\cite{Freitas:2002ja}, two--point
functions and vacuum bubbles are essentially all we need. Indeed, in order
to evaluate the Fermi coupling constant $G_{\ssF}$ from the muon life-time
we always work at zero momentum transfer and neglect terms proportional to
$\mm^2/\mw^2$; therefore, all diagrams contributing to this process (boxes
included) are simply equivalent to vacuum bubbles, i.e.\ generalized sunset
integrals evaluated at zero external momentum.

Feynman diagrams are built using propagators and vertices. In momentum
space, the former are represented by
\bq
\frac{N(p)}{p^2 + m^2 - i\,\delta} \, \, ,
\eq
where $\delta \to 0_+$, $m$ is the bare mass of the particle and $N(p)$ is
an expression depending on its spin. Our general approach towards the
numerical evaluation of an arbitrary, multi-scale, Feynman diagram $G$ is to
use a Feynman parameter representation and to obtain, diagram-by-diagram,
an integral representation of the following form:
\bq
G = \sum\left[\frac{1}{B_{\ssG}}\,\int_{\ssS}\,dx\,{\cal G}(x)\right],
\label{generalclass}
\eq
where $x$ is a vector of Feynman parameters, $S$ is a simplex, ${\cal G}$
is an integrable function (in the limit $\delta \to 0_+$) and $B_{\ssG}$ is
a function of masses and external momenta whose zeros correspond to true
singularities of the diagram $G$, if any. The Bernstein-Tkachov (hereafter
BT) functional relations~\cite{Tkachov:1997wh} are one realization of
\eqn{generalclass}, but in our previous work we considered different
possibilities.

Smoothness requires that the kernel in \eqn{generalclass}, together with its
first $N$ derivatives, should be a continuous function, with $N$ as large as
possible. However, in most cases we will be satisfied with absolute
convergence, e.g.\ with logarithmic singularities of the kernel. This is
particularly true around the zeros of $B_{\ssG}$, where the large number of
terms, induced requiring continuous derivatives of higher orders, leads to
large numerical cancellations.

As we stressed earlier, this article is by its own nature rather technical,
but we tried to avoid as much as possible a layout which overwhelmingly
privileges long lists of formulae in favor of interleaving the indispensable
amount of technical details with examples. For completeness, however, we
inserted Appendices where the reader can find a complete summary of the
results occurring in the reduction procedure.

The results presented in this paper are intermediate steps in any physical
calculation; although the presentation is organized through a series of 
concatenated formulae that can be used recursively, further derivations on 
the part of the reader are  required in order to obtain analytic 
or numerical results for a physical quantity. 

The outline of the paper is as follows: in \sect{conv} we recall our
notation and conventions. In \sect{tigc} we review the problem of gauge
cancellations and the use of Nielsen identities, while in \sect{prered} we
illustrate all preliminary steps that should be undertaken in any realistic
calculation (like projector techniques). The reduction of two-loop two-point
functions is discussed in \sect{stltpf} and a complete list of the results
is given in \sect{fulll}. The role of integration-by-part
identities is discussed in \sect{ibpoai}. In Sections~\ref{Sde}--\ref{SETLV}
we present the full body of our results for two-loop tensor integrals. (Rank
three tensors for three-point functions are shown in \sect{IRTIRT}.)
Conclusions are drawn in \sect{conclu}. Additional material is discussed in
the Appendices; in particular, the treatment of generalized one-loop
functions is discussed in \appendx{RGOLF}. A concatenated set of easy-to-use 
formulae for the reduction of two-loop three-point functions is summarized 
in \appendx{SummaR}; symmetries of diagrams are presented in \appendx{symme}.

\section{Conventions and notation \label{conv}}
Our conventions for arbitrary two-loop diagrams were introduced in Sect.~2
of I. Specific conventions for three-point functions were introduced in
Sect.~2 of III; vertex topologies were classified in III and are reproduced,
for the reader's convenience, in \figs{TLvertaba}{TLvertbbb}. Here we
briefly recall the terminology.

A generalized one-loop diagram will be denoted by
\bq
G_{\mu_1\,,\,\cdots\,,\,\mu_{\ssL}}(\{\alpha\}_{\ssN}\,;\,
\{p\}_{\ssN-1}\,,\,\{m\}_{\ssN}) =
\frac{\tHs^{4-n}}{i\,\pi^2}\,\int\,d^nq\,
q_{\mu_1}\,\cdots\,q_{\mu_{\ssL}}\,\prod_{i=1}^{N}\,[i]^{-\alpha_i}_{\ssG},
\label{Gold}
\eq
where $n = 4 - \ep$, $n$ is the space-time dimension,\footnote{In our
metric, space-like $p$ implies $p^2={\vec{p}}\,^2 + p_4^2 >0$. Also, it is
$p_4 = i\,p_0$ with $p_0$ real for a physical four-momentum.} $\tHs$ is the
arbitrary unit of mass, $N$ is the number of vertices, and
\bq
\{\alpha\}_{\ssN} = \alpha_1\,,\,\cdots\,,\,\alpha_{\ssN} \, ,
\qquad
[i]_{\ssG} = (q+\sum_{j=0}^{i-1}\,p_j)^2 + m^2_i, \quad p_0 = 0 \, .
\eq
The one-loop two-, three-,...\ point functions will be denoted by $G =
B,C,\cdots$.

A generalized two-loop diagram is defined with arbitrary, non-canonical,
powers of its propagators; it can be cast in the following form
\bqa
{}&{}&
G^{\{\alpha\}_a\,|\,\{\beta\}_b\,|\,\{\gamma\}_c}(\mu_1\,,\,\cdots\,,\,
\mu_{\ssR}\,|\,\nu_1\,,\,\cdots\,,\,\nu_{\ssS}\,;\,
\{\eta^1\,p\}\,,\,\{\eta^{12}\,p\}\,,\,\{\eta^2\,p\}\,,\,
\{m\}_{a+b+c}) =
\nl
{}&{}&
\frac{\mu^{2 (4-n)}}{\pi^4}\,
\intmomsii{n}{q_1}{q_2}\,\prod_{r=1}^{R}\,q_{1\mu_r}\,
\prod_{s=1}^{S}\,q_{2\nu_s}\,
\prod_{i=1}^{a}\,(k^2_i+m^2_i)^{-\alpha_i}\,
\!\!\!\!\prod_{j=a+1}^{a+c}\,(k^2_j+m^2_j)^{-\gamma_j}\,
\!\!\!\!\prod_{l=a+c+1}^{a+c+b} \!\!\!\!(k^2_l+m^2_l)^{-\beta_l},
\label{Gdiag}
\eqa
where $a, b$ and $c$ indicate the number of lines 
in the $q_1, q_2$ and $q_1-q_2$ loops, respectively. 
For generalized functions we use $\alpha = \sum_{i=1}^a \alpha_i$ etc, 
while for standard functions (i.e.\ those where all the propagators have 
canonical power $-1$), $\alpha = a, \beta = b$ and $\gamma = c$ and we will
write $G^{\alpha\beta\gamma}$. Furthermore, 
\[
\ba{ll}
k_i = q_1+\sum_{j=1}^{\ssN}\,\eta^1_{ij}\,p_j \, , \;&\; i=1,\dots,a \,  , \\
k_i = q_1-q_2+\sum_{j=1}^{\ssN}\,\eta^{12}_{ij}\,p_j \, , \;&\;
i=a+1,\dots,
a+c \,  , \\
k_i = q_2+\sum_{j=1}^{\ssN}\,\eta^2_{ij}\,p_j \, , \;&\;
i=a+c+1,\dots,
a+c+b \, ,
\label{matrixe}
\ea
\]
where $\eta^s = \pm 1,$ or $0$, and $\{p\}$ is the set of external
momenta. Diagrams which can be reduced to combinations of other diagrams
with a smaller number of internal lines will not receive a particular name.
Otherwise, a two-loop diagram will be denoted by 
$G^{\alpha\beta\gamma}$, where $G = S, V, B\,$ etc. stands for two-, three-, 
four-point etc. For scalar integrals we will use the symbol 
$G^{\alpha\beta\gamma}_0 = G^{\alpha\beta\gamma}(0|0;\ldots)$.
Following \eqn{Gdiag} diagrams are further classified according to non empty
entries in the matrices $\eta^s$ and in the list of internal masses. 

\begin{description}
\item{\bf Integrals:} To keep our results as compact as possible, we
introduce the following notation ($x_0 = y_0 = 1$) where $C$ stands for
(hyper)cube and $S$ for simplex,
\bqa
\dcubs{\{x\}}{\{y\}}f(x_1,\cdots,x_{n_1},y_1,\cdots,y_{n_2}) &=&
\int_0^1\prod_{i=1}^{n_1}dx_i
\prod_{j=1}^{n_2}\!\int_0^{y_{j-1}}\!\!\!\!\!\!\! dy_j
f(x_1,\cdots,x_{n_1},y_1,\cdots,y_{n_2}),
\nl
\dsimp{n}(\{x\})\,f(x_1,\cdots,x_n) &=& 
\prod_{i=1}^{n}\,\int_0^{x_{i-1}}\,dx_i\,f(x_1,\cdots,x_n),
\nl
\dcub{n}(\{x\}) \, \, \!\!f(x_1,\cdots,x_n) &=& \int_0^1 \prod_{i=1}^{n}\,dx_i
f(x_1,\cdots,x_n).
\eqa
Also, the so-called $'+'$-distribution will be extensively used, e.g.\
\bqa
\dcub{n}(\{z\})\,\intfx{x}\,\frac{f(x,\{z\})}{x}\bmid_+ &=&
\dcub{n}(\{z\})\,\intfx{x}\,\frac{f(x,\{z\}) - f(0,\{z\})}{x},
\nl
\dcub{n}(\{z\})\,\intfx{x}\,\frac{f(x,\{z\})}{x-1}\bmid_+ &=&
\dcub{n}(\{z\})\,\intfx{x}\,\frac{f(x,\{z\}) - f(1,\{z\})}{x-1},
\nl
\dcub{n}(\{z\})\,\intfx{x}\,\frac{f(x,\{z\})\,\ln^n x}{x}\bmid_+ &=&
\dcub{n}(\{z\})\,\intfx{x}\,\frac{\Bigl[f(x,\{z\}) - 
f(0,\{z\})\Bigr]\,\ln^n x}{x}.
\label{plusdist}
\eqa
The last relation in \eqn{plusdist} is used when evaluating integrals
of the following type:
\bqas
\intfx{x}\,\frac{f(x)}{x^{1-\ep}} &=& \frac{f(0)}{\ep} +
\intfx{x}\,\frac{f(x)}{x}\bmid_+ -
\ep\,\intfx{x}\,\frac{f(x)\,\ln x}{x}\bmid_+ + \ord{\ep^2}.
\eqas
\item{\bf Lists of arguments:}
To avoid long lists of arguments we introduce the symbol
\bq
\{m\}_{i\,j\,\cdots\,k} = m_i\,,\,m_j\,,\,\cdots\,,\,m_k,
\qquad \mbox{exactly in this order}.
\eq
\item{\bf Miscellanea:} We often need combinations of squared masses and
momenta,
\bqa
l_{ijk} &=& p^2_i - m^2_j + m^2_k, \qquad 
l_{\ssP jk} = P^2 - m^2_j + m^2_k, \qquad 
l_{pjk} = p^2 - m^2_j + m^2_k,  
\nl 
m^2_{ijk} &=& m^2_i - m^2_j + m^2_k, \qquad 
m^2_{ij} = m^2_i - m^2_j, \qquad
p_{ij} = \spro{p_i}{p_j},
\nl
(G)_{ij} &=& p_{ij}, \quad D = {\mbox{det}}\,G = p^2_1\,p^2_2 -
(\spro{p_1}{p_2})^2, \quad D_1 = p^2_1\,p^2_2, \quad D_2 = p_{12}\,p^2_2,
\quad D_3 = p_{12}\,p^2_1,
\label{defmisc}
\eqa
and of Feynman parameters,
\bq
\ox = 1 - x, \qquad \ox_i = 1 - x_i, \qquad \oy_i = 1 - y_i, \quad \mbox{etc,}
\quad
X = \frac{1-x_1}{1-x_2} = 1 - \bX\, 
\eq
\bq 
Y_i = -(1 - y_i + y_3\,X) \, , 
\quad
\bY_2 = 1 - y_2\,\bX,
\quad
H_i = 1 - x_1 - x_2\,Y_i \, , \quad i=1,2
\, . \label{HY1HY2}
\eq
\bq
F(x,y) = p^2_1\,x^2 + 2\,p_{12}\,x\,y + p^2_2\,y^2,
\qquad
m_x^2 = \frac{m^2_1}{x} + \frac{m^2_2}{1-x}.
\label{defGfun}
\eq
\item{\bf Symmetrized tensors:}
We define (partially) symmetrized tensors as follows ($\delta_{\alpha
  \beta}$ is the Kronecker delta function),
\bqa
\{p\,k\}_{\mu\nu} &=& p_{\mu}\,k_{\nu} + p_{\nu}\,k_{\mu},
\qquad
\{\delta\,p\}_{\alpha\beta\gamma} = \delta_{\alpha \beta} p_{\gamma} + 
\delta_{\alpha \gamma} p_{\beta} + \delta_{\beta \gamma} p_{\alpha},
\nl 
\{p p k\}_{\alpha\beta\gamma} &=&
p_{\alpha}\,p_{\beta}\,k_{\gamma} + 
p_{\alpha}\,p_{\gamma}\,k_{\beta} + 
p_{\gamma}\,p_{\beta}\,k_{\alpha}\,
\nl
\{p p k\}_{\alpha\gamma\,|\,\beta} &=& p_{\alpha} p_{\beta} 
k_{\gamma} + p_{\gamma} p_{\beta} k_{\alpha},
\qquad
\{\delta\,p\}_{\alpha\beta\,|\,\gamma} = \delta_{\alpha\gamma}\,
p_{\beta} + \delta_{\beta\gamma}\,p_{\alpha}.
\label{symmete}
\eqa
\item{\bf Contraction:}
If $p$ is a vector and $f$ is a function, we introduce the symbol
\bq
f(\cdots\,,\,p\,,\,\cdots) = p^{\mu}\,f(\cdots\,,\,\mu\,,\,\cdots), \qquad
f(\cdots\,\,\mu\mu\,,\,\cdots) = 
\delta^{\mu\nu}\,f(\cdots\,\,\mu\nu\,,\,\cdots).
\eq
\item{\bf {\boldmath $\MSB$} factors:}
Finally we remind the reader of the definition of $\MSB$ factors,
\bq
\Delta_{\ssU\ssV} = \gamma + \ln\pi - \ln\frac{\tHss}{\mid P^2\mid},
\qquad
\oDUV = \frac{1}{\ep} - \Delta_{\ssU\ssV},
\qquad
\omega = \frac{\mu^2}{\pi} \ ,
\label{defDUV}
\eq
where $\gamma= 0.577216\cdots$ is the Euler constant. In one-loop
calculations the definition $\oDUV = 2/\ep -\Delta_{\ssU\ssV}$ is often
employed. Finally some authors prefer to define $n = 4 - 2\,\ep$.
\end{description}
\subsection{Definition of one-loop generalized functions}
Products of one-loop functions occur in the reduction of two-loop diagrams;
generalized one-loop functions are defined in \eqn{Gold}, specific examples
of one- and two-point (scalar) functions are
\bqa
A_{\Z}(\alpha;m) = \frac{\mu^{\ep}}{i\,\pi^2}\,
\int\,\frac{d^nq}{(q^2 + m^2)^{\alpha}} &,&
\qquad
A_{\Z}(\alpha;[m_i,m_j]) = A_{\Z}(\alpha;m_i) - A_{\Z}(\alpha;m_j),
\nl
B_{\Z}(\alpha,\beta\,;\,p,m_1,m_2) &=& \frac{\mu^{\ep}}{i\,\pi^2}\,
\int\,\frac{d^nq}{(q^2+m^2_1)^{\alpha}((q+p)^2+m^2_2)^{\beta}}.
\label{GABfun}
\eqa
Note that we always drop strings like $1,1,\cdots$ in the argument of
standard functions, namely, we write $A_{\Z}(m)$ for $A_{\Z}(1,m)$ etc.
Tensor integrals are:
\bqa
\frac{\mu^{\ep}}{i\,\pi^2}\,\int\,
\frac{d^nq\,q_{\mu}}{(q^2 + m^2_1)^{\alpha}\,((q+p)^2+m^2_2)^{\beta}} &=& 
B_{\sa}(\alpha,\beta\,;\,p,\lstm{12})\,p_{\mu},
\label{GBV}
\eqa
\bqa
\frac{\mu^{\ep}}{i\,\pi^2}\,\int\,
\frac{d^nq\,q_{\mu}q_{\nu}}{(q^2 + m^2_1)^{\alpha}\,((q+p)^2+m^2_2)^{\beta}} 
&=& 
B_{\sba}(\alpha,\beta\,;\,p,\lstm{12})\,p_{\mu}\,p_{\nu} + 
B_{\sbb}(\alpha,\beta\,;\,p,\lstm{12})\,\delta_{\mu\nu},
\label{GBT}
\eqa
their reduction is given in \sect{fulll}. Generalized one-loop three-point 
functions are introduced as follows:
\bq 
C_{\mu_1,\cdots\mu_l}(\{\alpha\}_3\,;\,p_1,p_2,\lstm{123}) = 
\frac{\mu^{\ep}}{i\,\pi^2}\,\int\,d^nq\,
\prod_{j=1}^{l}\,q_{\mu_j}\,\prod_{i=1}^{3}\,[i]^{-\alpha_i},
\label{GCfun}
\eq
with $[i] = Q^2_i + m^2_i$ and $Q_i= q + p_0 + \cdots + p_{i-1}$, $p_0 = 0$.
In particular,
\bq 
C_{\mu}(p_1,p_2,\lstm{123}) = \frac{\mu^{\ep}}{i\,\pi^2}\,\int\,d^nq
\frac{q_{\mu}}{[q^2 + m_1^2 ][(q + p_1)^2 + m_2^2]
[(q + p_1 + p_2)^2 + m_3^2]}.
\label{GCfunV}
\eq
The integrals of \eqn{GCfun} can be reduced, for example,
\bq
C_{\mu}(\{\alpha\}_3\,;\,p_1,p_2,\lstm{123}) =
C_{\saa}(\{\alpha\}_3\,;\,\cdots)\,p_{1\mu} + 
C_{\sab}(\{\alpha\}_3\,;\,\cdots)\,p_{2\mu},
\eq 
\bqa
C_{\mu\nu}(\{\alpha\}_3\,;\,p_1,p_2,\lstm{123}) &=&
C_{\sba}(\{\alpha\}_3\,;\,\cdots)\,p_{1\mu} p_{1\nu} + 
C_{\sbb}(\{\alpha\}_3\,;\,\cdots)\,p_{2\mu} p_{2\nu} 
\nl
{}&+& 
C_{\sbc}(\{\alpha\}_3\,;\,\cdots)\,\{p_1 p_2\}_{\mu\nu} + 
C_{\sbd}(\{\alpha\}_3\,;\,\cdots)\,\delta_{\mu\nu}, 
\eqa
where the symmetrized product is defined by \eqn{symmete}; for the reduction
we refer to \appendx{RGOLF}. 

For completeness we also define generalized one-loop four-point functions,
although they are not needed in this article :
\bq 
D_{\mu_1,\cdots,\mu_l}(\{\alpha\}_4\,;\,p_1,p_2,p_3,\lstm{1234}) = 
\frac{\mu^{\ep}}{i\,\pi^2}\,\int\,d^nq\,
\prod_{j=1}^{l}\,q_{\mu_j}\,\prod_{i=1}^{4}\,[i]^{-\alpha_i},
\label{GDfun}
\eq
etc. Once again $[i]= Q^2_1 + m^2_i$, $Q_i= q + p_0 + \cdots + p_{i-1}$ with 
$p_0 = 0$.
\subsection{Alphameric classification of graphs}
In our conventions any scalar two-loop diagram is identified by a capital
letter ($S, V$, etc.) indicating the number of external legs, and by a
triplet of numbers $(\alpha, \beta$ and $\gamma$) giving the number of
internal lines (carrying internal momenta $q_1, q_2$ and $q_1 - q_2$,
respectively).  There is a compact way of representing this triplet: assume
that $\gamma \ne 0$, i.e.\ that we are dealing with non-factorizable
diagrams, then we introduce
\bq
\kappa = \gamma_{\rm max}\,\Bigl[ \alpha_{\rm max}\,(\beta - 1) +
\alpha - 1\Bigr] + \gamma
\eq
for each diagram. For $G = V$ we have $\alpha_{\rm max} = 2$ and
$\gamma_{\rm max} = 2$. Furthermore, we can associate a letter of the
English alphabet to each value of $\kappa$. Therefore, the following
correspondence holds:
\bq
121 \to E, \quad 131 \to I, \quad 141 \to M, \quad
221 \to G, \quad 231 \to K, \quad 222 \to H.
\label{americv}
\eq
For $G = S$ we have $\alpha_{\rm max} = 2$ and $\gamma_{\rm max} = 1$,
therefore
\bq
111 \to A, \quad 121 \to C, \quad 131 \to E, \quad 221 \to D.
\label{americs}
\eq
This classification is extensively used throughout the paper and motivated
by the unavoidable proliferation of indices; the reader not familiar with it
should remember that storing the elements of a matrix into a vector is a
well-known coding procedure (e.g.\ in Fortran).  Note that in II and in III
this convention was not yet used and the correspondence of results is simply
provided by \eqns{americv}{americs}.

\section{Tensor integrals and gauge cancellations \label{tigc}}
Any Feynman integral with a tensor structure can be written as a combination 
of form factors
\bq
G_{\mu_1\,\cdots\,\mu_{\ssN}} = \sum_{i=1}^{i_{\rm max}}\,
G^i_{\ssS}\,F_{i\,;\,\mu_1\,\cdots\,\mu_{\ssN}},
\eq
where the $F_{i\,;\,\mu_1\,\cdots\,\mu_{\ssN}}$ are tensor structures made
up of external momenta, Kronecker delta functions, $\ep$-tensors (which will
cancel in any CP-even observable) and elements of the Dirac algebra; the
scalar projections $G^i_{\ssS}$ admit a parametric representation which is
equivalent to the one for the corresponding scalar diagram but with
polynomials of Feynman parameters occurring in the numerator. Once we have
an integral representation for the primary scalar diagram, with the desired
properties of smoothness, then, analogous representations, with the same
properties, also follow for the induced scalar projections. Therefore, from
the point of view of numerical evaluation there is really little difference
between scalar and tensor integrals.

However, there is a problem due to the fact that we are dealing with gauge
theories with inherent gauge cancellations which do not support a blind
application of the procedure just described. A very simple example will be
useful to illustrate the roots of this problem. Consider the one-loop
photon self-energy in QED and express the result in terms of scalar one-loop
form factors~\cite{Passarino:1979jh}; we obtain
\bq
\Pi^{f}_{\mu\nu} = \Pi^f_1\delta_{\mu\nu} + \Pi^f_2p_{\mu}p_{\nu},
\eq
\bq
\Pi^f_1 = -4\,e^2\Bigl\{ (2 - n)B_{\sbb} - p^2\Bigl[ B_{\sba} + B_{1}\Bigr] - 
m^2_f\,B_{\Z}\Bigr\},
\qquad
\Pi^f_2 = -8\,e^2\Bigl[B_{\sba} + B_1\Bigr],
\eq
where $e$ is the bare electric charge and $B_{\Z}$ etc.\ are the standard
one-loop functions of~\cite{Passarino:1979jh}, all with arguments
$(p^2;m_f,m_f)$. The gauge invariance of the theory is controlled by a set
of Ward--Slavnov--Taylor identities~\cite{Taylor:ff} (hereafter WST), one of
which requires $\Pi^f_{\mu\nu}$ to be transverse; this hardly follows from
expressing the form factors in parametric space followed by some numerical
integration.  Rather, it follows from a set of identities that one can write
among the standard one-loop functions ($B_{\sbb}$ etc.) directly in momentum
space, the so-called reduction procedure (``scalarization'' in jargon). This
procedure, in its original design, is plagued by the occurrence of inverse
powers of Gram determinants whose zeros are unphysical but sometimes
dangerous for the numerical stability of the result.

There is another example where gauge cancellations play a crucial role.
Suppose that we decide to work in the so-called $R_{\xi}$ gauges with one or
more gauge parameters which we will collectively indicate by $\xi$: the
expected $\xi$ independence is seen at the level of $S$-matrix elements and
not for individual contributions to Green functions. From this point of
view, any procedure that computes single diagrams and sums the corresponding
numerical results, without controlling gauge cancellations analytically, is
bound to have its own troubles.

These two rather elementary considerations suggest the 
following strategy: first impose all the requirements dictated by
WST identities and see that they are satisfied.
At this point organize the calculation according to building blocks that are,
by construction, gauge-parameter independent.

The first step requires some form of scalarization (which, as we saw, may be
numerically unstable), but the perspective is different: scalarization is
now needed only to prove that certain combinations of form factors are zero,
and any occurrence of Gram determinants does not therefore pose a problem.

In the second step we need to control the $\xi$ behavior of individual Green
functions; a possible tool is represented by the use of the Nielsen
identities (hereafter NI)~\cite{Grassi:2001bz}.
Typically we will consider the transverse propagator of a gauge field:
\bq
D_{\mu\nu} = \frac{1}{\tpfi}\,
\frac{\delta_{\mu\nu} - p_{\mu}p_{\nu}/s}{s - M^2_0 + \Pi(\xi,s)},
\eq
where $p^2 = - s$, $M_0$ is the bare mass of the particle and $\Pi(\xi,s)$
is the self-energy. The corresponding NI reads as follows:
\bq
\frac{\partial}{\partial\,\xi}\,\Pi(\xi,s) = \Lambda(\xi,s)\,
\Pi(\xi,s),
\eq
where $\Lambda$ is a complex, amputated, $1$PI, two-point Green function and
the complex pole is defined by
\bq
{\bar s} - M^2_0 + \Pi(\xi,{\bar s}) = 0, \qquad
\partial_{\xi}\,{\bar s} = 0.
\eq
Let us consider now the amplitude for $i \to V \to f$, where $V$ is an
unstable gauge boson and $i/f$ are initial/final states. The overall
amplitude becomes
\bq 
A_{fi}(s) = \frac{\delta_{\mu\nu}}{s - {\bar s}}\,
\frac{V^{\mu}_f({\bar s})\,V^{\nu}_i({\bar s})}{1 + \Pi'({\bar
s})} \quad + \quad \mbox{non-resonant terms}, \eq
where it is understood that the vertex functions $V^{\mu}_f$ and $V^{\nu}_i$
include the wave-function renormalization factors of the external,
on-shell, particles. It has been proved that
\bq
\frac{d}{d\,\xi}\,\Bigl[1 + \Pi'({\bar s})\Bigr]^{-1/2}\,
V^{\mu}_f({\bar s}) = 0;
\eq
this combination is the prototype of one of the gauge-parameter independent
building blocks that are needed to assemble our calculation of a physical
observables. All gauge-parameter independent blocks will then be mapped into
one (multi-dimensional) integral to be evaluated numerically.

\section{Projector techniques \label{prered}}
Any realistic calculation requires several steps to be performed before we
can actually start to compute diagrams or sums of diagrams; in all of them,
some action can be taken in order to simplify the structure of the amplitude
in some efficient way. Much work has been invested in this area and we refer
to recent work of Glover~\cite{Glover:2004si} for an exhaustive list of
references.

Here we focus on few examples. Consider, for instance, the matrix element
for the decay of a vector particle $V$ into a fermion-antifermion pair,
$V(P_+) \to \bar f(p_+) f(p_-)$ (all particles are on their mass-shell);
instead of decomposing all tensor integrals into form factors, we first
decompose the vertex $V_{\mu}$ into the following structures,
\bqa
{\cal M} &=& {\bar u}(p_-)\,\,\spro{\ep}{V}\,v(p_+) = 
{\bar u}(p_-)\,\Bigl[ 
F_{\ssV}\,\sla{\ep} +
F_{\ssA}\,\sla{\ep}\,\gfd +
F_{\ssS}\,\spro{P_-}{\ep} + 
F_{\ssP}\,\spro{P_-}{\ep}\,\gfd 
\Bigr]\,v(p_+),
\label{decompo}
\eqa
where $\ep_{\mu} = \ep_{\mu}(P_+)$ is the polarization vector for the $V$
particle, subject to the constraint $\spro{\ep}{P_+} = 0$ ($P_{\pm}= p_+ \pm
p_-$). We also introduce projectors to extract the form factors appearing in
\eqn{decompo}~\cite{Glover:2004si}
\bqa
\sum_{\rm spin}\,P_{\ssI}\,{\cal M} &=& F_{\ssI},
\eqa
where $I = V,A,S$ or $P$.
The explicit solution for the projectors is obtained considering four
auxiliary quantities
\[
\ba{ll}
P_1 = {\bar v}(p_+)\,\sla{\ep}\,u(p_-), \;&\; 
P_2 = {\bar v}(p_+)\,\sla{\ep}\,\gfd\,u(p_-), \\
P_3 = \spro{\ep}{P_-}\,{\bar v}(p_+)\,u(p_-),  \;&\;
P_4 = \spro{\ep}{P_-}\,{\bar v}(p_+)\,\gfd\,u(p_-).
\ea
\]
Let us define $\beta_{\ssM} = M^2 - 4\,m^2$, where $M$ is the mass of the 
vector boson $V$ and $m$ is the mass of the fermion $f$: we get
\bqa
P_{\ssV} &=& -\,\frac{1}{2\,(2-n)\,M^2}\,\Bigl[ P_1 +
2\,i\,\frac{m}{\beta_{\ssM}}\,P_3\Bigr],
\qquad
P_{\ssA} = -\,\frac{1}{2\,(2-n)\,\beta_{\ssM}}\,P_2,
\nl
P_{\ssP} &=& \frac{1}{2\,M^2\,\beta_{\ssM}}\,P_4,
\qquad
P_{\ssS} = -\,i\,\frac{m}{M^2\,(2-n)\,\beta_{\ssM}}\,\Bigl\{
P_1 +\,\frac{i}{2\,m\,\beta_{\ssM}}\,\Bigl[4\,m^2 + 
(n - 2)\,M^2\Bigr]\,P_3\Bigr\},
\eqa
thus providing the scalar coefficients $F_{\ssI}$. For example,
\bqa
F_{\ssV} = \frac{1}{n-2}\,{\mbox{Tr}}\,{\cal F}_{\ssV},
\qquad
{\cal F}_{\ssV} =
&-&\,\frac{1}{2\,M^2}\,\gapu{\mu}\,\Lambda_-\,V_{\mu}\,\Lambda_+
-\,\frac{i}{2\,M^4}\,\Lambda_+\,\Lambda_-\,\spro{P_+}{V}\,\Lambda_+
\nl
{}&-& \,\frac{ i\,m}{M^4}\,\Lambda_-\,\spro{P_+}{V}\,\Lambda_+
-\,\frac{ i \, m}{M^2\,\beta_{\ssM}}\,\Lambda_-\,\spro{P_-}{V}\,\Lambda_+,
\label{vector}
\eqa
where $\Lambda_+ = -i\,\sla{p_+}-m$ and $\Lambda_- = -i\,\sla{p_-}+m$.
This procedure completely saturates indices and allows us to consider only
integrals with positive powers of scalar products in the numerators. Then a
reduction procedure follows and we will show that the final answer contains
only generalized scalar integrals. For a discussion on projector techniques
in conventional dimensional regularization or in the 't Hooft-Veltman
scheme~\cite{'tHooft:fi} we refer again to~\cite{Glover:2004si}.

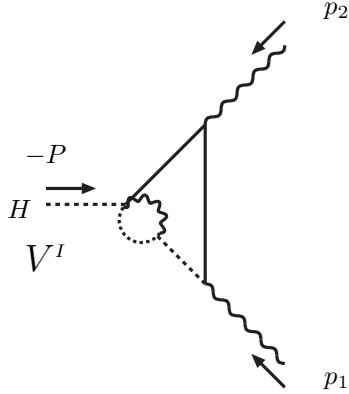
\begin{figure}[th]
\vspace{2.0cm}
\bqas  
{}&{}&
  \vcenter{\hbox{
  \begin{picture}(150,0)(0,0)
  \SetScale{0.6}
  \SetWidth{2.}
  \DashLine(0,0)(50,0){3}
  \Line(50,0)(100,50)
  \Photon(100,50)(150,100){2}{5}
  \Photon(100,-50)(150,-100){2}{5}
  \Line(100,-50)(100,50)
  \PhotonArc(60,-10)(14,-35,145){2}{5}
  \DashCArc(60.,-10.)(14.,-215,-35){2}
  \DashLine(70,-20)(100,-50){3}
  \Text(0,-25)[cb]{\Large $V^{\aca}$}
  \LongArrow(0,10)(30,10)   \Text(0,15)[cb]{$-P$}
  \LongArrow(150,115)(130,95)   \Text(110,70)[cb]{$p_2$}
  \LongArrow(150,-115)(130,-95)     \Text(110,-70)[cb]{$p_1$}  
  \Text(-10,-5)[cb]{$H$}
  \end{picture}}}
\eqas
\vspace{1.5cm}
\caption[]{$I$-family contribution to the decay $H \to \gamma\gamma$.
Internal dotted lines represent a Higgs-Kibble $\phi$-field, while solid 
ones indicate a $W$-field.}
\label{exafig}
\end{figure}

Another example we want to consider is the amplitude for $s \to
\gamma\gamma$, where $s$ is a generic scalar particle; for this case we
follow the procedure of Binoth, Guillet and Heinrich~\cite{Binoth:1999sp}
and introduce the vectors
\bq
r_{i\mu} = \sum_{j=1}^{i}\,p_{j\mu},
\qquad
R_{i\mu} = \sum_{j=1}^{2}\,G^{-1}_{ij}\,r_{j\mu}, \quad
G_{ij} = 2\,\spro{r_i}{r_j}.
\eq
The square of the $s \to \gamma\gamma$ vertex is further decomposed into
\bq
V_{\mu\nu} = F_{\ssD}\,\delta_{\mu\nu} + \sum_{i,j=1}^{2}\,F_{\ssP,ij}\,
r_{i\mu}\,r_{j\nu}.
\eq
The form factors of this decomposition are expressed through the action of
projectors,
\bq
F_{\ssD} = P^{\mu\nu}_{\ssD}\,V_{\mu\nu}, \qquad
F_{\ssP,ij} = P^{\mu\nu}_{\ssP,ij}\,V_{\mu\nu},
\eq
\bq
P^{\mu\nu}_{\ssD} = \frac{1}{n-2}\,\Bigl[ \delta^{\mu\nu} - 2\,
r^{\mu}\,G^{-1}\,r^{\nu}\Bigr],
\qquad
P^{\mu\nu}_{\ssP,ij} = 4 \, \Bigl[ R^{\mu}_i\,R^{\nu}_j - 
\frac{1}{2}\,G^{-1}_{ij}\,
P^{\mu\nu}_{\ssD}\Bigr].
\eq
These projectors have the following properties:
\bq
\spro{R_i}{r_j} = \frac{1}{2}\delta_{ij}, \qquad
P^{\mu\nu}_{\ssD}\,r_{i\nu} = 0, \qquad
P^{\mu\nu}_{\ssD}\,P_{\ssD,\mu\nu} = \frac{1}{n-2}.
\eq
The whole procedure is better illustrated in terms of an example,
an $I$-family contribution to the decay $H \to \gamma\gamma$, 
see \fig{exafig}. The corresponding integral is 
\bqa
V^{\mu\nu} &=& g^5\,s^4_{\theta}\,\frac{\mw}{2}\,\tHs^{2\ep}\,
\int\,d^nq_1\,d^nq_2\,
\Bigl\{\Bigl[ \spro{(q_2+q_1)}{(p_2-p_1)}\, - \,q_2^2\, - \,
\spro{q_1}{q_2}\Bigr]\,\delta^{\mu\nu}
\nl
{}&+& 
2\,\Bigl[ 
  q_1^{\mu}\,q_2^{\nu} 
+ (q_1 + q_2)^{\mu}\,p_1^{\nu} 
- (q_1 + q_2)^{\nu}\,p_2^{\mu}\Bigr]
+ q_2^{\mu}\,(q_2 - q_1)^{\nu} 
+ (q_1 + q_2)^{\mu}\,p_2^{\nu} 
- (q_1 + q_2)^{\nu}\,p_1^{\mu} 
\Bigr\}
\nl
{}&\times& \Bigl[ (q^2_1+\mw^2)\,(q_1-q_2)^2\,(q^2_2+\mw^2)\,
((q_2+p_1)^2+\mw^2)\,
((q_2+P)^2+\mw^2)\Bigr]^{-1},
\eqa
where $s_{\theta} \, (c_{\theta})$ is the sine (cosine) of the weak-mixing 
angle. Terms containing $q^2_2,\,\spro{q_1}{q_2}$ and $\spro{q_2}{p_1}$ are 
immediately eliminated from the final answer. Consider now terms with 
$q^{\mu}_2$ (for those with $q^{\mu}_1$ there is an analogous argument); with 
straightforward substitutions we obtain
\bq
\int\,F\,q^{\mu}_2 \quad \to \quad F_1\,p^{\mu}_1 +
F_2\,p^{\mu}_2 \quad \to \quad {\cal F}_1\,r^{\mu}_1 +
{\cal F}_2\,r^{\mu}_2,
\eq
where with $F$, etc we indicate some combination of form factors of the
$\cV^{\aca}$ family whose explicit expression is not relevant for our 
discussion at this stage.
When we project with $P^{\mu\nu}_{\ssD}$ or with $R_{i\nu}$ it follows that
\bqa
{}&{}& \int\,F\, P^{\mu\nu}_{\ssD}\,q_{2\mu} \quad \to \quad
\sum_i\,{\cal F}_i\,P^{\mu\nu}_{\ssD}\,r_{i\mu} = 0,
\nl
{}&{}& \int\,F\, R^{\mu}_i\,q_{2\mu} \quad \to \quad
\sum_j\,{\cal F}_j\,\spro{R_i}{r_j} \quad \to \quad
{\cal F}_i.
\eqa
When we have a term with $q^{\mu}_2\,q^{\nu}_2$ and project with
$P^{\mu\nu}_{\ssD}$ it follows
\bqa
\int\,F\, P^{\mu\nu}_{\ssD}\,q^{\mu}_2\,q^{\nu}_2 \quad &\to& \quad
\Bigl[\sum_{ij}\,{\cal F}_{ij}\,r^{\mu}_i\,r^{\nu}_j +
{\cal F}_d\,\delta^{\mu\nu}\Bigr]\,
P_{\ssD,\mu\nu} \; \to \; {\cal F}_d\,P_{\ssD,\mu\mu} =
{\cal F}_d.
\eqa
The number of form factors may be further reduced by 
requiring that on-(off-)shell 
WST identities hold.

The procedure that we just illustrated can be easily generalized to other
situations; consider, for instance, the off-shell vertex corresponding to
$V_1 \to V_2 + V_3$ where the $V_i$ are gauge bosons. By off-shell we mean
that the sources $J^{\mu}_{\ssV}$ emitting/absorbing the vector bosons are
not physical (therefore $\partial_{\mu}\,J^{\mu}_{\ssV} = 0$ is not assumed)
and are not on their mass-shell; this choice is also needed when two of the
particles correspond to (idealized) stable, physical, vector bosons and we
want to check a WST identity. In full generality we write the following
decomposition of the vertex:
\bqa
V^{\mu\alpha\beta} &=& \sum_{i=1}^{2}\,\Bigl[
A_i\,\delta^{\mu\alpha}\,r^{\beta}_i +
B_i\,\delta^{\mu\beta}\,r^{\alpha}_i +
C_i\,\delta^{\alpha\beta}\,r^{\mu}_i\Bigr] +
\sum_{ijk=1}^{2}\,D_{ijk}\,r^{\mu}_i\,r^{\alpha}_j\,r^{\beta}_k.
\label{fullG}
\eqa
Using the relations
\bq
\spro{R_i}{r_j} = \frac{1}{2} \delta_{ij}, \qquad
\spro{R_i}{R_j} = \frac{1}{2}\,G^{-1}_{ij}, \qquad
{\cal G}^{\mu\nu} = R^{\mu}\,G\,R^{\nu} = r^\mu\,G^{-1}\,r^\nu,
\eq
we introduce the following projectors:
\bqa
{\cal P}^{\mu\alpha\beta}_{\ssA l} &=& \delta^{\mu\alpha}\,R^{\beta}_l -
2\,{\cal G}^{\mu\alpha}\,R^{\beta}_l,
\quad
{\cal P}^{\mu\alpha\beta}_{\ssB l} = \delta^{\mu\beta}\,R^{\alpha}_l -
2\,{\cal G}^{\mu\beta}\,R^{\alpha}_l,
\quad
{\cal P}^{\mu\alpha\beta}_{\ssC l} = \delta^{\alpha\beta}\,R^{\mu}_l -
2\,{\cal G}^{\alpha\beta}\,R^{\mu}_l,
\nl
{\cal P}^{\mu\alpha\beta}_{\ssD ijl} &=&
R^{\mu}_i\,R^{\alpha}_j\,R^{\beta}_l - \frac{1}{2(n-2)}\,\Bigl[
G^{-1}_{ij}\,{\cal P}^{\mu\alpha\beta}_{\ssA l} +
G^{-1}_{il}\,{\cal P}^{\mu\alpha\beta}_{\ssB j} +
G^{-1}_{jl}\,{\cal P}^{\mu\alpha\beta}_{\ssC i}\Bigr].
\eqa
Their action can be represented as
\bq
A_i = \frac{2}{n-2}\,{\cal P}^{\mu\alpha\beta}_{\ssA i}\,V_{\mu\alpha\beta},
\quad
B_i = \frac{2}{n-2}\,{\cal P}^{\mu\alpha\beta}_{\ssB i}\,V_{\mu\alpha\beta},
\quad
C_i = \frac{2}{n-2}\,{\cal P}^{\mu\alpha\beta}_{\ssC i}\,V_{\mu\alpha\beta},
\quad
D_{ijl} = 8\,{\cal P}^{\mu\alpha\beta}_{\ssD ijl}\,V_{\mu\alpha\beta}.
\label{scalcomp}
\eq
At the level of triple vector boson couplings we encounter an additional
complication, namely CP-odd form factors are absent only in the total amplitude
but not in single diagrams. Therefore, one should write a more general form for
the vertex, including CP-odd terms:
\bqa
V^{\mu\alpha\beta}_{\ep} = \sum_{i=1}^{2}\,\Bigl[
E_i\,\ep(\lambda,\sigma,\mu,\alpha)\,r^{\beta}_i +
F_i\,\ep(\lambda,\sigma,\mu,\beta)\,r^{\alpha}_i +
G_i\,\ep(\lambda,\sigma,\alpha,\beta)\,r^{\mu}_i\Bigr]\,
r_{1\lambda}\,r_{2\sigma}.
\eqa
The following property holds:
${\cal P}^{\mu\alpha\beta}_{\ssI i}\,V_{\ep,\mu\alpha\beta} = 0$,
for $I = A,B,C$ and
${\cal P}^{\mu\alpha\beta}_{\ssD ijk}\,V_{\ep,\mu\alpha\beta} = 0$.

For external Proca fields (and also for Rarita-Schwinger fields), however, 
our preference will be for other methods~\cite{Passarino:1983zs} where the 
wave-functions for vector particles can be entirely expressed in terms of 
Dirac spinors with arbitrary polarization vectors allowing for the 
implementation of projector techniques for helicity 
amplitudes~\cite{Glover:2004si}. 
\section{Techniques for the reduction of two-loop two-point functions 
\label{stltpf}}
It is well-known that the reduction of two-loop tensor integrals can be
achieved up to two-point functions if we are ready to enlarge the class of
scalar functions. The original derivation is due to Weiglein, Scharf and
Bohm~\cite{Weiglein:hd}; for completeness we review here the necessary
technology and refer the reader to Appendix B for the full list of results.

Standard reduction to scalar integrals amounts to writing down the most
general decomposition of tensor integrals, and to transform this relation
into a system of linear equations whose unknowns are the form factors and
the known terms follow from algebraic reduction between saturated numerators
and denominators.

The well-known obstacle on the road to scalarization of multi-loop diagrams
is represented by the occurrence of irreducible numerators, i.e.\ those
cases where a $\spro{q_i}{p}$ term appears in the numerator, but no
parameterization of the diagram can be found where the inverse propagators
$q^2_i + m^2_j$ and $(q_i+p)^2 + m^2_l$ simultaneously occur. For any
two-loop self-energy diagram with $I$ propagators there are 
$5 - I$ irreducible scalar products.
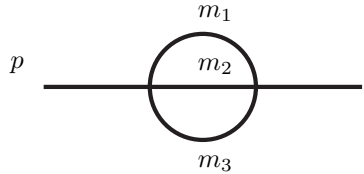
\begin{figure}[th]
\vspace{0.5cm}
\[
  \vcenter{\hbox{
  \begin{picture}(150,0)(0,0)
  \SetScale{0.8}
  \SetWidth{2.}
  \Line(0,0)(150,0)
  \CArc(75,0)(25,0,180)
  \CArc(75,0)(25,180,360)
  \Text(-10,5)[cb]{$p$}
  \Text(65,5)[cb]{$m_2$}
  \Text(65,25)[cb]{$m_1$}
  \Text(65,-32)[cb]{$m_3$}
  \end{picture}}}
\]
\vspace{0.3cm}
\caption[]{Scalar diagram of the $S^{\ssA}$-family, the so-called sunset 
(or sunrise) configuration.}
\label{thisisssa}
\end{figure}
To illustrate the procedure we start considering some simple example, e.g.\
a vector integral of the $S^{\ssA}$-family, depicted in \fig{thisisssa},
\bq
S^{\ssA}(\mu\,|\,0\,;\,p,\lstm{123}) =
\frac{\mu^{2\ep}}{\pi^4}\,\intmomsii{n}{q_1}{q_2}\,
\frac{q_{1\mu}}{[1]\,[2]_{\ssA}\,[3]_{\ssA}},
\label{exass}
\eq
where we introduced a shorthand notation for the inverse propagators:
\bq [1] =  q^2_1 + m^2_1, \quad [2]_{\ssA} = (q_1-q_2+p)^2 + m^2_2, \quad
[3]_{\ssA} = q^2_2 + m^2_3. 
\label{propsunset}
\eq
Apparently we meet an irreducible numerator, but we can generalize the
procedure considering an intermediate reduction with respect to sub-loops, a
technique originally introduced in~\cite{Weiglein:hd}. In the following
subsection we briefly illustrate this technique.
\subsection{Reduction to sub-loops \label{scalsubl}}
Consider \eqn{exass}: we may write
\bq
\int\,d^nq_1\,\frac{q_{1\mu}}{[1][2]_{\ssA}} = X_{\ssA}\,(q_2 - p)_{\mu}.
\label{boths}
\eq
If we multiply both sides of \eqn{boths} by $(q_2-p)_{\mu}$ and use the
identities
\bq
\frac{\spro{q_1}{p}}{[2]_{\ssA}} = \frac{\spro{q_1}{q_2}}{[2]_{\ssA}} +
\frac{1}{2}\,\Bigl[ 1 - \frac{q^2_1 + q^2_2 -
2\,\spro{q_2}{p} + m^2_2 + p^2}{[2]_{\ssA}}\Bigr],
\qquad
\frac{q^2_1}{[1]} = 1 - \frac{m^2_1}{[1]},
\eq
we can solve for $X_{\ssA}$ obtaining 
\bq
X_{\ssA} =  -\frac{1}{2}\,\frac{1}{[0]_{\ssA}}\,\int\,d^nq_1\,\Bigl[
    \frac{m^2_{12}-[0]_{\ssA}}{[1][2]_{\ssA}} + \frac{1}{[1]} - 
\frac{1}{[2]_{\ssA}}\Bigr],
\eq 
\label{af4}
where a new propagator has made its appearance, $[0]_{\ssA} = (q_2 - p)^2$.
We then use a second pair of identities,
\bq
\frac{\spro{q_2}{p}}{[0]_{\ssA}} = - \frac{1}{2}\,\Bigl[ 1 -
\frac{q^2_2 + p^2}{[0]_{\ssA}}\Bigr],
\qquad
\frac{q^2_2}{[3]_{\ssA}} = 1 - \frac{m^2_3}{[3]_{\ssA}},
\eq
to obtain the following result:
\bqa
S^{\ssA}(\mu\,|\,0\,;\,p,\lstm{123}) &=&
\frac{1}{2}\,S^{\ssA}(0\,|\,\mu\,;\,p,\lstm{123}) +
\frac{1}{4}\,S^{\ssA}_{a}\,p_{\mu},
\eqa
\bqa
S^{\ssA}_{a} &=&
m^2_{12}\,( \frac{m^2_3}{p^2} + 1 )\,S^{\ssC}_{\Z}(p,\lstm{12},0,m_3) +
( \frac{m^2_{12}}{p^2} - 2 )\,S^{\ssA}_{\Z}(p,\lstm{123})
\nl
{}&-& \frac{m^2_{12}}{p^2}\,S^{\ssA}_{\Z}(0,0,\lstm{12}) -
A_{\Z}([m_1,m_2])\,
\Bigl[ ( 1 + \frac{m^2_3}{p^2} )\,B_{\Z}(p,m_3,0)
+ \frac{1}{p^2}\,A_{\Z}(m_3)\Bigr].
\label{transfot}
\eqa
In \eqn{transfot} we used generalized one-loop scalar functions defined in
\eqn{GABfun} and
\bqa
S^{\ssC}_{\Z}(p,\lstm{1234}) &=&
\frac{\mu^{2\ep}}{\pi^4}\,\intmomsii{n}{q_1}{q_2}\,
\frac{1}{[1]\,[2]_{\ssC}\,[3]_{\ssC}\,[4]_{\ssC}},
\eqa
where the propagators are 
$[2]_{\ssC} = (q_1 - q_2)^2 + m^2_2$, 
$[3]_{\ssC} = q^2_2 + m^2_3$ and
$[4]_{\ssC} = (q_2+p)^2 + m^2_4$. Henceforth we continue our derivation for
$S^{\ssA}(0\,|\,\mu\,;\,p,\lstm{123})$ and write another equation,
\bq
\int\,d^nq_2\,\frac{q_{2\mu}}{[2]_{\ssA}[3]_{\ssA}} = Y_{\ssA}\,
(q_1 + p)_{\mu};
\eq
a solution for $Y_{\ssA}$ is obtained,
\bq
Y_{\ssA} = -\,\frac{1}{2}\,\frac{1}{[0]_{\ssA\ssA}}\,\int\,d^nq_2\,\Bigl[
\frac{m^2_{32}-[0]_{\ssA\ssA}}{[2]_{\ssA}[3]_{\ssA}} -
\frac{1}{[2]_{\ssA}} + \frac{1}{[3]_{\ssA}}\Bigr],
\eq
with a new propagator defined by $[0]_{\ssA\ssA} = (q_1+p)^2$. It follows that
\bqa
S^{\ssA}(0\,|\,\mu\,;\,p,\lstm{123}) &=&
\frac{1}{2}\,S^{\ssA}(\mu\,|\,0\,;\,p,\lstm{123}) +
\frac{1}{4}\,S^{\ssA}_{b}\,p_{\mu},
\eqa
\bqa
S^{\ssA}_{b} &=&
-m^2_{32}\,( \frac{m^2_1}{p^2} + 1 )\,S^{\ssC}_{\Z}(p,\lstm{32},0,m_1) -
( \frac{m^2_{32}}{p^2} - 2 )\,S^{\ssA}_{\Z}(p,\lstm{123})
\nl
{}&+& \frac{m^2_{32}}{p^2}\,S^{\ssA}_{\Z}(0,0,\lstm{32}) +
A_{\Z}([m_3,m_2])\,
\Bigl[ ( 1 + \frac{m^2_1}{p^2} )\,B_{\Z}(p,m_1,0)
+ \frac{1}{p^2}\,A_{\Z}(m_1)\Bigr].
\label{transfto}
\eqa
Therefore, using the system of \eqns{transfot}{transfto} we can solve for 
both vector integrals in terms of scalar functions.

The full list of results will be given in \sect{fulll}. Already from
this simple example we see the appearance of generalized scalar loop 
integrals in the reduction of tensor ones. In the next Section we present the 
strategy for their evaluation and discuss the general case based on a special
set of identities.
\section{Integration by parts identities and generalized recurrence relations 
\label{ibpoai}}
A popular and quite successful tool in dealing with multi-loop diagrams, in 
particular those containing powers of irreducible scalar products, is 
represented by the integration-by-parts identities (hereafter 
IBPI)~\cite{'tHooft:1972fi}. 
It is well-known that arbitrary integrals can be reduced~\cite{Laporta:2001dd}
to an handful of Master Integrals (MI) using IBPI~\cite{'tHooft:1972fi} 
and Lorentz-invariance identities~\cite{Gehrmann:1999as}.

For one-loop diagrams IBPI can be written as
\bq
\int\,d^nq\,\frac{\partial}{\partial q_{\mu}}\,\Bigl[ v_{\mu}\,
F(q,p,m_1\cdots)\Bigr] = 0 ,
\eq
where $v = q,p_1\cdots,p_\ssE$, and $E$ is the number of independent
external momenta. By careful examination of the IBPI one can show that all
one-loop diagrams can be reduced to a limited set of MI. Here we would like
to point out one drawback of this solution. Consider, for instance, the
following identity~\cite{Remiddi:1997ny},
\bqa
B_{\Z}(1,2\,;\,p,m_1,m_2) &=& \frac{1}{\lambda(-p^2,m^2_1,m^2_2)}\,\Bigl\{
(n - 3)\,(m^2_1 - m^2_2 - p^2)\,B_{\Z}(p,m_1,m_2) 
\nl
{}&+& (n - 2)\,\Bigl[ A_{\Z}(m_1) - \frac{p^2+m^2_1+m^2_2}{2\,m^2_2}\,
A_{\Z}(m_2)\Bigr]\Bigr\},
\label{squareb}
\eqa
where $\lambda(x,y,z)$ is the familiar K\"allen lambda function
$\lambda(x,y,z)=x^2 +y^2 +z^2 -2xy -2xz -2yz$.  The factor in front of the
curly bracket is exactly the BT-factor associated with the diagram; from the
general analysis of~\cite{Ferroglia:2002mz} we know that at the normal
threshold the leading behavior of $B_{\Z}(1,2)$ is $\lambda^{-1/2}$, so that
the reduction to MI apparently overestimates the singular behavior; of
course, by carefully examining the curly bracket in \eqn{squareb} one can
derive the right expansion at threshold, but the result, as it stands, is
again a source of cancellations/instabilities. Our experience, e.g.\ with
one-loop multi-leg diagrams~\cite{Ferroglia:2002mz}, shows that numerical
evaluation following smoothness algorithms (e.g.\ Bernstein-Tkachov
functional relations~\cite{Tkachov:1997wh}) does not increase the degree of
divergence when going from scalar to tensor integrals.

The IBPI for two-loop diagrams can be written as
\bq
\int\,d^nq_1\,d^nq_2\,\frac{\partial}{\partial a_{\mu}}\,\Bigl[
b^{\mu}\,F(q_1,q_2,\{p\},m_1\cdots)\Bigr] = 0,
\qquad
a = q_{i}, \quad b =  q_{i},p_{1}\cdots,p_{\ssE},
\eq
where $i = 1,2$, and $E$ is the number of independent external momenta.
Again, using IBPI, arbitrary two-loop integrals can be written in terms of a
restricted number of MI. The problem remains in the explicit evaluation of
the MI; in the following of this Section we want to show that the solution
is purely algebraic and, at the same time, we will discuss the relationship
with our approach. Consider again the scalar and the two vector integrals in
the $S^{\ssA}$-family: for them we have
\bqa
S^{\ssA}(0\,|\,0\,;\,p,\lstm{123}) &=& S^{\ssA}_{\Z},
\quad
S^{\ssA}(\mu\,|\,0\,;\,p,\lstm{123}) = S^{\ssA}_{\sa}\,p_{\mu},
\quad
S^{\ssA}(0\,|\,\mu\,;\,p,\lstm{123}) = S^{\ssA}_{\msb}\,p_{\mu}.
\eqa
Introducing the notation
\bq
\CIM{S}_{\ssA} = \intfxx{x}{y}\,\Bigl[ x\,(1-x) \Bigr]^{-\ep/2}\,y^{\ep/2-1},
\eq
we derive the parametric representation for the scalar and the two vector 
integrals:
\bq
S^{\ssA}_{i} = \omega^{\ep}\,\egam{\ep-1}\,
\CIM{S}_{\ssA}\,P^{\aaa}_{i}(x,y)\,\chiu{\aaa}^{1-\ep}(x,y),
\label{s111}
\eq
where $\egam{z}$ denotes the Euler gamma function, $\omega$ is defined in
\eqn{defDUV} and where we introduced the auxiliary polynomials
\bq
P^{\aaa}_0 = -1,
\qquad
P^{\aaa}_1 = x\,(1 - y),
\qquad
P^{\aaa}_2 = -\,y.
\label{tobem}
\eq
The quadratic form $\chiu{\aaa}$ in \eqn{s111} is given by
$\chiu{\aaa} = -p^2\,y^2 + (p^2 - m_3^2 + m_x^2)\,y + m_3^2$, 
with $m_x^2$ defined in \eqn{defGfun}.

The evaluation of the scalar integral was discussed
in~\cite{Passarino:2001wv} and can be easily extended to cover the two
remaining cases.  This simple example can be fully generalized, thus proving
that any smoothness algorithm designed for scalar integrals will also be
effective to deal with tensor ones; physical observables can be evaluated
without using a reduction procedure. Needless to say, however, that when
cancellations are at the basis of the result -- for instance when testing
the WST identities of the theory -- scalarization should be attempted;
indeed, in these cases the goodness of the result depends crucially on our
capability to express the whole set of graphs in terms of a minimal set of
integrals.

One way of deriving this result is purely algebraic: to achieve scalarization, 
which is equivalent to express irreducible tensor integrals in terms of truly 
scalar functions, we write down generalized functions
\bq
\cS^{\alpha_1 | \alpha_3 | \alpha_2}_{\ssA}(n\,;\,p,\lstm{123}) =
\frac{\mu^{2(4-n)}}{\pi^4}\,
\int\,d^nq_1\,d^nq_2\,\prod_{i=1}^{3}\,[i]^{-\alpha_i}_{\ssA},
\eq
with $[1]_{\ssA} = [1]$, which are defined for arbitrary space-time dimension 
$n$. Subsequently we fix $n$ to be $n = \sum_i\alpha_i + 1 - \ep$ and obtain
\bqa
\cS^{\alpha_1 | \alpha_3 | \alpha_2}_{\ssA}(n\,;\,p,\lstm{123}) &=&
-\,\frac{\egam{\ep-1}}{\prod_i\,\egam{\alpha_i}}\,
\omega^{3-\sum\alpha+\ep}\,
\dcub{2}\,x^{-\rho_1/2}\,
(1-x)^{-\rho_2/2}\,(1-y)^{\rho_3}\,
y^{\rho_4/2}\,\chiu{\aaa}^{1-\ep},
\label{workofT}
\eqa
where $\omega$ is defined in \eqn{defDUV} and where we introduced powers
\bq
\rho_1 = 1 + \alpha_1 - \alpha_2 - \alpha_3 + \ep,
\;
\rho_2 = 1 + \alpha_2 - \alpha_1 - \alpha_3 + \ep,
\;
\rho_3 = \alpha_3 - 1,
\;
\rho_4 = \alpha_1 + \alpha_2 - \alpha_3 - 3 + \ep.
\eq
According to the work of Tarasov~\cite{Tarasov:2000sf} the content of 
\eqn{workofT} can be interpreted by saying that we have a scalar integral in 
shifted space-time dimension and with non-canonical powers of propagators; 
equivalently, we may interpret it as an integral in the canonical $4-\ep$ 
dimensions, with all powers in the propagators equal to one but with 
polynomials of Feynman parameters in the numerator. To formally show this 
equivalence we write
\bq
S^{\ssA}_i = \sum_{j=1}^{2}\,\omega^{n_j-4+\ep}\,k_{ij}\,
\cS^{\alpha_{j}\,|\,\beta_{j}\,|\,\gamma_{j}}_{\ssA}(n_j\,;\,p,\lstm{123}),
\quad i = 1,2,
\eq
with $n_j = \alpha_j + \beta_j + \gamma_j + 1 - \ep$, and fix all
coefficients and exponents in order to match \eqn{tobem}. A solution is
therefore given by $\alpha_1 = 1, \beta_1 = 2,\gamma_1 = 2$, or by $\alpha_2
= 2,\beta_2 = 1,\gamma_2 = 2$, with coefficients $k_{11} = -1, \; k_{12} =
0$ and $k_{21} = 0, \; k_{22} = 1$.

Note that, starting with two-loop vertices and due to the presence of
irreducible scalar products, we should abandon the term ``scalarization'' in
favor of a more general concept, namely the reduction to a minimum number of
functions that are needed to classify the problem at hand. One can hence
adopt a reduction to scalar integrals in shifted dimensions, followed by a
solution of generalized recursion relations~\cite{Tarasov:2000sf} (which
include the IBPI method as a particular case) reducing the large set of
integrals to relatively few MI. Alternatively, we can decide to relate the
form factors to truly scalar integrals in the same number of dimensions and
belonging to contiguous families, and to integrals with contracted and
irreducible numerators for which a numerical solution is available; the
quality of this latter numerical solution is as good as the one for the
scalar configurations.

The two procedures are algebraically equivalent and preference is, to some
extent, a matter of taste, although the power of a procedure can only be
justified a posteriori by the goodness of the corresponding result. As a
matter of fact, a reduction to master integrals is notoriously difficult to
achieve when the problem is characterized by several scales.  For
completeness we stress that Davydychev~\cite{Davydychev:1991va} and, later
on, Bern, Dixon and Kosower~\cite{Bern:1993kr} gave expressions for one-loop
tensor integrals with shifted dimensions; Campbell, Glover and
Miller~\cite{Campbell:1996zw} discovered good numerical stability for
one-loop integrals in higher dimensions; and a simple formula expressing any
$N$-point integral in terms of integrals in higher dimensions was given by
Fleischer, Jegerlehner and Tarasov~~\cite{Fleischer:1999hq}.

An example of reduction of generalized functions with the help of IBP
techniques is provided by the well-known result that all generalized scalar
sunset diagrams with zero external momentum (i.e.\ vacuum-bubbles) can be
fully reduced. To see this we first introduce
\bq
S^{\ssA}_{\Z}(\{\alpha\}_3\,;\,0,\lstm{123}) =
\frac{\mu^{2\ep}}{\pi^4}\,\intmomsii{n}{q_1}{q_2}\,
\prod_{i=1}^{3}\,[i(p=0)]^{-\alpha_i}_{\ssA},
\eq
where $[i]_{\ssA}$ is defined in \eqn{propsunset} but with $p = 0$ and
$[1]_{\ssA} = [1]$. IBPI reduce all functions in this class to combinations
of $S^{\ssA}_{\Z}(1,1,2\,;\,0,\lstm{123})$ and products of one-loop integrals.
For instance we obtain
\bq
S^{\ssA}_{\Z}(\{\alpha\}_3\,;\,0,\lstm{123}) = \frac{1}{m^2_{132}}\,
{\cal S}^{\ssA}_{\Z}(\{\alpha\}_3\,;\,0,\lstm{123}),
\eq
etc, with
\bqa
{\cal S}^{\ssA}_{\Z}(0,\lstm{123}) &=&
\frac{\lambda(m^2_1,m^2_2,m^2_3)}{n-3}\,S^{\ssA}_{\Z}(1,1,2\,;\,0,\lstm{123}) +
\frac{n-2}{n-3}\,\Bigl[ A_{\Z}(m_1)\,A_{\Z}(m_2) 
\nl
{}&-&
\frac{1}{2}\,\Bigl(1 - \frac{m^2_1}{m^2_3} + \frac{m^2_2}{m^2_3}\Bigr)\,
A_{\Z}(m_1)\,A_{\Z}(m_3) -
\frac{1}{2}\,\Bigl(1 + \frac{m^2_1}{m^2_3} - \frac{m^2_2}{m^2_3}\Bigr)\,
A_{\Z}(m_2)\,A_{\Z}(m_3)\Bigr],
\nl
{\cal S}^{\ssA}_{\Z}(2,1,1\,;\,0,\lstm{123}) &=& 
m^2_{213}\,S^{\ssA}_{\Z}(1,1,2\,;\,0,\lstm{123}) 
+ \frac{n-2}{2}\,\Bigl[ \frac{1}{m^2_3}\,A_{\Z}(m_2)\,A_{\Z}(m_3)
\nl
{}&-& \frac{1}{m^2_1}\,A_{\Z}(m_1)\,A_{\Z}(m_2) +
\Bigl( \frac{1}{m^2_1} - \frac{1}{m^2_3}\Bigr)\,A_{\Z}(m_1)\,A_{\Z}(m_3)\Bigr], 
\nl
{\cal S}^{\ssA}_{\Z}(1,2,1\,;\,0,\lstm{123}) &=&
 m^2_{123}\,S^{\ssA}_{\Z}(1,1,2\,;\,0,\lstm{123}) 
+ \frac{n-2}{2}\,\Bigl[
\Bigl( \frac{1}{m^2_2} - \frac{1}{m^2_3}\Bigr)\,A_{\Z}(m_2)\,A_{\Z}(m_3)
\nl
{}&-& \frac{1}{m^2_2}\,A_{\Z}(m_1)\,A_{\Z}(m_2) +
\frac{1}{m^2_3}\,A_{\Z}(m_1)\,A_{\Z}(m_3)\Bigr],
\eqa
etc. The number of terms in the reduction tends to increase considerably for
higher powers in the propagators of the generalized sunset functions but, as
we said, all of them can be expressed through the $(1,1,2)$ sunset integral
and products of one-loop $A_{\Z}$-functions.

\section{Reduction for tensor two-point functions \label{fulll}}
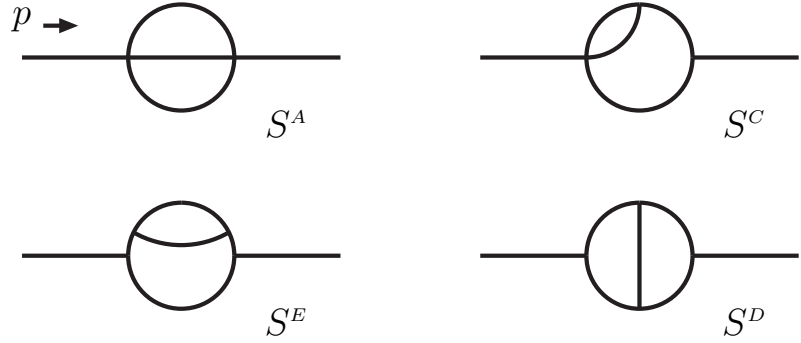
\begin{figure}[th]
\vspace{0.5cm}
\bqas  
{}&{}&
  \vcenter{\hbox{
  \begin{picture}(150,0)(0,0)
  \SetScale{0.8}
  \SetWidth{2.}
  \Line(0,0)(150,0)
  \CArc(75,0)(25,0,180)
  \CArc(75,0)(25,180,360)
  \LongArrow(10,15)(25,15)   
  \Text(0,10)[cb]{\Large $p$}
  \Text(100,-30)[cb]{\Large $S^{\ssA}$}
  \end{picture}}}
\qquad
  \vcenter{\hbox{
  \begin{picture}(150,0)(0,0)
  \SetScale{0.8}
  \SetWidth{2.}
  \Line(0,0)(50,0)
  \CArc(75,0)(25,0,90)
  \CArc(75,0)(25,-180,0)
  \CArc(75,0)(25,90,180)
  \CArc(50,25)(25,-90,0)
  \Line(100,0)(150,0)
  \Text(100,-30)[cb]{\Large $S^{\ssC}$}
  \end{picture}}}
\nl\nl\nl\nl\nl
{}&{}&
  \vcenter{\hbox{
  \begin{picture}(150,0)(0,0)
  \SetScale{0.8}
  \SetWidth{2.}
  \Line(0,0)(50,0)
  \CArc(75,0)(25,0,90)
  \CArc(75,0)(25,-180,0)
  \CArc(75,0)(25,90,180)
  \CArc(75,50)(45,-120,-60)
  \Line(100,0)(150,0)
  \Text(100,-30)[cb]{\Large $S^{\ssE}$}
  \end{picture}}}
\qquad
  \vcenter{\hbox{
  \begin{picture}(150,0)(0,0)
  \SetScale{0.8}
  \SetWidth{2.}
  \Line(0,0)(50,0)
  \CArc(75,0)(25,0,90)
  \CArc(75,0)(25,-180,0)
  \CArc(75,0)(25,90,180)
  \Line(75,25)(75,-25)
  \Line(100,0)(150,0)
  \Text(100,-30)[cb]{\Large $S^{\ssD}$}
  \end{picture}}}
\eqas
\vspace{0.1cm}
\caption[]{Irreducible classes for two-loop two-point functions.}
\label{TLse}
\end{figure}

In this Section we give a full list of results following the method of
Weiglein, Scharf and Bohm~\cite{Weiglein:hd} as derived in \sect{stltpf}.
Scalar two-loop two-point functions are defined by
\bqa
S^{\ssA}_{\Z}(p,\lstm{123}) &=&
\frac{\mu^{2\ep}}{\pi^4}\,\intmomsii{n}{q_1}{q_2}\,
\frac{1}{[1]\,[2]_{\ssA}\,[3]_{\ssA}},
\nl
S^{\ssC}_{\Z}(p,\lstm{1234}) &=&
\frac{\mu^{2\ep}}{\pi^4}\,\intmomsii{n}{q_1}{q_2}\,
\frac{1}{[1]\,[2]_{\ssC}\,[3]_{\ssC}\,[4]_{\ssC}},
\nl
S^{\ssD}_{\Z}(p,\lstm{12345}) &=&
\frac{\mu^{2\ep}}{\pi^4}\,\intmomsii{n}{q_1}{q_2}\,
\frac{1}{[1]\,[2]_{\ssD}\,[3]_{\ssD}\,[4]_{\ssD}\,[5]_{\ssD}},
\nl
S^{\ssE}_{\Z}(p,\lstm{12345}) &=&
\frac{\mu^{2\ep}}{\pi^4}\,\intmomsii{n}{q_1}{q_2}\,
\frac{1}{[1]\,[2]_{\ssE}\,[3]_{\ssE}\,[4]_{\ssE}\,[5]_{\ssE}},
\label{Deftltp}
\eqa
with propagators
\bq
[1] = q^2_1 + m^2_1, \qquad
[2]_{\ssA} = (q_1 - q_2 + p)^2 + m^2_2, \quad
[3]_{\ssA} = q^2_2 + m^2_3,
\eq
\bq
[2]_{\ssC} = (q_1 - q_2)^2 + m^2_2, \quad
[3]_{\ssC} = q^2_2 + m^2_3, \quad
[4]_{\ssC} = (q_2 + p)^2 + m^2_4,
\eq
\bq
[2]_{\ssD} = (q_1 + p)^2 + m^2_2, \quad
[3]_{\ssD} = (q_1 - q_2)^2 + m^2_3, \quad
[4]_{\ssD} = q_2^2 + m^2_4, \quad
[5]_{\ssD} = (q_2 + p)^2 + m^2_5.
\eq
\bq
[2]_{\ssE} = (q_1 - q_2)^2 + m^2_2, \quad
[3]_{\ssE} = q^2_2 + m^2_3, \quad
[4]_{\ssE} = (q_2 + p)^2 + m^2_4, \quad
[5]_{\ssE} = q^2_2 + m^2_5,
\eq
Propagators $[i]_{\ssE}$ should not be confused with those appearing in
\eqn{enummaba} which refer to a three-point function.
These scalar diagrams were investigated in~\cite{Passarino:2001wv}, Eq.~(89)
and Eqs.~(146-147) for $S^{\ssA}_{\Z} \equiv S^{\saaa}_{\Z}$; 
in~\cite{Passarino:2001jd}, Sect.~(5.8) for $S^{\ssC}_{\Z} \equiv
S^{\saba}_{\Z}$, Sect.~(7.3) for $S^{\ssD}_{\Z} \equiv S^{\sbba}_{\Z}$ and 
Sect.~(7.8) for $S^{\ssE}_{\Z} \equiv S^{\saca}_{\Z}$.  Furthermore, we 
define form factors according to
\bqa
S^{\ssI}(\mu\,|\,0)
&=& S^{\ssI}_{\sa}\,p_{\mu},
\quad
S^{\ssI}(0\,|\,\mu)
= S^{\ssI}_{\msb}\,p_{\mu},
\quad
S^{\ssI}(\mu,\nu\,|\,0)
= S^{\ssI}_{\saab}\,\delta_{\mu\nu} +
S^{\ssI}_{\saaa}\,p_{\mu}\,p_{\nu},
\nl
S^{\ssI}(\mu\,|\,\nu)
&=& S^{\ssI}_{\sabb}\,\delta_{\mu\nu} +
S^{\ssI}_{\saba}\,p_{\mu}\,p_{\nu},
\quad
S^{\ssI}(0\,|\,\mu,\nu)
= S^{\ssI}_{\sbbb}\,\delta_{\mu\nu} +
S^{\ssI}_{\sbba}\,p_{\mu}\,p_{\nu},
\label{generalabg}
\eqa
with $I= A,C,D,E$; the irreducible classes for two-loop two-point functions 
are shown in \fig{TLse}.
Generalized one-loop functions are given in \eqns{GABfun}{GBT}:
after reduction,(with $\lambda_{ij} = \lambda(-p^2,m^2_i,m^2_j)$) we obtain
\bq
A_{\Z}(\alpha,m) = \frac{\mu^{\ep}}{i\,\pi^2}\,
\int\,\frac{d^nq}{(q^2+m^2)^{\alpha}} =\frac{1}{m^2}
\Bigl[ 1- \frac{4-\ep}{2\,(\alpha-1)}\Bigr]\,A_{\Z}(\alpha-1,m),
\quad \Reb\,\alpha > 1,
\eq
\bqa
B_{\sa}(2,1\,;\,p,\lstm{12}) &=& 
\frac{1}{2 \ p^2} \Bigl[ A_{\Z}(2,m_1) - B_{\Z}(p,\lstm{12}) -
l_{p 12}\,B_{\Z}(2,1,p,\lstm{12}) \Bigr],
\nl
B_{\sba}(2,1\,;\,p,\lstm{12}) &=& -\,\frac{1}{4}\,\frac{1}{(n-1) p^4}\,\Bigl[
n\,l_{p 12}\,A_{\Z}(2,m_1) + n\,A_{\Z}([m_1,m_2]) 
+ 2\,(2\,p^2 - n\,l_{p 12})\,B_{\Z}(p,m_1,m_2) \nl
&+& 
(4\,p^2\,m_1^2\,(n-1)-n\,\lambda_{12})\,B_{\Z}(2,1\,;\,p,m_1,m_2)
\Bigr],
\nl
B_{\sbb}(2,1\,;\,p,\lstm{12}) &=& -\frac{1}{4}\,\frac{1}{(n-1) p^2} 
\Bigl[A_{\Z}([m_1,m_2]) + l_{p 12}\,A_{\Z}(2,m_1)
+ 2\,l_{p 21}\,B_{\Z}(p,\lstm{12}) 
\nl
&+& \lambda_{12}\,B_{\Z}(2,1\,;\,p,\lstm{12})\Bigr],
\nl
B_{\sa}(1,2;\,p,\lstm{12}) &=& B_{\sa}(2,1\,;\,-p,\lstm{21}) -
B_{\Z}(2,1\,;\,-p,\lstm{21}), \quad \mbox{etc.}
\eqa
In this Section it is always understood that the space-time dimension is
$n = 4 - \ep$. Whenever reducibility is at hand we apply standard methods and
obtain the following list of results:
\bqa
S^{\ssC}_{\saaa} &=&
-\frac{1}{p^2}\,\Bigl[
 A_{\Z}(m_2)\,B_{\Z}(p,\lstm{34}) +  n\,S^{\ssC}_{\saab}(p,\lstm{1234}) +
 m^2_1\,S^{\ssC}_{\Z}(p,\lstm{1234})\Bigr],
\nl\nl
S^{\ssC}_{\saba} &=& \frac{1}{2}\,\frac{1}{(n-1)\,p^4}\,\Bigl\{
 -\,p^2\,A_{\Z}([m_1,m_2])\,B_{\Z}(p,\lstm{34}) + 
p^2\,m_{123}^2\,S^{\ssC}_{\Z}(p,\lstm{1234}) \nl
&-& p^2\,S^{\ssA}_{\Z}(p,\lstm{1234}) - n\,
\Bigl[ l_{\ssP 34}\,S^{\ssC}_{\sa}(p,\lstm{1234}) + 
S^{\ssA}_{\sa}(p,\lstm{124})\Bigr] 
\Bigr\} \, ,
\eqa
\bqa
S^{\ssC}_{\sbba} &=& \frac{1}{4}\,\frac{1}{(n-1)\,p^4}\,\Bigl\{
\lambda_{34}\,S^{\ssC}_{\Z}(p,\lstm{1234}) - 
l_{\ssP 43}\,\Bigl[ S^{\ssA}_{\Z}(p,\lstm{124})
+ S^{\ssA}_{\Z}(0,\lstm{123})\Bigr] - 2\,p^2\,S^{\ssA}_{\msb}(p,\lstm{124})
\nl
{}&+& (n-1)\,\Bigl[ (\lambda_{34} - 4\,p^2\,m^2_3)\,S^{\ssC}(p,\lstm{1234})
+ (3\,p^2 - m^2_3 + m^2_4)\,S^{\ssA}_{\Z}(p,\lstm{124})
\nl
{}&-& l_{\ssP 34}\,S^{\ssA}_{\Z}(0,\lstm{123})
- 2\,p^2\,S^{\ssA}_{\msb}(p,\lstm{124})\Bigr]\Bigr\} \, ,
\eqa
\bqa
S^{\ssC}_{\sabb} &=& \frac{1}{2}\,\frac{1}{n-1}\,\Bigl[ 
A_{\Z}([m_1,m_2])\,B_{\Z}(p,\lstm{34}) - m^2_{132}\,S^{\ssC}_{\Z}(p,\lstm{123}) + 
S^{\ssA}_{\Z}(p,\lstm{124}) 
\nl
{}&+& l_{\ssP 34}\,S^{\ssC}_{\sa}(p,\lstm{1234}) + 
S^{\ssA}_{\sa}(p,\lstm{124})\Bigr]\, ,
\eqa
\bqa
S^{\ssC}_{\sbbb} &=&
\frac{1}{4}\!\frac{1}{(n-1) p^2}\!\Bigl\{ 
\!-\lambda_{34}S^{\ssC}_{\Z}(p,\lstm{1234}) \!+\! 
l_{\ssP 43}\Bigl[S^{\ssA}_{\Z}(p,\lstm{124}) \!+\!
S^{\ssA}_{\Z}(0,\lstm{123})\Bigr]\! 
\!+ \!p^2 S^{\ssA}_{\msb}(p,\lstm{124})\! \Bigr\},
\eqa
\bqa
S^{\ssC}_{\msb} &=& \frac{1}{2\,p^2}\, \Bigl[-l_{\ssP 34}\,
S^{\ssC}_{\Z}(p,\lstm{1234}) - S^{\ssA}_{\Z}(p,\lstm{124}) + 
S^{\ssA}_{\Z}(0,\lstm{124})\Bigr] \, ,
\eqa
\bqa
S^{\ssE}_{\saaa} &=& -\frac{1}{p^2}\,\Bigl[
 A_{\Z}(m_2)\,B_{\Z}(2,1\,;\,p,\lstm{34}) +  
n\,S^{\ssE}_{\saab}(p,\lstm{12343}) +
m^2_1\,S^{\ssE}_{\Z}(p,\lstm{12343})\Bigr],
\eqa
\bqa
S^{\ssE}_{\saba} &=&
\frac{1}{2}\,\frac{1}{(n-1) p^2}\, \Bigl[
A_{\Z}([m_1,m_2])\,C_{\Z}(p,-p,\lstm{343}) + 
m^2_{123}\,S^{\ssE}_{\Z}(p,\lstm{12343}) - S^{\ssC}_{\Z}(p,\lstm{1234}) 
\nl 
&-& n\,l_{\ssP 34}\,S^{\ssE}_{\sa}(p,\lstm{12343})\, -
n\,S^{\ssC}_{\sa}(p,\lstm{1234})\Bigr] \, ,
\eqa
\bqa
S^{\ssE}_{\sbba} &=&
\frac{1}{4}\,\frac{1}{(n-1) p^4}\, \Bigl\{
( n\,l_{p 34}^2  + 4\,p^2\,m_3^2 )\,
S^{\ssE}_{\Z}(p,\lstm{12343}) 
+ 2\,( n\,l_{p 34}  - 2\,p^2 )\,S^{\ssC}_{\Z}(p,\lstm{1234})
\nl
{}&-& n\,l_{p 34} \,S^{\ssC}_{\Z}(0,\lstm{1233}) 
+ n\,\Bigl[ S^{\ssA}_{\Z}(p,\lstm{124})  - 
S^{\ssA}_{\Z}(0,\lstm{123})\Bigr] \Bigr\} \, , 
\eqa
\bqa
S^{\ssE}_{\sabb} &=&
\frac{1}{4}\,\frac{1}{(n-1)}\,\Bigr[
A_{\Z}([m_1,m_2])\,B_{\Z}(2,1\,;\,p,\lstm{34}) 
- m^2_{123}\,S^{\ssE}_{\Z}(p,\lstm{12343}) +
S^{\ssC}_{\Z}(p,\lstm{1234}) \nl
&+& l_{p 34}\,S^{\ssE}_{\sa}(p,\lstm{12343}) + 
S^{\ssC}_{\sa}(p,\lstm{1234}) 
\Bigl]\, , 
\eqa
\bqa
S^{\ssE}_{\sbbb} &=&
\frac{1}{4}\,\frac{1}{(n-1) p^2}\,\Bigr[
-\lambda_{34}\,S^{\ssE}_{\Z}(p,\lstm{12343}) 
+2\,l_{p43}\,S^{\ssC}_{\Z}(p,\lstm{1234}) \nl
&+&\,l_{p34}\,S^{\ssC}_{\Z}(0,\lstm{1233}) -
 S^{\ssA}_{\Z}(p,\lstm{124}) +  S^{\ssA}_{\Z}(0,\lstm{123})
\Bigl] \, , 
\eqa
\bqa
S^{\ssE}_{\msb} &=&
\frac{1}{2\,p^2}\,\Bigr[
- l_{p34}\,S^{\ssE}_{\Z}(p,\lstm{12343}) -
S^{\ssC}_{\Z}(p,\lstm{1234}) + S^{\ssC}_{\Z}(0,\lstm{1233})\Bigl] \, , 
\eqa
\bqa
S^{\ssD}_{\saaa} &=& \frac{1}{4}\,\frac{1}{(n-1)\,p^4}\,\Bigl\{
( n\,l^2_{p12} + 4\,m_1^2\,p^2 )\,S^{\ssD}_{\Z}(p,\lstm{12345}) 
- n\,l_{p 12}\,S^{\ssC}_{\Z}(p,\lstm{1345}) 
\nl
{}&-& 
\Bigr[ 4\,p^2 -\,n\,(3\,p^2 - m_{12}^2) \Bigl]\,S^{\ssC}_{\Z}(p,\lstm{2354}) 
+ 2\,n\,p^2\,\Bigl[ S^{\ssC}_{\sa}(p,\lstm{1345}) + 
S^{\ssC}_{\sa}(p,\lstm{2354})\Bigr]
\Bigr\} \, ,
\eqa
\bqa
S^{\ssD}_{\saba} &=& \frac{1}{4}\,\frac{1}{(n-1)\,p^4}\,\Bigl\{
-2\,p^2\,B_{\Z}(p,\lstm{12})\,B_{\Z}(p,\lstm{45}) \nl
&+& ( n\,(p^4 + p^2\,m^2_{245} - p^2\,m^2_1  + m^2_{12}\,m^2_{45})
+ 2\,m_{134}^2\,p^2 )\,S^{\ssD}_{\Z}(p,\lstm{12345}) \nl 
&-& n\,l_{p 45}\,
S^{\ssC}_{\Z}(p,\lstm{1345}) - ( 2\,p^2 - n\,l_{p 45})\,
S^{\ssC}_{\Z}(p,\lstm{2354}) 
-\,n\,l_{p 12}\, S^{\ssC}_{\Z}(p,\lstm{4312}) \nl
&-& ( 2\,p^2 - n\,l_{p 12} )\,S^{\ssC}_{\Z}(p,\lstm{5321})
-\,n\,\Bigl[S^{\ssA}_{\Z}(p,\lstm{432}) + S^{\ssA}_{\Z}(p,\lstm{531})\Bigr]\, \nl
&+& n\,\Bigl[ S^{\ssA}_{\Z}(0,\lstm{431}) +S^{\ssA}_{\Z}(0,\lstm{532}) \Bigr]
\Bigr\} \, ,
\eqa
\bqa
S^{\ssD}_{\sbba} &=&\frac{1}{4}\,\frac{1}{(n-1)\,p^4}\,\Bigl\{
( 4\,m_4^2\,p^2  + n\,l^2_{p 45} )\,S^{\ssD}_{\Z}(p,\lstm{12345}) 
- n\,l_{p 54}\,S^{\ssC}_{\Z}(p,\lstm{4312}) 
\nl
{}&-& \Bigl[ 4\,p^2 - n\,(3\,p^2 - m_{45}^2)\Bigr]\
S^{\ssC}_{\Z}(p,\lstm{5321}) 
+ 2\,n\,p^2\,S^{\ssC}_{\sa}(p,\lstm{4312}) + 
2\,n\,S^{\ssC}_{\sa}(p,\lstm{5321})
\Bigr\} \, ,
\eqa
\bqa
S^{\ssD}_{\saab} &=& \frac{1}{4}\,\frac{1}{(n-1)\,p^2}\,\Bigl[
-\,\lambda_{12}\,S^{\ssD}_{\Z}(p,\lstm{12345}) 
+ l_{p 12}\,S^{\ssC}_{\Z}(p,\lstm{1345}) 
+ l_{p 21}\,S^{\ssC}_{\Z}(p,\lstm{2354}) \nl
&-&2\,p^2\, S^{\ssC}_{\sa}(p,\lstm{1345}) + 
2\,p^2  S^{\ssC}_{\sa}(p,\lstm{2354})
\Bigr] \, ,
\eqa
\bqa
S^{\ssD}_{\sabb} &=& \frac{1}{4}\,\frac{1}{(n-1)\,p^2}\,\Bigl[
2\,p^2\,B_{\Z}(p,\lstm{12})\,B_{\Z}(p,\lstm{45}) 
\nl
&+& \Bigl[ m^2_{12}\,m^2_{54} -p^2\,(m_1^2 +m_2^2 - 2 m_3^2 + m_4^2 +
m_5^2) - p^4\Bigr]\,S^{\ssD}_{\Z}(p,\lstm{12345}) 
\nl 
&+& l_{p 45}\,S^{\ssC}_{\Z}(p,\lstm{1345}) + 
l_{p 54}\,S^{\ssC}_{\Z}(p,\lstm{2354}) + 
l_{p 12}\,S^{\ssC}_{\Z}(p,\lstm{4312}) +
l_{p 21}\,S^{\ssC}_{\Z}(p,\lstm{5321}) 
\nl
&+& S^{\ssA}_{\Z}(p,\lstm{432}) +  S^{\ssA}_{\Z}(p,\lstm{531}) -
 S^{\ssA}_{\Z}(0,\lstm{431}) - S^{\ssA}_{\Z}(0,\lstm{532})\Bigr] \, , 
\eqa
\bqa
S^{\ssD}_{\sbbb} &=& \frac{1}{4}\,\frac{1}{(n-1)\,p^2}\,\Bigl[
- \lambda_{45}\,S^{\ssD}_{\Z}(p,\lstm{12345}) + 
l_{p 45}\,S^{\ssC}_{\Z}(p,\lstm{4312}) + 
l_{p 54}\,S^{\ssC}_{\Z}(p,\lstm{5321}) 
\nl
&-& 2\,p^2\,S^{\ssC}_{\sa}(p,\lstm{4312}) -
2\,p^2\,S^{\ssC}_{\sa}(p,\lstm{5321})\Bigl] \, ,
\eqa
\bqa
S^{\ssD}_{\sa} &=& \frac{1}{2 p^2}\,\Bigl[ 
-\,l_{p 12}\,S^{\ssD}_{\Z}(p,\lstm{12345}) + 
S^{\ssC}_{\Z}(p,\lstm{1345}) - S^{\ssC}_{\Z}(p,\lstm{2354})\Bigr] \, ,
\eqa
\bqa
S^{\ssD}_{\msb} &=& \frac{1}{2 p^2}\,\Bigl[ 
-\,l_{p 45}\,S^{\ssD}_{\Z}(p,\lstm{12345}) + 
S^{\ssC}_{\Z}(p,\lstm{4312}) - S^{\ssC}_{\Z}(p,\lstm{5321})\Bigr] \, . 
\eqa
The standard reduction procedure does not work for the
$S^{\ssC}_{\sa}$ and $S^{\ssE}_{\sa}$ form factors, since for them the scalar 
product $\spro{p}{q_1}$ is irreducible.
In order to express these form factors in term of other scalar
functions, it is then necessary to employ the procedure outlined in Section
\ref{stltpf}, i.e.\ considering first the scalarization with respect to the
sub-loops. Employing \eqn{faf23} one obtains the following relations
\bqa
S^{\ssC}_{\sa} &=& \frac{1}{4}\,\frac{1}{m_3^2 p^2} \Bigl\{
A_{\Z}([m_1,m_2])\,\Bigl[ A_{\Z}(m_3) 
- B_{\Z}(p,\lstm{34}) \, l_{p 34} 
+ B_{\Z}(p,0,m_4) \, ( p^2 + m_4^2 )\Bigr] 
\nl
{}&-& S^{\ssC}_{\Z}(p,\lstm{1234}) \, l_{p 34}\,   m^2_{123}
+ S^{\ssC}_{\Z}(p,\lstm{12},0,m_4) \, m^2_{12}\,( p^2 + m_4^2 )
- S^{\ssA}_{\Z}(p,\lstm{124}) \,m_3^2 
\nl
{}&+& S^{\ssA}_{\Z}(0,\lstm{123}) \, m^2_{123}
- S^{\ssA}_{\Z}(0,\lstm{12},0) \, m^2_{12} \Bigr\}\,, 
\eqa
\bqa
S^{\ssE}_{\sa} &=&  \frac{1}{4}\,\frac{1}{m_3^4 p^2} \Bigl\{
A_{\Z}([m_1,m_2])\,\Bigl[
\frac{4 - n}{2}\,\,A_{\Z}(m_3) 
 - B_{\Z}(p,\lstm{34}) \, ( p^2 + m_4^2 )
 + B_{\Z}(p^2,0,m_4) \, ( p^2 + m_4^2 )
\nl
{}&-& m_3^2\,l_{p 34}\,B_{\Z}(2,1\,;\,p,\lstm{34}) \Bigr]
- m_3^2\,l_{p 34}\,m^2_{123}\,S^{\ssE}_{\Z}(p,\lstm{12343}) 
\nl 
&-& \Bigl[ (m^2_{12}\,( p^2 + m_4^2) + m_3^4 \Bigr]\,
S^{\ssC}_{\Z}(p,\lstm{1234})
\nl 
&+& m^2_{12}\,(p^2 + m_4^2)\, S^{\ssC}_{\Z}(p,\lstm{12},0,m_4) \, 
+ m_3^2\,m^2_{123}\,S^{\ssC}_{\Z}(0,\lstm{1233}) 
\nl 
&+& m^2_{12}\,\Bigl[ S^{\ssA}_{\Z}(0,\lstm{123}) 
- S^{\ssA}_{\Z}(0,\lstm{12},0)  \Bigr]\Bigr\}\, .
\eqa
\section{Strategies for the evaluation of two-loop self-energies \label{Sde}}
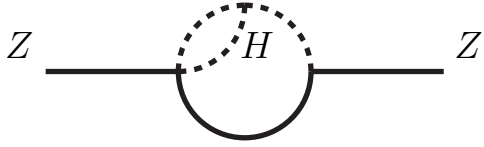
\begin{figure}[th]
\vspace{0.5cm}
\[
  \vcenter{\hbox{
  \begin{picture}(150,0)(0,0)
  \SetWidth{2.}
  \Line(0,0)(50,0)
  \Line(100,0)(150,0)
  \CArc(75,0)(25,-180,0)
  \DashCArc(75,0)(25,0,90){3.}
  \DashCArc(75,0)(25,90,180){3.}
  \DashCArc(50,25)(25,-90,0){3.}
  \Text(-10,5)[cb]{\Large $\zb$}
  \Text(160,5)[cb]{\Large $\zb$}
  \Text(80,5)[cb]{\Large $\hb$}
  \end{picture}}}
\]
\vspace{0.5cm}
\caption[]{Example of a diagram belonging to the $S^{\ssC}$-family and
contributing to the $\zb$ self-energy. Dashed lines represent a $\hb$-field.}
\label{exaZse}
\end{figure}

In this Section we provide an explicit example of possible strategies to
evaluate diagrams with a non-trivial spin structure. Consider the diagram in
\fig{exaZse}, representing one of the two-loop contributions to the
$\zb$-boson self-energy (the diagram may be needed to assemble the
components of a scattering amplitude or to compute a doubly-contracted WST
identity, in which case we have to multiply the corresponding expression by
$p_{\mu} p_{\nu}$).
In the $R_{\xi}$-gauge, with $[\xi] = (q_2+p)^2+\xi^2\,\mz^2$ the diagram is 
be written as
\bqa
S_{\mu\nu} &=& -\frac{3}{8}\,\frac{g^4\,\mhs}{c^4_{\theta}}\,
\mu^{2\ep}\,\int\,d^nq_1\,d^nq_2\,
\Bigl\{ \frac{\delta_{\mu\nu}}{[\xi=1]} + \,
\frac{(q_2+p)_{\mu}(q_2+p)_{\nu}}{\mz^2} 
\Bigl[\frac{1}{[\xi=1]} - \frac{1}{[\xi_{\ssZ}]}\Bigr]\}
\nl
{}&\times&\,
\frac{1}{(q_1^2+\mh^2) \, [(q_1-q_2)^2+\mh^2] \, (q_2^2+\mh^2)} =
-\frac{3}{8}\,\frac{g^4\,\pi^4\,\mhs}{c^4_{\theta}\,\mz^2}\,
\Bigl( \Pi^d\,\delta_{\mu\nu} + \Pi^p\,p_{\mu}\,p_{\nu}
\Bigr).
\eqa
After multiplication by $p_{\mu} p_{\nu}$ we can perform all the algebraic
manipulations, like rewriting $\spro{q_2}{p}$ and $q^2_2$ in terms of
propagators, or we can use \eqn{generalabg} and the results of
\sect{fulll} in order to obtain a fully scalarized expression.
Alternatively, again using \eqn{generalabg}, we can write
\bqa
\Pi^d &=& S^{\ssC}_{\sbbb}(1) - S^{\ssC}_{\sbbb}(\xi_{\ssZ}) + 
\mz^2\,S^{\ssC}_{\Z}(1),
\nl
\Pi^p &=& S^{\ssC}_{\sbba}(1) - S^{\ssC}_{\sbba}(\xi_{\ssZ}) + 
2\,S^{\ssC}_{\msb}(1) - 2\,S^{\ssC}_{\msb}(\xi_{\ssZ}) + 
S^{\ssC}_{\Z}(1)  - S^{\aba}_{\Z}(\xi_{\ssZ}),
\eqa
where we explicitly indicated the dependence on the gauge parameter
$\xi_{\ssZ}$.  To derive an explicit expression for the form factors we
decompose the diagram according to $S = S_{\ssD\ssP} + S_{\ssS\ssP} +
S_{\ssF}$, where the subscripts refer to double and single ultraviolet poles
and to the finite ultraviolet part. Note, however, that the splitting is
defined only modulus constants. The three components of the result (with a
presentation limited here to the $\xi = 1$ part) are given in the following
list:
\bqa
S^{\ssC}_{\Z\,;\,\ssD\ssP} &=& -\,\frac{1}{\ep^2} 
       -\,\oDUV^2,
\qquad
S^{\ssC}_{\msb\,;\,\ssD\ssP} = -\,\frac{1}{2}\,S^{\ssC}_{\Z\,;\,\ssD\ssP},
\qquad
S^{\ssC}_{\sbba\,;\,\ssD\ssP} =  \frac{1}{3}\,S^{\ssC}_{\Z\,;\,\ssD\ssP},
\nn
\eqa
\bqa
S^{\ssC}_{\sbbb\,;\,\ssD\ssP} &=&
   ( \frac{1}{6}\,p^2 + \frac{49}{6}\,\mhs + \frac{1}{2}\,\mzs )\,
 \oDUV\,\frac{1}{\ep} +
  ( \frac{1}{12}\,p^2 - \frac{5}{4}\,\mhs + \frac{1}{4}\,\mzs )\,
       \Delta^2_{\ssU\ssV},
\label{smoothproofDP}
\eqa
%
%
\bqa
S^{\ssC}_{\Z\,;\,\ssS\ssP} &=&
 2\,\oDUV\,\Bigl[ 
       \intfx{x}\,\ln\chi(x) - \frac{1}{2}\Bigr],
\qquad
S^{\ssC}_{\msb\,;\,\ssS\ssP} =
       \frac{1}{4}\,\oDUV\,\Bigl[ 1 - 
       8\,\intfx{x}\,x\,\ln\chi(x)\Bigr],
\nn
\eqa
\bqa
S^{\ssC}_{\sbba\,;\,\ssS\ssP} &=&
       2\,\oDUV\,\Bigl[
       \intfx{x}\,x^2\,\ln\chi(x) - \frac{1}{36}\Bigr],
\nn
\eqa
\bqa
S^{\ssC}_{\sbbb\,;\,\ssS\ssP} &=&
       \frac{20}{9}\,\mhs \,\Delta_{\ssU\ssV}  
       - (  \frac{5}{72}\,p^2 + \frac{97}{8}\,\mhs + \frac{13}{24}\,\mzs )\,
         \oDUV
+  \frac{23}{3}\,\mhs\, \ln\mu^2_H\,\oDUV
\nl
{}&+& \oDUV\,\dcubs{x}{y,z}\,\Bigl\{
-  \Bigl[ 3\,y\,\mhs + 3\,(p^2 - \mhs + \mzs)\,z - 4\,z^2\,p^2 \Bigr]
      \,\ln\chi(x,y,z)
\nl
{}&+&\,\mhs\, ( 3\,y - 2)\,\Bigl[  \frac{\ln\chi(x,y,z)}{x}\bmid_+ -
         \frac{\ln\chi(x,y,z)}{x-1}\bmid_+\Bigr]\Bigr\},       
\label{smoothproofSP}
\eqa
%
%
\bqa
S^{\ssC}_{\ssF} &=&
       \dcubs{x}{y,z} \frac{\ln\chi(x,y,z)}{1-y}\bmid_+
       + \intfx{x}\,\ln\chi(x)\,\Bigl[ L_1(x) + 2\Bigr]
       - \frac{3}{2} - \frac{1}{2}\,\zeta(2),
\nn
\eqa
\bqa
S^{\ssC}_{\msb\,;\,\ssF} &=&
       - \dcubs{x}{y,z}\,z \frac{\ln\chi(x,y,z)}{1-y}\bmid_+
       -\,\intfx{x}\,x\,\ln\chi(x)\,\Bigl[ L_1(x) + 2\Bigr]
+ \frac{11}{16} + \frac{1}{4}\,\zeta(2),
\nl
S^{\ssC}_{\sbba\,;\,\ssF} &=&
       \dcubs{x}{y,z}\,z^2\, \frac{\ln\chi(x,y,z)}{1-y}\bmid_+
       +\intfx{x}\,x^2\,\ln\chi(x)\,\Bigl[ L_1(x) + 2\Bigr]
- \frac{97}{216} - \frac{1}{6}\,\zeta(2),
\nl
S^{\ssC}_{\sbbb\,;\,\ssF} &=&  \dcubs{x}{y,z}\,\Bigl\{
    \mhs\,( 1 - \frac{3}{2}\,y ) \ln(1 -
    x)\,\left(\frac{\ln\chi(x,y,z)}{x}\bmid_+\right) 
\nl
{}&+& \frac{3}{2}\,\Bigl[ y\,\mhs + (p^2 - \mhs + \,\mzs)\,z - 
6\,z^2\,p^2 \Bigr]\,\ln\chi(x,y,z)\,L_2(x,y,z) 
\nl
{}&-& \frac{5}{2}\,\Bigl[  y\,\mhs + (p^2 - \mhs + \mzs)\,z - z^2\,p^2 \Bigr]
         \ln\chi(x,y,z)  
\nl
{}&+& 
\mhs\,( \frac{3}{2}\,y - 1 ) \ln^2(1 - x)\,\left(\frac{\ln\chi(x,y,z)}{x}\bmid_+
\right)
+ \frac{5}{2}\,\mhs\,( 1 - y )\, \Bigl[
      \frac{\ln\chi(x,y,z)}{x-1}\bmid_+ - 
      \frac{\ln\chi(x,y,z)}{x}\bmid_+ \Bigr] 
\nl
{}&+& \mhs\,( \frac{3}{2}\,y - 1)\,\Bigl[
      \frac{\ln\chi(x,y,z)\,L_2(x,y,z)}{x-1}\bmid_+ -
      \frac{\ln\chi(x,y,z)\,L_2(x,y,z)}{x}\bmid_+\Bigr]
  \Bigr\}
  + \frac{5}{36}\,\mhs\,\ln\mu^2_h  
\nl
{}&+& \Bigl[ \frac{1}{24}\,p^2 - \frac{43}{24}\,\mhs + \frac{1}{8}\,\mzs
\Bigr]\,\zeta(2) + 
     \frac{145}{864}\,p^2 - \frac{1}{3}\,\zeta(3)\,\mhs + 
    \frac{9247}{864}\,\mhs + \frac{251}{288}\,\mzs.
\label{smoothproofF}
\eqa
Here $\zeta(n)$ denotes the Riemann zeta function. To derive our result we
introduced the auxiliary functions
\bqa
\chi(x) &=& x\,(1-x)\,s_p + \mu_{\ssH}^2\,(1-x) + \mu_{\ssZ}^2\,x,
\quad
\mu^2_x= \mu_{\ssH}^2\,\Bigl( \frac{1}{x} + \frac{1}{1-x}\Bigr),
\nl
\chi(x,y,z) &=& x\,(1-x)\,\Bigl[ z\,(1-z)\,s_p + \mu_{\ssH}^2\,(y-z) + 
\mu_{\ssZ}^2\,z\Bigr] + \mu^2_x\,(1-y),
\eqa
where $s_p = {\rm sign}\,(p^2)$ and $\mu_{\ssH}^2 = \mhs/\mid p^2\mid$,
$\mu_{\ssZ}^2 = \mzs/\mid p^2\mid$.
We define $L_{1,2}$,
\bq
L_1(x) = \ln (1-x) -\ln \chi(x)\,,
\qquad
L_2(x,y,z) = \ln \chi(x,y,z) - \ln (1-y) - \ln x - \ln (1-x).
\eq
The $'+'$-distribution is used according to the definitions of \eqn{plusdist}.

In conclusion we may say that \eqns{smoothproofDP}{smoothproofF} prove that
whenever we can find an algorithm of smoothness for scalar integrals, then
the same algorithm can be generalized to handle tensor integrals. The whole
diagram, or even sets of diagrams, can be successively mapped into one
multi-dimensional integral. The only additional complication is represented
by those cases where the scalar diagram is ultraviolet convergent while
tensors of the same family diverge; in this case one cannot set $\ep = 0$
from the very beginning but, apart from this caveat, the procedure will be
essentially the same.

\section{Reduction of tensor two-loop three-point functions \label{svert}}
In this Section we move to the complex environment of three-point
functions. Two independent external momenta induce seven scalar products
containing $q_1$ and/or $q_2$, and the number of irreducible ones is $7-I$
where $I$ is the number of internal lines in the diagram (note that $4 \le I
\le 6$); the choice of the set of irreducible scalar products has, of
course, some arbitrariness. In any case, for two-loop diagrams we never have
complete reducibility with respect to both $q_1$ and $q_2$.  Actually, in
evaluating observables for physical processes, we encounter a more general
situation: massive SM gauge bosons are unstable particles and final states
are always made up of stable fermions and/or photons.  Referring to
\fig{Moreirr} we have propagators that depend on $p_1$ and $p_2+p_3$,
therefore losing full reducibility in the $q_2$ sub-loop.
\begin{figure}[ht]
\begin{center}
\begin{picture}(150,75)(0,0)
 \SetWidth{2.}
 \DashLine(0,0)(40,0){3}         
 \Line(128,-53)(100,-35)  
 \Line(128,53)(100,35)    
 \Line(70,-17.5)(40,0)             
 \Line(70,17.5)(40,0)              
 \ArrowLine(100,-35)(100,35)            
 \ArrowLine(100,35)(70,17.5)            
 \ArrowLine(70,17.5)(70,-17.5)          
 \ArrowLine(70,-17.5)(100,-35)          
 \ArrowLine(158,63)(128,53)
 \ArrowLine(128,53)(158,43)
 \Text(160,72)[cb]{$p_3$}
 \Text(160,30)[cb]{$p_2$}
 \Text(138,-70)[cb]{$p_1$}
 \Text(138,-55)[cb]{$\zb$}
 \Text(-11,7)[cb]{$-P$}
 \Text(-11,-12)[cb]{$\hb$}
\end{picture}
\end{center}
\vspace{2cm}
\caption[]{A contribution of the $V^{\bca}$ family to $H \to \zb^* \zb \to
\zb \barf f$. External momenta flow inwards.}
\label{Moreirr}
\end{figure}
The whole procedure will be developed on a diagram-by-diagram basis with the
double goal of writing explicit integral representations for all form
factors and of deriving a suitable algorithm to express them in terms of
ordinary and generalized scalar functions. Therefore, for each set of
graphs, we will show that all integrals can be expressed in terms of
generalized scalar functions, part of which should be subsequently treated
within the context of generalized recurrence relations~\cite{Tarasov:2000sf}; 
the final answer will contain a limited number of master integrals.

Alternatively, and this represents our preferred solution, all the integrals 
not belonging to ${\cal S}_4$ -- the class of ordinary scalar functions 
in $n= 4 - \ep$ dimensions -- can be evaluated according to the given integral 
representation, following the same lines that we have already adopted in II and
in III for solving the problem in the ${\cal S}_4$ class.

For each diagram there are many equivalent ways to assign loop momenta; we
will make a specific choice for the matrix $\eta$ of \eqn{matrixe} (the 
defining parametric representation of the graph) and stick to it also when 
diagrams of a given family appear in the result of the reduction of the tensor
integrals of other families. In these cases the necessary permutations of 
momenta should be performed, as it will be shown in \sect{TopFFPerm}.

In our presentation the different families are ordered according to the
choice made in~\cite{Ferroglia:2003yj}, where the scalar members were
computed explicitly  and where the ordering was dictated by a criterion of
increasing complexity in the evaluation and by the fact that three graphs
belong to the same class $V^{\ono}$. Therefore, in the following
subsections we present and discuss techniques for treating $\{V^{\aba},
V^{\aca}, V^{\ada}\} \in V^{\ono}$ and the more complicated ones, $V^{\bba}, 
V^{\bca}$ and $V^{\bbb}$. A complete summary of all results for reduction 
of three-point functions is provided in \appendx{SummaR}.
\subsection{The $V^{\aba}$-family ($\alpha = 1, \beta = 2, \gamma = 1$)
\label{abafam}}
We start our analysis considering the scalar member of the 
$V^{\aba}$-family of \fig{TLvertaba}, which is representable as 
\bq
\pi^4\,\cV^{\aba}_{\Z}(p_2,P,\lstm{1234}) =
\mu^{2\ep}\,
\intmomii{n}{q_1}{q_2}\,\frac{1}{[1][2]_{\aba}[3]_{\aba}[4]_{\aba}},
\label{DEFcVaba}
\eq
with propagators defined by
\bqa
[1] &\equiv& q^2_1+m^2_1, \quad [2]_{\aba} \equiv (q_1-q_2)^2+m^2_2,
\quad
[3]_{\aba} \equiv (q_2+p_2)^2+m^2_3, \quad [4]_{\aba} \equiv (q_2+P)^2+m^2_4.
\label{enummaba}
\eqa
Note the symmetry property $\cV^{\aba}_{\Z}(p_2,P,\lstm{1234}) = 
\cV^{\aba}_{\Z}(P,p_2,\lstm{1243})$, besides the one shown in \eqn{sym1} of 
\appendx{symme}. 

The scalar diagram is overall divergent and so is the $(\alpha,\gamma)$
sub-diagram. Vector and tensor integrals for all classes usually show
additional ultraviolet divergences which have been transferred from the
momentum integration to the parametric one. Also for this reason we will
keep the $n$ dependence explicit, i.e. $n \not= 4 \, (\ep \not =0)$ in all
parametrizations. In the following we will discuss vector and rank two
tensor integrals for all families. Rank three tensor are fully analyzed in
\sect{IRTIRT}.
\begin{figure}[ht]
\begin{center}
\begin{picture}(150,75)(0,0)
 \SetWidth{1.5}
 \Line(0,0)(40,0)         
\LongArrow(0,8)(20,8)          \Text(-11,7)[cb]{$-P$}
 \Line(128,-53)(100,-35)  
\LongArrow(128,-63)(114,-54)   \Text(138,-70)[cb]{$p_1$}
 \Line(128,53)(100,35)    
\LongArrow(128,63)(114,54)     \Text(138,62)[cb]{$p_2$}
 \CArc(100,-35)(70,90,150)     \Text(64,31)[cb]{$1$}
 \CArc(40,70)(70,270,330)      \Text(80,-2)[cb]{$2$}
 \Line(100,-35)(40,0)          \Text(70,-30)[cb]{$4$}
 \Line(100,-35)(100,35)        \Text(107,-3)[cb]{$3$}
 \Text(0,-25)[cb]{\Large $V^{\aba}$}
\end{picture}
\end{center}
\vspace{2cm}
\caption[]{The irreducible two-loop vertex diagrams $V^{\aba}$. External 
momenta flow inwards. Internal masses are enumerated according to the
parametrization of \eqn{enummaba}.} 
\label{TLvertaba}
\end{figure}
\subsubsection{Vector integrals in the $V^{\aba}$ family \label{rankoneaba}}
We also consider the $V^{\aba}$ vector integrals and introduce the 
following decomposition in terms of the ${p_1\,,\,p_2}$ basis: 
\bqa
\cV^{\aba}(\mu\,|\,0\,;\,p_2,P,\lstm{1234}) &=&
\sum_{i=1,2}\,\cV^{\aba}_{\sagen}(p_2,P,m_{1234})\,p_{i\mu},
\nl
\cV^{\aba}(0\,|\,\mu\,;\,p_2,P,\lstm{1234}) &=&
\sum_{i=1,2}\,\cV^{\aba}_{\sbgen}(p_2,P,\lstm{1234})\,p_{i\mu}.
\label{ffaba}
\eqa
Note that we always use the convention $\cV^{\ssX}(p\,|\,0\,; \cdots) = 
p^{\mu}\,\cV^{\ssX}(\mu\,|\,0\,; \cdots)$.

$\cV^{\aba}$ will often appear in the reduction of the form factors
belonging to other families and special care should be applied in writing the 
correct list of arguments.
To help understanding this list we rewrite $\cV^{\aba}_{\sijgen}$ according to the 
following equation:
\bqa
\frac{\mu^{2\ep}}{\pi^4}\int\,d^nr_1 d^nr_2\,r_{i\mu}\,
\prod_{l=a}^{d}\,D^{-1}_l &\equiv&  
\cV^{\aba}_{\sgena}(k_c,k_d,\lstm{abcd})\,(k_d - k_c)_{\mu}\, +
\cV^{\aba}_{\sgenb}(k_c,k_d,\lstm{abcd})\,k_{c\mu},
\label{explist}
\eqa
where the propagators are now generically written as
\bq
D_a = r_1^2 + m_a^2\, , 
\quad 
D_b = (r_1 - r_2)^2 + m_b^2\, , 
\quad
D_c = (r_2 + k_c)^2 + m_c^2\, , 
\quad 
D_d = (r_2 + k_d)^2 + m_d^2. 
\label{SF}
\eq
Here $i = 1,2$ and $m_a, \ldots, m_d$ are generic masses, $k_c$ and $k_d$ are
the external momenta appearing in the propagators $D_c$ and 
$D_d$, respectively, and $r_1$, $r_2$ are the loop momenta.
Note that the following identities hold:
\bq
\cV^{\aba}_{\sgena}(c,d) = - \cV^{\aba}_{\sgena}(d,c) + \cV^{\aba}_{\sgenb}(d,c),
\qquad
\cV^{\aba}_{\sgenb}(c,d) = \cV^{\aba}_{\sgenb}(d,c),
\eq
where $(c,d) = (k_c,k_d,\lstm{abcd})$ etc.
Therefore, \eqns{explist}{SF} tell us how to identify the proper list of 
arguments when these integrals appear as the result of a reduction of 
tensor integrals belonging to other classes and a permutation has been
applied in order to conform to the convention of \eqn{SF}.

As we explained earlier, all these form factors could be computed directly,
without having to perform a reduction. For this reason it is important to
list their integral representation. The explicit expression for the vector
form factors of this family is
\bq 
\cV^{\aba}_{\sijgen} = -\,\egam{\ep}\,\CIM{V}_{\aba}\,
P_{ij;\aba}\,\chiu{\aba}^{-\ep}(x,y,z),
\eq
\bq
P_{\szz;\aba} = 1, \quad P_{\saa;\aba} = -x\,z, \quad
P_{\sab;\aba} = -x\,y, \quad P_{\sba;\aba} = -z, \quad 
P_{\sbb;\aba} = -y,
\label{Pffaba}
\eq
with an integration measure defined as follows:
\bq
\CIM{V}_{\aba} = \omega^{\ep}\,\dcubs{x}{y,z}\,
\Bigl[x\,(1-x)\Bigr]^{-\ep/2}\,(1-y)^{\ep/2-1},
\eq
where $\omega$ is defined in \eqn{defDUV} and where, with our choice for the 
Feynman parameters, the polynomial $\chiu{\aba}$ is given by
\bq 
\chiu{\aba} (x,y,z) = -\,F(z,y) + (p^2_2 - m^2_x + m^2_3)\,y + 
(2\,p_{12} + l_{134})\,z + m_x^2, 
\label{Defchiaba}
\eq
where we used \eqn{defGfun}. All these functions can be manipulated according 
to the procedure introduced in III and they will give rise to smooth integral 
representations. 

The generic scalar function in this family is
\bqa
{}&{}&
\cV^{\alpha_1 | \alpha_3,\alpha_4 | \alpha_2}_{\aba}(n = \sum_{i=1}^4 \alpha_i 
- \ep) =
\pi^{-4} (\mu^2)^{4-n} \intmomii{n}{q_1}{q_2} 
\prod_{i=1}^{4} [i]^{-\alpha_i}_{\aba}
\nl
{}&=& - \frac{\egam{\ep}}{\prod_{i=1}^4 \egam{\alpha_i}} 
\omega^{\rho_0} \!\!\! 
\dcubs{x}{y,z} x^{\rho_1} (1-x)^{\rho_2} (1-y)^{\rho_3} (y-z)^{\rho_4} 
z^{\rho_5} \chiu{\aba}^{-\ep}(x,y,z),
\label{genaba}
\eqa
where $[1]_{\aba} \equiv [1]$, $\omega$ is defined in \eqn{defDUV}, with 
powers $\rho_0 = 4 - \sum_{j=1}^4\alpha_j+\ep$ and
\bqa
\rho_1 &=& \frac{1}{2}\,(-\alpha_1+\alpha_2+\alpha_3+\alpha_4) - 1 -
\frac{\ep}{2},
\qquad
\rho_2 = \frac{1}{2}\,(\alpha_1-\alpha_2+\alpha_3+\alpha_4) - 1 -
\frac{\ep}{2},
\nl
\rho_3 &=& \frac{1}{2}\,(\alpha_1+\alpha_2-\alpha_3-\alpha_4) -1 +
\frac{\ep}{2},
\qquad
\rho_4 = \alpha_3 - 1,
\qquad
\rho_5 = \alpha_4 - 1.
\eqa
Henceforth, for the form factors of \eqn{ffaba} we can write
\bq
\cV^{\aba}_{\sijgen} = \sum_{l=1}^{4}\,\omega^{n_{lij}-4+\ep}\,k_{lij}\,
\cV^{\alpha_{1lij}\,|\,\alpha_{3lij},\alpha_{4lij}\,|\,\alpha_{2lij}}_{\aba}
(n_{lij}),
\eq
with $\omega$ defined in \eqn{defDUV} and $n_{lij} = \sum_{k=1}^4\,
\alpha_{klij} - \ep$. The coefficients $k_{lij}$ and the exponents 
$\alpha_{lij}$, can be easily read out of the following explicit expressions:
\bqa
\cV^{\aba}_{\saa} &=& -\omega^2\,\cV^{1\,|\,1,2\,|\,2}_{\aba}, \quad
\cV^{\aba}_{\sab} =  - \omega^2\,\Bigl[ \cV^{1\,|\,2,1\,|\,2}_{\aba} +
\cV^{1\,|\,1,2\,|\,2}_{\aba}\Bigr], \quad
\cV^{\aba}_{\sba} = - \omega^2\,\Bigl[ \cV^{2|1,2|1}_{\aba} + 
\cV^{1|1,2|2}_{\aba}\Bigr], \nl
\cV^{\aba}_{\sbb} &=&  - \omega^2\,\Bigl[ \cV^{2|2,1|1}_{\aba} +
\cV^{2|1,2|1}_{\aba} + \cV^{1|2,1|2}_{\aba} + \cV^{1|1,2|2}_{\aba}\Bigr],
\label{alltheseIb}
\eqa
all to be evaluated for $n= 6 - \ep$.  As usual, there still is the problem
of evaluating the integrals of \eqn{alltheseIb} by means of recurrence
relations or, in other words, to link all of them to MI and to develop an
algorithm to evaluate the master integrals.

Based on our experience with
one-loop multi-leg diagrams, we propose an alternative: an algorithm for the
evaluation of tensor integrals  offering the same stability characteristics as
for scalar integrals.
More precisely, we mean a result which is,
from a numerical point of view, of the same degree of stability for all
integrals and where the real nature of any singularity, apparent or not, is
independent of the rank of the integral under consideration. Let us start
with $\cV^{\aba}(\mu\,|\,0)$, where sub-loop reduction techniques may be 
applied giving
\bq
\int\,d^nq_1\,\frac{q_{1\mu}}{[1][2]_{\aba}} = X_{\aba}\,q_{2\mu}, 
\label{faf31}
\eq
and where $X_{\aba}$ by standard methods is computed to be
\bq 
X_{\aba} = \frac{1}{2}\,\int\,d^n{q_1}\,\Bigl\{ \frac{1}{[1][2]_{\aba}} + 
\frac{1}{[0]_{\aba}}\,
\Bigl[ \frac{m^2_{21}}{[1][2]_{\aba}} - \frac{1}{[1]} + 
\frac{1}{[2]_{\aba}}\Bigr]\Bigr\},
\label{faf23}
\eq
with $[0]_{\aba} = q^2_2$. As a consequence of this result we obtain
\bqa 
\cV^{\aba}(\mu\,|\,0\,;\,p_2,P,\lstm{1234}) &=&
\frac{1}{2}\,\Bigl[ m^2_{21}\, 
\cV^{\aca}(0\,|\,\mu\,;\,p_2,P,\lstm{12},0,\lstm{34}) 
+ \,\cV^{\aba}(0\,|\,\mu\,;\,p_2,P,\lstm{1234}) 
\nl
{}&-& \,C_{\mu}(p_2,p_1,0,\lstm{34})\,A_{\Z}([m_2,m_1])\Bigr]\, ,
\label{anothereqa}
\eqa
so that the $q_{1\mu}$ vector integral in the $E$-family is related to the
$q_{2\mu}$ vector integrals of the $I-E$ families. The function $C_{\mu}$ in
\eqn{anothereqa} is defined in \eqn{GCfunV}, the $I$ family will be
discussed in \sect{acafam}.  For the $V^{\aba}$-family we have partial
reducibility, i.e.  $\cV^{\aba}(0|\,p_1)$ can be expressed in term of known
quantities:
\bq
\cV^{\aba}(0|\,p_1;p_2,P,\lstm{1234}) =
-\frac{1}{2}\Bigl[ (l_{134} + 2p_{12})
\cV^{\aba}_{\Z}(p_2,P,\lstm{1234}) - S^{\ssA}_{\Z}(p_2,\lstm{123}) +
S^{\ssA}_{\Z}(P,\lstm{124})\Bigr].
\eq
Thus we can write
\bqa 
\cV^{\aba}(0\,|\,p_1\,;\,p_2,P,\lstm{1234}) &=& 
p^2_1\,I_{\aban{zy}} + \spro{p_1}{P}\,\Bigl[ I_{\aban{y}} -
\cV^{\aba}_{\Z}(p_2,P,\lstm{1234})\Bigr],
\label{unknownone}
\eqa
where two new quantities were introduced,
\bqa
I_{\aban{zy}} &=& \egam{\ep}\,\CIM{V}_{\aba}\,(z-y)\,\chiu{\aba}^{-\ep} , 
\qquad
I_{\aban{y}} = -\,\egam{\ep}\,\CIM{V}_{\aba}\,(1-y)\,\chiu{\aba}^{-\ep} . 
\eqa
Similarly, we derive
\bqa 
\cV^{\aba}(0\,|\,p_2\,;\,p_2,P,\lstm{1234}) &=& 
\spro{p_1}{p_2}\,I_{\aban{zy}} + \spro{p_2}{P}\,\Bigl[ I_{\aban{y}} -
\cV^{\aba}_{\Z}(p_2,P,\lstm{1234})\Bigr].
\label{unknowntwo}
\eqa
Assuming $p^2_1 \ne 0$, we can eliminate one of the two unknowns
from \eqn{unknownone} obtaining
\bqa
p^2_1\,I_{\aban{zy}} &=&  -\,\spro{p_1}{P}\,\Bigl[ I_{\aban{y}}   
-\cV^{\aba}_{\Z}(p_2,P,\lstm{1234})\Bigr] +
\cV^{\aba}(0\,|\,p_1\,;\,p_2,P,\lstm{1234}),
\eqa
and express $\cV^{\aba}(0\,|\,p_2)$ in terms of standard functions and
$I_{\aban{y}}$, which is the integral of \eqn{genaba} with $\alpha = \beta =
2$ and $\gamma = \delta = 1$, corresponding to $n = 6 -\ep$. Hence, one
generalized scalar function in shifted space-time dimension suffices in this
class, although we certainly prefer to use $\cV^{\aba}(0|p_2)$ in $n= 4 -
\ep$ dimensions, for which we can derive a smooth integral representation.
\subsubsection{Rank two tensor integrals in the $V^{\aba}$ family
\label{ranktwoaba}}
Tensor integrals with two powers of momenta in the numerator can be treated
in a similar way. New ultraviolet divergences arise; for instance with a
$q_{1\mu}\,q_{2\nu}$ numerator also the $(\alpha,\beta)$ sub-diagram is 
divergent. We define a decomposition according to
\bqa 
\cV^{\aba}(\mu,\nu\,|\,0\,;\,\cdots) &=& 
\cV^{\aba}_{\saaa}\,p_{1\mu}\,p_{1\nu} + 
\cV^{\aba}_{\saab}\,p_{2\mu}\,p_{2\nu} +
\cV^{\aba}_{\saac}\,\{p_1\,p_2\}_{\mu\nu} + 
\cV^{\aba}_{\saad}\,\delta_{\mu\nu}, 
\label{tensoraba} 
\eqa
where the list of arguments has been suppressed and where
$\{p\,k\}_{\mu\nu}$ is defined in \eqn{symmete}.  Strictly analogous
definitions hold for the $q_{1\mu} q_{2\nu}$ tensor integrals
($\cV^{\aba}_{12i}$ form factors) and for the $q_2^{\mu} q_2^{\nu}$ ones
($\cV^{\aba}_{22i}$ form factors).  Consider first the form factors in the
$22i$ series; taking the trace in \eqn{tensoraba} gives
\bqa 
n\,\cV^{\aba}_{\sbbd} + p^2_1\,\cV^{\aba}_{\sbba} +
2\,p_{12} \cV^{\aba}_{\sbbc} + p^2_2\,
\cV^{\aba}_{\sbbb} = -\,(p^2_2 + m^2_3)\,\cV^{\aba}_{\Z} - 
2\,\Bigl[ p_{12}\,\cV^{\aba}_{\sba} + p^2_2\,
\cV^{\aba}_{\sbb}\Bigr] + S^{\ssA}_{\Z}(P,\lstm{124}). 
\eqa
As a second step we multiply \eqn{tensoraba} by $p_{1\nu}$ and obtain
\bqa 
\cV^{\aba}_{\sbbd} + p^2_1\,\cV^{\aba}_{\sbba} + 
p_{12}\,\cV^{\aba}_{\sbbc} &=& \frac{1}{2}\,\Bigl[ -\,(l_{134} 
+ 2\,p_{12} )\, \cV^{\aba}_{\sba} +  
S^{\ssA}_{\Z}(P,\lstm{124}) + S^{\ssA}_{\msb}(P,\lstm{124})\Bigr], 
\nl 
p_{12}\,\cV^{\aba}_{\sbbb} + p^2_1\,
\cV^{\aba}_{\sbbc} &=& \frac{1}{2}\,\Bigl[-\,(l_{134} + 
2\,p_{12})\,\cV^{\aba}_{\sbb} 
\nl
{}&+& \,S^{\ssA}_{\Z}(P,\lstm{124}) - 
S^{\ssA}_{\Z}(p_2,\lstm{123}) + S^{\ssA}_{\msb}(P,\lstm{124}) - 
S^{\ssA}_{\msb}(p_2,\lstm{123})\Bigr]. 
\label{tensorabat} 
\eqa
\eqns{tensoraba}{tensorabat} give a system of three equations for four 
unknowns; for one of the form factors we can write a combination of two 
generalized scalar functions, e.g.
\bqa
\cV^{\aba}_{\sbbd} &=& \frac{1}{2}\,\omega^2\,\Bigl[
\cV^{2|1,1|1}_{\aba} + \cV^{1|1,1|2}_{\aba}\Bigr]\,
\bmid_{n = 6-\ep},
\eqa
where $\omega$ is defined in \eqn{defDUV}, and solve 
\eqns{tensoraba}{tensorabat} in terms of the generalized scalar functions. 
An alternative procedure is based on the following integral representations:
\bqa
\cV^{\aba}_{\sbbgen} &=& -\egam{\ep}\,\CIM{V}_{\aba}\,
P_{\sbbgen;\aba}\,\chiu{\aba}^{-\ep}(x,y,z), \quad i \ne 4,
\qquad
\cV^{\aba}_{\sbbd} = -\,\frac{1}{2}\,\egam{\ep-1}\,\CIM{V}_{\aba}\,
\chiu{\aba}^{1-\ep}(x,y,z) \, ,
\nn
\eqa
\bq
P_{\sbba;\aba} = z^2, \qquad
P_{\sbbb;\aba} = y^2, \qquad
P_{\sbbc;\aba} = y\,z.
\eq
The three integrals can be computed via their representation, following the
general strategy already adopted for scalar integrals.  Similar
representations hold for the remaining tensor integrals, for instance we
obtain
\bqa
\cV^{\aba}_{\sabd} &=& -\,\frac{1}{2}\,
\egam{\ep-1}\,\CIM{V}_{\aba}\,x\,\chiu{\aba}^{1-\ep}(x,y,z) ,
\nl
\cV^{\aba}_{\saad} &=& -\,\egam{\ep}\,\CIM{V}_{\aba}
\Bigl\{ -\frac{x(1-x)}{2-\ep}
\Bigl[ \frac{1}{2}\,\frac{4-\ep}{\ep-1}\,\chiu{\aba} 
+  R_{\aba} \Bigr]
+\frac{1}{2}\,\frac{x^2}{\ep-1}\,\chiu{\aba}\Bigr\}\,\chiu{\aba}^{-\ep} ,
\nl
P_{\sabgen;\aba} &=& x\,P_{\sbbgen;\aba}, \qquad
P_{\saagen;\aba} =  x^2\,P_{\sbbgen;\aba}\, \qquad \mbox{for} 
\quad i \ne 4,
\qquad
R_{\aba} = F(z,y) + m^2_x,
\eqa
with $F$ defined in \eqn{defGfun}.  Note the singularity hidden in the
$x$-integration in the formulae above. The $q_{1\mu}q_{2\nu}$ tensor
integrals are easily reduced as the following relation holds:
\bqa 
\cV^{\aba}(\mu\,|\,\nu\,;\,p_2,P,\lstm{1234}) &=&
\frac{1}{2}\,\Bigl[ m^2_{21}\, 
\cV^{\aca}(0\,|\,\mu\,,\nu\,;\,p_2,P,\lstm{12},0,\lstm{34}) 
+ \,\cV^{\aba}(0\,|\,\mu\,,\nu\,;\,p_2,P,\lstm{1234}) 
\nl
{}&+&\,C_{\mu\nu}(p_2,p_1,0,\lstm{34})\,A_{\Z}([m_2,m_1])\Bigr],
\eqa
while those with $q_{1\mu}q_{1\nu}$ require some additional work.
To derive the corresponding result we start with
\bq
\int\,d^nq_1\,\frac{q_{1\mu}\,q_{1\nu}}{[1][2]_{\aba}} =
B_{\aban{\sbb}}\,\delta_{\mu\nu} + 
B_{\aban{\sba}}\,q_{2\mu}\,q_{2\nu},
\eq
where the sub-loop form factors are
\bq
B_{\aban{\sbb}} = \frac{1}{n-1}\,(X_{\aban{1}} - X_{\aban{2}}),
\qquad
B_{\aban{\sba}} = \frac{1}{q^2_2}\,(X_{\aban{2}} - B_{\aban{\sbb}}),
\eq
and also
\bq
X_{\aban{1}} = \int\,d^nq_1\,\Bigl[ \frac{1}{[2]_{\aba}} - 
\frac{m^2_1}{[1][2]_{\aba}}\Bigr],
\eq
\bq
X_{\aban{2}} = \frac{1}{4}\,\int\,d^nq_1\,\Bigl[ \frac{q^2_2 + 2\,m^2_{21}}
{[1][2]_{\aba}} + \frac{m^4_{12}}{[0]_{\aba}[1][2]_{\aba}} + 
\frac{3}{q^2_1 + m_2^2} - \frac{1}{[1]}  + 
m^2_{12}\,\Bigl( \frac{1}{[0]_{\aba}[1]} - 
\frac{1}{[0]_{\aba}[2]_{\aba}}\Bigr)\Bigr],
\eq
with $[0]_{\aba} = q^2_2$.
The complete result reads as follows:
\bq
\cV^{\aba}(\mu,\nu\,|\,0\,;\,\cdots) =
\frac{1}{4\,(n-1)}\,\Bigl[
\cV^{\aba}_{\ssA}\,\delta_{\mu\nu} + \cV^{\aba}_{\ssB,\mu\nu}\Bigr],
\eq
\bqa
\cV^{\aba}_{\ssA} &=&
-\,m^4_{12}\,\cV^{\aca}_{\Z}(p_2,P,\lstm{12},0,\lstm{34}) -
2\, (m^2_1 + m^2_2)\,\cV^{\aba}_{\Z}(p_2,P,\lstm{1234})-
\cV^{\aba}(0|\mu,\mu;p_2,P,\lstm{1234})
\nl
{}&-& A_{\Z}(m_1)\,\Bigl[ m^2_{21}\,C_{\Z}(p_2,p_1,0,\lstm{34}) +
B_{\Z}(\,p_1,\lstm{34})\Bigr]
\nl
{}&-& A_{\Z}(m_2)\,\Bigl[ m^2_{12}\,C_{\Z}(p_2,p_1,0,\lstm{34}) +
B_{\Z}(\,p_1,\lstm{34})\Bigr],
\nl
\cV^{\aba}_{\ssB,\mu\nu} &=&
n\,m^4_{12}\,\cV^{\ada}(0\,|\,\mu,\nu\,;\,p_1,P,\lstm{12},0,\lstm{34},0)
\nl
{}&+& 2\,( n\,m^2_{21} + 2\,m^2_1)\,
\cV^{\aca}(0\,|\,\mu,\nu\,;\,p_2,P,\lstm{12},0,\lstm{34}) + 
n\,\cV^{\aba}(0\,|\,\mu,\nu\,;\,p_2,P,\lstm{1234})
\nl
{}&-& n\,A_{\Z}(m_1)\,\Bigl[ m^2_{12}\,
C_{\mu\nu}(2,1,1\,;\,p_2,p_1,0,\lstm{34}) -
C_{\mu\nu}(p_2,p_1,0,\lstm{34})\Bigr]
\nl
{}&-& A_{\Z}(m_2)\,\Bigl[ (3\,n - 4)\,C_{\mu\nu}(p_2,p_1,0,\lstm{34}) +
n\,m^2_{21}\,C_{\mu\nu}(2,1,1\,;\,p_2,p_1,0,\lstm{34})\Bigr],
\label{generC}
\eqa
which concludes our analysis of the $V^{\aba}$-family; 
note that $\cV^{\aba}_{\ssB,\mu\nu}$ can be further decomposed following
the standard procedure and also contributes to the $\delta_{\mu\nu}$ part
of $\cV^{\aba}(\mu,\nu\,|\,0\,;\,\cdots)$. Also for the
$q_{2\mu}q_{2\nu}$ tensor integrals we can write down a system of equations
and solve it, or we can use their explicit representations.  In \eqn{generC}
we used generalized $C$-functions; since these functions refer to one-loop
diagrams we have full reducibility of tensors while the scalars can be
expressed in terms of standard $C_{\Z}(1,1,1)$ and $B_{\Z}(1,1)$ functions by
repeated applications of IBP identities; one should only be aware of the
appearance of denominators vanishing at the anomalous threshold. Once again,
we could use their explicit parametric representations treated with the
BT-algorithm.

Results for this family are summarized in \appendx{summaaba}. $V^{\aba}_{\Z}
\equiv V^{121}_{\Z}$ is discussed in Sect.~5.1 of III, the evaluation of the
corresponding form factors is addressed in \sect{exaaba}. Note that
$\chiu{\aba}$ (\eqn{Defchiaba}) is defined in Eq.~(62) of III by rescaling by
$1/|P^2|$, a normalization which is better suited for numerical integration and
that we  have used for all the $\chi$ functions of III.  The same comment
(rescaling $\chi$ by $1/|P^2|$ in III) applies to all  families of diagrams.
The $\nu_i$ of Eqs.~(62) --  (63) of III, defined in Eq.~(7) of the same paper,
coincide with the $\nu_i$  quantities defined in \eqn{scaledq} of the present
paper.
\subsection{The $V^{\aca}$-family ($\alpha = 1, \beta = 3, \gamma = 1$)
\label{acafam} }
We continue our derivation considering the scalar function in the
$V^{\aca}$-family of \fig{TLvertaca}, where only the $(\alpha,\gamma)$
sub-diagram is ultraviolet divergent. This function is representable as
\bq
\pi^4\,\cV^{\aca}_{\Z}(p_1,P,\lstm{12345}) =
\mu^{2\ep}\,\intmomii{n}{q_1}{q_2}\,
\frac{1}{[1][2]_{\aca}[3]_{\aca}[4]_{\aca}[5]_{\aca}},
\label{DEFcVaca}
\eq
with propagators
\bqa
[1] &\equiv& q^2_1+m^2_1, \qquad\qquad [2]_{\aca} \equiv (q_1-q_2)^2+m^2_2,
\qquad
[3]_{\aca} \equiv q^2_2+m^2_3,
\nn
\eqa
\bqa
[4]_{\aca} &\equiv& (q_2+p_1)^2+m^2_4, \quad [5]_{\aca} \equiv (q_2+P)^2+m^2_5.
\label{enummaca}
\eqa
Note the symmetry property $\cV^{\aca}_{\Z}(p_1,P,\lstm{12345}) =
\cV^{\aca}_{\Z}(P,p_1,\lstm{12354})$, besides the one of \eqn{sym1}. 
\begin{figure}[ht]
\begin{center}
\begin{picture}(150,75)(0,0)
 \SetWidth{1.5}
 \Line(0,0)(42,0)         
\LongArrow(0,8)(20,8)          \Text(-11,7)[cb]{$-P$}
 \Line(128,-53)(100,-35)  
\LongArrow(128,-63)(114,-54)   \Text(138,-70)[cb]{$p_1$}
 \Line(128,53)(100,35)    
\LongArrow(128,63)(114,54)     \Text(138,62)[cb]{$p_2$}
 \CArc(55,-9)(15,0,360)         \Text(75,-3)[cb]{$1$}\Text(38,-27)[cb]{$2$}
 \Line(100,-35)(67,-15.75)      \Text(80,-37)[cb]{$3$}
 \Line(100,-35)(100,35)         \Text(107,-3)[cb]{$4$}
 \Line(100,35)(45,3)            \Text(68,23)[cb]{$5$}
 \Text(0,-25)[cb]{\Large $V^{\aca}$}
\end{picture}
\end{center}
\vspace{2cm}
\caption[]{The irreducible two-loop vertex diagrams $V^{\aca}$. 
External momenta flow inwards. Internal masses are enumerated according 
to \eqn{enummaca}.} 
\label{TLvertaca}
\end{figure}
\subsubsection{Vector integrals in the $V^{\aca}$ family \label{rankoneaca}}
By standard methods we write a decomposition of the vector integrals into 
form factors 
\bqa
\cV^{\aca}(\mu\,|\,0\,;\,p_1,P,\lstm{12345}) &=&
\sum_{i=1,2}\,\cV^{\aca}_{\sagen}(p_1,P,\lstm{12345})\,p_{i\mu},
\nl
\cV^{\aca}(0\,|\,\mu\,;\,p_1,P,\lstm{12345}) &=&
\sum_{i=1,2}\,\cV^{\aca}_{\sbgen}(p_1,P,\lstm{12345})\,p_{i\mu}.
\label{ffaca}
\eqa
As we mentioned earlier, special care must be used when $\cV^{\aca}_{\sijgen}$
appears in the reduction of other form factors and one has to bring the
integrand in a form adhering to \eqn{DEFcVaca}; this can be done using the
definition,
\bq
\frac{\mu^{2\ep}}{\pi^4}\,\int\,d^nr_1\,d^nr_2\,r_{i\mu}\,
\prod_{l=a}^{e}\,D^{-1}_l \equiv
\cV^{\aca}_{\sgena}(k_d,k_e,\lstm{abcde})\,k_{d\mu} + 
\cV^{\aca}_{\sgenb}(k_d,k_e,\lstm{abcde})\,(k_{e} - k_{d})_{\mu} \, ,
\eq
where, with an obvious notation
\[
\ba{lll}
D_a = r_1^2 + m_a^2\, , \quad & \quad
D_b = (r_1 - r_2)^2 + m_b^2\,, \quad & \quad
D_c = r_2^2 + m_c^2\,, \\
D_d = (r_2 + k_d)^2 + m_d^2\,,  \quad & \quad
D_e = (r_2 + k_e)^2 + m_e^2 .\quad & \quad\, 
\ea
\]
Note that the following identities hold:
\bq
\cV^{\aca}_{\sgena}(d,e) =  \cV^{\aca}_{\sgena}(e,d) ,
\qquad
\cV^{\aca}_{\sgenb}(d,e) = \cV^{\aca}_{\sgena}(e,d) - \cV^{\aca}_{\sgenb}(e,d),
\eq
where $(d,e) = (k_d,k_e,\lstm{abcde})$ etc.
The explicit expression for the form factors of \eqn{ffaca} is
\bqa
\cV^{\aca}_{\sijgen} &=&
-\,\egam{1+\ep}\,\CIM{V}_{\aca}\,P_{\acan{ij}}\,\chiu{\aca}^{-1-\ep},
\nl
P_{\acan{\szz}} &=& 1, \quad
P_{\acan{\sagen}} = x\,P_{\acan{\sbgen}}, \quad
P_{\acan{\sba}} = -z_1, \quad
P_{\acan{\sbb}} = -z_2,
\eqa
where $P_{\szz}$ is the factor corresponding to the scalar integral and,
with our choice for the Feynman parameters, the polynomial $\chiu{\aca}$ is
\bqa
\chiu{\aca}(x,y,z_1,z_2) &=& -\,F(z_1,z_2)\,+\,l_{134}\,z_1\,+ 
\,(l_{245}\,+\,2\,p_{12})\,z_2\,+
\,(m_3^2\,-\,m_x^2)\,y\, + \,m_x^2,
\label{Defchiaca}
\eqa
where $F$ and $m_x^2$ are defined in \eqn{defGfun}. Finally, the integration 
measure is
\bq
\CIM{V}_{\aca} = \omega^{\ep}\,
\dcubs{x}{y,z_1,z_2}\,\Bigl[ x\,(1-x)\Bigr]^{-\ep/2}\,(1-y)^{\ep/2-1},
\eq
with $\omega$ defined in \eqn{defDUV}.  All these functions can be
manipulated according to the procedure introduced in III with
correspondingly smooth integral representations.

This family is the first example of a vertex with  full $q_2$
reducibility. Consider the $q_1$ vector integral: for the case $m_3 \ne 0$
and by methods similar to the ones used in \sect{abafam} for $V^{\aba}$ we 
obtain
\bqa
\cV^{\aca}(\mu\,|\,0\,;\,p_1,P,\lstm{12345}) &=&
\frac{1}{2}\,\frac{m^2_{123}}{m^2_3}\,
\cV^{\aca}(0\,|\,\mu\,;\,p_1,P,\lstm{12345}) - 
\frac{1}{2}\,\frac{m^2_{12}}{m^2_3}\,
\cV^{\aca}(0\,|\,\mu\,;\,p_1,P,\lstm{12},0,\lstm{45}) \nl
{}&-& \frac{1}{2 m_3^2}\,A_{\Z}([m_1,m_2])\,
\Bigl[C_{\mu}(p_1,p_2,\lstm{345}) - C_{\mu}(p_1,p_2,0,\lstm{45}) \Bigr].
\eqa
Furthermore, the $q_2$ vector integral can be reduced according to
the following relation:
\bqa
\cV^{\aca}(0|p_1;p_1,P,\lstm{12345}) &=&
\frac{1}{2}\Bigl[ -l_{134}\cV^{\aca}_{\Z}(p_1,P,\lstm{12345})
-\cV^{\aba}_{\Z}(p_1,P,\lstm{1245}) + \cV^{\aba}_{\Z}(0,P,\lstm{1235})\Bigr],
\nl
\cV^{\aca}(0|p_2;p_1,P,\lstm{12345}) &=&
\frac{1}{2}\Bigl[ (l_{154} \!\! -\!\! P^2)V^{\aca}_{\Z}(p_1,P,\lstm{12345})
\!+\!\cV^{\aba}_{\Z}(0,p_1,\lstm{1234}) -\!\cV^{\aba}_{\Z}(0,P,\lstm{1235})\Bigr].
\label{redI2}
\nl
\eqa
\subsubsection{Rank two tensor integrals in the $V^{\aca}$ family
\label{ranktwoaca}}
All tensor integrals in this class are overall ultraviolet divergent, with a
divergent $(\alpha,\gamma)$ sub-diagram.
Adopting the same notation employed in the analysis of the $\cV^{\aba}$
functions, we introduce the form factors $\cV^{\aca}_{ij m}$, 
($m = 1,\ldots,4$):
\bqa 
\cV^{\aca}(0|\mu,\nu\,;\,\cdots) &=& 
\cV^{\aca}_{\sbba}\,p_{1\mu}\,p_{1\nu} + 
\cV^{\aca}_{\sbbb}\,p_{2\mu}\,p_{2\nu} + 
\cV^{\aca}_{\sbbc}\,\{p_1 p_2\}_{\mu\nu} + 
\cV^{\aca}_{\sbbd}\,\delta_{\mu\nu}, 
\label{tensoraca}
\eqa
where the symmetrized product is given by \eqn{symmete}.
Taking the trace of both sides in \eqn{tensoraca} gives
\bq 
p^2_1\,\cV^{\aca}_{\sbba} + 2\,p_{12}\,
\cV^{\aca}_{\sbbc} +p^2_2\,\cV^{\aca}_{\sbbb} +\,
n \, \cV^{\aca}_{\sbbd} = \cV_{\Z}^{\aba}(p_1,P,\lstm{1245}) -
m_3^2\,\cV_{\Z}^{\aca}(p_1,P,\lstm{12345}), 
\label{af3}
\eq
while, multiplying both sides of \eqn{tensoraca} by $p_{1 \nu}$, we
have the relations
\bqa
p^2_1\cV^{\aca}_{\sbba} + p_{12}\cV^{\aca}_{\sbbc} +
\cV^{\aca}_{\sbbd} &=&
\frac{1}{2}\Bigl[ -\cV^{\aba}_{\sba}(p_1,P,\lstm{1245}) + 
\cV^{\aba}_{\sbb}(0,P,\lstm{1235}) 
- l_{134}\cV^{\aca}_{\sbb}(p_1,P,\lstm{12345})\Bigr] , 
\nl
p_{12}\cV^{\aca}_{\sbbb} + p^2_1\cV^{\aca}_{\sbbc} &=&
\frac{1}{2}\Bigl[-\cV^{\aba}_{\sba}(p_1,P,\lstm{1245}) + 
\cV^{\aba}_{\sba}(0,P,\lstm{1235}) -
l_{134}\cV^{\aca}_{\sba}(p_1,P,\lstm{12345})\Bigr].
\label{af2}
\eqa
Similarly, contracting \eqn{tensoraca} with $p_{2\nu}$, it is possible to 
write additional identities:
\bqa
p_{12}\,\cV^{\aca}_{\sbba} + p^2_2\,\cV^{\aca}_{\sbbc} &=&
-\,\frac{1}{2}\,\Bigl[ \cV^{\aba}_{\sba}(0,P,\lstm{1235}) - 
\,\cV^{\aba}_{\sba}(0,p_1,\lstm{1234}) 
+\,(l_{\ssP 45} - p^2_1)\,\cV^{\aca}_{\sba}(p_1,P,\lstm{12345})\Bigr], 
\nl
p^2_2\,\cV^{\aca}_{\sbbb} + p_{12}\,\cV^{\aca}_{\sbbc} +
\cV^{\aca}_{\sbbd} &=&
\frac{1}{2}\,\Bigl[-\,\cV^{\aba}_{\sba}(0,P,\lstm{1235}) 
-\,(l_{\ssP 45}-p^2_1)\,\cV^{\aca}_{\sbb}(p_1,P,\lstm{12345})\Bigr].
\label{af1}
\eqa
Solving the system formed by \eqns{af3}{af1} it is then possible to express
the $\cV^{\aca}_{\sbbgen}$ form factor in terms of functions
$\cV^{\aca}_{\sbgen}$ and form factors belonging to the $V^{\aba}$ family.

The integral representation for the $\cV^{\aca}_{\sbbgen}$ functions is the
following:
\bqa
\cV^{\aca}_{\sbbgen} &=&
-\,\egam{1+\ep}\,\CIM{V}_{\aca}\,R_{\acan{\sbbgen}}\,\chiu{\aca}^{-1-\ep},
\qquad
\cV^{\aca}_{\sbbd} =
-\frac{1}{2}\,\egam{\ep}\,\CIM{V}_{\aca}\,\chiu{\aca}^{-\ep},
\nl
R_{\acan{\sbba}} &=& z_1^2, \quad
R_{\acan{\sbbb}} = z_2^2, \quad
R_{\acan{\sbbc}} = z_1\,z_2,
\eqa
showing for instance a double ultraviolet pole for $\cV^{\aca}_{\sbbd}$.
Similar integral representations can be found for the form factors
$\cV^{\aca}_{\sabgen}$:
\bqa
\cV^{\aca}_{\sabgen} &=&
-\,\egam{1+\ep}\,\CIM{V}_{\aca}\,R_{\acan{\sabgen}}\,\chiu{\aca}^{-1-\ep},
\quad
\cV^{\aca}_{\sabd} =
-\frac{1}{2}\,\egam{\ep}\,\CIM{V}_{\aca}\,x\,\chiu{\aca}^{-\ep},
\quad
R_{\acan{\sabgen}} = x\,R_{\acan{\sbbgen}}.
\eqa
The $q_{1\mu}\,q_{2\nu}$ or $q_{1\nu}\,q_{2\mu}$ tensor integrals
can be written in terms of form factors $\cV^{\aca}_{\sbbgen}$ employing 
the following relation, valid for $m_3 \neq 0$:
\bq
\frac{1}{[0]_{\aca}[3]_{\aca}} = \frac{1}{m_3^2}\,
\Bigl( \frac{1}{[0]_{\aca}} - \frac{1}{[3]_{\aca}}\Bigr)
\,, 
\label{af5}
\eq
where $[0]_{\aca} = q_2^2$; in this way one obtains
\bqa
\cV^{\aca}(\mu|\nu\,;\,p_1,P,\lstm{12345}) &=& \frac{m^2_{123}}{2 m_3^2}\,
\cV^{\aca}(0|\mu,\nu\,;\,p_1,P,\lstm{12345}) + \frac{m^2_{21}}{2 m_3^2}\,
\cV^{\aca}(0|\mu ,\nu\,;\,p_1,P,\lstm{12},0,\lstm{45})
\nl
&-&\,\frac{1}{2 m_3^2}\,A_{\Z}([m_1,m_2])
\Bigl[\,C_{\mu \nu}(p_1,p_2,\lstm{345}) -
C_{\mu \nu}(p_1,p_2,0,\lstm{45}) \Bigr]. 
\eqa
The integral representation of the $\cV^{\aca}_{\saagen}$ form factors is
\bqa
\cV^{\aca}_{\saagen} &=&
-\,\egam{1+\ep}\,\CIM{V}_{\aca}\,R_{\acan{\saagen}}\,\chiu{\aca}^{-1-\ep},
\nl
\cV^{\aca}_{\saad} &=& -\,\egam{\ep}\,\CIM{V}_{\aca}\,
\chiu{\aca}^{-\ep}\,\Bigl\{ -\frac{x(1-x)}{2-\ep}\,
\Bigl[ \frac{4-\ep}{2} +\,\ep\,\chiu{\aca}^{-1}\,R_{\aca}\Bigr]
+\,\frac{1}{2}\,x^2\,\Bigr\},
\nl
R_{\acan{\saagen}} &=& x^2\,R_{\acan{\sbbgen}}, \qquad
R_{\aca} = F(z_1,z_2) + m^2_x,
\eqa
with $F$ defined in \eqn{defGfun}, The form factors $\cV^{\aca}_{\saagen}$
can be reduced using $q_1$ sub-loop techniques, similarly to what we did for
the $\cV^{\aba}_{\saagen}$ functions, and employing \eqn{af5}. One
obtains
\bq
\cV^{\aca}(\mu,\nu\,|\,0\,;\,\cdots) =
\frac{1}{4\,(n-1)}\,\Bigl[
\cV^{\aca}_{\ssA}\,\delta_{\mu\nu} + \cV^{\aca}_{\ssB,\mu\nu}\Bigr],
\eq
\bqa
\cV^{\aca}_{\ssA} &=& \frac{1}{m_3^2}\,\Bigl\{ -A_{\Z}(m_1)\,\Bigl[ 
m^2_{123}\,C_{\Z}(p_1,p_2,\lstm{345}) + 
m^2_{21}\,C_{\Z}(p_1,p_2,0,\lstm{45})\Bigr] 
\nl 
{}&-&  A_{\Z}(m_2)\,\Bigl[ 
m^2_{213}\,C_{\Z}(p_1,p_2,\lstm{345}) + 
m^2_{12}\,C_{\Z}(p_1,p_2,0,\lstm{45}) \Bigr] 
\nl
{}&-& m_3^2\,\cV^{\aca}(0|\mu, \mu\,;\,p_1,P,\lstm{12345}) +
\Bigl[ m^4_{12} - 2\,m_3^2\,(m_1^2 + m_2^2)\Bigr]\,
\cV^{\aca}_{\Z}(p_1,P,\lstm{12345})  
\nl
{}&-& m^4_{12}\,\cV^{\aca}_{\Z}(p_1,P,\lstm{12},0,\lstm{45}) \Bigr\} ,
\nl
\cV^{\aca}_{\ssB,\mu\nu} &=& 
\frac{1}{m^2_3}\,\Bigl\{
n\,m_3^2\,m^4_{12}\,\Bigl[
\cV^{\ada}(0|\mu ,\nu\,;\,p_1,P,\lstm{12},0,\lstm{45},0)  
-\cV^{\ada}(0|\mu ,\nu\,;\,p_1,P,\lstm{12345},0)\Bigr]
\nl
{}&-& n\,m^2_{3}\,A_{\Z}(m_1)\,\Bigl[
C_{\mu \nu}\,(p_1,p_2,\lstm{345})-
C_{\mu \nu}\,(p_1,p_2,0,\lstm{45}) \Bigr]
\nl
&-& n\,m^2_{12}\,A_{\Z}([m_1,m_2])\,\Bigl[
m^2_3\,C_{\mu \nu}(2,1,1\,;\,p_1,p_2,0,\lstm{45})- 
C_{\mu \nu}\,(p_1,p_2,0,\lstm{45}) 
\nl
{}&+&
C_{\mu \nu}\,(p_1,p_2,\lstm{345})\Bigr] 
+ (3\,n-4)\,A_{\Z}(m_2)\,
\Bigl[ C_{\mu \nu}\,(p_1,p_2,\lstm{345}) - 
C_{\mu \nu}\,(p_1,p_2,0,\lstm{45})\Bigr] 
\nl 
{}&+& m_3^2\,(2\,n\,m_{12}^2 + n\,m_3^2 - 4\,m_1^2)\,
\cV^{\aca}(0|\mu ,\nu\,;\,p_1,P,\lstm{12345})
\nl
{}&-& \,m_3^2\,\Bigl[(n-4)\,m_1^2 -2\,n\,m_2^2\Bigr]\,
\cV^{\aca}(0|\mu ,\nu\,;\,p_1,P,\lstm{12},0,\lstm{45}) \Bigr\} \, .   
\label{justabove}
\eqa
Note that $\cV^{\aca}_{\ssB,\mu\nu}$ will be further decomposed into
$\delta_{\mu\nu}$ and $p_{i\mu} p_{j\nu}$ terms.
Results for this family are summarized in \appendx{summaaca}.
$V^{\aca}_{\Z} \equiv V^{131}_{\Z}$ is discussed in Sect.~6.1 of III
(see comment at the end of \sect{ranktwoaba}),
evaluation of form factors in \sect{exaacaada}.
Note that $\chiu{\aca}(x,1,y,z)$ does not depend on $x$ and, in III,
we used $\chiu{\aca}(y,z) \equiv \chiu{\aca}(x,1,y,z)$.
In the following Section we move to the discussion of the $V^{\ada}$ class of 
diagrams.
\subsection{The $V^{\ada}$-family ($\alpha = 1, \beta = 4, \gamma = 1$)
\label{adafam} }
The scalar $V^{\ada}$ function of \fig{TLvertada} is overall ultraviolet 
convergent with the $(\alpha,\gamma)$ sub-diagram divergent and is 
representable as follows: 
\bq 
\pi^4\,\cV^{\ada}_{\Z}(p_1,P,\lstm{123456}) =
\mu^{2\ep}\,\intmomii{n}{q_1}{q_2}\,
\frac{1}{[1][2]_{\ada}[3]_{\ada}[4]_{\ada}[5]_{\ada}[6]_{\ada}},
\eq
with propagators
\bqa
[1] &\equiv& q^2_1+m^2_1, \qquad\qquad [2]_{\ada} \equiv (q_1-q_2)^2+m^2_2,
\qquad
[3]_{\ada} \equiv q^2_2+m^2_3,
\nn
\eqa
\bqa
[4]_{\ada} &\equiv& (q_2+p_1)^2+m^2_4, \qquad\qquad
[5]_{\ada} \equiv (q_2+P)^2+m^2_5, \qquad\qquad 
[6]_{\ada} \equiv q^2_2+m^2_3 \, .
\label{enummada}
\eqa
Note the symmetry property $\cV^{\ada}_{\Z}(p_1,P,\lstm{123456}) =
\cV^{\ada}_{\Z}(P,p_1,\lstm{123546})$, as shown in \eqn{sym1} of
\appendx{symme}.  Scalar, vector and rank two tensor integrals have an
$(\alpha,\gamma)$ sub-diagram which is ultraviolet divergent.  As it was
pointed out in III, we need to consider just the case $m_3 = m_6$ and, as a
consequence, $m_6$ drops from the list of arguments; in fact, when these two
masses are different it is possible to rewrite the integral as a difference
of $V^{\aca}$-functions.
\begin{figure}[ht]
\begin{center}
\begin{picture}(150,75)(0,0)
 \SetWidth{1.5}
 \Line(0,0)(40,0)         
\LongArrow(0,8)(20,8)          \Text(-11,7)[cb]{$-P$}
 \Line(128,-53)(100,-35)  
\LongArrow(128,-63)(114,-54)   \Text(138,-70)[cb]{$p_1$}
 \Line(128,53)(100,35)    
\LongArrow(128,63)(114,54)     \Text(138,62)[cb]{$p_2$}
 \Line(57,-10)(40,0)            \Text(46,-16)[cb]{$6$}
 \CArc(70,-17.5)(15,0,360)      \Text(80,0)[cb]{$1$}\Text(58,-40)[cb]{$2$}
 \Line(100,-35)(83,-25)         \Text(89,-42)[cb]{$3$}
 \Line(100,-35)(100,35)         \Text(107,-3)[cb]{$4$}
 \Line(100,35)(40,0)            \Text(68,23)[cb]{$5$}
 \Text(0,-25)[cb]{\Large $V^{\ada}$}
\end{picture}
\end{center}
\vspace{2cm}
\caption[]{The irreducible two-loop vertex diagrams $V^{\ada}$. 
External momenta flow inwards. Internal masses are enumerated according 
to \eqn{enummada}.} 
\label{TLvertada}
\end{figure}
\subsubsection{Vector integrals in the $V^{\ada}$ family \label{rankoneada}}
As usual, we introduce form factors for the vector integrals according to the
equations
\bqa
\cV^{\ada}(\mu\,|\,0\,;\,p_1,P,\lstm{12345}) &=&
\sum_{i=1,2}\,\cV^{\ada}_{\sagen}(p_1,P,\lstm{12345})\,p_{i\mu},
\nl
\cV^{\ada}(0\,|\,\mu\,;\,p_1,P,\lstm{12345}) &=&
\sum_{i=1,2}\,\cV^{\ada}_{\sbgen}(p_1,P,\lstm{12345})\,p_{i\mu},
\label{ffada}
\eqa
where the form factors $\cV^{\ada}_{\sijgen}(p,k,\lstm{a \cdots e})$ refer to the
basis $p_{\mu}$ and $(k - p)_{\mu}$.

The integral representation of the form factors introduced in \eqn{ffada}
is obtained employing standard methods:
\bqa
\cV^{\ada}_{\sijgen} &=&
-\,\egam{2+\ep}\,\CIM{V}_{\ada}\,P_{\adan{ij}}\,\chiu{\ada}^{-2-\ep},
\eqa
\bqa
P_{\adan{\szz}} &=& 1, \quad
P_{\adan{\sagen}} = x\,P_{\adan{\sbgen}}, \quad
P_{\adan{\sba}} = - z_1, \quad
P_{\adan{\sbb}} = - z_2,
\eqa
where the integration measure is
\bq
\CIM{V}_{\ada} = \omega^{\ep}\,\dcubs{x}{y,z_1,z_2}\,(y-z_1)\,
\Bigl[ x\,(1-x)\Bigr]^{-\ep/2}\,(1-y)^{\ep/2-1},
\label{measada}
\eq
and $\omega$ is defined in \eqn{defDUV}; $P_{\szz}$ is the factor
corresponding to the integral representation of the scalar function. The
polynomial $\chiu{\ada}$ is equal to $\chiu{\aca}$.

It is possible to rewrite the $q_1$ vector integral in terms of the $q_2$ 
vector integral; when $m_3 \neq 0$ one finds
\bqa
{}&{}&
\cV^{\ada}(\mu|0\,;\,p_1,P,\lstm{12345}) =
\frac{m^2_{123}}{2 m^2_3}\,
\cV^{\ada}(0\,|\,\mu\,;\,p_1,P,\lstm{12345}) +
\,\frac{m^2_{12}}{2 m^4_3}\,[
\cV^{\aca}(0\,|\,\mu\,;\,p_1,P,\lstm{12345})
\nl 
{}&-&
\cV^{\aca}(0\,|\,\mu\,;\,p_1,P,\lstm{12},0,\lstm{45})] -
\frac{A_{\Z}([m_1,m_2])}{2 m_3^4}\,\Bigl[
C_{\mu}(2,1,1\,;\,p_1,p_2,\lstm{345})\,m_3^2 
\nl 
{}&+& C_{\mu}(p_1,p_2,\lstm{345}) - 
C_{\mu}(p_1,p_2,0,\lstm{45}) \Bigr].
\eqa
The $q_2$ vector integrals can be reduced to a linear combination of scalar
factors as follows:
\bqa
\cV^{\ada}(0|p_1\,;\,p_1,P,\lstm{12345}) &=&\frac{1}{2}\,\Bigl[
        -\,\cV^{\ada}_{\Z}\,(p_1,P,\lstm{12345})\, l_{134} \nl
&-& \cV^{\aca}_{\Z}\,(p_1,P,\lstm{12345})\,
       + \cV^{\aca}_{\Z}\,(0,P,\lstm{12335}) \Bigr]\, , \nl
\cV^{\ada}(0|p_2\,;\,p_1,P,\lstm{12345}) &=&\frac{1}{2}\,\Bigl[
       \cV^{\ada}_{\Z}\,(p_1,P,\lstm{12345})\, ( l_{154} - P^2 ) \nl
&+& \cV^{\aca}_{\Z}\,(0,p_1,\lstm{12334})\,
- \cV^{\aca}_{\Z}\,(0,P,\lstm{12335})\, \Bigr]\, . \quad\quad
\label{omada}
\eqa
\subsubsection{Rank two tensor integrals in the $V^{\ada}$ family
\label{ranktwoada}}
The tensor integrals with two powers of the integration momenta in
the numerator can be treated analogously to the case of $\cV^{\aca}$. 
Only the $(\alpha,\gamma)$ sub-diagram is ultraviolet divergent.
Using \eqn{symmete} we define the corresponding decomposition as
\bqa 
\cV^{\ada}(0 |\mu,\nu\ ;\,\cdots) &=& 
\cV^{\ada}_{\sbba}\,p_{1\mu}\,p_{1\nu} + 
\cV^{\ada}_{\sbbb}\,p_{2\mu}\,p_{2\nu} + 
\cV^{\ada}_{\sbbc}\,\{p_1 p_2\}_{\mu\nu} + 
\cV^{\ada}_{\sbbd}\,\delta_{\mu\nu}, 
\label{tensorada}
\eqa
and the corresponding form factors $\cV^{\ada}_{\saagen}$ and
$\cV^{\ada}_{\sabgen}$. 
The symmetrized product is given by \eqn{symmete}.
Consider the form factors of the $\cV_{\sbbgen}$ family: taking the trace 
in both sides of \eqn{tensorada} one obtains the relation
\bq 
p^2_1\,\cV^{\ada}_{\sbba} + 2\,p_{12}\,\cV^{\ada}_{\sbbc} +
p^2_2\,\cV^{\ada}_{\sbbb} + n\,\cV^{\ada}_{\sbbd} = 
\cV_{\Z}^{\aca}(p_1,P,\lstm{12345}) -
m_3^2\,\cV_{\Z}^{\ada}(p_1,P,\lstm{12345}) \,
\label{af13}.
\eq
Similarly, contracting \eqn{tensorada} with $p_{1\nu}$ we have
\bq 
p^2_1\cV^{\ada}_{\sbba} + p_{12}\cV^{\ada}_{\sbbc} + 
\cV^{\ada}_{\sbbd} = 
\frac{1}{2}\Bigl[ \!-\cV^{\aca}_{\sba}(p_1,P,\lstm{12345}) \!+
\cV^{\aca}_{\sbb}(0,P,\lstm{12335}) \!- 
l_{134}\cV^{\ada}_{\sba}(p_1,P,\lstm{12345})\Bigr],
\eq
\bq
p_{12}\cV^{\ada}_{\sbbb} + p^2_1\cV^{\ada}_{\sbbc} =
 \frac{1}{2}\Bigl[ -\cV^{\aca}_{\sbb}(p_1,P,\lstm{12345}) + 
\cV^{\aca}_{\sbb}(0,P,\lstm{12335}) - 
l_{134}\cV^{\ada}_{\sbb}(p_1,P,\lstm{12345})\Bigr].
\label{af12} 
\eq
Contracting \eqn{tensorada} with $p_{2\nu}$ we get instead
\bqa 
p_{12}\cV^{\ada}_{\sbba} + p^2_2\cV^{\ada}_{\sbbc} &=&
 \frac{1}{2}\Bigl[\cV^{\aca}_{\sbb}(0,p_1,\lstm{12334})  -
 \cV^{\aca}_{\sbb}(0,P,\lstm{12335}) +
(p^2_1 - l_{\ssP 45})\cV^{\ada}_{\sba}(p_1,P,\lstm{12345})\Bigr], 
\nl 
p^2_2\cV^{\ada}_{\sbbb} + p_{12}\cV^{\ada}_{\sbbc} + 
\cV^{\ada}_{\sbbd} &=& \frac{1}{2}\Bigl[ -
\cV^{\aca}_{\sbb}(0,P,\lstm{12335}) + 
(p^2_1 - l_{\ssP 45})\cV^{\ada}_{\sbb}(p_1,P,\lstm{12345})\Bigr].
\label{af11}
\eqa
Solving the system given by \eqns{af13}{af11}, we can reduce the 
$\cV^{\ada}_{\sbbgen}$ form factors to linear combinations of vector and 
scalar integrals. The integral representation of these form factors is the 
following:
\bqa 
\cV^{\ada}_{\sbbgen} &=&
-\,\egam{2+\ep}\,\CIM{V}_{\ada}\,
R_{\adan{\sbbgen}}\,\chiu{\aca}^{-2-\ep}, \quad i \neq 4 \,,
\qquad
\cV^{\ada}_{\sbbd} =
-\frac{1}{2}\,\egam{1+\ep}\,\CIM{V}_{\ada}\,\chiu{\ada}^{-1-\ep},
\nn
\eqa
\bq
R_{\adan{\sbba}} = z_1^2, \quad R_{\adan{\sbbb}} = z_2^2, \quad
R_{\adan{\sbbc}} = z_1\,z_2.
\eq
The form factors of the $\cV^{\ada}_{\sabgen}$ family can be
written in terms of those of the $\cV^{\ada}_{\sbbgen}$ family: in fact we 
have that
\bqa 
\cV^{\ada}(\mu|\nu\,;\,p_1,P,\lstm{12345}) &=& 
\cV^{\ada}(0|\mu,\nu\,;\,p_1,P,\lstm{12345}) \,\frac{m^2_{312}}{2 m_3^2} + 
\frac{m_{12}^2}{2\,m_3^4}\,\Bigl[\cV^{\aca}(0|\mu,\nu\,;\,p_1,P,\lstm{12345})
\nl
&-& \cV^{\aca}(0|\mu,\nu\,;\,p_1,P,\lstm{12},0,\lstm{45})\Bigr]\,   
-\,\frac{A_{\Z}([m_1,m_2])}{2 m_3^4}\,\Bigl[
C_{\mu\nu}(p_1,p_2,\lstm{345}) \nl
&-& 
C_{\mu \nu}(p_1,p_2,0,\lstm{45}) 
+
m_3^2\,C_{\mu \nu}(2,1,1\,;\,p_1,p_2,\lstm{345})\Bigr].
\eqa
The integral representation of the same form factors is the following:
\bqa 
\cV^{\ada}_{\sabgen} &=&
-\,\egam{2+\ep}\,\CIM{V}_{\ada}\,R_{\adan{\sabgen}}\,\chiu{\ada}^{-2-\ep}, 
\quad i \neq 4, 
\quad
\cV^{\ada}_{\sabd} =
-\frac{1}{2}\,\egam{1+\ep}\,\CIM{V}_{\ada}\,
x\,\chiu{\ada}^{-1-\ep}, 
\eqa
with $R_{\adan{\sabgen}} = x\,R_{\adan{\sbbgen}}$.
Similarly to the case of the $V^{\aba}$ and $V^{\aca}$ families, the reduction 
of $q_{1\mu} q_{1\nu}$ tensor integrals leads to expressions which are more 
involved. Introducing the definitions
\bq 
\cV^{\ada}(\mu,\nu\,|\,0\,;\,\cdots) =
\frac{1}{4\,(n-1)}\,\Bigl[ \cV^{\ada}_{\ssA}\,\delta_{\mu\nu}
+ \cV^{\ada}_{\ssB, \, \mu\nu}\Bigr], 
\eq
and employing standard techniques one finds
\bqa 
\cV^{\ada}_{\ssA} &=& \frac{1}{m_3^4}\,\Bigl\{ 
\cV^{\ada}_{\Z}(p_1,P,\lstm{12345}) \, m_3^2\,\Bigl[
m_{12}^4 - 2\,(m_1^2 + m_2^2)\,m_3^2 \Bigr] -
m_3^4\,\cV^{\ada}(0|\mu,\mu\,;\, p_1,P,\lstm{12345})
\nl 
{}&+& m^4_{12}\,\Bigl[ \cV^{\aca}_{\Z}(p_1,P,\lstm{12345})  -
\cV^{\aca}_{\Z}(p_1,P,\lstm{12},0,\lstm{45})\Bigr] 
\nl 
{}&-& m^2_{12}\,A_{\Z}([m_1,m_2])\, 
\Bigl[ C_{\Z}(p_1,p_2,\lstm{345}) - 
C_{\Z}(p_1,p_2,0,\lstm{45})\Bigr] 
\nl 
{}&-&
m^2_3\,C_{\Z}(2,1,1\,;\,p_1,p_2,\lstm{345})\,\Bigl[
m^2_{123}\,A_{\Z}(m_1) + m^2_{213}\,A_{\Z}(m_2)\Bigr]\Bigr\},
\nl 
\cV^{\ada}_{\ssB, \, \mu\nu} &=&
\frac{1}{m_3^4}\,\Bigl\{ ( n\,m_{123}^4 - 4\,m_1^2\,m_3^2 ) 
\,\cV^{\ada}(0|\mu,\nu\,;\,p_1,P,\lstm{12345}) +
\cV^{\ada}(0|\mu,\nu\,;\,p_1,P,\lstm{12},0,\lstm{45}) \, n\,m^4_{12} 
\nl 
{}&+&
\frac{2}{m_3^2}\,( 2\,m_1^2\,m_3^2 - n\,m_{12}^2\,m_{123}^2 ) \,
\Bigl[ \cV^{\aca}(0|\mu ,\nu\,;\,p_1,P,\lstm{12},0,\lstm{45}) - 
\cV^{\aca}(0|\mu,\nu\,;\,p_1,P,\lstm{12345})\Bigr] 
\nl 
{}&+& n\,\Bigl(2\,\frac{m^2_{123}}{m_3^2}-1\Bigr)\,A_{\Z}([m_1,m_2])\,\Bigl[ 
C_{\mu \nu}(p_1,p_2,\lstm{345}) -
C_{\mu \nu}(p_1,p_2,0,\lstm{45})\Bigr] 
\nl 
&-& n\,A_{\Z}([m_1,m_2])\,\Bigl[m_{123}^2  
\,C_{\mu \nu}(2,1,1\,;\,p_1,p_2,\lstm{345}) + 
m_{12}^2\,C_{\mu \nu}(2,1,1\,;\,p_1,p_2,0,\lstm{45})\Bigr] 
\nl 
{}&+&2\,(n-2)\,A_{\Z}(m_2)\,\Bigl[ 
C_{\mu \nu}(p_1,p_2,\lstm{345}) -
C_{\mu \nu}(p_1,p_2,0,\lstm{45})
\nl
{}&+& m^2_3\,C_{\mu \nu}(2,1,1\,;\,p_1,p_2,\lstm{345}) \Bigr\}.
\eqa
Note that $\cV^{\ada}_{\ssB,\mu\nu}$ will be further decomposed into
$\delta_{\mu\nu}$ and $p_{i\mu} p_{j\nu}$ terms.
The integral representations for the form factor $\cV^{\ada}_{\saagen}$ are 
the following:
\bqa \cV^{\ada}_{\saagen} &=&
-\,\egam{2+\ep}\,\CIM{V}_{\ada}\,
R_{\adan{\sabgen}}\,\chiu{\ada}^{-2-\ep}, \quad i\neq 4\, ,
\nl 
\cV^{\ada}_{\saad} &=& -\,\egam{1+\ep}\,\CIM{V}_{\ada}\,
\Bigl\{ -\frac{x(1-x)}{2-\ep}\,\Bigl[ \frac{4-\ep}{2}\,
+\,(1+\ep)\,\chiu{\ada}^{-1}\,R_{\ada}\Bigr]
+\,\frac{1}{2}\,x^2\,\Bigr\}\,\chiu{\ada}^{-1-\ep},
\nn
\eqa
\bq
R_{\adan{\saagen}} = x^2\,R_{\adan{\sbbgen}}, \qquad
R_{\ada} =  F(z_1,z_2) + m_x^2, 
\eq
with $F$ and $m^2_x$ defined in \eqn{defGfun}.
Consider now the generalized $\cV^{\alpha_1 | \alpha_2 , \alpha_3 ,
\alpha_4| \alpha_5 }$ function where the propagator carrying mass $m_i$ is
raised to the $\alpha_i$ power:
\bqa
\cV^{\alpha_1 | \alpha_3  , \alpha_4 , \alpha_5| \alpha_2 }_{\ada} (n) &=& - 
\frac{\egam{2 +\ep}}{\Pi_{i=1}^5\,\egam{\alpha_i}}\, 
\omega^{6-\sum_{j=1}^5\alpha_j+\ep}\,\dcubs{x}{y,z_1,z_2} 
\nl 
&\times& (1-y)^{\rho_1}\,(y-z_1)^{\rho_2}\,(z_1-z_2)^{\rho_3}\,
x^{\rho_4}\,(1-x)^{\rho_5}\,z_2^{\rho_6}\,\chiu{\ada}^{-2-\ep}.
\label{genada}
\eqa
The space-time dimension is $n = \sum_{i=1}^5 \alpha_i - 2 - \ep$ and the
various powers appearing in \eqn{genada} are:
\bqa
\rho_1 &=& \alpha_1 + \alpha_2 - \frac{1}{2}\,( \sum \alpha  - \ep),
\qquad
\rho_2 = \alpha_3-1,
\qquad
\rho_3 = \alpha_4 - 1,
\nl
\rho_4 &=& \frac{1}{2}\,\sum \alpha - \alpha_1 - 2 -  \frac{1}{2}\,\ep,
\qquad
\rho_5 = \frac{1}{2}\,\sum \alpha - \alpha_2 - 2 - \frac{1}{2}\,\ep,
\qquad
\rho_6 = \alpha_5 - 1.
\eqa
$\omega$ is defined in \eqn{defDUV}.
Results for this family are summarized in \appendx{summaada}.
$V^{\ada}_{\Z} \equiv V^{141}_{\Z}$ is discussed in Sects.~8.1--8.2 of III
(see comment at the end of \sect{ranktwoaba}), evaluation of form factors 
in \sect{exaacaada}.
\subsection{The $V^{\bba}$-family ($\alpha = 2, \beta = 2, \gamma = 1$)
\label{bbafam} }
We continue our analysis considering the scalar function in the 
$V^{\bba}$-family of \fig{TLvertbba} which is ultraviolet convergent with all 
its sub-diagrams and is representable as
\bq
\pi^4\,\cV^{\bba}_{\Z}(p_1,p_1,P,\lstm{12345}) =
\mu^{2\ep}\,\intmomii{n}{q_1}{q_2}\,
\frac{1}{[1][2]_{\bba}[3]_{\bba}[4]_{\bba}[5]_{\bba}},
\eq
with propagators
\bqa
[1] &\equiv& q^2_1 + m^2_1,  \quad
[2]_{\bba} \equiv (q_1+p_1)^2+ m^2_2,  \quad
[3]_{\bba} \equiv (q_1-q_2)^2 + m^2_3,
\nn
\eqa
\bqa
[4]_{\bba} &\equiv& (q_2+p_1)^2 + m^2_4,  \quad
[5]_{\bba} \equiv (q_2+P)^2 + m^2_5.
\label{def221}
\eqa
This family represents the first case where the scalar configuration is 
ultraviolet finite while tensor integrals are divergent.
\begin{figure}[ht]
\begin{center}
\begin{picture}(150,75)(0,0)
 \SetWidth{1.5}
 \Line(0,0)(40,0)         
\LongArrow(0,8)(20,8)          \Text(-11,7)[cb]{$-P$}
 \Line(128,-53)(100,-35)  
\LongArrow(128,-63)(114,-54)   \Text(138,-70)[cb]{$p_1$}
 \Line(128,53)(100,35)    
\LongArrow(128,63)(114,54)     \Text(138,62)[cb]{$p_2$}
 \Line(100,-35)(40,0)           \Text(70,-33)[cb]{$1$}
 \Line(100,0)(100,35)           \Text(107,-21)[cb]{$2$}
 \Line(100,0)(40,0)             \Text(83,5)[cb]{$3$}
 \Line(100,-35)(100,0)          \Text(107,15)[cb]{$4$}
 \Line(100,35)(40,0)            \Text(70,25)[cb]{$5$}
 \Text(0,-25)[cb]{\Large $V^{\bba}$}
\end{picture}
\end{center}
\vspace{2cm}
\caption[]{The irreducible two-loop vertex diagrams $V^{\bba}$. External 
momenta flow inwards. Internal masses are enumerated according to 
\eqn{def221}.} 
\label{TLvertbba}
\end{figure}
\subsubsection{Vector integrals in the $V^{\bba}$ family \label{rankonebba}}
Decomposition of vector integrals follows in the usual way:
\bqa 
\cV^{\bba}(\mu\,|\,0\,;\,p_1,p_1,P,\lstm{12345}) &=&
\sum_{i=1}^{2}\cV^{\bba}_{\sagen}(p_1,p_1,P,\lstm{12345})\,p_{i\mu} , 
\nl 
\cV^{\bba}(0\,|\,\mu\,;\,p_1,p_1,P,\lstm{12345}) &=& 
\sum_{i=1}^{2}\cV^{\bba}_{\sbgen}(p_1,p_1,P,\lstm{12345})\,p_{i\mu} , 
\label{ffbba} 
\eqa
where the form factors $\cV^{\bba}_{\sijgen}(p,p,k,\lstm{a \cdots e})$ refer to 
the basis $p_{\mu}$ and $(k - p)_{\mu}$. Their explicit expression is
\bqa
\cV^{\bba}_{\sijgen} &=&
-\,\egam{1+\ep}\,\CIM{V}_{\bba}\,P_{\bban{ij}}\,\chiu{\bba}^{-1-\ep},
\nl
\CIM{V}_{\bba} &=& \omega^{\ep}\,
\dsimp{2}(\{x\})\,\dsimp{2}(\{y\})\,
\Bigl[ x_2\,(1-x_2)\Bigr]^{-1-\ep/2}\,y^{\ep/2}_2,
\nl
P_{\bban{\szz}} &=& 1, \quad
P_{\bban{\saa}} =  - 1 + x_1 - x_2\,(1 - y_2) - x_2\,y_2\,X,
\quad
P_{\bban{\sab}} = x_2\,(y_1 - 1),
\nl
P_{\bban{\sba}} &=&  - 1 + y_2\,\bX, \quad
P_{\bban{\sbb}} = y_1 -1,
\label{elimi}
\eqa
where $\omega$ is defined in \eqn{defDUV} and $X = (1-x_1)/(1-x_2) = 1 - \bX$.
The polynomial $\chiu{\bba}$ is given by
\bqa
\chiu{\bba} &=& \left[x_2(1-x_2)\right]^{-1}\Bigl\{
-\,\frac{1}{\ox_2}\,F(\ox\,y_2\,,\,\ox_2\,y_1 )
+ \ox_2\,l_{254}\,y_1 
\nl
{}&+& \Bigl[ \ox_2\,(M_x^2 - m_4^2 - p^2_2 + P^2) - 
\ox_1\,(p^2_1 - p^2_2 + P^2)\Bigr]\,y_2 + \ox_2\,M_x^2 \Bigr\},
\label{Defchibba}
\eqa
with $F$ defined in \eqn{defGfun} and
\bq
x_2\,\ox_2\,M_x^2 = \ox\,m_1^2 + \ox_1\,m_2^2
 + x_2\,m_3^2 + x_1\,\ox_1\,p^2_1,
\eq
where $\ox_i = 1- x_i$, $\ox = x_1 - x_2$.
All these functions can be manipulated according to the procedure
introduced in III. The generic scalar function in this family is
\bqa
\cV^{\alpha_1,\alpha_2|\alpha_3,\alpha_4|\alpha_5}_{\bba}
(n) &=& 
-\,\frac{\egam{1+\ep}}{\prod_{i=1}^5\egam{\alpha_i}}\,
\omega^{5-\sum_{j=1}^5\alpha_j+\ep}\,
\dsimp{2}(\{x\})\,\dsimp{2}(\{y\})\,\chiu{\bba}^{-1-\ep}
\nl
{}&\times& (x_1-x_2)^{\rho_1}\,(1-x_1)^{\rho_2}\,
(y_1-y_2)^{\rho_3}\,(1-y_1)^{\rho_4}\,(1-x_2)^{\rho_5}\, 
y^{\rho_6}_2\,x^{\rho_7}_2,
\label{genbba}
\eqa
with $\omega$ defined in \eqn{defDUV}, the dimension $n = \sum_{i=1}^5\,
\alpha_i - 1 - \ep$ and powers
\bqa
\rho_i &=& \alpha_i - 1, \quad i = 1,\cdots,4,
\qquad
\rho_5 = \frac{1}{2}\,\sum\alpha - \alpha_5 - \alpha_1 - \alpha_2 -
\frac{1}{2}\,(1 + \ep),
\nl
\rho_6 &=& \alpha_5 + \alpha_1 + \alpha_2 - \frac{1}{2}\,(\sum\alpha + 1 - \ep),
\qquad
\rho_7 = \frac{1}{2}\,\sum\alpha - \alpha_1 -  \alpha_2 - 
\frac{1}{2}\,( 3 + \ep).
\eqa
Also for this case we have partial reducibility,
\bqa
\cV^{\bba}(p_1\,|\,0\,;\,p_1,p_1,P,\lstm{12345}) &=&
 \frac{1}{2}\,\Bigl[ -\,l_{112}\,
\cV^{\bba}_{\Z}(p_1,p_1,P,\lstm{12345}) + 
\cV^{\aba}_{\Z}(p_1,P,\lstm{1345})
\nl
{}&-&
\cV^{\aba}_{\Z}(0,p_2,\lstm{2345})\Bigr],
\nl
\cV^{\bba}(0\,|\,p_2\,;\,p_1,p_1,P,\lstm{12345}) &=&
\frac{1}{2}\,\Bigl[(- l_{245} - 2\,p_{12})\,
\cV^{\bba}_{\Z}(p_1,p_1,P,\lstm{12345})
-\,\cV^{\aba}_{\Z}(-p_2,-P,\lstm{5321})
\nl
{}&+&\,\cV^{\aba}_{\Z}(0,-p_1,\lstm{4321})\Bigr]. 
\label{ombba}
\eqa
Note that $\cV^{\aba}_{\Z}(0,p)$ is equivalent to two-point functions of the 
$S^{\ssC}$ family, \eqn{Deftltp}. The system of equations that we obtain is
\bqa
\cV^{\bba}(p_1\,|\,0) &=&
p^2_1\,\cV^{\bba}_{\saa} + p_{12}\,\cV^{\bba}_{\sab},
\qquad
\cV^{\bba}(0\,|\,p_2) =
p_{12}\,\cV^{\bba}_{\sba} + p^2_2\,\cV^{\bba}_{\sbb}.
\label{systembba}
\eqa
Assuming that $p^2_1 \ne 0$ we can eliminate the integral with $P_{\saa}$
in \eqn{elimi} in favor of the integral with $P_{\sab}$ which contains the
factor $x_2(y_1-1)$ and obtain the generalized function with
$\alpha_1 = \alpha_2 = 1$, $\alpha_3 = 1, \alpha_4 = 2$, $\alpha_5= 2$
corresponding to $n = 6-\ep$, i.e.\
\bq
\cV^{\bba}_{\sab} = - \omega^2\,\cV^{1,1|1,2|2}_{\bba}(n = 6 - \ep).
\eq
Under the same assumption we eliminate the integral with $P_{\sba}$ in favor
of the integral with $P_{\sbb}$ which contains a factor $1-y_1$ and obtain a
combination of three generalized functions
\bq
\cV^{\bba}_{\sbb} = -\,\omega^2\,\Bigl[ \cV^{1,2|1,2|1}_{\bba} +
\cV^{2,1|1,2|1}_{\bba} + \,\cV^{1,1|1,2|2}_{\bba}\Bigr]\,\bmid_{n = 6-\ep}.
\label{GStt}
\eq
\subsubsection{Rank two tensor integrals in the $V^{\bba}$ family
\label{ranktwobba}}
Tensor integrals become ultraviolet divergent; with $q_{i\mu}\,q_{i\nu}$
in the numerator the integrals are overall divergent with $(\alpha,\gamma)$
or $(\beta,\gamma)$ sub-diagram divergent. With $q_{1\mu}\,q_{2\nu}$
the function is overall divergent (but sub-diagrams are convergent).

Henceforth we want to analyze the tensor integrals with two powers of 
$q_2$ in the numerator: adopting the usual decomposition in form factors we 
have that
\bqa 
\cV^{\bba}(0 |\mu,\nu\ ;\,\cdots) &=& 
\cV^{\bba}_{\sbba}\,p_{1\mu}\,p_{1\nu} + 
\cV^{\bba}_{\sbbb}\,p_{2\mu}\,p_{2\nu} + 
\cV^{\bba}_{\sbbc}\,\{p_1 p_2\}_{\mu\nu} + 
\cV^{\bba}_{\sbbd}\,\delta_{\mu\nu} 
\label{tensorbba} \, ,
\eqa
with the symmetrized product of \eqn{symmete}.
The integral representation for the form factors introduced in \eqn{tensorbba}
is given by
\bqa 
\cV^{\bba}_{\sbbgen} &=&
-\,\egam{1+\ep}\,\CIM{V}_{\bba}\,
R_{\bban{\sbbgen}}\,\chiu{\bba}^{-1-\ep}, \quad i \neq 4,
\qquad
\cV^{\bba}_{\sbbd} = -\,\frac{1}{2}\,\egam{\ep}\,\CIM{V}_{\bba}\,
\chiu{\bba}^{-\ep},
\nl 
R_{\bban{\sbba}} &=& \bY_2^2, \quad 
R_{\bban{\sbbb}} = (1- y_1)^2, \quad
R_{\bban{\sbbc}} = (1-y_1)\,\bY_2, 
\eqa
where $X = (1-x_1)/(1-x_2)$ and $\bY_2 = 1 - y_2\,\bX$.
The reduction of the form factors of the $\cV^{\bba}_{\sbbgen}$
family proceeds as follows: at first we take the trace in both sides
of \eqn{tensorbba} and obtain
\bqa 
{}&{}&
p^2_1\,\cV^{\bba}_{\sbba} + 2\,p_{12}\,
\cV^{\bba}_{\sbbc} +p^2_2\,\cV^{\bba}_{\sbbb} + n\,
\cV^{\bba}_{\sbbd} = -(p^2_1 + m_4^2)
\cV_{\Z}^{\bba}(p_1,p_1,P,\lstm{12345}) 
\nl 
&-& 2\,p^2_1\,\cV_{\sba}^{\bba}(p_1,p_1,P,\lstm{12345}) -
2\,p_{12}\,\cV_{\sbb}^{\bba}(p_1,p_1,P,\lstm{12345}) + 
\cV_{\Z}^{\aba}(-p_2,-P,\lstm{5321}) 
\label{af013}.
\eqa
Similarly, contracting \eqn{tensorbba} with $p_{2\nu}$, we get
additional relations
\bqa 
p_{12}\,\cV^{\bba}_{\sbba} + p^2_2\,\cV^{\bba}_{\sbbc} &=&
 \frac{1}{2}\,\Bigl[ ( l_{154} - P^2 )\,
\cV^{\bba}_{\sba}(p_1,p_1,P,\lstm{12345})
\nl
&+& \,\cV^{\aba}_{\saa}(-p_2,-P,\lstm{5321}) -
\,\cV^{\aba}_{\saa}(0,-p_1,\lstm{4321}) 
\nl
&+& \,\cV^{\aba}_{\Z}(-p_2,-P,\lstm{5321}) -
\,\cV^{\aba}_{\Z}(0,-p_1,\lstm{4321})\Bigr], 
\nl
p^2_2\,\cV^{\bba}_{\sbbb} + p_{12}\,
\cV^{\bba}_{\sbbc} + \cV^{\bba}_{\sbbd} &=& \,  
\frac{1}{2}\,\Bigl[ (l_{154} - P^2 )\,
\cV^{\bba}_{\sbb}(p_1,p_1,P,\lstm{12345})
\nl
&+& \,\cV^{\aba}_{\sab}(-p_2,-P,\lstm{5321}) +
\,\cV^{\aba}_{\Z}(0,-p_1,\lstm{4321})\Bigr].
\label{af012} 
\eqa
In \eqn{af012}, where necessary, the momenta have been permuted
to bring the integrand in the standard form of \eqn{explist}.

\eqns{af013}{af012} can be solved for the form factors with $i < 4$ 
when we use one generalized scalar function, 
\bq
\cV^{\bba}_{\sbbd} = \frac{1}{2}\,\omega^2\,\cV^{1,1|1,1|2}(n = 6 -\ep) \, .
\eq
The $q_{1\mu}\,q_{2\nu}$ tensor integral can be expressed in terms of 
form factors as follows:
\bqa 
\cV^{\bba}(\mu|\nu\ ;\,\cdots) &=& 
\cV^{\bba}_{\saba}\,p_{1\mu}\,p_{1\nu} + 
\cV^{\bba}_{\sabb}\,p_{2\mu}\,p_{2\nu} + 
\cV^{\bba}_{\sabc}\,p_{1\mu}\,p_{2\nu} +\cV^{\bba}_{\sabe}\,
p_{1\nu}\,p_{2\mu} + \cV^{\bba}_{\sabd}\,\delta_{\mu\nu}
\label{tensorbba2} \, .
\eqa
$\cV^{\bba}(\mu|\nu\ ;\,\cdots)$ is not symmetric in $\mu$ and $\nu$, and we 
have to distinguish between $\cV^{\bba}_{\sabc}$ and $\cV^{\bba}_{\sabe}$. 
The integral representation for the form factors of \eqn{tensorbba2} is:
\bqa 
\cV^{\bba}_{\sabgen} &=& -\,\egam{1+\ep}\,\CIM{V}_{\bba}\,
R_{\bban{\sabgen}}\,\chiu{\bba}^{-1-\ep}, \quad i \neq 4, 
\qquad 
\cV^{\bba}_{\sabd} = -\,\frac{1}{2}\,\egam{\ep}\,\CIM{V}_{\bba}\,
x_2\,\chiu{\bba}^{-\ep},
\nl
R_{\bban{\saba}} &=& \bY_2\,( 1-x_1+x_2\,\bY_2)\,, 
\quad
R_{\bban{\sabb}} = x_2\,(1-y_1)^2 ,
\nl 
R_{\bban{\sabc}} &=& (1-y_1)\,( 1-x_1+x_2\,\bY_2), 
\quad 
R_{\bban{\sabe}} = (1-y_1)\,x_2\,\bY_2 . 
\label{indep}
\eqa
Since the $q_1$ sub-diagram involves three propagators, it is not
possible to rewrite $\cV^{\bba}(\mu|\nu)$ in terms of 
$\cV^{\bba}(0|\mu ,\nu)$. We can, however, express the five 
form factors of \eqn{tensorbba2} in terms of scalar function employing the 
same technique adopted for the $\cV^{\bba}_{\sbbgen}$ form factors. In fact,
taking the trace of both sides of \eqn{tensorbba2} one obtains the relation
\bqa 
{}&{}&
p^2_1\,\cV^{\bba}_{\saba} + p_{12}\,(
\cV^{\bba}_{\sabc} + \cV^{\bba}_{\sabe}) + p^2_2\,\cV^{\bba}_{\sabb}
+ n\,\cV^{\bba}_{\sabd} = \frac{1}{2}\Bigl[\,m^2_{31}\,
\cV_{\Z}^{\bba}(p_1,p_1,P,\lstm{12345}) 
\nl 
&+& \cV^{\bba}(0|\mu, \mu ; p_1,p_1,P,\lstm{12345}) +
\cV_{\Z}^{\aba}(0,p_2,\lstm{2345}) + B_{\Z}(p_1,\lstm{12})\,
B_{\Z}(p_2,\lstm{45})\Bigr]
\label{aff1}; 
\eqa
contracting \eqn{tensorbba2} with $p_{1\mu}$ we have
\bqa 
p^2_1\,\cV^{\bba}_{\saba} + p_{12}\,\cV^{\bba}_{\sabe} + 
\cV^{\bba}_{\sabd} &=& 
\frac{1}{2}\,\Bigl[ -\,l_{112}\,\cV^{\bba}_{\sba}(p_1,p_1,P,\lstm{12345})
\nl
&+& \,\cV^{\aba}_{\sbb}(p_1,P,\lstm{1345}) -
 \,\cV^{\aba}_{\Z}(0,p_2,\lstm{2345})\Bigr], 
\nl
p^2_1\,\cV^{\bba}_{\sabc} + p_{12}\,\cV^{\bba}_{\sabb}  &=& 
\frac{1}{2}\,\Bigl[ -\,\,l_{112}\,\cV^{\bba}_{\sbb}(p_1,p_1,P,\lstm{12345})
\nl
&+& \,\cV^{\aba}_{\sba}(p_1,P,\lstm{1345}) -
\,\cV^{\aba}_{\sba}(0,p_2,\lstm{2345})\Bigr];
\label{aff2} 
\eqa
finally, contracting \eqn{tensorbba2} with $p_{2\nu}$ we have
\bqa
p^2_2\,\cV^{\bba}_{\sabc} + p_{12}\,\cV^{\bba}_{\saba} &=& 
\frac{1}{2}\,\Bigl[ (l_{145} - P^2 )
\,\cV^{\bba}_{\saa}(p_1,p_1,P,\lstm{12345})
- \,\cV^{\aba}_{\saa}(0,-p_1,\lstm{4321}) 
\nl
{}&+& \cV^{\aba}_{\sab}(-p_2,-P,\lstm{5321}) 
- \,\cV^{\aba}_{\Z}(0,-p_1,\lstm{4321}) +
\,\cV^{\aba}_{\Z}(-p_2,-P,\lstm{5321})\Bigr] 
\, , 
\nl
 p^2_2\,\cV^{\bba}_{\sabb} + p_{12}
\cV^{\bba}_{\sabe} + \cV^{\bba}_{\sabd} &=& 
\,\frac{1}{2}\,\Bigl[ (l_{145} - P^2 )\,
\cV^{\bba}_{\sab}(p_1,p_1,P,\lstm{12345})
\nl
&+& \,\cV^{\aba}_{\sbb}(-p_2,-P,\lstm{5321}) +
\,\cV^{\aba}_{\Z}(-p_2,-P,\lstm{5321})\Bigr].
\label{aff3} 
\eqa
Furthermore we have
\bq
\cV^{\bba}_{\sabd}= \frac{1}{2}\,\omega^2\,\cV^{1,1|1,1| 2}_{\bba}\,
\bmid_{n= 6-\ep}.
\eq
The system composed by \eqns{aff1}{aff3} gives the form factors with $i \ne 4$.

It is now necessary to analyze the form factors of the 
$\cV^{\bba}_{\saagen}$ family, which are defined through the relation
\bqa 
\cV^{\bba}(\mu ,\nu|0\,;\,\cdots) &=& 
\cV^{\bba}_{\saaa}\,p_{1\mu}\,p_{1\nu} + 
\cV^{\bba}_{\saab}\,p_{2\mu}\,p_{2\nu} + 
\cV^{\bba}_{\saac}\,\{p_1 p_2\}_{\mu\nu} + 
\cV^{\bba}_{\saad}\,\delta_{\mu\nu} 
\label{tensorbba3} \, ,
\eqa
with $\{p_1 p_2\}$ defined in \eqn{symmete}.
The integral representation for these form factors can be obtained
with standard techniques:
\bqa 
\cV^{\bba}_{\saagen} &=& -\egam{1+\ep}\,\CIM{V}_{\bba}\,
R_{\bban{\saagen}}\,\chiu{\bba}^{-1-\ep}, \qquad i \neq 4 \,, 
\nl
R_{\bban{\saaa}} &=& (1-x_1 + x_2\,\bY)^2, \quad
R_{\bban{\saab}} = x_2^2\,(1-y_1)^2 \quad 
R_{\bban{\saac}} = (1-y_1)\,x_2\,( 1-x_1+x_2\,\bY_2), \,  
\nl 
\cV^{\bba}_{\saad} &=& -\frac{1}{2}\,\egam{\ep}\,\CIM{V}_{\bba}\,
\frac{x_2}{y_2}\,( 1-x_2 + x_2\,y_2 )\,\chiu{\bba}^{-\ep}\, . 
\label{hidden}
\eqa
In order to reduce the form factors to linear combination of
scalar functions, we start by taking the trace in both sides of
\eqn{tensorbba3} so that we obtain
\bqa 
p^2_1\,\cV^{\bba}_{\saaa} + 2\,p_{12}\,\cV^{\bba}_{\saac} +
p^2_2\,\cV^{\bba}_{\saab} + n\,\cV^{\bba}_{\saad} &=& -m_1^2\,
\cV_{\Z}^{\bba}(p_1,p_1,P,\lstm{12345}) 
+ \cV^{\aba}_{\Z}(0,p_2,\lstm{2345}) 
\label{aff11}. 
\eqa
Contracting \eqn{tensorbba3} with $p_{1\mu}$ we have additional
relations,
\bqa 
p^2_1\,\cV^{\bba}_{\saaa} + p_{12}\cV^{\bba}_{\saac} + 
\cV^{\bba}_{\saad} &=& 
\frac{1}{2}\,\Bigl[ -\,l_{112}\,\cV^{\bba}_{\saa}(p_1,p_1,P,\lstm{12345})
\nl
&+& \,\cV^{\aba}_{\sab}(p_1,P,\lstm{1345}) +
\,\cV^{\aba}_{\Z}(0,p_2,\lstm{2345})\Bigr], 
\nl
p^2_1\,\cV^{\bba}_{\saac} + p_{12}\,\cV^{\bba}_{\saab} &=& 
\frac{1}{2}\,\Bigl[ -\,l_{112}\,\cV^{\bba}_{\sab}(p_1,p_1,P,\lstm{12345})
\nl
&+& \,\cV^{\aba}_{\saa}(p_1,P,\lstm{1345}) -
\,\cV^{\aba}_{\saa}(0,p_2,\lstm{2345})\Bigr].
\label{aff21} 
\eqa
It is not possible to obtain more equations by multiplying both
sides of \eqn{tensorbba2} by $p_{2\mu}$, since the scalar product 
$\spro{q_1}{p_2}$ is irreducible. 
\eqns{aff11}{aff21} give the form factors with $i \ne 2$ when we 
introduce one generalized scalar function,
\bq 
\cV^{\bba}_{\saab} = 4\,\omega^4\,
\cV^{1 , 1 | 1 , 3 | 3}_{\bba} (n = 8 - \ep)\,.
\eq
There are other equivalent solutions.
Results for this family are summarized in \appendx{summabba}.
$V^{\bba}_{\Z} \equiv V^{221}_{\Z}$ is discussed in Sect.~7.1 of III
(see comment at the end of \sect{ranktwoaba}), evaluation of form factors 
in \sect{exabba}.
\subsection{The $V^{\bca}$-family ($\alpha = 2, \beta = 3, \gamma = 1$)
\label{bcafam} }
Next we consider the scalar diagram in the $V^{\bca}$-family of
\fig{TLvertbca}, which is overall ultraviolet convergent (with all
sub-diagrams convergent) and which is representable as
\bq
\pi^4\,\cV^{\bca}_{\Z}(P,p_1,P,\lstm{123456}) =
\mu^{2\ep}\,\intmomii{n}{q_1}{q_2}\,
\frac{1}{[1][2]_{\bca}[3]_{\bca}[4]_{\bca}[5]_{\bca}[6]_{\bca}},
\eq
with propagators
\bqa
[1] &\equiv& q^2_1 + m^2_1,  \quad
[2]_{\bca} \equiv (q_1+P)^2+ m^2_2,  \quad
[3]_{\bca} \equiv (q_1-q_2)^2 + m^2_3,
\nn
\eqa
\bqa
[4]_{\bca} &\equiv& q^2_2 + m^2_4,  \quad
[5]_{\bca} \equiv (q_2+p_1)^2 + m^2_5,  \quad
[6]_{\bca} \equiv (q_2+P)^2 + m^2_6.
\label{defbca}
\eqa
\begin{figure}[bh]
\begin{center}
\begin{picture}(150,75)(0,0)
 \SetWidth{1.5}
 \Line(0,0)(40,0)         
\LongArrow(0,8)(20,8)          \Text(-11,7)[cb]{$-P$}
 \Line(128,-53)(100,-35)  
\LongArrow(128,-63)(114,-54)   \Text(138,-70)[cb]{$p_1$}
 \Line(128,53)(100,35)    
\LongArrow(128,63)(114,54)     \Text(138,62)[cb]{$p_2$}
 \Line(70,-17.5)(40,0)             \Text(53,-21)[cb]{$1$}
 \Line(70,17.5)(40,0)              \Text(53,14)[cb]{$2$}
 \Line(70,-17.5)(70,17.5)          \Text(77,-3)[cb]{$3$}
 \Line(100,-35)(70,-17.5)          \Text(82,-39)[cb]{$4$}
 \Line(100,-35)(100,35)            \Text(107,-3)[cb]{$5$}
 \Line(100,35)(70,17.5)            \Text(82,31)[cb]{$6$}
 \Text(0,-25)[cb]{\Large $V^{\bca}$}
\end{picture}
\end{center}
\vspace{2cm}
\caption[]{The irreducible two-loop vertex diagrams $V^{\bca}$. 
External momenta flow inwards. Internal masses are enumerated 
according to \eqn{defbca}.} 
\label{TLvertbca}
\end{figure}
\subsubsection{Vector integrals in the $V^{\bca}$ family \label{rankonebca}}
The form factors for the vector integrals are defined by
\bqa
\cV^{\bca}(\mu\,|\,0\,;\,P,p_1,P,\lstm{123456}) &=&
\sum_{i=1}^2 \cV^{\bca}_{\sagen}(P,p_1,P,\lstm{123456})\,p_{i\mu} ,
\nl
\cV^{\bca}(0\,|\,\mu\,;\,P,p_1,P,\lstm{123456}) &=&
\sum_{i=1}^2 \cV^{\bca}_{\sbgen}(P,p_1,P,\lstm{123456})\,p_{i\mu} ,
\label{ffbca}
\eqa
where the form factors $\cV^{\bca}_{\sijgen}(k,p,k,\lstm{a \cdots f})$ refer 
to the basis $p_{\mu}$ and $(k - p)_{\mu}$. Their explicit expression is
\bqa
\cV^{\bca}_{\sijgen} &=& -\,\egam{2+\ep}\,\CIM{V}_{\bca}\,P_{\bcan{ij}}\,
\chiu{\bca}^{-2-\ep}\,
,
\nl
\CIM{V}_{\bca} &=& \omega^{\ep}\,
\dsimp{2}(\{x\})\,\dsimp{3}(\{y\})\,
\Bigl[ x_2\,(1-x_2)\Bigr]^{-1-\ep/2}\,y^{\ep/2}_3,
\nl
P_{\bcan{\szz}} &=& 1, \quad\!\!\!
P_{\bcan{\saa}} = -\,H_2,\quad\!\!\!
P_{\bcan{\sab}} = -\,H_1,\quad\!\!\!
P_{\bcan{\sba}} = Y_2, \quad\!\!\!
P_{\bcan{\sbb}} = Y_1,\quad
\label{varaibca}
\eqa
where $\omega$ is defined in \eqn{defDUV}, and the quantities $Y_i$ and
$H_i$ are given in \eqn{HY1HY2}.
The polynomial $\chiu{\bca}$ is given by
\bqa
\chiu{\bca} &=& -\,F(y_2 - X\,y_3\,,\,y_1 - X\,y_3)
+ l_{265}\,y_1 +( P^2 - l_{245} )\,y_2
-  ( 2\,X\,P^2 - m_{xx}^2 + m_4^2 )\,y_3 + m^2_6,
\label{Defchibca}
\eqa
with $F$ defined in \eqn{defGfun} and
\bqa
\nn 
m_{xx}^2 = \frac{-P^2  x_1^2 + x_1  (P^2 + m_{12}^2) + 
x_2  m_{312}^2 }{x_2 (1-x_2)},
\eqa
with $X = (1-x_1)/(1-x_2)$. The generalized function in this family is
\bqa
{}&{}&
\cV^{\alpha_1,\alpha_2|\alpha_4,\alpha_5,\alpha_6|\alpha_3}_{\bca}
(n = \sum_{i=1}^6\alpha_i-2-\ep) =
\pi^{-4}\,(\mu^2)^{4-n}\,\int\,d^nq_1\,d^nq_2\,\prod_{i=1}^{6}\,
[i]^{-\alpha_i}_{\bca}
\nl
{}&=& -\,
\frac{\egam{2+\ep}}{\prod_{i=1}^6\egam{\alpha_i}}\,
\omega^{6-\sum_{j=1}^6\alpha_j+\ep}\,
\dsimp{2}(\{x\})\,\dsimp{3}(\{y\})\,
(1-x_1)^{\rho_1}
\nl
{}&\times&\,(x_1-x_2)^{\rho_2}\, x^{\rho_3}_2\,(1-x_2)^{\rho_4}\,(1-y_1)^{\rho_5}\,
(y_1-y_2)^{\rho_6}\,(y_2-y_3)^{\rho_7}\,y^{\rho_8}_3\,
\chiu{\bca}^{-2-\ep},
\label{genbca}
\eqa
where $[1]_{\bca} \equiv [1]$, $\omega$ is defined in \eqn{defDUV} and the 
powers $\rho_i$ are
\bqa
\rho_1 &=& \alpha_2 -1, \quad
\rho_2 = \alpha_1 -1, \quad
\rho_3 = \frac{1}{2}\,(\sum \alpha -2\,\alpha_1 -2\,\alpha_2 - 4 - \ep),
\nl
\rho_4 &=& \frac{1}{2}\,(\sum \alpha -2\,\alpha_1 -2\,\alpha_2 -2\,\alpha_3- 
2 - \ep), \quad
\rho_5 = \alpha_6 - 1, \quad
\rho_6 = \alpha_5 -1,
\nl
\rho_7 &=& \alpha_4 -1, \quad
\rho_8 = \frac{1}{2}\,
(\ep - \sum \alpha + 2\,\alpha_1 +2\,\alpha_2 +2\,\alpha_3).
\eqa
There is partial reducibility with respect to $q_1$ and complete reducibility
with respect to $q_2$. We obtain
\bqa
\cV^{\bca}(P|0;P,p_1,P,\lstm{123456}) &=&
-\frac{1}{2}\Bigl[ l_{\ssP 12} 
\!\cV^{\bca}_{\Z}(P,p_1,P,\lstm{123456}) \!-\! 
\cV^{\aca}_{\Z}(p_1,P,\lstm{13456}) 
\nl
{}&+& \cV^{\aca}_{\Z}(-p_2,-P,\lstm{23654})\!\Bigr],
\nl
\cV^{\bca}(0|p_1;P,p_1,P,\lstm{123456}) &=&
\!-\frac{1}{2}\Bigl[l_{145}\cV^{\bca}_{\Z}(P,p_1,P,\lstm{123456}) \!+\! 
\cV^{\bba}_{\Z}(P,P,p_1,\lstm{12365}) 
\nl
{}&-& \cV^{\bba}_{\Z}(P,P,0,\lstm{12364})\!\Bigr], 
\nl
\cV^{\bca}(0|P;P,p_1,P,\lstm{123456}) &=&
-\,\frac{1}{2}\Bigl[ l_{\ssP 46}\cV^{\bca}_{\Z}(P,p_1,P,\lstm{123456}) + 
\cV^{\bba}_{\Z}(P,P,p_1,\lstm{12365})
\nl
{}&-& \cV^{\bba}_{\Z}(-P,-P,-p_2,\lstm{21345})\Bigr].
\label{af30}
\eqa
We can write
\bqa
\cV^{\bca}(P\,|\,0\,;\,P,p_1,P,\lstm{123456}) &=&
\spro{p_1}{P}\,\cV^{\bca}_{\saa} +
\spro{p_2}{P}\,\Bigl[\cV^{\bca}_{\saa} - I_{\bcan{\ssR}}\Bigr],
\label{af30bis}
\eqa
\bqa
I_{\bcan{\ssR}} &=& \egam{2+\ep}\,\CIM{V}_{\bca}\,
x_2\,(y_1 - y_2)\,\chiu{\bca}^{-2-\ep} = \omega^2\,
\cV^{1,1|1,2,1|2}_{\bca}(n = 6 -\ep),
\label{Igen}
\eqa
which gives the reduction of the $11$-component. Reduction of the
$12$-component follows from
\bq
\cV^{\bca}_{\sab} = \cV^{\bca}_{\saa} - I_{\bcan{\ssR}}.
\label{af30tris}
\eq
A similar argument holds for the $2i$ components with the same
$I_{\bcan{\ssR}}$, although the reduction of the $\cV^{\bca}_{\sbgen}$
components can also be obtained solving the system composed by the last two
equations of \eqn{af30}.
\subsubsection{Rank two tensor integrals in the $V^{\bca}$ family
\label{ranktwobca}}
Only the $q_{1\mu}\,q_{1\nu}$ tensor integral has an ultraviolet divergent 
$(\alpha,\gamma)$ sub-diagram.

Henceforth we consider the $q_{2\mu}\,q_{2\nu}$ tensor integral:
we introduce the form factors of the $\cV^{\bca}_{\sbbgen}$ family through
the relation
\bqa 
\cV^{\bca}(0|\mu ,\nu\,;\,\cdots) &=& \cV^{\bca}_{\sbba}\,
p_{1\mu}\,p_{1\nu} + \cV^{\bca}_{\sbbb}\,p_{2\mu}\,p_{2\nu} 
+ \cV^{\bca}_{\sbbc}\,\{p_1\,p_2\}_{\mu\nu} + 
\cV^{\bca}_{\sbbd}\,\delta_{\mu\nu} 
\label{tensorbca3} \, ,
\eqa
with the symmetrized product of \eqn{symmete}. Their integral representation
is given by
\bqa 
\cV^{\bca}_{\sbbgen} &=& -\egam{2+\ep}\CIM{V}_{\bca}
R_{\bcan{\sbbgen}}\,\chiu{\bca}^{-2-\ep}, \quad i \neq 4,
\qquad 
\cV^{\bca}_{\sbbd} = -\frac{1}{2}\egam{1+\ep}
\CIM{V}_{\bca}\,\chiu{\bca}^{-1-\ep},
\nn
\eqa
\bq
R_{\bcan{\sbba}} = Y_2^2 , \qquad
R_{\bcan{\sbbb}} = Y_1^2  ,  \qquad
R_{\bcan{\sbbc}} = Y_1\,Y_2.
\eq
We want to express the $\cV^{\bca}_{\sbbgen}$ form factors as linear
combinations of scalar functions. Taking the trace of both sides of
\eqn{tensorbca3} one obtains
\bqa 
p^2_1\,\cV^{\bca}_{\sbba} + 2\,p_{12}\,\cV^{\bca}_{\sbbc} +
p^2_2\,\cV^{\bca}_{\sbbb} + n\,\cV^{\bca}_{\sbbd}
&=& \cV_{\Z}^{\bba}(P,P,p_1,\lstm{12365}) 
- m_4^2\,\cV_{\Z}^{\bca}(P,p_1,P,\lstm{123456}).
\label{af413}
\eqa
Contracting both sides of \eqn{tensorbca3} with $p_{2\mu}$ we get
\bqa 
p_{12}\,\cV^{\bca}_{\sbba} + p^2_2\,\cV^{\bca}_{\sbbc} &=& 
\frac{1}{2}\,\Bigl[ (l_{165} - P^2)\,
\cV^{\bca}_{\sba}(P,p_1,P,\lstm{123456}) -
\,\cV^{\bba}_{\sba}(P,P,0,\lstm{12364}) 
\nl
&-& \,\cV^{\bba}_{\sba}(-P,-P,-p_2,\lstm{21345}) +
\,\cV^{\bba}_{\sbb}(P,P,0,\lstm{12364}) 
\nl 
&+& \,\cV^{\bba}_{\sbb}(-P,-P,-p_2,\lstm{21345}) -
\,\cV^{\bba}_{\Z}(-P,-P,-p_2,\lstm{21345})\Bigr], 
\nl 
p^2_2\,\cV^{\bca}_{\sbbb} + p_{12}\,\cV^{\bca}_{\sbbc} + 
\cV^{\bca}_{\sbbd} &=& 
\frac{1}{2}\,\Bigl[ (l_{165} - P^2 )\,
\cV^{\bca}_{\sbb}(p_1,p_1,P,\lstm{123456}) -
\,\cV^{\bba}_{\sba}(P,P,0,\lstm{12364}) 
\nl
&-& \,\cV^{\bba}_{\sba}(-P,-P,-p_2,\lstm{21345})
+\,\cV^{\bba}_{\sbb}(P,P,0,\lstm{12364}) 
\nl
&-& \,\cV^{\bba}_{\Z}(-P,-P,-p_2,\lstm{21345}) \Bigr]
\label{af412}. 
\eqa
Once again, one should be particularly careful in shifting the integration
momenta in order to bring the integrand of the $V^{\bba}$ functions in the
chosen standard form:
\bqa
\frac{\mu^{2\ep}}{\pi^4}\int d^n r_1 d^n r_2 
\frac{r_{i\mu}}{D_a D_b D_c 
D_d D_e } &\equiv&  
\cV^{\bba}_{\sgena}(k_b,k_b,k_e,\lstm{abcde})\,k_{b\mu}\, 
+ \cV^{\bba}_{\sgenb}(k_b,k_b,k_e,\lstm{abcde})\,(k_e -  k_b)_{\mu},
\eqa
\[
\ba{lll}
D_a = r_1^2 + m_a^2\, , \quad & \quad  
D_b = (r_1 +  k_b)^2 + m_b^2\, , \quad & \quad
D_c = (r_2 - r_2)^2 + m_c^2\, , \\
D_d = (r_1 + k_b)^2 + m_d^2\, , \quad & \quad
D_e = (r_2 + k_e)^2 + m_e^2\,. &
\label{SFG}
\ea
\]
Contracting both sides of \eqn{tensorbca3} with $p_{1\mu}$ we get
\bqa 
p_{12}\,\cV^{\bca}_{\sbbb} + p^2_1\,\cV^{\bca}_{\sbbc} &=& 
\frac{1}{2}\,\Bigl[ -\,l_{145}\,\cV^{\bca}_{\sbb}(P,p_1,P,\lstm{123456}) +
\,\cV^{\bba}_{\sba}(P,P,0,\lstm{12364})
\nl
{}&-& \,\cV^{\bba}_{\sbb}(P,P,0,\lstm{12364})\Bigr], 
\nl
p^2_1\,\cV^{\bca}_{\sbba} + p_{12}\,\cV^{\bca}_{\sbbc} + 
\cV^{\bca}_{\sbbd} &=& 
\frac{1}{2}\,\Bigl[ -\,\,l_{145}\,\cV^{\bca}_{\sba}(P,p_1,P,\lstm{123456}) - 
\,\cV^{\bba}_{\sba}(P,P,p_1,\lstm{12365})
\nl 
{}&+& \,\cV^{\bba}_{\sba}(P,P,0,\lstm{12364})
- \,\cV^{\bba}_{\sbb}(P,P,0,\lstm{12364})\Bigr]. 
\label{af414} 
\eqa
A solution of \eqns{af413}{af414} give the form factors in the $22$ group. 
We can now analyze the $q_{1\mu}\,q_{2 \nu}$ tensor integrals. As 
for the $\cV^{\bba}$ case, this tensor integral is not symmetric 
in $\mu\nu$, so that we need to introduce five form factors:
\bqa 
\cV^{\bca}(\mu |\nu\ ;\,\cdots) &=&
\cV^{\bca}_{\saba}\,p_{1\mu}\,p_{1\nu} +
\cV^{\bca}_{\sabb}\,p_{2\mu}\,p_{2\nu} 
+ \cV^{\bca}_{\sabc}\,p_{1\mu}\,p_{2\nu} +
\cV^{\bca}_{\sabe}\,p_{1\nu}\,p_{2\mu} +
\cV^{\bca}_{\sabd}\,\delta_{\mu\nu}.
\label{tensorbca2}
\eqa
The integral representation of these form factors is the following:
\bqa 
\cV^{\bca}_{\sabgen} &=& -\,\egam{2+\ep}\,\CIM{V}_{\bca}\,
R_{\bcan{\sabgen}}\,\chiu{\bca}^{-2-\ep}, \quad i \neq 4,
\qquad 
\cV^{\bca}_{\sabd} = -\,\frac{1}{2}\,\egam{1+\ep}\,
\CIM{V}_{\bca}\,x_2\,\chiu{\bca}^{-1-\ep},
\nl 
R_{\bcan{\saba}} &=&
-Y_2\,H_2 , \qquad 
R_{\bcan{\sabb}} = -Y_1\,H_1\,  ,  
\qquad
R_{\bcan{\sabc}} = -Y_1\,H_2\, ,
\qquad
R_{\bcan{\sabe}} = -Y_2\,H_1. 
\eqa
Employing the usual procedure we can reduce the form factors. Contracting
\eqn{tensorbca2} with $\delta_{\mu\nu}$, $p_{1\nu}$ and $p_{2\nu}$ we obtain
\bqa 
{}&{}&
p^2_1\,\cV^{\bca}_{\saba} +p_{12}\,(\cV^{\bca}_{\sabc} + 
\cV^{\bca}_{\sabe}) + p^2_2\,\cV^{\bca}_{\sabb}
+ n\,\cV^{\bca}_{\sabd} = -\,\frac{1}{2}\Bigl[\,m^2_{134}\,
\cV_{\Z}^{\bca}(P,p_1,P,\lstm{123456}) 
\nl 
&-& \cV^{\bba}_{\Z}(P,P,p_1,\lstm{12365}) -
\cV_{\Z}^{\aca}(-p_2,-P,\lstm{23654}) - B_{\Z}(P,\lstm{12})\,
C_{\Z}(p_1,p_2,\lstm{456})\Bigr]
\label{aff51}, 
\eqa
%
%
\bqa 
p^2_1\,\cV^{\bca}_{\saba} + p_{12}\,
\cV^{\bca}_{\sabc} + \cV^{\bca}_{\sabd} &=& 
\frac{1}{2}\,\Bigl[ -\,l_{145}\,\cV^{\bca}_{\saa}(P,p_1,P,\lstm{123456}) + 
\,\cV^{\bba}_{\saa}(P,P,0,\lstm{12364}) 
\nl 
&-& \,\cV^{\bba}_{\saa}(P,P,p_1,\lstm{12365}) 
-\,\cV^{\bba}_{\sab}(P,P,0,\lstm{12364})\Bigr],
\nl
p^2_1\,\cV^{\bca}_{\sabe} + p_{12}\,\cV^{\bca}_{\sabb} &=& 
\frac{1}{2}\,\Bigl[ -\,l_{145}\,\cV^{\bca}_{\sab}(P,p_1,P,\lstm{123456}) + 
\,\cV^{\bba}_{\saa}(P,P,0,\lstm{12364}) 
\nl 
&-& \,\cV^{\bba}_{\saa}(P,P,p_1,\lstm{12365}) 
-\,\cV^{\bba}_{\sab}(P,P,0,\lstm{12364}) 
\nl
&+& \,\cV^{\bba}_{\sab}(P,P,p_1,\lstm{12365})\Bigr],
\label{aff52} 
\eqa
%
\bqa 
p^2_2\,\cV^{\bca}_{\sabb} + p_{12}
\cV^{\bca}_{\sabe} + \cV^{\bca}_{\sabd} &=& 
\frac{1}{2}\,\Bigl[ (l_{165} -P^2 )\,
\cV^{\bca}_{\sab}(P,p_1,P,\lstm{123456}) +
\,\cV^{\bba}_{\sab}(P,P,0,\lstm{12364}) 
\nl 
&-& \,\cV^{\bba}_{\saa}(P,P,0,\lstm{12364}) -
\,\cV^{\bba}_{\saa}(-P,-P,-p_2,\lstm{21345})
\nl
&-& \,\cV^{\bba}_{\Z}(-P,-P,-p_2,\lstm{21345})\Bigr],
\nl
p^2_2\,\cV^{\bca}_{\sabc} + p_{12}\,\cV^{\bca}_{\saba} &=& 
\frac{1}{2}\,\Bigl[ (l_{165} -P^2 )\,
\cV^{\bca}_{\saa}(P,p_1,P,\lstm{123456}) 
-\,\cV^{\bba}_{\saa}(P,P,0,\lstm{12364}) 
\nl
&-& \,\cV^{\bba}_{\saa}(-P,-P,-p_2,\lstm{21345}) + 
\,\cV^{\bba}_{\sab}(P,P,0,\lstm{12364}) 
\nl
&+& \,\cV^{\bba}_{\sab}(-P,-P,-p_2,\lstm{21345}) - 
\,\cV^{\bba}_{\Z}(-P,-P,-p_2,\lstm{21345})\Bigr].
\label{aff53} 
\eqa
The solution of \eqns{aff51}{aff53} gives the form factors in the $12$ group.

Finally, we consider the  $q_{1\mu}\,q_{1\nu}$ tensor integral for which 
we introduce the form factors $\cV^{\bca}_{\saagen}$:
\bqa 
\cV^{\bca}(\mu, \nu|0\,;\,\cdots) &=& \cV^{\bca}_{\saaa}\,
p_{1\mu}\,p_{1\nu} + \cV^{\bca}_{\saab}\,p_{2\mu}\,p_{2\nu}  
+ \cV^{\bca}_{\saac}\,\{p_1 p_2\}_{\mu\nu} + 
\cV^{\bca}_{\saad}\,\delta_{\mu\nu} 
\label{tensorbca1} \, ,
\eqa
where the symmetrized product is given in \eqn{symmete}. Their integral 
representation is given by
\bqa 
\cV^{\bca}_{\saagen} &=& -\,\egam{2+\ep}\,\CIM{V}_{\bca}\,
R_{\bcan{\saagen}}\,\chiu{\bca}^{-2-\ep}, \qquad i \neq 4 \,, 
\nl 
R_{\bcan{\saaa}} &=& H_2^2\, , \quad
R_{\bcan{\saab}} = H_1^2\, ,  \quad
R_{\bcan{\saac}} = H_1\,H_2 ,
\nl 
\cV^{\bca}_{\saad} &=& -\,\frac{1}{2}\,\egam{1+\ep}\,\CIM{V}_{\bca}\,
\Bigl[\frac{x_2\,(1-x_2)}{y_3} + x_2^2\Bigr]\,\chiu{\bca}^{-1-\ep}\, . 
\label{thisdiv}
\eqa
$H_1$ and $H_2$ were defined in \eqn{HY1HY2}.  Contracting both sides of
\eqn{tensorbca1} first with $\delta_{\mu\nu}$ and then with $P_{\nu}$, it is
possible to obtain the following set of three equations:
\bqa 
p^2_1\,\cV^{\bca}_{\saaa} +2\,p_{12}\,\cV^{\bca}_{\saac}+
p^2_2\,\cV^{\bca}_{\sabb}
+ n\,\cV^{\bca}_{\sabd} &=& -m_1^2\,\cV_{\Z}^{\bca}(P,p_1,P,\lstm{123456})  
+ \cV_{\Z}^{\aca}(-p_2,-P,\lstm{23654})
\label{haff51}, 
\eqa
\bqa 
\spro{p_1}{P}\,\cV^{\bca}_{\saaa} + \spro{p_2}{P}\cV^{\bca}_{\saac} + 
\cV^{\bca}_{\saad} &=& 
\frac{1}{2}\,\Bigl[ -\,l_{\ssP 12}\,\cV^{\bca}_{\saa}(P,p_1,P,\lstm{123456}) +
\,\cV^{\aca}_{\saa}(p_1,P,\lstm{13456}) 
\nl 
&+& \,\cV^{\aca}_{\sab}(-P,-p_2,\lstm{23645}) +
\,\cV^{\aca}_{\Z}(-P,-p_2,\lstm{23645})\Bigr],
\nl
\spro{p_1}{P}\,\cV^{\bca}_{\saac} + \spro{p_2}{P}\,\cV^{\bca}_{\saab}  + 
\cV^{\bca}_{\saad} &=& 
\frac{1}{2}\,\Bigl[ -\,l_{\ssP 12}\,\cV^{\bca}_{\sab}(P,p_1,P,\lstm{123456}) +
\,\cV^{\aca}_{\sab}(p_1,P,\lstm{13456}) 
\nl 
&+& \,\cV^{\aca}_{\saa}(-P,-p_2,\lstm{23645})  
+ \,\cV^{\aca}_{\Z}(-P,-p_2,\lstm{23645})\Bigr].
\label{haff52} 
\eqa
We have then three equations and four unknown form factors, so that
we should look for relations between form factors and generalized scalar 
function; for example we have that
\bqa
\cV^{\bca}_{\saaa} \!-\! 2\cV^{\bca}_{\saac} \!+\! \cV^{\bca}_{\saab} &=& 
 -\egam{2+\ep}\!\CIM{V}_{\bca}\,
x_2^2 (y_1 - y_2)^2\,\chiu{\bca}^{-2-\ep} =  4\omega^4
\cV^{1,1 | 1 , 3 , 1 | 3}_{\bca}\,\bmid_{n = 8 - \ep} .
\eqa
Results for this family are summarized in \appendx{summabca}.
$V^{\bca}_{\Z} \equiv V^{231}_{\Z}$ is discussed in Sects.~9.1~-~9.2 of III
(see comment at the end of \sect{ranktwoaba}), evaluation of form factors 
in \sect{exabcabbb}.
\subsection{The $V^{\bbb}$-family ($\alpha = 2, \beta = 2, \gamma = 2$)
\label{bbbfam} }
Finally, we consider the non-planar diagram of the $V^{\bbb}$-family, given
in \fig{TLvertbbb}, which is representable as 
\bq
\pi^4\,\cV^{\bbb}_{\Z}(-p_2,p_1,-p_2,-p_1,\lstm{123456}) =
\mu^{2\ep}\,\intmomii{n}{q_1}{q_2}\,
\frac{1}{[1][2]_{\bbb}[3]_{\bbb}[4]_{\bbb}[5]_{\bbb}[6]_{\bbb}},
\eq
with propagators
\bqa
[1] &\equiv& q^2_1 + m^2_1,  \quad
[2]_{\bbb} \equiv (q_1-p_2)^2+ m^2_2,  \quad
[3]_{\bbb} \equiv (q_1-q_2 +p_1)^2 + m^2_3,
\nn
\eqa
\bqa
[4]_{\bbb} &\equiv& (q_1-q_2-p_2)^2 + m^2_4, \quad
[5]_{\bbb} \equiv q^2_2 + m^2_5,  \quad
[6]_{\bbb} \equiv (q_2-p_1)^2 + m^2_6.
\label{def222}
\eqa
The basis for the form factors 
$\cV^{\bbb}_{i \cdots j}(k,p,k,-p,\lstm{a \cdots f})$ is chosen to be
$p_{\mu}$ and $-k_{\mu}$.
\begin{figure}[ht]
\begin{center}
\begin{picture}(150,75)(0,0)
 \SetWidth{1.5}
 \Line(0,0)(40,0)         
\LongArrow(0,8)(20,8)          \Text(-11,7)[cb]{$-P$}
 \Line(128,-53)(100,-35)  
\LongArrow(128,-63)(114,-54)   \Text(138,-70)[cb]{$p_1$}
 \Line(128,53)(100,35)    
\LongArrow(128,63)(114,54)     \Text(138,62)[cb]{$p_2$}
 \Line(70,-17.5)(100,35)                            \Text(97,10)[cb]{$2$}
 \Line(100,35)(70,17.5)                             \Text(82,31)[cb]{$1$}
 \Line(70,-17.5)(40,0)                              \Text(53,-21)[cb]{$4$}
 \Line(70,17.5)(40,0)                               \Text(53,14)[cb]{$3$}
 \Line(100,-35)(82,-3.5)\Line(78,3.5)(70,17.5)      \Text(97,-16)[cb]{$6$}
 \Line(100,-35)(70,-17.5)                           \Text(82,-39)[cb]{$5$}
 \Text(0,-25)[cb]{\Large $V^{\bbb}$}
\end{picture}
\end{center}
\vspace{2cm}
\caption[]{The irreducible two-loop vertex diagrams $V^{\bbb}$. External 
momenta flow inwards. Internal masses are enumerated according to
\eqn{def222}.} 
\label{TLvertbbb}
\end{figure}
All members of this family, including rank-two tensors, are overall ultraviolet
convergent with all sub-diagrams convergent.

Adopting the parametrization presented in Sect.~10.2 of III, the integral 
representation for the scalar integral of the $\cV^\bbb$ family, with 
arbitrary powers for the propagators, is
\bqa
{}&{}&
\cV^{\alpha_1,\alpha_2|\alpha_5,\alpha_6| \alpha_3,\alpha_4}_{\bbb}
(n = \sum_{i=1}^6\alpha_i-2-\ep) =
\pi^{-4}\,(\mu^2)^{4-n}\,\int\,d^nq_1\,d^nq_2\,\prod_{i=1}^{6}\,
[i]^{-\alpha_i}_{\bbb}
\nl
{}&=& -\,
\frac{\egam{2+\ep}}{\prod_{i=1}^6\egam{\alpha_i}}\,
\omega^{6-\sum_{j=1}^6\alpha_j+\ep}\,
\dcub{2}(x,y)\,\dcub{3}(\{z\}) \,
\nl
{}&\times&\,(1-z_1)^{\rho_1}\,z_1^{\rho_2}\,(1-z_2)^{\rho_3}\,
z_2^{\rho_4}\,(1-z_3)^{\rho_5}\,z_3^{\rho_6}\,
(1-y)^{\rho_7}\,y^{\rho_8}\,(1-x)^{\rho_9}\,x^{\rho_{10}}\,
\chiu{\bbb}^{-2-\ep},
\label{genbbb}
\eqa
where $[1]_{\bbb} \equiv [1]$, where $\omega$ is defined in \eqn{defDUV} and 
the powers $\rho_i$ ($i = 1, \ldots, 10$) are
\bqa
\rho_1 &=& \alpha_1 -1, \quad
\rho_2 = \alpha_2 -1, \quad
\rho_3 =\alpha_4 -1, \quad \rho_4 = \alpha_3 -1, \nl
\rho_5 &=& \alpha_5 -1, \quad \rho_6 = \alpha_6 -1, \quad
\rho_7 = \alpha_5+ \alpha_6 - 1, \quad
\rho_8 = \frac{1}{2}(\ep -\sum \alpha + 2\,\alpha_1 + 2\,\alpha_2
+ 2\,\alpha_3  + 2\,\alpha_4),
\nl
\rho_9 &=& -\frac{1}{2}(4 + \ep  - \sum \alpha + 2\,\alpha_1  + 2\,\alpha_2 ),
 \, \, \, \,
\rho_{10} = 
-\frac{1}{2}\,(4 + \ep  - \sum \alpha + 2\,\alpha_3 + 2\,\alpha_4). \quad \,
\eqa
The polynomial $\chiu{\bbb}$ is given by 
$\chiu{\bbb} = \,- Q^2\,y^2 + ( M^2_x - M^2 + Q^2)\,y + M^2$
where:
\bqa
\nn
Q_{\mu} = K_{1\mu} - K_{2\mu} - K_{3\mu}, \qquad M^2 = R^2_3 - K^2_3,  \qquad
M^2_x = \frac{x\,(R^2_1 - K^2_1) + (1-x)\,(R^2_2-K^2_2)}{x\,(1-x)} \, ,
\eqa
\bqa
\nn
R^2_1 &=& l_{212}\,z_1 + m^2_1,
\quad
R^2_2 = z_2\,(p^2_1+m^2_3) + (1-z_2)\,(p^2_2+m^2_4),
\quad
R^2_3 = l_{156}\,z_3 + m^2_5,
\nl 
\nn
K_{1\mu} &=& - z_1\,p_{2\mu}, \quad K_{2\mu} = z_2\,p_{1\mu} - 
(1 - z_2)\,p_{2\mu}, \quad K_{3\mu} = - z_3\,p_{1\mu}.
\eqa
\subsubsection{Vector integrals in the $V^{\bbb}$ family \label{rankonebbb}}
The form factors for the vector integrals are defined by the relations
\bqa
\cV^{\bbb}(\mu\,|\,0\,;\,-p_2,p_1,-p_2,-p_1,\lstm{123456}) &=&
\sum_{i=1}^2 \cV^{\bbb}_{\sagen}(-p_2,p_1,-p_2,-p_1,\lstm{123456})\,p_{i\mu} ,
\nl
\cV^{\bbb}(0\,|\,\mu\,;\,-p_2,p_1,-p_2,-p_1,\lstm{123456}) &=&
\sum_{i=1}^2 \cV^{\bbb}_{\sbgen}(-p_2,p_1,-p_2,-p_1,\lstm{123456})\,p_{i\mu} .
\label{ffbbb}
\eqa
Their explicit expression, in terms of integrals over the Feynman parameters,
is
\bqa
\cV^{\bbb}_{\sijgen} &=& -\,\egam{2+\ep}\,\CIM{V}_{\bbb}\,
P_{\bbbn{ij}}\,\chiu{\bbb}^{-2-\ep},
\nl
\CIM{V}_{\bbb} &=& \omega^{\ep}\,\dcub{5}(x,y,\{z\})\,
\Bigl[ x\,(1-x)\Bigr]^{-1-\ep/2}\,y^{1+\ep/2}\,(1-y),
\nl
P_{\bbbn{\szz}} &=& 1, \quad
P_{\bbbn{\saa}} = -(z_2-z_3)\,(1-x)\,(1-y), \quad
P_{\bbbn{\sab}} = (1-z_1-z_2)\,(1-x)\,(1-y),
\nl
P_{\bbbn{\sba}} &=& y\,(z_2-z_3), \quad
P_{\bbbn{\sbb}} = -y\,(1-z_1-z_2),
\label{measbbb}
\eqa
where $\omega$ is defined in \eqn{defDUV} and
$P_{\szz}$ is the factor that arises in the calculation of the scalar
integral. The form factors for the vector integrals can be reduced as follows:
first it is possible to simplify the scalar products $\spro{q_1}{p_2}$ and
$\spro{q_2}{p_1}$, respectively, obtaining the relations
\bqa
\cV^{\bbb}(p_2\,|\,0\,;\, -p_2,p_1,-p_2,-p_1,\lstm{123456}) &=&
\frac{1}{2}\,\Bigl[ l_{212}\,\cV^{\bbb}_{\Z}(-p_2,p_1,-p_2,-p_1,\lstm{123456})
\nl
{}&-& \,\cV^{\bba}_{\Z}(p_1,p_1,-p_2,\lstm{56134})
\nl
{}&+& \,\cV^{\bba}_{\Z}(-P,-P,-p_2,\lstm{34256})\Bigr],
\nl
\cV^{\bbb}(0\,|\,p_1\,;\,-p_2,p_1,-p_2,-p_1,\lstm{123456}) &=&
\frac{1}{2}\,\Bigl[ l_{156}\,\cV^{\bbb}_{\Z}(-p_2,p_1,-p_2,-p_1,\lstm{123456})
\nl
{}&+& \,\cV^{\bba}_{\Z}(p_2,p_2,-p_1, \lstm{21634})
\nl
{}&-& \,\cV^{\bba}_{\Z}(p_2,p_2,-p_1, \lstm{12543})\Bigr].
\label{afy30}
\eqa
Since there are no other reducible scalar products we must find relations 
that link the form factors of the vector integrals to a linear combination 
of generalized scalar functions. The following identities hold:
\bqa
\cV^{\bbb}_{\sbb}(-p_2,p_1,-p_2,-p_1,\lstm{123456}) &=&
\omega^2\,\Bigl[ \cV^{1,2 | 1 , 1 | 2 , 1 }_{\bbb} -
\cV^{2+ 1 | 1 , 1 | 1 , 2 }_{\bbb}\Bigr]\,\bmid_{n = 6-\ep},  \nl
\cV^{\bbb}_{\saa}(-p_2,p_1,-p_2,-p_1,\lstm{123456}) &=&
\omega^2\,\Bigl[ \cV^{1 , 1 | 1 , 2 | 1 , 2 }_{\bbb} -
\cV^{1 , 1 | 2 , 1 | 2 , 1 }_{\bbb}\Bigr]\,\bmid_{n = 6-\ep} \, .
\eqa
\subsubsection{Rank two tensor integrals in the $V^{\bbb}$ family
\label{ranktwobbb}}
It is then necessary to consider the tensor integrals that have two
momenta of integration with free Lorentz indices. We
start from the the $\cV^{\bbb}(0 | \mu, \nu)$ integral and introduce the
relevant form factors through the relation
\bqa 
\cV^{\bbb}(0|\mu, \nu\,;\,\cdots) &=& 
\cV^{\bbb}_{\sbba}\,p_{1\mu}\,p_{1\nu} + 
\cV^{\bbb}_{\sbbb}\,p_{2\mu}\,p_{2\nu} + 
\cV^{\bbb}_{\sbbc}\,\{ p_1 p_2\}_{\mu\nu} + 
\cV^{\bbb}_{\sbbd}\,\delta_{\mu\nu} 
\label{tensorbbb3}.
\eqa
The integral representation of these form factors is given by
\bqa 
\cV^{\bbb}_{\sbbgen} &=& -\,\egam{2+\ep}\,\CIM{V}_{\bbb}\,
y^2\,R_{\bbbn{\sbbgen}}\,\chiu{\bbb}^{-2-\ep}, \quad i \neq 4, 
\qquad 
\cV^{\bbb}_{\sbbd} = -\,\frac{1}{2}\,\egam{1+\ep}\,\CIM{V}_{\bbb}\,
\chiu{\bbb}^{-1-\ep},
\nl
R_{\bbbn{\sbba}} &=& (z_2-z_3)^2, \quad 
R_{\bbbn{\sbbb}} = (1-z_1 - z_2)^2\,  ,  \quad
R_{\bbbn{\sbbc}} = -\,(1-z_1 - z_2)\,(z_2-z_3)\, .
\eqa
Multiplying \eqn{tensorbbb3} by $\delta_{\mu\nu}$ and $p_{1\mu}$,
respectively, we obtain the relations
\bqa
p^2_1\cV^{\bbb}_{\sbba}\! + p^2_2 \cV^{\bbb}_{\sbbb}
\!+\! 2p_{12}\cV^{\bbb}_{\sbbc} \! +\! n\cV^{\bbb}_{\sbbd}\! &=&
\cV^{\bba}_{\Z}(p_2,p_2,-p_1,\lstm{21634}) - m_5^2\,
\cV^{\bbb}_{\Z}(-p_2,p_1,-p_2,-p_1,\lstm{123456})\, , 
\nl
p^2_1\,\cV^{\bbb}_{\sbba}  + p_{12}\,\cV^{\bbb}_{\sbbc}
+\cV^{\bbb}_{\sbbd} &=& 
\frac{1}{2}\,\Bigl[ l_{156}\,
\cV^{\bbb}_{\sba}(-p_2,p_1,-p_2,-p_1,\lstm{123456}) 
-\,\cV^{\bba}_{\sab}(p_2,p_2,-p_1,\lstm{21634}) 
\nl
{}&-& \,\cV^{\bba}_{\sab}(-p_2,-p_2,p_1,\lstm{12543}) 
+\,\cV^{\bba}_{\sbb}(p_2,p_2,-p_1,\lstm{21634}) 
\nl
{}&+& \,\cV^{\bba}_{\sbb}(-p_2,-p_2,p_1,\lstm{12543}) +
\,\cV^{\bba}_{\Z}(p_2,p_2,-p_1,\lstm{21634})\Bigr], 
\nl
p^2_1\,\cV^{\bbb}_{\sbbc}  + p_{12}\,\cV^{\bbb}_{\sbbb} &=& 
\frac{1}{2}\,\Bigl[ l_{156}\,
\cV^{\bbb}_{\sbb}(-p_2,p_1,-p_2,-p_1,\lstm{123456})  
+\,\cV^{\bba}_{\saa}(p_2,p_2,-p_1,\lstm{21634})
\nl
{}&+& \,\cV^{\bba}_{\saa}(-p_2,-p_2,p_1,\lstm{12543}) 
- \,\cV^{\bba}_{\sab}(p_2,p_2,-p_1,\lstm{21634})
\nl
{}&-& \,\cV^{\bba}_{\sab}(-p_2,-p_2,p_1,\lstm{12543}) 
- \,\cV^{\bba}_{\sba}(p_2,p_2,-p_1,\lstm{21634})
\nl 
&-& \,\cV^{\bba}_{\sba}(-p_2,-p_2,p_1,\lstm{12543}) 
+ \,\cV^{\bba}_{\sbb}(p_2,p_2,-p_1,\lstm{21634})
\nl 
&+& \,\cV^{\bba}_{\sbb}(-p_2,-p_2,p_1,\lstm{12543})\Bigr].
\eqa
We then have a set of three equations that can be solved for $i \ne 4$
when we express one of the form factors in terms of a generalized scalar 
function; for example we have that
\bqa
\cV^{\bbb}_{\sbbd} &=& \frac{\omega^2}{2}\,\Bigl[
\cV^{1 , 1 | 1 , 1 | 1 , 2 }_{\bbb}  +
\cV^{1 , 1 | 1 , 1 | 2  , 1  }_{\bbb} +
\cV^{2 , 1 | 1 , 1|1 , 1 }_{\bbb}  +
\cV^{1 , 2 | 1 , 1 | 1 , 1 }_{\bbb}\, \Bigr]\,\bmid_{n = 6-\ep}.
\eqa
We can proceed in a completely analogous way for the $q_{1\mu}\,q_{1\nu}$
tensor integral. The relevant form factors are defined through the relation
\bqa 
\cV^{\bbb}(\mu, \nu\,|0;\,\cdots) &=&
\cV^{\bbb}_{\saaa}\,p_{1\mu}\,p_{1\nu} +
\cV^{\bbb}_{\saab}\,p_{2\mu}\,p_{2\nu} +
\cV^{\bbb}_{\saac}\,\{ p_1 p_2\}_{\mu\nu} +
\cV^{\bbb}_{\saad}\,\delta_{\mu\nu}
\label{tensorbbb2}.
\eqa
The integral representation of these form factors is given by
\bqa 
\cV^{\bbb}_{\saagen} &=& -\,\egam{2+\ep}\,\CIM{V}_{\bbb}\,
(1-x)^2\,(1-y)^2\,R_{\bbbn{\saagen}}\,\chiu{\bbb}^{-2-\ep}, \quad i \neq 4, 
\nl 
\cV^{\bbb}_{\saad} &=& -\,\frac{1}{2}\,\egam{1+\ep}\,\CIM{V}_{\bbb}\,
R_{\bbbn{\saad}}\,\chiu{\bbb}^{-1-\ep}, 
\nl
R_{\bbbn{\saagen}} &=& R_{\bbbn{\sbbgen}}, \qquad
\quad
R_{\bbbn{\saad}} = (1-x)\,(1 - x + \frac{x}{y} ) \, .
\eqa
Contracting \eqn{tensorbbb2} by $\delta_{\mu\nu}$ and $p_{2\mu}$,
respectively, we obtain the relations
\bqa
p^2_1\,\cV^{\bbb}_{\saaa} + p^2_2\,\cV^{\bbb}_{\saab}
+ 2\,p_{12}\,\cV^{\bbb}_{\saac}  + n\, \cV^{\bbb}_{\saad} &=&
\cV^{\bba}_{\Z}(-P,-P,-p_2,\lstm{34256}) 
\nl
{}&-& m_1^2\,\cV^{\bbb}_{\Z}(-p_2,p_1,-p_2,-p_1,\lstm{123456})\, , 
\eqa
\bqa
p^2_2\,\cV^{\bbb}_{\saab}  + p_{12}\,\cV^{\bbb}_{\saac}
+\cV^{\bbb}_{\saad} &=& 
\frac{1}{2}\,\Bigl[ l_{212}\,
\cV^{\bbb}_{\sab}(-p_2,p_1,-p_2,-p_1,\lstm{123456})
- \,\cV^{\bba}_{\saa}(-P,-P,-p_2,\lstm{34256})
\nl 
&+& \,\cV^{\bba}_{\sba}(-P,-P,-p_2,\lstm{34256})
+ \,\cV^{\bba}_{\Z}(-P,-P,-p_2,\lstm{34256})
\nl 
&+& \,\cV^{\bba}_{\sbb}(p_1,p_1,-p_2,\lstm{56234})
- \,\cV^{\bba}_{\sab}(p_1,p_1,-p_2,\lstm{56234})\Bigr], 
\nl
p^2_2\,\cV^{\bbb}_{\saac}  + p_{12}\,\cV^{\bbb}_{\saaa} &=&  
\frac{1}{2}\,\Bigl[ l_{212}\,
\cV^{\bbb}_{\saa} (-p_2,p_1,-p_2,-p_1,\lstm{123456})
+ \,\cV^{\bba}_{\saa}(p_1,p_1,-p_2,\lstm{56234})
\nl 
&-& \,\cV^{\bba}_{\sba}(p_1,p_1,-p_2,\lstm{56234}) 
+ \,\cV^{\bba}_{\sbb}(p_1,p_1,-p_2,\lstm{56234})
\nl 
&-& \,\cV^{\bba}_{\sab}(p_1,p_1,-p_2,\lstm{56234})\Bigr].
\eqa
Once again, we can rewrite one of the form factors of the
$\cV^{\bbb}_{\saagen}$ family as a linear combination of generalized
scalar functions and solve the system for the others: or instance
\bqa
\cV^{\bbb}_{\saaa} &=& 4\,\omega^4\,\Bigl[
\cV^{1 , 1 | 3 , 1 | 3 , 1 }_{\bbb} +
\cV^{1 , 1 | 1 , 3 | 1 , 3  }_{\bbb} -
\frac{1}{2}\,\cV^{1 , 1 | 2 , 2 | 2 , 2  }_{\bbb}\,\Bigr]\,
\bmid_{n = 8-\ep} \quad \mbox{or}
\nl
\cV^{\bca}_{\saad} &=& \frac{1}{2}\,\omega^2\,\Bigl[
\cV^{1,1 | 1,2 | 1,1}_{\bbb} +
\cV^{1,1 | 2,1 | 1,1}_{\bbb} +
\cV^{1,1 | 1,1 | 1,2}_{\bbb} +
\cV^{1,1 | 1,1 | 2,1}_{\bbb}\Bigr]\,\bmid_{n = 6 - \ep}.
\eqa
The $q_{1\mu}\,q_{2\nu}$ tensor integrals are symmetric in the exchange of 
$\mu$ and $\nu$;  
this fact can be understood noticing that the integral with respect to $q_1$
is proportional to $a_1\,q_2^{\mu} + a_2\,Q^{\mu}$, where $a_1$ and $a_2$
are scalar factors and $Q^{\mu}$ is a linear combination of the external
momenta. Therefore, after the $q_1$ integration, the integrand will
split into a part proportional to $q_2^\mu q_2^\nu$, obviously
symmetric with respect to $\mu \leftrightarrow \nu$, and into a part 
proportional to $q_2^{\nu} Q^{\mu}$; also the latter is symmetric since
the vector integral with $q_2^\nu$ in the numerator is proportional to $Q^\nu$.

To describe their tensor structure it is necessary to introduce four 
form factors:
\bq
\cV^{\bbb}(\mu | \nu ;\,\cdots) = 
\cV^{\bbb}_{\saba}\,p_{1\mu}\,p_{1\nu} + 
\cV^{\bbb}_{\sabb}\,p_{2\mu}\,p_{2\nu} + 
\cV^{\bbb}_{\sabc}\,\{p_1\,p_2\}_{\mu\nu} + 
\cV^{\bbb}_{\sabd}\,\delta_{\mu\nu} 
\label{tensorbbb1},
\eq
with $\{p_1 p_2\}$ given in \eqn{symmete} and with corresponding integral 
representations given by
\bqa 
\cV^{\bbb}_{\sabgen} &=& -\,\egam{2+\ep}\,\CIM{V}_{\bbb}\,
y\,(1-x)\,(1-y)\,R_{\bbbn{\sabgen}}\,\chiu{\bbb}^{-2-\ep} , \; i \neq 4\,,  
\nl
\cV^{\bbb}_{\sabd} &=& -\,\frac{1}{2}\,\egam{1+\ep}\,\CIM{V}_{\bbb}\,
(1-x)\,\chiu{\bbb}^{-1-\ep}\, , 
\qquad
R_{\bbbn{\sabgen}}= -\,R_{\bbbn{\sbbgen}}.
\eqa
Contracting both sides of \eqn{tensorbbb1} with $p_{1\nu}$ and $p_{2\mu}$ 
we obtain the following set of four equations:
\bqa 
p^2_1\,\cV^{\bbb}_{\saba} + p_{12}\
\cV^{\bbb}_{\sabc} + \cV^{\bbb}_{\sabd} &=& 
\frac{1}{2}\,\Bigl[ l_{156}\,
\cV^{\bbb}_{\saa}(-p_2,p_1,-p_2,-p_1,\lstm{123456}) 
\nl 
&-& \,\cV^{\bba}_{\sab}(p_2,p_2,-p_1,\lstm{21634})
- \,\cV^{\bba}_{\sab}(-p_2,-p_2,p_1,\lstm{12543})\Bigr], 
\nl
p^2_1\,\cV^{\bbb}_{\sabc} + p_{12}\,\cV^{\bbb}_{\sabb} &=& 
\frac{1}{2}\,\Bigl[ l_{156}\,
\cV^{\bbb}_{\sab} (-p_2,p_1,-p_2,-p_1,\lstm{123456}) 
+ \,\cV^{\bba}_{\saa}(p_2,p_2,-p_1,\lstm{21634})
\nl 
&+& \,\cV^{\bba}_{\saa}(-p_2,-p_2,p_1,\lstm{12543}) 
- \,\cV^{\bba}_{\sab}(p_2,p_2,-p_1,\lstm{21634})
\nl 
&-& \,\cV^{\bba}_{\sab}(-p_2,-p_2,p_1,\lstm{12543}) 
+ \,\cV^{\bba}_{\Z}(p_2,p_2,-p_1,\lstm{21634})\Bigr],
\nl
p^2_2\,\cV^{\bbb}_{\sabc} + p_{12}\,\cV^{\bbb}_{\saba} &=& 
\frac{1}{2}\,\Bigl[ l_{212}\,
\cV^{\bbb}_{\sba} (-p_2,p_1,-p_2,-p_1,\lstm{123456}) 
+ \,\cV^{\bba}_{\saa}(p_1,p_1,-p_2,\lstm{56234})
\nl 
&+& \,\cV^{\bba}_{\Z}(-P,-P,-p_2,\lstm{34256})
- \,\cV^{\bba}_{\sab}(p_1,p_1,-p_2,\lstm{56234})\Bigr],
\nl
p^2_2\,\cV^{\bbb}_{\sabb} + p_{12}\,
\cV^{\bbb}_{\sabc} + \cV^{\bbb}_{\sabd} &=& 
\frac{1}{2}\,\Bigl[
l_{212}\, \cV^{\bbb}_{\sbb}(-p_2,p_1,-p_2,-p_1,\lstm{123456}) 
+ \,\cV^{\bba}_{\sba}(-P,-P,-p_2,\lstm{34256})
\nl 
&+& \,\cV^{\bba}_{\Z}(-P,-P,-p_2,\lstm{34256})
- \,\cV^{\bba}_{\sab}(p_1,p_1,-p_2,\lstm{56234})\Bigr]. 
\nl 
\eqa
Note that the form factor $\cV^{\bbb}_{\sabd}$ could be expressed in terms of 
generalized scalar functions; indeed we have
\bq 
\cV^{\bbb}_{\sabd} = \frac{\omega^2}{2}\,
\Bigl[ \cV^{1 , 1| 1,1 | 1 , 2}_{\bbb}(n = 6 -\ep) +
\cV^{1 , 1 | 1 , 1 | 2 , 1 }_{\bbb}(n = 6 - \ep)\Bigr], 
\eq
while the remaining ones can be obtained solving the corresponding system
of equations.
Results for this family are summarized in \appendx{summabbb}.
$V^{\bbb}_{\Z} \equiv V^{222}_{\Z}$ is discussed in Sect.~10.4 of III
(see comment at the end of \sect{ranktwoaba}), evaluation of form factors 
in \sect{exabcabbb}.
\subsection{Integral representation for tensor integrals of rank three
\label{IRTIRT}}
Our aim in this work was to derive all the ingredients needed for the
two-loop renormalization of the standard model (or of any other
renormalizable theory) and to discuss all the tensor integrals that are
relevant for the calculation of physical observables related to processes of
the type $V(S) \to \overline{f} f$. For the classes of diagrams involving at
least one four-point vertex, it is sufficient to analyze tensor integral
that include up to two integration momenta in the numerator.  However, for
the remaining classes, $\cV^{\ada}$, $\cV^{\bca}$, and $\cV^{\bbb}$ it is
necessary to consider in addition tensor integrals that include up to three
momenta. As specified in the Introduction, we make use of the following
shorthand notation: $\ox = 1 - x, \ox_i = 1 - x_i, \oy_i = 1 - y_i$, etc.
\subsubsection{$\cV^{\ada}$ family }
For general definitions see \sect{adafam}.
We start by considering the integral with three uncontracted $q_2$ momenta
in the numerator:
\bqa
\cV^{\ada}(0 | \alpha ,\beta ,\gamma ; \cdots) &=& 
\cV^{\ada}_{\sbbba}\,\{\delta\,p_1\}_{\alpha\beta\gamma} + 
\cV^{\ada}_{\sbbbb}\, \{\delta\,p_2\}_{\alpha\beta\gamma} 
+ \cV^{\ada}_{\sbbbc}\,\{p_1 p_1 p_2\}_{\alpha\beta\gamma} 
+ \cV^{\ada}_{\sbbbd}\,\{p_2 p_2 p_1\}_{\alpha\beta\gamma}
\nl
{}&+&
 \cV^{\ada}_{\sbbbe}\, p_{1 \alpha}\,p_{1 \beta}\,p_{1 \gamma} +
\cV^{\ada}_{\sbbbf}\, p_{2 \alpha}\,p_{2 \beta}\,p_{2 \gamma} \,,
\label{ff1}
\eqa
where we used the definitions of \eqn{symmete}.
The various form factors have the following integrals representations
(with integration measure defined in \eqn{measada}):
\bqa
\cV^{\ada}_{\sbbbgen} &=& -\,\egam{2+\ep}\,\CIM{V}_{\ada}\,
P_{\adan{\sbbbgen}}\,\chiu{\ada}^{-2-\ep}, \quad i > 2 ,
\nl
P_{\adan{\sbbbc}} &=& -z_1^2\,z_2, \quad
P_{\adan{\sbbbd}} = -z_1\,z_2^2, 
\quad 
P_{\adan{\sbbbe}} = -z_1^3, \quad
P_{\adan{\sbbbf}} = -z_2^3, 
\nl
\cV^{\ada}_{\sbbbgen} &=& -\,\frac{\egam{1+\ep}}{2}\,\CIM{V}_{\ada}\,
P_{\adan{\sbbbgen}}\,\chiu{\ada}^{-1-\ep}, \quad i = 1,2, 
\qquad
P_{\adan{\sbbbgen}} = -z_i\, . \,\,\,
\eqa
$\chiu{\ada} \equiv \chiu{\aca}$ is given in \eqn{Defchiaca}.
For the tensor integral with three uncontracted $q_1$ momenta in the numerator
we use a decomposition identical to the one of \eqn{ff1}.
The integral representation for the corresponding form factors is given by
\bqa
\cV^{\ada}_{\saaagen} &=& -\egam{2+\ep}\CIM{V}_{\ada} 
P_{\adan{\saaagen}}\,\chiu{\ada}^{-2-\ep} , \quad 
P_{\adan{\saaagen}} = x^3\,P_{\adan{\sbbbgen}},
\qquad i > 2 ,
\nl
\cV^{\ada}_{\sbbbgen} &=& - \frac{\egam{1+\ep}}{2} \CIM{V}_{\ada} 
x^2\,\left[ P_{\adan{\saaagen}}  + \frac{R_{\adan{\saaagen}}}{2-\ep} +
2\,\frac{1+\ep}{2-\ep}\,\chiu{\ada}^{-1}\,Q_{\adan{\saaagen}}\right]\, 
\chiu{\ada}^{-1-\ep},
\quad i = 1,2 , 
\nl
Q_{\adan{\saaagen}} &=&  \ox z_i [F(z_1,z_2) + m^2_x],
\qquad
R_{\adan{\saaagen}} = (6-\ep)\,\ox z_i,  
\quad
P_{\adan{\saaagen}} = -x\,z_i.
\eqa
Employing again definitions analogous to those of \eqn{ff1}, the form
factors for the $\cV^{\ada}_{\sabbgen}$ family are written as
\bqa
\cV^{\ada}_{\sabbgen} &=& - \egam{2+\ep} \CIM{V}_{\ada} 
P_{\adan{\sabbgen}}\,\chiu{\ada}^{-2-\ep}, 
\qquad P_{\adan{\sabbgen}} = x\,P_{\adan{\sbbbgen}},
\quad i > 2,
\nl
\cV^{\ada}_{\sabbgen} &=& - \frac{\egam{1+\ep}}{2} \CIM{V}_{\ada} 
P_{\adan{\sabbgen}}\,\chiu{\ada}^{-1-\ep}, 
\qquad
P_{\adan{\sabbgen}} = x\,P_{\adan{\sbbbgen}}
\quad i = 1,2\, .
\eqa
Since the tensor integral $\cV^{\ada}(\alpha, \beta | \gamma)$ is only 
symmetric with respect to the exchange of the first two indices,
a new decomposition in form factors is introduced:
\bqa
\cV^{\ada}(\alpha ,\beta | \gamma ; \cdots) &=& 
\cV^{\ada}_{\saaba}\,\{\delta\,p_1\}_{\alpha\beta\gamma} +
\cV^{\ada}_{\saabb}\,\{\delta\,p_2\}_{\alpha\beta\gamma} +
\cV^{\ada}_{\saabc}\,\{p_1 p_1 p_2\}_{\alpha\beta\gamma} +
\cV^{\ada}_{\saabd}\,\{p_2 p_2 p_1\}_{\alpha\beta\gamma} 
\nl 
&+& \cV^{\ada}_{\saabe}\, p_{1 \alpha}\,p_{1 \beta}\,p_{1 \gamma} +
\cV^{\ada}_{\saabf}\, p_{2 \alpha}\,p_{2 \beta}\,p_{2 \gamma} 
+
\cV^{\ada}_{\saabg}\,\{\delta\,p_1\}_{\alpha\beta\,|\,\gamma} +
\cV^{\ada}_{\saabh}\,\{\delta\,p_2\}_{\alpha\beta\,|\,\gamma}\, ,
\label{ff2}
\eqa
where all the symmetrized products were defined in \eqn{symmete}.  The
integral representation for the form factors in \eqn{ff2} is as follows:
\bqa
\cV^{\ada}_{\saabgen} &=& -\,\egam{2+\ep}\,\CIM{V}_{\ada}\,
P_{\adan{\saabgen}}\,\chiu{\ada}^{-2-\ep}, 
\qquad P_{\adan{\saabgen}} = x^2\,P_{\adan{\sbbbgen}},
\quad i \neq 1,2,7,8,
\nl
\cV^{\ada}_{\saabgen} &=& -\,\frac{\egam{1+\ep}}{2}\,\CIM{V}_{\ada}\,
x\,\left[ P_{\adan{\saabgen}}  + 
\frac{R_{\adan{\saabgen}}}{2-\ep} +
2\,\frac{1+\ep}{2-\ep}\,\chiu{\ada}^{-1}\,Q_{\adan{\saabgen}}\right]\, 
\chiu{\ada}^{-1-\ep}\,
\quad i = 1,2 ,
\nl
Q_{\adan{\saabgen}} &=& 
\ox\,z_i\,[F(z_1,z_2) + m^2_x ],
\qquad
R_{\adan{\saabgen}} = (6-\ep)\,\ox\,z_i, 
\quad
P_{\adan{\saabgen}} = -x\,z_i\,, 
\nl
\cV^{\ada}_{\saabgen} &=& -\,\frac{\egam{1+\ep}}{2-\ep}\,\CIM{V}_{\ada}\,
x\,\ox\,\chiu{\ada}^{-1-\ep}\,\Bigl[
\frac{1}{2}\,R_{\adan{\saabgen}} 
+ (1 + \ep)\,\chiu{\ada}^{-1}\,Q_{\adan{\saabgen}}\Bigr] , \quad i = 7,8 ,
\nl
Q_{\adan{\saabg}} &=& -\,z_1\,[ F(z_1,z_2) + m_x^2] ,
\qquad
Q_{\adan{\saabh}} = -\,z_2\,[ F(z_1,z_2) + m_x^2] ,
\nl
R_{\adan{\saabg}} &=& -(6-\ep)\,z_1, 
\quad
R_{\adan{\saabh}} = -(6-\ep)\,z_2 . 
\eqa
It is straightforward to show that the form factors $\cV^{\ada}_{\saaac}$,
$\cV^{\ada}_{\saaad}$, $\cV^{\ada}_{\saaae}$, and $\cV^{\ada}_{\saaaf}$,
are generalized integrals of the type
$\cV^{\alpha_1 | \alpha_2 , \alpha_3  , \alpha_4| \alpha_5 }_{\ada}$:
\bqa
\cV^{\ada}_{\saaae} &=& -36\,\omega^6\,
\cV^{1 |2,4,1| 4 }_{\ada}(n  = 10 -\ep) -
\cV^{\ada}_{\saaaf} +3\,\cV^{\ada}_{\saaad} - 3\,\cV^{\ada}_{\saaac}\,, 
\nl
\cV^{\ada}_{\saaac} &=& -12\,\omega^6\,
\cV^{1 |2,3,2| 4 }_{\ada}(n  = 10 -\ep) - \cV^{\ada}_{\saaaf} -2\, 
\cV^{\ada}_{\saaad}\,,
\nl
\cV^{\ada}_{\saaad} &=&  -12\,\omega^6\,  
\cV^{1 |2,2,3|4 }_{\ada}(n  = 10 -\ep) + \cV^{\ada}_{\saaaf}\, , 
\quad 
\cV^{\ada}_{\saaaf} = -36\,\omega^6\, 
\cV^{1 |2,1,4|4 }_{\ada}(n  = 10 -\ep) \,.
\eqa
\subsubsection{$\cV^{\bca}$ family}
For general definitions see \sect{bcafam}.
We start by considering the tensor integrals $\cV(\mu, \nu ,\alpha |
0;\cdots)$ and $\cV(0 | \mu ,\nu ,\alpha ;\cdots)$ which are obviously totally
symmetric and for which we can then adopt the same decomposition in form
factors already presented in \eqn{ff1}.
Employing the standard procedure, one finds that
\bqa 
\cV^{\bca}_{\sbbbgen} &=& -\,\egam{2+\ep}\,\CIM{V}_{\bca}\,
R_{\bcan{\sbbbgen}}\,\chiu{\bca}^{-2-\ep}, \qquad i > 2 \,, 
\nl 
R_{\bcan{\sbbbc}} &=& Y_1\,Y_2^2, 
\quad
R_{\bcan{\sbbbd}} = Y_1^2\,Y_2\,  ,  
\quad
R_{\bcan{\sbbbe}} = Y_2^3\, ,
\quad
R_{\bcan{\sbbbf}} = Y_1^3\, , 
\nl 
\cV^{\bca}_{\sbbbgen} &=& -\,\frac{\egam{1+\ep}}{2}\,\CIM{V}_{\bca}\,
R_{\bcan{\sbbbgen}}\,\chiu{\bca}^{-1-\ep}  \, , \quad i = 1,2 \, , 
\nl 
R_{\bcan{\sbbba}} &=& Y_2\, ,
\quad
R_{\bcan{\sbbbb}} = Y_1\,, 
\eqa
where we recall that the quantities $Y_1$, $Y_2$  
(see \eqn{HY1HY2}) are given by $Y_i = - 1 + y_i - y_3\,X$, with 
$X = (1-x_1)/(1-x_2)$. The integration measure is defined in \eqn{varaibca}.
For the $\cV^{\bca}_{\saaagen}$ form factors we have
\bqa 
\cV^{\bca}_{\saaagen} &=& -\,\egam{2+\ep}\,\CIM{V}_{\bca}\,
R_{\bcan{\saaagen}}\,\chiu{\bca}^{-2-\ep}, \qquad i > 2 \,, 
\nl 
R_{\bcan{\saaac}} &=& -H_1\,H_2^2 , 
\quad
R_{\bcan{\saaad}} = -H_1^2\,H_2  ,  
\quad
R_{\bcan{\saaae}} = -H_2^3\, ,
\quad
R_{\bcan{\saaaf}} = -H_1^3\,  ,
\nl 
\cV^{\bca}_{\saaagen} &=& -\,\frac{\egam{1+\ep}}{2}\,\CIM{V}_{\bca}\,
x_2\,\Bigl( x_2\,R_{\bcan{\saaagen}} - Q_{\bcan{\saaagen}}\Bigr)\,
\chiu{\bca}^{-1-\ep}
 \, , \quad i = 1,2 \, , 
\nl
R_{\bcan{\saaaa}} &=& -\,H_2  ,
\quad
R_{\bcan{\saaab}} = -\,H_1 ,  
\quad
Q_{\bcan{\saaaa}} = \frac{\ox_2}{y_3}  H_2 ,
\quad
Q_{\bcan{\saaab}} = \frac{\ox_2}{y_3} H_1 . 
\eqa
The quantities $H_1$ and $H_2$ were introduced in \eqn{HY1HY2}.
Consider now the integral $\cV^{\bca}(\mu | \nu ,\alpha)$; in this 
case we have symmetry in the last two indices and a larger number of 
form factors; with symmetrized products defined in \eqn{symmete} we have
\bqa
\cV^{\bca}(\mu |  \nu , \alpha ; \cdots) &=& 
\cV^{\bca}_{\sabba}\,
\{\delta\,p_1\}_{\nu\alpha\,|\,\mu} +
\cV^{\bca}_{\sabbb}\, 
\{\delta\,p_2\}_{\nu\alpha\,|\,\mu} +
\cV^{\bca}_{\sabbc}\, \delta_{\nu \alpha} p_{1 \mu} + 
\cV^{\bca}_{\sabbd}\, \delta_{\nu \alpha} p_{2 \mu}
\nl 
&+& 
\cV^{\bca}_{\sabbe}\, p_{1 \alpha}\,p_{1 \mu}\,p_{1 \nu} +
\cV^{\bca}_{\sabbf}\, p_{2 \alpha}\,p_{2 \mu}\,p_{2 \nu} +
\cV^{\bca}_{\sabbg}\, 
\{p_1\,p_1\,p_2\}_{\alpha\nu\,|\mu} 
\nl 
&+& 
\cV^{\bca}_{\sabbh}\, 
\{p_2\,p_2\,p_1\}_{\alpha\nu\,|\,\mu} +
\cV^{\bca}_{\sabbi}\, p_{1 \alpha}\,p_{1 \nu}\,p_{2 \mu} +
\cV^{\bca}_{\sabbj}\, p_{1 \mu}\,p_{2 \nu}\,p_{2 \alpha}\, .
\label{ff3}
\eqa
The integral representations of the form factors of \eqn{ff3} are given by
\bqa 
\cV^{\bca}_{\sabbgen} &=& -\,\egam{2+\ep}\,\CIM{V}_{\bca}\,
R_{\bcan{\sabbgen}}\,\chiu{\bca}^{-2-\ep}, \qquad i > 4\,, 
\nl 
R_{\bcan{\sabbe}} &=& -Y_{2}^2\,H_2 ,
\quad
R_{\bcan{\sabbf}} = -Y_{1}^2\,H_1 , 
\quad
R_{\bcan{\sabbg}} = -Y_1\,Y_2\,H_2 ,
\nl
R_{\bcan{\sabbh}} &=& -Y_1\,Y_2\,H_1 , 
\quad
R_{\bcan{\sabbi}} = -Y_2^2\,H_1 , 
\quad
R_{\bcan{\sabbj}} = -Y_1^2\,H_2 , 
\nl
\cV^{\bca}_{\sabbgen} &=& -\,\frac{\egam{1+\ep}}{2}\,\CIM{V}_{\bca}\,
R_{\bcan{\sabbgen}}\,\chiu{\bca}^{-1-\ep}\, , \quad i = 1,\,\cdots\,4 \, ,
\nl
R_{\bcan{\sabba}} &=& 1 - x_1 - H_2 ,
\quad
R_{\bcan{\sabbb}} =  1 - x_1 - H_1 ,
\quad
R_{\bcan{\sabbc}} = - H_2 ,
\quad
R_{\bcan{\sabbd}} = - H_1  . 
\eqa
The integral $\cV^{\bca}(\mu , \nu | \alpha; \cdots)$ is symmetric 
in the first two indexes; using the definitions of \eqn{symmete} we obtain 
\bqa
\cV^{\bca}(\mu , \nu | \alpha ; \cdots) &=& 
\cV^{\bca}_{\saaba} 
\{\delta\,p_1\}_{\mu\nu\,|\,\alpha} +
\cV^{\bca}_{\saabb}\, 
\{\delta\,p_2\}_{\mu\nu\,|\,\alpha} +
\cV^{\bca}_{\saabc}\, \delta_{\mu \nu} p_{1 \alpha} + 
\cV^{\bca}_{\saabd}\, \delta_{\mu \nu} p_{2 \alpha}
\nl 
&+& 
\cV^{\bca}_{\saabe}\, p_{1 \alpha}\,p_{1 \mu}\,p_{1 \nu} +
\cV^{\bca}_{\saabf}\, p_{2 \alpha}\,p_{2 \mu}\,p_{2 \nu} +
\cV^{\bca}_{\saabg}\, 
\{p_1 p_1 p_2\}_{\mu\nu\,|\,\alpha} 
\nl 
&+& \cV^{\bca}_{\saabh}\, 
\{p_2 p_2 p_1\}_{\nu\mu\,|\,\alpha} +
\cV^{\bca}_{\saabi}\, p_{1 \mu}\,p_{1 \nu}\,p_{2 \alpha} +
\cV^{\bca}_{\saabj}\, p_{1 \alpha}\,p_{2 \mu}\,p_{2 \nu}\, .
\label{ff4}
\eqa
The integral representation of the form factors in \eqn{ff4} is the 
following: 
\bqa 
\cV^{\bca}_{\saabgen} &=& -\,\egam{2+\ep}\,\CIM{V}_{\bca}\,
R_{\bcan{\saabgen}}\,\chiu{\bca}^{-2-\ep}, \qquad i > 4 \,, 
\nl 
R_{\bcan{\saabe}} &=& Y_2\,H_2^2\, ,
\quad
R_{\bcan{\saabf}} = Y_1\,H_1^2\, , 
\quad
R_{\bcan{\saabg}} = Y_2\,H_1\,H_2\,, 
\nl
R_{\bcan{\saabh}} &=& Y_1\,H_1\,H_2\,,
\quad
R_{\bcan{\saabi}} = Y_1 H_2^2\, ,
\quad
R_{\bcan{\saabj}} = Y_2\,H_1^2 , 
\nl
\cV^{\bca}_{\saabgen} &=& -\,\frac{\egam{1+\ep}}{2}\,\CIM{V}_{\bca}\,
x_2\,R_{\bcan{\saabgen}}\,\chiu{\bca}^{-1-\ep} \, , \quad i = 1,2 \, ,
\nl
R_{\bcan{\saaba}} &=& -\,H_2 \, ,
\quad
R_{\bcan{\saabb}} = -\,H_1\,, 
\nl
\cV^{\bca}_{\saabgen} &=& -\,\frac{\egam{1+\ep}}{2}\,\CIM{V}_{\bca}\,
x_2\,\Bigl[ R_{\bcan{\saabgen}} +
y_3^{-1}\,\ox_2 \, Q_{\bcan{\saabgen}}\Bigr]\,
\chiu{\bca}^{-1-\ep} \, , \quad i = 3,4 \, ,
\nl
R_{\bcan{\saabc}} &=& 1-x_1-H_2\, ,
\quad
R_{\bcan{\saabd}} = 1-x_1-H_1\, , 
\quad
Q_{\bcan{\saabc}} = Y_2\, ,
\quad
Q_{\bcan{\saabd}} = Y_1\, . 
\eqa
\subsubsection{$\cV^{\bbb}$ family}
For general definitions see \sect{bbbfam}.
We finally analyze the rank three tensor integrals in the family
$\cV^{\bbb}$. The tensor integrals $\cV^{\bbb}(\mu ,\nu ,\alpha | 0)$
and $\cV^{\bbb}(0 | \mu ,\nu ,\alpha)$ can be decomposed into
form factors in complete analogy with \eqn{ff1}. We provide here the integral
representations for these form factors,
\bqa 
\cV^{\bbb}_{\sbbbgen} &=& -\,\egam{2+\ep}\,\CIM{V}_{\bbb}\,
y^3\,R_{\bbbn{\sbbbgen}}\,\chiu{\bbb}^{-2-\ep}, \qquad i > 2 \,, 
\nl 
R_{\bbbn{\sbbbc}} &=& -\,(z_2-z_3)^2\,(1-z_1-z_2)\, , 
\quad
R_{\bbbn{\saaad}} = (z_2-z_3)\,(1-z_1-z_2)^2\, ,  \nl
R_{\bbbn{\sbbbe}} &=& (z_2-z_3)^3\, ,
\quad
R_{\bbbn{\sbbbf}} = -\,(1-z_1-z_2)^3\,,  
\nl 
\cV^{\bbb}_{\sbbbgen} &=& -\,\frac{1}{2}\,\egam{1+\ep}\,\CIM{V}_{\bbb}\,
y\,R_{\bbbn{\sbbbgen}}\,\chiu{\bbb}^{-1-\ep} , \quad i = 1,2 \, ,
\nl
R_{\bbbn{\sbbba}} &=& z_2-z_3\, ,
\quad
R_{\bbbn{\sbbbb}} = -\,(1-z_1-z_2)\,,
\eqa
where the integration measure is given in \eqn{measbbb}. Also
\bqa 
\cV^{\bbb}_{\saaagen} &=& -\,\egam{2+\ep}\,\CIM{V}_{\bca}\,
\ox^3\,\oy^3\,R_{\bbbn{\saaagen}}\,\chiu{\bbb}^{-2-\ep},\qquad  
R_{\bbbn{\saaagen}} = -\,R_{\bbbn{\sbbbgen}}, \quad
i > 2 \,,
\nl 
\cV^{\bbb}_{\saaagen} &=& -\,\frac{1}{2}\,\egam{1+\ep}\,\CIM{V}_{\bbb}\,
\ox^2\,\oy\,\Bigl[ \ox\,R_{\bbbn{\saaagen}} + \frac{x}{y}\,
Q_{\bbbn{\saaagen}}\Bigr]\,
\chiu{\bbb}^{-1-\ep}\,, \quad i = 1,2 ,
\nl
R_{\bbbn{\saaaa}} &=& Q_{\bbbn{\saaaa}} = z_3 - z_2 \,,
\quad
R_{\bbbn{\saaab}} = Q_{\bbbn{\saaab}} = 1-z_1-z_2 \,.
\eqa
For the tensor integral $\cV^{\bbb}(\mu | \nu ,\alpha)$ we employ 
another decomposition into form factors, based on the definitions of 
\eqn{symmete}:
\bqa
\cV^{\bbb}(\mu |  \nu,  \alpha ; \cdots) &=& 
\cV^{\bbb}_{\sabba} 
\{\delta\,p_1\}_{\nu\alpha\,|\,\mu} +
\cV^{\bbb}_{\sabbb}\, 
\{\delta\,p_2\}_{\nu\alpha\,|\,\mu} +
\cV^{\bbb}_{\sabbc}\, \delta_{\nu \alpha} p_{1 \mu} + 
\cV^{\bbb}_{\sabbd}\, \delta_{\nu \alpha} p_{2 \mu}
+ \cV^{\bbb}_{\sabbe}\, p_{1 \alpha}\,p_{1 \mu}\,p_{1 \nu}
\nl
{}&+&
\cV^{\bbb}_{\sabbf}\, p_{2 \alpha}\,p_{2 \mu}\,p_{2 \nu}\!+\!
\cV^{\bbb}_{\sabbg}\, \{p_{1}\,p_{1}\,p_{2}\}_{\mu \nu \alpha}\!+\! 
\cV^{\bbb}_{\sabbh}\, \{p_{1}\,p_{2}\,p_{2}\}_{\mu \nu \alpha}\,, \, \, \,
\label{nff3}
\eqa
obtaining the following parametrization:
\bqa 
\cV^{\bbb}_{\sabbgen} &=& -\,\egam{2+\ep}\,\CIM{V}_{\bbb}\,
y^2\,\ox\,\oy\,R_{\bbbn{\sabbgen}}\,\chiu{\bbb}^{-2-\ep}, 
\qquad i > 4 \,, 
\nl 
R_{\bbbn{\sabbe}} &=& -\,(z_2-z_3)^3 ,
\quad
R_{\bbbn{\sabbf}} = (1-z_1-z_2)^3, 
\nl
R_{\bbbn{\sabbg}} &=& (z_2-z_3)^2\,(1-z_1-z_2), 
\quad
R_{\bbbn{\sabbh}} = -\,(z_2-z_3)\,(1-z_1-z_2)^2, 
\nl
\cV^{\bbb}_{\sabbgen} &=& -\,\frac{1}{2}\,\egam{1+\ep}\,\CIM{V}_{\bbb}\,
\ox\,R_{\bbbn{\sabbgen}}\,\chiu{\bbb}^{-1-\ep} \, , \quad i \le 4 \, ,
\nl
R_{\bbbn{\sabba}} &=&  y\,(z_2-z_3) ,
\quad
R_{\bbbn{\sabbb}} = -\,y\,(1- z_1-z_2)  , 
\nl
R_{\bbbn{\sabbc}} &=& -\,\oy\,(z_2-z_3)  ,
\quad
R_{\bbbn{\sabbd}} = \oy\,(1- z_1-z_2)\, . 
\eqa
Finally, for the tensor integral $\cV^{\bbb}(\mu, \nu | \alpha)$
we adopt the decomposition 
\bqa
\cV^{\bbb}(\mu,  \nu | \alpha ; \cdots) &=& 
\cV^{\bbb}_{\saaba} 
\{\delta\,p_1\}_{\mu\nu\,|\,\alpha} +
\cV^{\bbb}_{\saabb}\, 
\{\delta\,p_2\}_{\mu\nu\,|\,\alpha} +
\cV^{\bbb}_{\saabc}\, \delta_{\mu \nu} p_{1 \alpha} + 
\cV^{\bbb}_{\saabd}\, \delta_{\mu \nu} p_{2 \alpha}
+\cV^{\bbb}_{\saabe}\, p_{1 \alpha}\,p_{1 \mu}\,p_{1 \nu}
\nl
{}&+&
\cV^{\bbb}_{\saabf}\, p_{2 \alpha}\,p_{2 \mu}\,p_{2 \nu}\!+\!
\cV^{\bbb}_{\saabg}\, \{p_{1}\,p_{1}\,p_{2}\}_{\mu \nu \alpha}\!+\! 
\cV^{\bbb}_{\saabh}\, \{p_{1}\,p_{2}\,p_{2}\}_{\mu \nu \alpha}\, ,
\label{ffbb4}
\eqa 
where symmetrized products are defined in \eqn{symmete}.
We obtain the corresponding expression for the form factors:
\bqa
\cV^{\bbb}_{\saabgen} &=& -\,\egam{2+\ep}\,\CIM{V}_{\bbb}\,
y\,\ox^2\,\oy^2\,
R_{\bbbn{\saabgen}}\,\chiu{\bbb}^{-2-\ep},\qquad  
R_{\bbbn{\saabgen}} = -\,R_{\bbbn{\sabbgen}} ,
\quad i > 4 \,,
\nl
\cV^{\bbb}_{\saabgen} &=& -\,\frac{1}{2}\,\egam{1+\ep}\,\CIM{V}_{\bbb}\,
\ox\,R_{\bbbn{\saabgen}}\,\chiu{\bbb}^{-1-\ep} \, , \quad i = 1 \cdots 4 \, ,
\nl
R_{\bbbn{\saaba}} &=& -\ox\,\oy\,(z_2-z_3)\, ,
\quad
R_{\bbbn{\saabb}} =  \ox\,\oy\,(1-z_1-z_2)\,, 
\nl
R_{\bbbn{\saabc}} &=& (1-\ox\,\oy)\,(z_2-z_3)\, ,
\quad
R_{\bbbn{\saabd}} =  -\,(1-\ox\,\oy)\,(1-z_1-z_2)\, . 
\eqa
Note that $\cV^{\bbb}(\mu | \nu ,\alpha)$ and also 
$\cV^{\bbb}(\mu ,\nu | \alpha)$ require a smaller number of form factors 
than $\cV^{\bca}(\mu | \nu ,\alpha)$ and $\cV^{\bca}(\mu, \nu | \alpha)$. 
One can check that this is indeed the case by repeating the arguments 
already used in discussing $\cV^{\bbb}(\mu |\nu)$.

We conclude observing that another way of parametrizing rank three tensors
is through \eqn{fullG}, after which the corresponding form factors are
obtained with the help of \eqn{scalcomp}; the two sets of form factors are
easily related but with this parametrization and for a singular Gram matrix
the inversion can be done with its pseudo-inverse, as pointed out
in~\cite{Binoth:1999sp}.
\subsection{Diagrammatic interpretation of the reduction procedure 
\label{dirp}}
All the manipulations discussed in the previous Sections, aimed at reducing
form factors to combinations of scalar integrals, have a diagrammatic
counterpart. Diagrams with reducible scalar products in the numerator give
rise to standard scalar functions of the same family and contractions
corresponding to diagrams with fewer internal lines, as illustrated in
\fig{contra} (there, the symbol $\otimes$ denotes insertion of a scalar
product into the numerator of the diagram). The figure is based on the
simple relation $2\,\spro{q_2}{p_1} = [5]_{\bca} - [4]_{\bca} -
l_{145}$. After permutation of momenta we obtain the first of
Eqs.~(\ref{af30}) where the form factors are expressed in their standard
form.
\begin{figure}[ht]
\vspace{0.5cm}
{\Large\bqas  
\spro{q_2}{p_1}
\,\,
\otimes
\,\,
 \vcenter{\hbox{
 \begin{picture}(130,0)(0,0)
 \SetScale{0.6}
 \SetWidth{2.}
 \Line(0,0)(40,0)
 \Line(128,-53)(100,-35)
 \Line(128,53)(100,35)
 \Line(70,-17.5)(40,0)
 \Line(70,17.5)(40,0)
 \Line(70,-17.5)(70,17.5)
 \Line(100,-35)(70,-17.5)
 \Line(100,-35)(100,35)
 \Line(100,35)(70,17.5)
 \end{picture}}}
\!\!\!\!\!\!\!\!\!\!\!
&=&
\quad
-\,\frac{1}{2}\,\,l_{145}\,\,
 \vcenter{\hbox{
 \begin{picture}(130,0)(0,0)
 \SetScale{0.6}
 \SetWidth{2.}
 \Line(0,0)(40,0)
 \Line(128,-53)(100,-35)
 \Line(128,53)(100,35)
 \Line(70,-17.5)(40,0)
 \Line(70,17.5)(40,0)
 \Line(70,-17.5)(70,17.5)
 \Line(100,-35)(70,-17.5)
 \Line(100,-35)(100,35)
 \Line(100,35)(70,17.5)
 \end{picture}}}
\nl\nl\nl\nl
-\,\frac{1}{2}\,\,
 \vcenter{\hbox{
 \begin{picture}(130,0)(0,0)
 \SetScale{0.6}
 \SetWidth{2.}
 \Line(0,0)(40,0)
 \Line(128,-53)(100,-35)
 \Line(128,53)(100,35)
 \Line(70,-17.5)(40,0)
 \Line(70,17.5)(40,0)
 \Line(70,-17.5)(70,17.5)
 \Line(70,-17.5)(100,35)
 \Line(100,-35)(70,-17.5)
 \Line(100,35)(70,17.5)
 \end{picture}}}
\!\!\!\!\!\!\!\!\!\!\!
&+&
\qquad\quad
\frac{1}{2}\,\,\,
 \vcenter{\hbox{
 \begin{picture}(140,0)(0,0)
 \SetScale{0.6}
 \SetWidth{2.}
 \Line(0,0)(40,0)
 \Line(140,-26)(100,0)
 \Line(140,26)(100,0)
 \CArc(70,0)(30,0,360)
 \Line(70,-30)(70,30)
 \end{picture}}}
\eqas}
\vspace{1cm}
\caption[]{Diagrammatic interpretation of the reduction induced by a 
reducible scalar product. Here $l_{145} = p^2_1 - m^2_4 + m^2_5$,
while the symbol $\otimes$ denotes insertion of a scalar
product into the numerator of the diagram.} 
\label{contra}
\end{figure}

There are $7-I$ irreducible scalar products for two-loop vertices, neglecting
additional branching of the external lines (as in \fig{Moreirr}), $I$ being
the number of internal lines in the graph; in the reduction procedure they 
give raise to both contractions, i.e. scalar diagrams with less propagators, 
and to ordinary/generalized scalar functions of the same family as 
illustrated in \fig{genscalfun}.
\begin{figure}[ht]
\vspace{0.5cm}
{\Large\bqas  
\qquad\quad
\spro{q_1}{p_1}
\,\,
\otimes
\,\,
 \vcenter{\hbox{
 \begin{picture}(130,0)(0,0)
 \SetScale{0.6}
 \SetWidth{2.}
 \Line(0,0)(40,0)
 \Line(128,-53)(100,-35)
 \Line(128,53)(100,35)
 \Line(70,-17.5)(40,0)
 \Line(70,17.5)(40,0)
 \Line(70,-17.5)(70,17.5)
 \Line(100,-35)(70,-17.5)
 \Line(100,-35)(100,35)
 \Line(100,35)(70,17.5)
 \end{picture}}}
\!\!\!\!\!\!\!\!\!\!\!\!\!\!
&=&
\quad
\frac{p_1^2\,p_2^2 - (\spro{p_1}{p_2})^2}{P^2}\,\,
\omega^2\,
 \vcenter{\hbox{
 \begin{picture}(130,0)(0,0)
 \SetScale{0.6}
 \SetWidth{2.}
 \Line(0,0)(40,0)
 \Line(128,-53)(100,-35)
 \Line(128,53)(100,35)
 \Line(70,-17.5)(40,0)
 \Line(70,17.5)(40,0)
 \Line(70,-17.5)(70,17.5)
 \Line(100,-35)(70,-17.5)
 \Line(100,-35)(100,35)
 \Line(100,35)(70,17.5)
 \GCirc(70,0){5}{1}
 \GCirc(100,0){5}{1}
 \end{picture}}}
\nl\nl\nl\nl
-\,\frac{\spro{p_1}{P}}{2\,P^2}\,\,
\Bigg[\,
l_{\ssP 12}\,
 \vcenter{\hbox{
 \begin{picture}(130,0)(0,0)
 \SetScale{0.6}
 \SetWidth{2.}
 \Line(0,0)(40,0)
 \Line(128,-53)(100,-35)
 \Line(128,53)(100,35)
 \Line(70,-17.5)(40,0)
 \Line(70,17.5)(40,0)
 \Line(70,-17.5)(70,17.5)
 \Line(100,-35)(70,-17.5)
 \Line(100,-35)(100,35)
 \Line(100,35)(70,17.5)
 \end{picture}}}
\!\!\!\!\!\!\!\!\!\!\!\!\!\!\!\!
&-&
\,\,
 \vcenter{\hbox{
 \begin{picture}(130,0)(0,0)
 \SetScale{0.6}
 \SetWidth{2.}
 \Line(0,0)(42,0)
 \Line(128,-53)(100,-35)
 \Line(128,53)(100,35)
 \CArc(55,-9)(15,0,360)
 \Line(100,-35)(67,-15.75)
 \Line(100,-35)(100,35)
 \Line(100,35)(45,3)
 \end{picture}}}
\!\!\!\!\!\!\!\!\!\!\!\!\!\!\!\!
+\,\,
 \vcenter{\hbox{
 \begin{picture}(130,0)(0,0)
 \SetScale{0.6}
 \SetWidth{2.}
 \Line(0,0)(42,0)
 \Line(128,-53)(100,-35)
 \Line(128,53)(100,35)
 \CArc(55,9)(15,0,360)
 \Line(100,35)(67,15.75)
 \Line(100,-35)(100,35)
 \Line(100,-35)(45,-3)
 \end{picture}}}
\!\!\!\!\!\!\!\!\!\!\!\!\!\!\!\!
\Bigg]
\eqas}
\vspace{1cm}
\caption[]{Diagrammatic interpretation of the reduction induced by an 
irreducible scalar product. In the first diagram of the RHS non-canonical 
powers $-2$ in propagators are explicitly indicated by a circle and the 
space-time dimension is $6 - \ep$. Here $l_{\ssP 12} = P^2 - m^2_1 + m^2_2$
and $\omega = \mu^2/\pi$ where $\mu$ is the unit of mass.
The symbol $\otimes$ denotes insertion of a scalar product into the numerator 
of the diagram.}
\label{genscalfun}
\end{figure}
The component with contractions and ordinary scalar functions is given in
the second row of \fig{genscalfun} while the irreducible component is
expressed through a generalized scalar function in $6 - \ep$ space-time
dimension, as depicted in the first row of \fig{genscalfun} (there, a circle
denotes a non-canonical power $2$ for the corresponding propagator). Note
that the irreducible component appears multiplied by the Gram determinant.

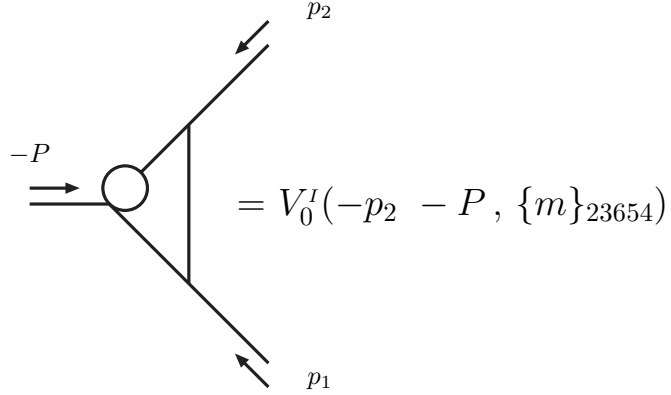
\begin{figure}[bh]
\vspace{1.9cm}
\bqas  
\qquad\qquad\qquad
\vcenter{\hbox{
  \begin{picture}(150,0)(0,0)
  \SetScale{0.6}
  \SetWidth{2.}
  \Line(0,0)(50,0)
  \Line(70,20)(150,100)
  \Line(50,0)(150,-100)
  \Line(100,-50)(100,50)
  \CArc(60.,10.)(14.,0.,360.)
  \LongArrow(0,10)(30,10)   \Text(0,15)[cb]{$-P$}
  \LongArrow(150,115)(130,95)   \Text(110,70)[cb]{$p_2$}
  \LongArrow(150,-115)(130,-95)     \Text(110,-70)[cb]{$p_1$}  
  \Text(160,-8)[cb]{\Large = $\cV^{\ssI}_0(-p_2\,\,-P\,,\,\lstm{23654})$}
  \end{picture}}}
\eqas
\vspace{1.5cm}
\caption[]{Rearrangement of arguments bringing the diagram in the l.h.s. to
the standard form of the $I$-family, see \sect{acafam}.} 
\label{tostandard}
\end{figure}

Whenever this relation, or similar ones, is used in the reduction procedure,
the last diagram on the r.h.s.\ of \fig{genscalfun} will be written as in
\eqn{af30} after a rearrangement of its arguments, see \fig{tostandard}.
In principle a generalized scalar function can be cast into the form of a
combination of ordinary scalar functions using IBPI techniques but, in
practice, these solutions are poorly known in the fully massive case; it is
somehow hard to accept that part of our present limitations are related to a
poor level of technical handling of large systems of linear equations;
however, this really represents the bottleneck of many famous approaches
(see~\cite{Smirnov:2003kc} for recent developments).

\section{Graphs, form factors and permutations \label{TopFFPerm}}
Diagrams of any renormalizable field theory, like the standard model, must
be generated according to the rules of the theory itself, they must be
assembled to construct some physical amplitude and a reduction must be
performed. There are many technical details hidden in this procedure, in
particular some efficient way of handling the different topologies while
assembling the grand total of diagrams.

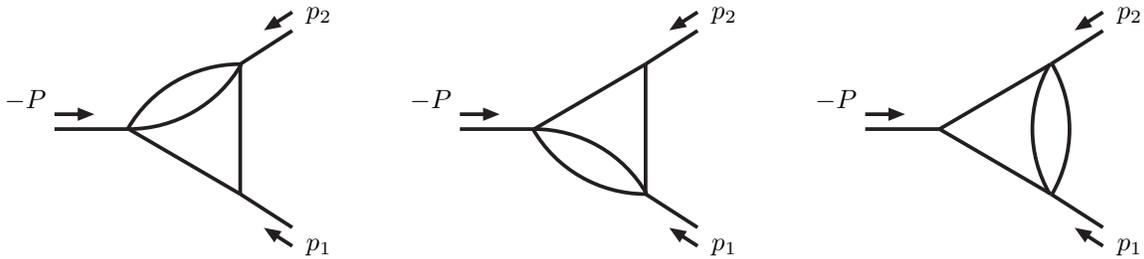
\begin{figure}[ht]
\vspace{2.0cm}
\bqas  
\qquad
{}&{}&
  \vcenter{\hbox{
  \begin{picture}(130,0)(0,0)
  \SetScale{0.7}
  \SetWidth{2.}
  \Line(0,0)(40,0)         
\LongArrow(0,8)(20,8)          \Text(-11,7)[cb]{$-P$}
  \Line(128,-53)(100,-35)  
\LongArrow(128,-63)(114,-54)   \Text(100,-48)[cb]{$p_1$}
  \Line(128,53)(100,35)    
\LongArrow(128,63)(114,54)     \Text(100,40)[cb]{$p_2$}
  \CArc(100,-35)(70,90,150)
  \CArc(40,70)(70,270,330)
  \Line(100,-35)(40,0)
  \Line(100,-35)(100,35)
  \end{picture}}}
\qquad
  \vcenter{\hbox{
  \begin{picture}(130,0)(0,0)
  \SetScale{0.7}
  \SetWidth{2.}
  \Line(0,0)(40,0)         
\LongArrow(0,8)(20,8)          \Text(-11,7)[cb]{$-P$}
  \Line(128,-53)(100,-35)  
\LongArrow(128,-63)(114,-54)   \Text(100,-48)[cb]{$p_1$}
  \Line(128,53)(100,35)    
\LongArrow(128,63)(114,54)     \Text(100,40)[cb]{$p_2$}
  \CArc(100,35)(70,210,270)
  \CArc(40,-70)(70,30,90)
  \Line(100,35)(40,0)
  \Line(100,-35)(100,35)
  \end{picture}}}
\qquad
  \vcenter{\hbox{
  \begin{picture}(130,0)(0,0)
  \SetScale{0.7}
  \SetWidth{2.}
  \Line(0,0)(40,0)         
\LongArrow(0,8)(20,8)          \Text(-11,7)[cb]{$-P$}
  \Line(128,-53)(100,-35)  
\LongArrow(128,-63)(114,-54)   \Text(100,-48)[cb]{$p_1$}
  \Line(128,53)(100,35)    
\LongArrow(128,63)(114,54)     \Text(100,40)[cb]{$p_2$}
  \CArc(160,0)(70,150,210)
  \CArc(40,0)(70,-30,30)
  \Line(100,-35)(40,0)
  \Line(100,35)(40,0)
  \end{picture}}}
\eqas
\vspace{1cm}
\caption[]{The $V^{\aba}$-family. External momenta flow inwards.} 
\label{Taba}
\end{figure}

\begin{figure}[ht]
\vspace{2.0cm}
\bqas  
\qquad
{}&{}&
  \vcenter{\hbox{
  \begin{picture}(130,0)(0,0)
  \SetScale{0.7}
  \SetWidth{2.}
  \DashLine(0,0)(40,0){3}              
  \LongArrow(0,8)(20,8)               \Text(-11,7)[cb]{$-P$}
  \Photon(128,-53)(100,-35){2}{4} 
  \LongArrow(128,-63)(114,-54)        \Text(100,-48)[cb]{$p_1$}
  \Photon(128,53)(100,35){2}{4}   
  \LongArrow(128,63)(114,54)          \Text(100,40)[cb]{$p_2$}
  \CArc(100,-35)(70,90,150)
  \DashCArc(40,70)(70,270,330){3}
  \Line(100,-35)(40,0)
  \DashLine(100,-35)(100,35){3}
  \Text(-15,-10)[cb]{$\hb$}
  \Text(80,0)[cb]{$\phi$}
  \Text(55,-10)[cb]{$\hb$}
  \Text(40,-48)[cb]{a)}
  \end{picture}}}
\qquad
  \vcenter{\hbox{
  \begin{picture}(130,0)(0,0)
  \SetScale{0.7}
  \SetWidth{2.}
  \DashLine(0,0)(40,0){3}              
  \LongArrow(0,8)(20,8)               \Text(-11,7)[cb]{$-P$}
  \Photon(128,-53)(100,-35){2}{4} 
  \LongArrow(128,-63)(114,-54)        \Text(100,-48)[cb]{$p_1$}
  \Photon(128,53)(100,35){2}{4}   
  \LongArrow(128,63)(114,54)          \Text(100,40)[cb]{$p_2$}
  \PhotonArc(160,0)(70,150,210){2}{8}
  \CArc(40,0)(70,-30,30)
  \Line(100,-35)(40,0)
  \Line(100,35)(40,0)
  \Text(40,-48)[cb]{b)}
  \end{picture}}}
\qquad
  \vcenter{\hbox{
  \begin{picture}(130,0)(0,0)
  \SetScale{0.7}
  \SetWidth{2.}
  \Photon(0,0)(40,0){2}{4}              
  \LongArrow(0,8)(20,8)               \Text(-10,7)[cb]{$p_2$}
  \DashLine(128,-53)(100,-35){3} 
  \LongArrow(128,-63)(114,-54)        \Text(102,-48)[cb]{$-P$}
  \Photon(128,53)(100,35){2}{4}   
  \LongArrow(128,63)(114,54)          \Text(100,40)[cb]{$p_1$}
  \CArc(100,-35)(70,90,150)
  \PhotonArc(40,70)(70,270,330){2}{8}
  \Line(100,-35)(40,0)
  \Line(100,-35)(100,35)
  \Text(40,-48)[cb]{c)}
  \end{picture}}}
\eqas
\vspace{1cm}
\caption[]{A $V^{\aba}$-family contribution to $H(-P) + \gamma(p_1) + 
\gamma(p_2) \to 0$. External momenta flow inwards.} 
\label{Hgg}
\end{figure}
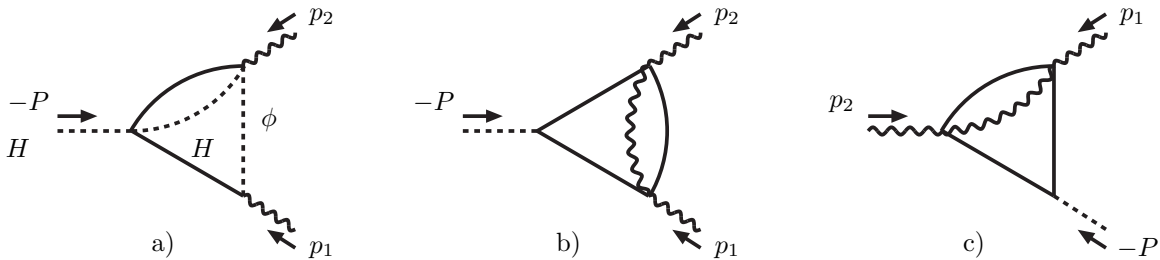

We briefly illustrate our approach: for the sake of clarity we refer to the
$V^{\aba}$-family. In principle, for a fixed choice of the external momenta
we should consider three kind of diagrams, as shown in \fig{Taba}.
However, in our automatized procedure, we will only compute the first diagram 
of \fig{Taba} since the remaining two are obtainable through permutation of 
the external momenta.
To illustrate the procedure we consider a specific example, the process
$H(-P) + \gamma(p_1) + \gamma(p_2) \to 0$; in the standard model there will be
diagrams like the one of \fig{Hgg}(a) which can be expressed as a
combinations of functions
$\cV^{\aba}_{i \cdots j}(p_2,P,\mw,\mh,\mw,\mw),$
but also diagrams like in \fig{Hgg}(b) which are always evaluated according
to the conventions of \fig{Hgg}(c).
Therefore they correspond to combinations of functions
\bq
\cV^{\aba}_{i \cdots j}(p_1,-p_2,\mw,0,\mw,\mw).
\eq
Similarly the decomposition into form factors will be as follows:
\bqa
\cV^{\aba}(\mu\,|\,0\,;\,p_1,-p_2,\mw,0,\mw,\mw) &=&
-\,\cV^{\aba}_{\saa}(p_1,-p_2,\,\cdots)\,P_{\mu}
+\,\cV^{\aba}_{12}(p_1,-p_2,\,\cdots)\,p_{1\mu},
\eqa
etc, showing that the consistent basis to expand the form factors is
$(-P\,,\,p_1)$. After permutation, the results of our paper follow
automatically. For a correct treatment of the combinatorial factors we refer
the reader to \appendx{symme}.
\section{Strategies for the evaluation of two-loop vertices \label{SETLV}}
Scalar configurations for irreducible two-loop vertices were considered and
evaluated in III, where tables of numerical results were presented.  The
techniques include several variations of the standard
BT-algorithm~\cite{Tkachov:1997wh} and the introduction of
parameter-dependent $C$-functions (for which we refer the reader to
Appendix~E of III, where they are introduced and their numerical evaluation
is discussed in Eqs.~(291-294)).

The same set of procedures can be easily generalized to cover a non-trivial 
theory (i.e.\ one with spin) using the defining parametric representations and 
the reduction formalism derived in this article. 

A few relevant examples will be shown and discussed in the
following sections. We will place special emphasis on proving that new
ultraviolet poles, not present in scalar configurations, do not prevent the
derivation of representations of the class \eqn{generalclass} 
for tensor integrals.

The three diagrams (with non-trivial numerators) belonging to the 
$V^{\ono}$-family, namely $V^{\aba}, V^{\aca}$ and $V^{\ada}$, are evaluated by
repeated applications of the BT algorithm~\cite{Tkachov:1997wh}; in this
case the procedure remains the same as for scalar configurations, since the
BT-algorithm works independently of the presence of additional polynomials
of Feynman parameters in the numerator. We only have to pay some attention
to the limit $\ep \to 0$, which cannot be taken from the very beginning for
tensor configurations that are ultraviolet divergent. The introduction of
parameter-dependent $C$-functions for tensor integrals of the remaining
families is also shown.
\subsection{Examples in the $V^{\aba}$-family \label{exaaba}}
A typical example is given by the $V^{\aba}$-family where we can easily 
provide integral representations for the scalar representative and for
the form factors, e.g.\
\bqa
\cV^{\aba}_{\Z} &=&  
-\,\frac{1}{\ep^2} - \oDUV^2 
+ \oDUV\,\Bigl[
2\,\dcub{2}\,\ln \chiu{\aba}(x,1,y) - 1\Bigr]
\nl
{}&+& \dcubs{x}{y,z}\,\frac{\ln \chiu{\aba}(x,y,z)}{1-y}\bmid_+
     + \dcub{2}\,\ln \chiu{\aba}(x,1,y)\,L_{\aba}(x,y)
     - \frac{3}{2} - \frac{1}{2}\,\zeta(2) ,
\nl
\cV^{\aba}_{a} &=&  
-\,\frac{1}{2}\,\Bigl[ \frac{1}{\ep^2} - \oDUV^2 
\Bigr] - \,\frac{3}{16} - \frac{1}{4}\,\zeta(2)
+  2\,\oDUV\,\Bigl[ \dcub{2}\,(1-y)\,
\ln \chiu{\aba}(x,1,y) + \frac{1}{8}\Bigr]
\nl
{}&-& \dcubs{x}{y,z}\,\Bigl[ \ln \chiu{\aba}(x,y,z) 
+ (1-z)\,\frac{\ln \chiu{\aba}(x,y,z)}{1-y}\bmid_+
\nl
{}&+& \dcub{2}\,( 1 - y )\,\ln \chiu{\aba}(x,1,y)\,L_{\aba}(x,y)  ,
\nl
\cV^{\aba}_{b} &=&  
\frac{1}{\ep^2} + \oDUV^2 - 2\,\oDUV\,
\dcub{2}\,\ln \chiu{\aba}(x,1,y)
  +\,\dcubs{x}{y,z}\,\ln \chiu{\aba}(x,y,z) 
\nl
{}&-& \dcubs{x}{y,z}\,\frac{\ln \chiu{\aba}(x,y,z)}{1-y}\bmid_+
   - \dcub{2}\,\ln \chiu{\aba}(x,1,y)\,L_{\aba}(x,y) 
   + 1 + \frac{1}{2}\,\zeta(2),
\eqa
\bq
\cV^{\aba}_{a} = \cV^{\aba}_{\sba} - \cV^{\aba}_{\sbb},
\qquad
\cV^{\aba}_{b} = \cV^{\aba}_{\sbb},  
\eq
where the l.h.s.\ of the last equation refers to the $(p_1,P)$ basis.
Furthermore, $L_{\aba}(x,y) = \ln (1-y) - \ln x - \ln(1-x) - \ln
\chiu{\aba}(x,1,y)$ and $\chiu{\aba}$ is obtained from \eqn{Defchiaba} by
rescaling by $1/\mid P^2\mid$.

Smooth integral representations for higher tensors can be classified according
to
\bqa
\cV^{\aba}_i &=& K_i +
a_i\,\oDUV\,\dcub{2}\,y^{\alpha_i}\,\ln\chiu{\aba}(x,1,y) 
+ b_i\,\dcubs{x}{y,z}\,z^{\beta_i}\,\frac{\ln\chiu{\aba}(x,y,z)}{1-y}\bmid_+
\nl
{}&+&
\dcubs{x}{y,z}\,P_i(y,z)\,\ln\chiu{\aba}(x,y,z) 
+ c_i\,\dcub{2}\,y^{\gamma_i}\,\ln\chiu{\aba}(x,1,y)\,L_{\aba}(x,y),
\label{fftable}
\eqa
and coefficients and exponents for the first few cases are reported in
\tabn{FFtable}.
\small
\begin{table}[ht]\centering
\renewcommand\arraystretch{1.2}
\begin{tabular}{|c|c|c|c|c|c|c|c|c|}
\hline
& & & & & & & & \\
$i$ & $K$ & $a$ & $\alpha$ & $b$ & $\beta$ & $P$ & $c$ & $\gamma$ \\ 
& & & & & & & & \\
\hline
& & & & & & & & \\
$221$ & $-\,\frac{2}{3}\,\oDUV\,\ep^{-1} - \frac{1}{18}\,\oDUV - \frac{1}{3}\,
\DUV^2 - \frac{151}{216} - \frac{1}{6}\,\zeta(2)$ &
$2$ & $2$ & $1$ & $2$ & $0$ & $1$ & $2$ \\
& & & & & & & & \\
\hline
& & & & & & & & \\
$222$ & $-2\,\oDUV\,\ep^{-1} + \frac{1}{2}\,\oDUV - \DUV^2
- \frac{7}{8} - \frac{1}{2}\,\zeta(2)$ &
$2$ & $0$ & $1$ & $0$ & $-1-y$ & $1$ & $0$ \\
& & & & & & & & \\
\hline
& & & & & & & & \\
$223$ & $-\,\oDUV\,\ep^{-1}  - \frac{1}{2}\,\DUV^2
- \frac{3}{4} - \frac{1}{4}\,\zeta(2)$ &
$2$ & $1$ & $1$ & $1$ & $-z$ & $1$ & $1$ \\
& & & & & & & & \\
\hline
\hline
\end{tabular}
\vspace*{3mm}
\caption[]{Parameters for the $V^{\aba}$ form factors according to
\eqn{fftable}.}
\label{FFtable}
\end{table}
\normalsize
\subsection{Examples in the $V^{\aca}, V^{\ada}$ families \label{exaacaada}}
Consider the $V^{\aca}$-family as a second example. All form factors can be
expressed as linear combinations of integrals of the following kind
($\chiu{\aca}$ is obtained from \eqn{Defchiaca} by rescaling by 
$1/\mid P^2\mid$):
\bqa
I_{\acan{0}} &=& \CIM{V}_{\aca}\,\chiu{\aca}^{-1-\ep},
\quad
I_{\acan{1x}} = \CIM{V}_{\aca}\,\chiu{\aca}^{-1-\ep}\,x, 
\quad
I_{\acan{2xx}} = \CIM{V}_{\aca}\,\chiu{\aca}^{-1-\ep}\,x^2, \qquad
\mbox{etc.}
\label{expacaone}
\eqa
As an illustration we derive the ultraviolet finite part for the first few
integrals of the list, introducing
\bqa
I^{\rm fin}_{\acan{n}} &=& \frac{1}{M^2\,b_{\aca}}\,\Bigl[
\int_0^1\,dx dy\int_0^{1-y}\,dz_1\,\int_0^{z_1}\,dz_2\,I^4_{\acan{n}} +
\int_0^1\,dx dy\int_0^{1-y}\,dz\,I^3_{\acan{n}} +
\dcub{2}\,I^2_{\acan{n}} + I^0_{\acan{n}} \Bigr].
\label{expacatwo}
\eqa
Notation follows closely that of Sect.~6.1 of III (see also 
Eq.~(12) of III for the definition of $[y,z,u]_i$)
and $b_{\aca}$ is the BT-factor of the function (see Eq.~(64) of III); 
therefore we have
\bq
f(\{x\}\,;\,[y\,z\,u]_i) = \left\{
\begin{array}{ll}
f(\{x\}\,;\,y,z) & \mbox{for $i = 0$} \\
f(\{x\}\,;\,z,z) & \mbox{for $i = 1$} \\
f(\{x\}\,;\,z,u) & \mbox{for $i = 2$} 
\end{array}
\right. .
\eq
\bqa
b_{\aca} &=& (\nu_1^2+\mu_3^2-\mu_4^2)^2
- \mu_{35}^2\,(1+\nu_1^2-\nu_2^2)\,(\nu_1^2+\mu_3^2-\mu_4^2) + 
\mu_{35}^4\,\nu_1^2 + \lambda_{\aca} \mu_3^2,
\eqa
where we have set $\lambda_{\aca} = \lambda(1,\nu_1^2,\nu_2^2)$ and where
\bqa
\mu^2_i &=& \frac{m^2_i}{\mid P^2\mid}, \quad i=1,\dots, N,
\qquad
\nu^2_j = \mid \frac{p^2_j}{P^2}\mid, \quad j=1,2,
\qquad
\mu_{ij}^2 = 1 + \mu_i^2 - \mu_j^2.
\label{scaledq}
\eqa
BT co-factors are
\bqa
Z_{\acan{0}} &=& -\,\lambda_{\aca}, 
\qquad
Z_{\acan{3}} = 0,
\qquad
Z_{\acan{1}} =   (1-\nu_1^2-\nu_2^2)\,(\nu_1^2-\mu_{45}^2) + 
2 \, (\nu_1^2+\mu_3^2-\mu_4^2)\,\nu_2^2 
\nl
Z_{\acan{2}} &=&  - (1-\nu_1^2-\nu_2^2)\,(\nu_1^2+\mu_3^2-\mu_4^2) - 
2 \, (\nu_1^2-\mu_{45}^2)\,\nu_1^2,
\quad
Z^-_{\acan{i}} = Z_{\acan{i}} - Z_{\acan{i+1}}.
\eqa
We define additional auxiliary functions:  
\bqa
\xiu{\aca}(x,y,z_1,z_2) &=& \chiu{\aca}(x,1-y,z_1,z_2),
\nl
L_{\aca}(x,y,z_1,z_2) &=& \ln(1-y)-\ln(x)-\ln(1-x) - 
\ln\xiu{\aca}(x,y,z_1,z_2).
\eqa
Our results are as follows:
\bqa
I^4_{\acan{0}} &=&
- \frac{\ln\xiu{\aca}(x,y,z_1,z_2)}{y}\bmid_+,
\nl
I^3_{\acan{0}} &=&
\Bigl[ 1 -\,L_{\aca}(x,0,1-y,z)\Bigr]\,\ln\,\xiu{\aca}(x,0,1-y,z)
+\frac{1}{2}\,\ln \xiu{\aca}(x,y,[1-y,z,0]_0)
\nl
{}&+&\,\frac{1}{2}\,\sum_{i=0}^{2}\,Z^-_{\acan{i}}\,
\frac{\ln\xiu{\aca}(x,y,[1-y,z,0]_i)}{y}\bmid_+,
\nl
I^2_{\acan{0}} &=&   
\frac{1}{2}\,\sum_{i=0}^{2}\,Z^-_{\acan{i}}\,
\ln\xiu{\aca}(x,0,[1-y,z,0]_i)\,L_{\aca}(x,0,[1-y,z,0]_i),
\qquad
I^0_{\acan{0}} = -\,\frac{1}{4}\,S_1(1) + \frac{1}{8},
\nl\nl
I^i_{\acan{1x}} &=& x\,I^i_{\acan{0}}, \quad (i \not= 0),
\qquad
I^0_{\acan{1x}} = -\,\frac{1}{8}\,S_1(2) + \frac{1}{8},
\qquad
I^4_{\acan{1y}} = -\,\ln\xiu{\aca}(x,y,z_1,z_2),
\nl\nl
I^3_{\acan{1y}} &=&
\frac{1}{2}\,\sum_{i=0}^{2}\,Z^-_{\acan{i}}\,
\ln\xiu{\aca}(x,y,[1-y,z,0]_i) +
\frac{1}{2}\,y\,\ln\xiu{\aca}(x,y,[1-y,z,0]_0),
\nl
I^3_{\acan{1y}} &=& I^2_{\acan{1y}} = 0,
\qquad
I^0_{\acan{1y}} = -\,1,
\qquad
I^4_{\acan{1z_1}} =
\frac{1}{2}\,(Z_1 - 3\,z_1)\,\frac{\ln\xiu{\aca}(x,y,z_1,z_2)}{y}\bmid_+,
\nl\nl
I^3_{\acan{1z_1}} &=&
\frac{1}{2}\,(Z_1 - 3\,z)\,L_{\aca}(x,0,1-y,z)\,\ln\xiu{\aca}(x,0,1-y,z)
+\frac{1}{2}\,(Z_1 - y)\,\ln\xiu{\aca}(x,y,[1-y,z,0]_0)
\nl
{}&+& z\,\ln\xiu{\aca}(x,0,1-y,z)
+\frac{1}{2}\,Z^-_{\acan{0}}\,\frac{\ln\xiu{\aca}(x,y,[1-y,z,0]_0)}{y}\bmid_+
+\frac{z}{2}\,Z^-_{\acan{1}}\,
\frac{\ln\xiu{\aca}(x,y,[1-y,z,0]_1)}{y}\bmid_+
\nl
{}&+& \frac{z}{2}\,Z^-_{\acan{2}}\,\frac{\ln\xiu{\aca}(x,y,[1-y,z,0]_2)}
{y}\bmid_+,
\nl
I^2_{\acan{1z_1}} &=&
\frac{1}{2}\,Z^-_{\acan{0}}\,L_{\aca}(x,0,[1,1-y,0]_0)\,
\ln\xiu{\aca}(x,0,[1,1-y,0]_0)
\nl
{}&+& \,\frac{1-y}{2}\,\sum_{i=1,2}\,Z^-_{\acan{i}}\,
L_{\aca}(x,0,[1,1-y,0]_i)\,\ln\xiu{\aca}(x,0,[1,1-y,0]_i),
\nl
I^0_{\acan{1z_1}} &=& -\,\frac{1}{6}\,S_1(1) + \frac{5}{36},
\qquad
I^4_{\acan{1z_2}} =
\frac{1}{2}\,(Z_2 - 3\,z_2)\,\frac{\ln\xiu{\aca}(x,y,z_1,z_2)}{y}\bmid_+,
\nl\nl
I^3_{\acan{1z_2}} &=&
\frac{1}{2}\,(Z_2 - 3\,z)\,L_{\aca}(x,0,1-y,z)\,\ln\xiu{\aca}(x,0,1-y,z)
+\,\frac{z_1}{2}\,\sum_{i=0}^{1}\,Z^-_{\acan{i}}\,
\frac{\ln\xiu{\aca}(x,y,[1-y,z,0]_i)}{y}\bmid_+
\nl
{}&+&\frac{z}{2}\,\ln\xiu{\aca}(x,y,[1-y,z,0]_0)
+z\,\ln\xiu{\aca}(x,0,1-y,z),
\nl
I^2_{\acan{1z_2}} &=&
\frac{1-y}{2}\,\sum_{i=0}^{1}\,Z^-_{\acan{i}}\,
L_{\aca}(x,0,[1,1-y,0]_i)\,\ln\xiu{\aca}(x,0,[1,1-y,0]_i),
\quad
I^0_{\acan{1z_2}} = -\,\frac{1}{12}\,S_1(1) + \frac{5}{72},
\label{expacathree}
\eqa
where $S_n(k) = \sum_{l=1}^{k}\,l^{-n}$. 
Similar expressions can be written also for higher order form factors
showing, once more, that scalar and tensor integrals give similar results
and can be treated in one single stroke.

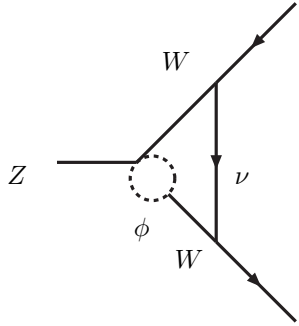
\begin{figure}[th]
\vspace{1.5cm}
\bqas  
{}&{}&
  \vcenter{\hbox{
  \begin{picture}(150,0)(0,0)
  \SetScale{0.6}
  \SetWidth{2.}
  \Line(0,0)(50,0)
  \Line(50,0)(100,50)
  \ArrowLine(150,100)(100,50)
  \Line(70,-20)(100,-50)
  \ArrowLine(100,-50)(150,-100)
  \ArrowLine(100,50)(100,-50)
  \DashCArc(60.,-10.)(14.,0.,360.){3}
  \Text(-15,-8)[cb]{$\zb$}
  \Text(70,-8)[cb]{$\nu$}
  \Text(45,35)[cb]{$\wb$}
  \Text(50,-40)[cb]{$\wb$}
  \Text(32,-30)[cb]{$\phi$}
  \end{picture}}}
\eqas
\vspace{1.5cm}
\caption[]{Diagram of the $V^{\aca}$-family contributing to $\zb \to l^+l^-$.} 
\label{zffvertaca}
\end{figure}

Once again the whole procedure can be described in terms of a specific
example.  Consider the diagram of \fig{zffvertaca} which contributes to the
on-shell decay amplitude $\zb \to l^+l^-$. The on-shell vertex, including
external wave-functions, is decomposed according to \eqn{decompo} and the
corresponding coefficients are subsequently evaluated; for instance we
consider the contribution coming from the diagrams of \fig{zffvertaca} and
derive the vector coefficient in the limit $m_f = 0$.
Using \eqn{vector} and taking the trace we obtain the following expression
\bqa
F_{\ssV} &=& \, \frac{i \pi^4 \gb^5 \stw^2}{16 \ctw} 
\Bigl[\,\mzs \, \cV^{\aca}_{\Z} \, (p_1,P,\mz,\mw,\mw,0,\mw) \, 
+ \, \mws \, \cV^{\aca}_{\saa} \, (p_1,P,\mz,\mw,\mw,0,\mw) \, 
\nl
{}&+& (\, \mws \, + \mzs \,) \,\cV^{\aca}_{12} \, (p_1,P,\mz,\mw,\mw,0,\mw) \, 
+\,A_{\Z}([\mw,\mz])\,C_{\Z}(p_1,p_2,\mw,0,\mw) \, 
\nl
{}&-& \, \cV^{\aba}_{\Z} \, (p_1,P,\mz,\mw,0,\mw) \,   
   +  \, \cV^{\aba}_{\Z} \, (0,P,\mz,\mw,\mw,\mw) \,  
\nl
{}&+& \, 2 \,  \cV^{\aba}_{\saa} \, (0,P,\mz,\mw,\mw,\mw) \,   
   -   \cV^{\aba}_{12} \, (p_1,P,\mz,\mw,0,\mw) \,  \Bigr].
\label{onesexa}
\eqa
The form factors of the $V^{\aba}$-family in \eqn{onesexa} can be further
reduced according to the results of \sect{abafam} or, more conveniently,
they can be computed according to \eqns{ffaba}{Pffaba}. A similar situation
appears for the form factors of \eqn{onesexa} of the $V^{\aca}$-family for
which we use the reduction techniques of \sect{acafam} or an explicit
evaluation using \eqns{expacaone}{expacathree}.

Results in the $V^{\ada}$-family are very similar in their structure and
will not be reported here. Furthermore, the graph corresponds to a one-loop
self-energy insertion which should be Dyson--re-summed.
\subsection{Examples in the $V^{\bba}$-family \label{exabba}}
Coming back to the strategy to evaluate tensor integrals, we observe that
two other scalar diagrams, $V^{\bba}$ and $V^{\bca}$, were expressed in
III in terms of integrals of $C$-functions (Appendix~E of III).

It is very easy to extend the derivation to tensors. Consider, for instance,
the $V^{\bba}$ case: starting from \eqn{systembba} the appropriate strategy
will be as follows. If we need to prove a WST identity, where the presence
of Gram determinants is inessential, we simply invert the system and derive
$\cV^{\bba}_{\sijgen}$ with $i,j = 1,2$ in terms of known quantities. If 
instead we need to use these form factors to compute some physical observable,
then a possible strategy is the following: suppose that $p^2_1 \not= 0$, then
$\cV^{\bba}_{\saa}$ is eliminated and $\cV^{\bba}_{\sab}$ is either given in
terms of $\cV^{1,1|1,2|2}$ at $n = 6 - \ep$ or explicitly evaluated.

If we choose the second strategy then $\chiu{\bba}$ is a quadratic form in 
$y_1,y_2$, with $x$-dependent coefficients and we can use the results of 
Appendix~E of III to write
\bq
\cV^{\bba}_{\sab} = -\,\frac{1}{M^2}\,\dsimp{2}(\{x\})\,x_2\,\Bigl[
C_{\saa}(0) - C_{\Z}(0)\Bigr],
\eq
with $|P^2|= M^2$. Similarly, if $\spro{p_1}{p_2} \not= 0$ we can eliminate 
$\cV^{\bba}_{\sba}$ and express $\cV^{\bba}_{\sbb}$ in terms of generalized 
scalars as in \eqn{GStt}, or explicitly derive
\bq
\cV^{\bba}_{\sbb} = -\,\frac{1}{M^2}\,\dsimp{2}(\{x\})\,\Bigl[
C_{\saa}(0) - C_{\Z}(0)\Bigr].
\eq
For this family the rank two tensor integrals are ultraviolet divergent.
For the form factors of the $22i$-family defined in \eqn{tensorbba} the
relevant quantity is $\cV^{\bba}_{\sbbd} = \,\cV^{2,1|1,1|1}$ evaluated at
$n = 6 -\ep$ which, with $\chiu{\bba}$ obtained from \eqn{Defchibba} by
rescaling by $1/\mid P^2\mid$, can be rewritten according to
\bqa
\cV^{\bba}_{\sbbd} &=& -\frac{1}{2}\,\Bigl(\frac{\omega}{M^2}\Bigr)^\ep
\egam{\ep}\,\dsimp{2}(\{x\})
\Bigl[ x_2\,(1 - x_2)\Bigr]^{-1-\ep}\,\dsimp{2}(\{y\})y^{\ep/2}_2\,
\chiu{\bba}^{-\ep}.
\label{showingthepole}
\eqa
\eqn{showingthepole} shows the expected ultraviolet poles; indeed the integral 
is overall ultraviolet divergent and so is the $(\beta,\gamma)$ sub-diagram.
With $\omega$ defined in \eqn{defDUV} and $\chiu{\bba\,;\,0} = 
\chiu{\bba}(x_2 = 0)$ we obtain
\bqa
\cV^{\bba}_{\sbbd} &=& \frac{1}{2}\,\dsimp{2}(x_1,x_2)\,\dsimp{2}(y_1,y_2)\,
(1- x_2)^{-1}\,\cV^{\bba\,;\,4}_{\sbbd} +
\frac{1}{2}\,\dcubs{x_1}{y_1,y_2}\,\cV^{\bba\,;\,3}_{\sbbd} + K,
\nl
\cV^{\bba\,;\,4}_{\sbbd} &=& \ln\chiu{\bba\,;\,0} +
\frac{\ln\chiu{\bba}}{x_2}\bmid_+,
\qquad
\cV^{\bba\,;\,3}_{\sbbd} = -\,\ln\chiu{\bba\,;\,0}\,
\Bigl( \oDUV - \ln x_1 + \frac{1}{2}\,\ln y_2 -
\frac{1}{2}\,\ln\chiu{\bba\,;\,0}\Bigr),
\nl
K &=& \frac{1}{8}\,\Bigl(\frac{1}{\ep^2} + 
\oDUV^2\Bigr) - \frac{3}{16}\, \oDUV + \frac{7}{64} - \frac{1}{8}\,
\zeta(2),
\eqa
where, as usual, $\oDUV = 1/\ep - \Delta_{\ssU\ssV}$ and
$\Delta_{\ssU\ssV} = \gamma - \ln\omega/M^2$,
with $M^2 = \mid P^2\mid$. Similarly $\cV^{\bba}_{\saad}$ will develop a double
ultraviolet pole being overall divergent with $(\alpha,\gamma)$ divergent. In
this case the additional pole is hidden in the $y_2$-integration, as shown in
\eqn{hidden}.
\subsection{Examples in the $V^{\bca}, V^{\bbb}$-families \label{exabcabbb}}
Also the $V^{\bca}$-family can be expressed in terms of well-behaved
integrals of $C$-functions, introduced in Appendix~E of III. From \eqn{Igen}
we see that one of the relevant objects to be evaluatet is $I_{\bcan{\ssR}}
= \cV^{1,1|1,2,1|2}$ for $n = 6 -\ep$: the important result is that this
quantity can be computed along the same lines of the corresponding scalar
integral.

The derivation is straightforward: starting from \eqn{varaibca} we will
adopt the same technique as in Sect.~9.1 of III; with $X = (1-x_1)/(1-x_2) =
1 - \bX$ we change variables according to $y_1 = y'_1 + X\,y_3, y_2 = y'_2 +
X\,y_3$ and $y_3 = y'_3$. Next we perform the $y_3$ integration
analytically; after that the $y_1 - y_2$ interval is mapped into the
standard triangle $0 \le y_2 \le y_1 \le 1$ and the net result is a
combination of $10$ integrals of $C$ functions with $\{x\}$ dependent
parameters, as defined in Tab.~2 of III.  Therefore, we obtain expressions
for both standard and generalized scalar as
\bqa
\cV^{\bca}_{\Z} &=& - \frac{1}{M^4}\,\dsimp{2}(x_1,x_2)\,
\frac{x_2}{\Delta(x_1,x_2)}
\nl
{}&\times&
\Bigl[ \ox_1^2\,C_{\Z}([1-2]) - \ox_1\,\ox_2\,C_{\Z}([3-4]) - 
\ox_1\,\ox_2 \,C_{\Z}([5-6]) - \ox^2\,C_{\Z}([7-8]) + \ox_1\,\ox\,C_{\Z}([9-10]
\Bigr], 
\nl
I_{\bcan{\ssR}} &=& - \frac{1}{M^4}\,\dsimp{2}(x_1,x_2)\,
\frac{x^2_2}{\Delta(x_1,x_2)}
\nl
{}&\times&
\Bigl\{
- \frac{\ox^3_1}{\ox_2} \, \Bigl[ C_{\saa}([1-2]) - C_{12}([1-2]) \Bigr]
- \ox_1\,\ox_2\, \Bigl[ C_{\saa}([5-6]) + C_{12}([3-4]) \Bigr]
\nl
{}&+& \ox^2_1 \, \Bigl[ C_{\saa}([3-4]) + C_{12}([5-6]) \Bigr]
- \frac{\ox^3}{\ox_2} \, \Bigl[  C_{\saa}([7-8]) - C_{12}([7-8]) \Bigr]
\nl
{}&-& \frac{\ox_1\,\ox^2}{\ox_2} \, C_{12}([9-10])
+ \ox_1\,\ox \, C_{\Z}([9-10])
- \frac{\ox^2_1\,\ox}{\ox_2} C_{\saa}([9-10)]
\Bigr\},
\eqa
where $C_n([i-j]) = C_n(i) - C_n(j)$ and where $\ox_i = 1 - x_i$, 
$\ox = x_1 - x_2$. Furthermore we have
\bqa
\Delta(x_1,x_2) &=& \nu^2_x - x_2\,(1-x_2)\,\mu^2_4 + x_2\,(1 - x_1)\,
(\mu^2_4 - \mu^2_6 + s_p),
\nl
\nu^2_x &=& - s_p\,x^2_1 + x_1\,( - s_p + \mu^2_1 - \mu^2_2) +
x_2\,( \mu^2_3 - \mu^2_1) + \mu^2_2,
\eqa
where, according to III, we introduced $P^2 = -\,s_p\,M^2$,
$\mu^2_i = m^2_i/\mid P^2\mid$ and $\nu^2_j = \mid p^2_j/P^2\mid$.

In this family we can show another example of ultraviolet divergent
form factor in a situation where the corresponding scalar integral is
convergent. Consider $\cV^{\bca}_{\saad}$ defined in \eqn{thisdiv}. Since the
$q_1$ sub-loop diverges we expect a simple pole at $\ep = 0$. Let us define
\bq
x_2\,(1 - x_2)\,\chiu{\bca}(x_1,x_2,y_1+X\,y_3,y_2+X\,y_3,y_3) = 
\xiu{\bca}(\{x\},\{y\}) \equiv \xiu{\bca}(y_1,y_2,y_3),
\eq
where $\chiu{\bca}$ is obtained from \eqn{Defchibca} by rescaling by 
$1/\mid P^2\mid$; $\xiu{\bca}$ is a quadratic form in $y_1,y_2$, linear 
in $y_3$, with $x$-dependent coefficients. The procedure of extracting the 
ultraviolet pole (a subtraction, as introduced in III), followed by a mapping 
of the $y_1,y_2$ integration regions into the standard triangle $0 \le y_2 
\le y_1 \le 1$, will introduce several new quadratic forms which will be 
enumerated as follows:
\bqa
\xi(1-y_1,1-y_2) &=& \xi_1,
\qquad
\xi(1-X\,y_1,1-X\,y_2) = \xi_2,
\nl
\xi(1-X\,y_1,1-y_2) &=& \xi_3,
\qquad
\xi(1-y_1,1-X\,y_2) = \xi_4,
\nl
\xi(\bX\,y_1,\bX\,y_2) &=& \xi_5,
\qquad
\xi(1-X\,y_2,1-y_1) = \xi_6,
\nl
\xi(1-y_2,1-X\,y_1) &=& \xi_7,
\qquad
\xi(1-X\,y_3\,y_1,1-X\,y_3\,y_2) = \xi_8,
\nl
\xi(1-X\,y_1,\bX\,y_2) &=& \xi_9,
\qquad
\xi(1-X\,y_3\,y_1,1-y_2) = \xi_{10},
\nl
\xi(1-X\,y_3\,y_2,1-y_1) &=& \xi_{11},
\qquad
\xi(\bX\,y_3\,y_2,\bX\,y_3\,y_1) = \xi_{12},
\nl
\xi(1-y_1,\bX\,y_2\,y_3) &=& \xi_{13},
\qquad
\xi(1-X\,y_3\,y_2,\bX\,y_3\,y_1) = \xi_{14},
\nl
\xi(1-X\,y_3\,y_1,\bX\,y_3\,y_2) &=& \xi_{15},
\qquad
\xi(1-y_2,\bX\,y_3\,y_1) = \xi_{16},
\nl
\xi(1-X\,y_3\,y_1,1-X\,y_3\,y_2,y_3) &=& \xi_{17},
\qquad
\xi(1-X\,y_3\,y_1,1-y_2,y_3) = \xi_{18},
\nl
\xi(1-X\,y_3\,y_1,\bX\,y_3\,y_2,y_3) &=& \xi_{19},
\qquad
\xi(1-y_1,\bX\,y_3\,y_2,y_3) = \xi_{20},
\nl
\xi(1-y_2\,y_3\,X,1-y_1,y_3) &=& \xi_{21},
\qquad
\xi(1-y_2\,y_3\,X,y_1\,y_3\,\bX,y_3) = \xi_{22},
\nl
\xi(1-y_2,y_1\,y_3\,\bX,y_3) &=& \xi_{23},
\qquad
\xi(y_2\,y_3\,\bX,y_1\,y_3\,\bX,y_3) = \xi_{24},
\nl
\xi(1-y_2,1-y_1) &=& \xi_{25},
\qquad
\xi(1-y_2,1-y_1,y_3) = \xi_{26},
\eqa
where $\xiu{\bca}(y_1,y_2) \equiv \xiu{\bca}(y_1,y_2,0)$. We introduce new
functions corresponding to well-defined integrals of the $C$-class:
\bq
\dsimp{2}(y_1,y_2)\,\xi^{-1-\ep}_l = \cC^0_{\Z}(l) - 
\frac{\ep}{2}\,\cC^1_{\Z}(l) + \ord{\ep^2},
\qquad
\cC^l_{1i}(l) = \dsimp{2}(y_1,y_2)\,\xi^{-1}_l\,\ln y_i.
\eq
All of them can be evaluated with the same algorithm described in Appendix~E of
III. Collecting all the ingredients we obtain
\bq
\cV^{\bca}_{\saad} = -\frac{1}{2}\,
\Bigl(\frac{\omega}{M^2}\Bigr)^{\ep}\,\frac{\egam{1+\ep}}{M^4}\,
\dsimp{2}(\{x\})\,x_2\,
\Bigl[ \cV^{\ssS\ssP}_{\bcan{114}}\,\frac{1}{\ep} + 
\cV^{\ssA}_{\bcan{114}} + \intfx{y_3}\,\cV^{\ssB}_{\bcan{114}}\Bigr],
\eq
\bq
\cV^{\ssP}_{\bcan{114}} = 2\,\ox_1\,\Bigl[ \cC^0_{\Z}(3) + \cC^0_{\Z}(6)\Bigr] -
2\,\frac{\ox^2_1}{\ox_2}\,\cC^0_{\Z}(2) +
2\,\frac{\ox^2}{\ox_2}\,\cC^0_{\Z}(5), 
\eq
\bqa
\cV^{\ssA}_{\bcan{114}} &=&
- \frac{\ox\,\ox_1}{\ox_2}\,\Bigl[  \cC^l_{\saa}(9) - \cC^l_{12}(9) \Bigr]
+ \ox_1\,\Bigl\{ 
\Bigl[ \cC^0_{\Z}(3) + \cC^0_{\Z}(6)\Bigr]\,\ln x_2\,(1-x_2) 
\nl
{}&-& \cC^1_{\Z}(3) - \cC^1_{\Z}(6) + \cC^l_{\saa}(3) + \cC^l_{12}(6) \Bigr\} 
- \frac{\ox^2_1}{\ox_2}\,\Bigl[  \cC^0_{\Z}(2)\,\ln x_2\,(1-x_2) - 
  \cC^1_{\Z}(2) + \cC^l_{\saa}(2) \Bigr]
\nl
{}&+& \frac{\ox^2}{\ox_2}\,\Bigl[ \cC^0_{\Z}(5)\,\ln x_2\,(1-x_2) -  
  \cC^1_{\Z}(5) + \cC^l_{12}(5) \Bigr],
\nl
\cV^{\ssB}_{\bcan{114}} &=&
- y_3\,x_2\,\frac{\ox_1}{\ox_2}\,\Bigl[  \cC^0_{\Z}(18) + \cC^0_{\Z}(21) \Bigr]
- y_3\,x_2\,\frac{\ox}{\ox_2}\,\Bigl[  \cC^0_{\Z}(20) + \cC^0_{\Z}(23) \Bigr]
\nl
{}&-& y_3\,\frac{\ox\,\ox_1}{\ox_2}\,\Bigl[ \cC^0_{\Z}(14) + \cC^0_{\Z}(15) - 
      \cC^0_{\Z}(19) - \cC^0_{\Z}(22) \Bigr]
- y_3\,\frac{\ox^2_1}{\ox_2}\,\Bigl[  \cC^0_{\Z}(8) - \cC^0_{\Z}(17) \Bigr]
\nl
{}&+& y_3^2\,x_2\,\frac{\ox\,\ox_1}{\ox^2_2}\,\Bigl[ \cC^0_{\Z}(19) + 
   \cC^0_{\Z}(22) \Bigr]
+ y_3^2\,x_2\,\frac{\ox^2_1}{\ox^2_2}\, \cC^0_{\Z}(17) 
+ y_3^2\,x_2\,\frac{\ox^2}{\ox^2_2}\, \cC^0_{\Z}(24) 
\nl
{}&+& x_2\, \cC^0_{\Z}(26) 
+ \ox_1\,\Bigl[ \cC^0_{\Z}(10) + \cC^0_{\Z}(11) - \cC^0_{\Z}(18) - 
\cC^0_{\Z}(21) \Bigr]
- \,\frac{\ox_2}{y_3}\,\Bigl[  \cC^0_{\Z}(25) - \cC^0_{\Z}(26) \Bigr]
\nl
{}&+& \ox\,\Bigl[ \cC^0_{\Z}(13) + \cC^0_{\Z}(16) - \cC^0_{\Z}(20) - 
\cC^0_{\Z}(23) \Bigr]
- y_3\,\frac{\ox^2}{\ox_2}\,\Bigl[  \cC^0_{\Z}(12) - \cC^0_{\Z}(24) \Bigr].
\eqa
Note that in $\cV_{\bcan{114\,;\,\ssB}}$ the $C$-functions have parameters 
which depend on $x_1,x_2$ and also on $y_3$.

The $V^{\bbb}$-family is characterized by having ultraviolet finite 
components up to rank four tensors. Therefore, the techniques introduced in 
Sect.~10 of III can be transferred in an integral manner to all relevant 
form factors discussed in this article.

\section{Conclusions \label{conclu}}
Any realistic calculation of physical observables in the framework of
quantum field theory is remarkably more demanding than simply having at our
disposal techniques to evaluate few special scalar diagrams. There are of
course different strategies to compute complex diagrams but, to a large
extent, they all amount to reducing a large number of integrals to some
minimal set of master (irreducible) integrals.

As a starting procedure, one always saturates the Lorentz indices in the
Green functions so that the numerator of the Feynman integrals contains
powers of scalar products. The novelty in the analysis of two-loop vertices
consists in the presence of so-called irreducible scalar products, namely,
configurations in which the available propagators are not sufficient to
algebraically simplify the numerator. Note that irreducible scalar products
already occur in two-loop self-energies; there, however, the technique of
reduction in sub-loops~\cite{Weiglein:hd} alleviates their irreducibility
(see our presentation in \sect{stltpf}).

We showed that tensor integrals can be first of all decomposed into a
combination of form factors, many of which can be reduced to scalar
integrals (either of the same family or of families with a smaller number of
propagators), while few irreducible integrals remain. It is then possible to
relate these latter ones to generalized scalar integrals of the same family,
i.e.\ integrals in shifted space-time dimensions and with non-canonical
powers of the propagators. The number of these generalized scalar integrals
can be further reduced using generalized recurrence relation techniques
introduced by Tarasov in~\cite{Tarasov:2000sf}.

Alternatively, we developed our favorite strategy: following the findings of
our work on one-loop multi-leg diagrams, we sought for a procedure where all
integrals occurring in a realistic calculation can be written in a form
analogous to \eqn{generalclass}.  The practicality of this approach was
strengthened in \sect{SETLV} by the explicit treatment of several form
factors, paying particular attention to those cases where new or additional
ultraviolet poles arise. In a line, we assembled the bases for extending a
diagram generator to an evaluator of physical observables.

In our opinion, the optimal algorithm puts tensor integrals on the same
footing as scalar ones and should not, therefore, introduce any
multiplication of the tensor integrals by negative powers of Gram
determinants.  The numerical quality of tensor integrals should also not be
worsened, as a consequence of the adopted reduction algorithm, by expressing
them as linear combinations of master integrals; in this case, the kinematic
coefficients have zeros corresponding to real singularities of the diagram,
but their behavior around the singularity is always badly overestimated.

These shortcomings are not severe in the (almost) massless world of QED/QCD,
but they turn into serious disadvantages in the massive world of the
full-fledged Standard Model (SM). We found it more convenient to interpret
irreducible configurations as integrals in the canonical $4 - \ep$
dimensions with polynomials of Feynman parameters in the numerator; they can
be computed -- numerically -- as well as the scalar ones. Several explicit
examples were presented in \sect{SETLV}.

Once we have reduced all obviously-reducible structures, we may as well 
compute all remaining quantities numerically. We must of course avoid 
situations where cancellations are expected: this may happen when the final 
result contains a very large number of terms, when apparent singularities
are present (see Sect. D of III for a discussion) or when inherent gauge 
cancellations do not support a blind application of the procedure. 
We do not expect our approach to suffer from problems more severe than those 
encountered in other methods, but this remains to be fully tested in explicit 
two-loop applications. Comfortingly, our findings in numerical one-loop 
analysis (but also independent work~\cite{PC}) support this claim.

In conclusion, we collected in one single place all the formulae needed to
reduce fully massive tensor integrals, diagram-by-diagram up to three-point
functions, to generalized scalar integrals. One may then choose how to
proceed; for instance, using explicit integral representations for these
functions and evaluating them with the same algorithms of smoothness (or
with some of their generalizations) introduced in~\cite{Ferroglia:2003yj}
for ordinary scalar functions.

Although we believe that there is no substitute for writing linearly, and
that any article should be read linearly as well, we inserted several
Appendices to be consulted as a reference.

The collection of results of this article contains all the ingredients needed
to renormalize the SM (or any other renormalizable field theory) at the
two-loop level, and to calculate the two-loop gauge boson complex poles as
well as physical observables related to processes of the type $V(S) \to
\overline{f} f$, the decay of vector or scalar particles into
fermion--anti-fermion pairs. The use of projector
techniques~\cite{Glover:2004si}, augmented by the explicit reduction
formulae that we collected, with the supplement of suitable integral
representations for irreducible components, are the main tools to carry out
the program.
\Acknowledgments The contribution of Chiara Arina to the general project of
algebraic-numerical evaluation of Feynman diagrams and to several
discussions concerning this article is gratefully acknowledged.  The work of
A.~F.\ was supported by the DFG-Forschergruppe ``\emph{Quantenfeldtheorie,
Computeralgebra und Monte-Carlo-Simulation}''.  The work of M.~P.\ was
supported by the MECyD fellowship SB2002-0105, the MCyT grant FPA2002-00612,
and by the Swiss National Science Foundation.
\clearpage
\appendix
\section{Reduction for generalized one-loop functions \label{RGOLF}}
Generalized one-loop functions can be treated according to the BT-algorithm
discussed in~\cite{Ferroglia:2002mz}. 
For $B_{\Z}(\alpha,\beta\,;\,p,m_1,m_2)$ one may use the results of Sect.~3
of~\cite{Ferroglia:2002mz}, in particular Eqs.~(23-24).

For the $C$-family there is full reducibility and, moreover, different
scalar integrals can be related among each other and expressed in terms of
standard scalar functions.  The most convenient approach is based on the
fact that all $C$-functions can be evaluated according to the
BT-algorithm. We illustrate the procedure for functions of weight $4$, where
the weight is defined to be the sum of the (positive) powers in the
propagators; all $C$ functions can be written as
\bq
C[w=4] = \dsimp{2}\,P(x_1,x_2)\,V^{-2-\ep/2}(x_1,x_2),
\quad
P(x_1,x_2) = \sum_{n=0}^{\ssN}\,\sum_{m=0}^{\ssM}\,a_{nm}\,x^n_1\,x^m_2,
\eq
where $w = \alpha_1+\alpha_2+\alpha_3$ and where the polynomial $P$ depends
on the specific case under consideration. For standard ($w = 3$) form factors
(from $C_{\saa}$ to $C_{\sbd}$) the corresponding polynomials are given in 
Eq.~(41) of~\cite{Ferroglia:2002mz}.
For scalar functions of weight $4$ the $P$ are $1-x_1$ for $C_{\Z}(2,1,1)$,
$x_1-x_2$ for $C_{\Z}(1,2,1)$ and $x_2$ for $C_{\Z}(1,1,2)$.

Higher weights can be evaluated recursively, e.g.
\bqa
C[w=4] &=& \sum_{n=0}^{\ssN}\,\sum_{m=0}^{\ssM}\,\frac{a_{nm}}{(2+\ep)\,B_3}\,
C_{nm}[w=4],
\nl
C_{nm}[w=4] &=&
\dsimp{2}\,\,V^{-1-\ep/2}(x_1,x_2)\,\Bigl[
n\,X_1\,x^{n-1}_1\,x^m_2 +
m\,X_2\,x^n_1\,x^{m-1}_2 +
(\ep - n - m)\,x^n_1\,x^m_2\Bigr] 
\nl
{}&+& \dcub{1}\,\Bigl[
(1-X_1)\,x^m_1\,V^{-1-\ep/2}(1,x_1) + 
(X_1-X_2)\,x^{n+m}_1\,V^{-1-\ep/2}(x_1,x_1)
\nl
{}&+&\,\delta_{m,0}\,X_2\,V^{-1-\ep/2}(x_1,0)\Bigr],
\eqa
where, with the definition of \eqn{GCfun}, the quadratic form $V$ is
\bqa
V(x_1,x_2) &=& x^t\,G\,x + 2\,K^t\,x + L ,
\qquad
G_{ij} = -\,\spro{p_i}{p_j}, \quad L = m^2_1,
\nl
K_1 &=& \frac{1}{2} \, ( p^2_1 + m_2^2 - m_1^2 ),
\quad
K_2 = \frac{1}{2} \, ( P^2 - p^2_1 + m_3^2 - m_2^2 ),
\label{Konetwo}
\eqa
with $P = p_1 + p_2$. Furthermore, $B_3 = L - K^t\,G^{-1}\,K$ and 
$X = -\,G^{-1}\,K$. For $w = 3$ we finally have
\bq
C_{nm}[w=3] = C_{mn}^0 - \frac{1}{2}\,\dcub{1}\,C_{mn}^1 -
\frac{1}{2}\,\dsimp{2}\,x_1^{n-1}\,x_2^{m-1}C_{mn}^2
\Bigr] ,
\eq
where the coefficients are
\bqa
C_{mn}^0 &=& \frac{1}{(2+n+m)\,(1+m)},
\nl
\nl
C_{mn}^1 &=&
(X_1-X_2)\,x^{n+m}_1\,\ln\,V(x_1,x_1) + 
\delta_{m,0}\,X_2\,x_1^n\,\ln\,V(x_1,0) + (1-X_1)\,x^m_1\,\ln\,V(1,x_1),
\nl
\nl
C_{mn}^2 &=&
( m\,X_2\,x_1 - (2 + n + m)\,x_1\,x_2 + n\,X_1\,x_2)\,\ln\,V(x_1,x_2).
\eqa
$D$-family functions of weight $5$ can be reduced recursively with three
iterations of the BT-algorithm; see Sect.~(6.2) of~\cite{Ferroglia:2002mz}
for a discussion, in particular Eqs.~(140-142).  These functions are not
needed in this paper.
\section{Summary of the results for the reduction of three-point functions
\label{SummaR}}
In this Section we present a summary of the results obtained for the
reduction of two-loop three-point functions and derived in
Sections~\ref{abafam}--\ref{bbbfam}.  Tensor integrals are defined by
having powers of momenta in the numerator; they are further decomposed into
form factors $\,\otimes\,$ tensor structures and, for completeness, the full
collection of results is presented for the form factors as well as for
tensor integrals with saturated indices; the latter are perhaps the most
important objects when one computes physical amplitudes in the framework of
the projector techniques introduced in \sect{prered}.

The presentation is organized through a series of concatenated formulae that
can be easily coded with any of the well-known packages for symbolic
manipulation; the formulae below can thus be used as they stand or they can
be used recursively. Each object of the list contains scalar functions or
form factors corresponding to tensors of lower rank and with fewer
propagators that can be found earlier in the list and, if needed, the
procedure can be iterated until the chain of reductions stops with a fully
scalarized expression.

Formulae for ordinary scalar vertex functions can be found in III where,
however, the alphameric convention was not yet used and therefore the
correspondence is based on \eqn{americv}. In particular, $V^{\aba}_{\Z} \equiv
V^{121}_{\Z}$ in Sect.~5.1, $V^{\aca}_{\Z} \equiv V^{131}_{\Z}$ in Sect.~6.1,
$V^{\bba}_{\Z} \equiv V^{221}_{\Z}$ in Sect.~7.1, $V^{\ada}_{\Z} \equiv 
V^{141}_{\Z}$ in Sects.~8.1 -- 8.2, $V^{\bca}_{\Z} \equiv V^{231}_{\Z}$ in 
Sects.~9.1 -- 9.2 and $V^{\bbb}_{\Z} \equiv V^{222}_{\Z}$ in Sect.~10.4 of III.
Additional material, with the extension to generalized scalar functions, is 
presented in this paper in \sectm{exaaba}{exabcabbb}. Finally, reduction of 
one-loop generalized form factors has been discussed in \appendx{RGOLF}.

Additional notation, relevant for this Appendix, was given in the
Introduction but is repeated here: we denote by $G$ the Gram matrix arising
in the context of a vertex function and use
\bq
G_{ij} = \spro{p_i}{p_j}, \quad D = {\mbox{det}}\,G = p^2_1\,p^2_2 -
(\spro{p_1}{p_2})^2, \quad D_1 = p^2_1\,p^2_2, \quad D_2 = p_{12}\,p^2_2,
\quad D_3 = p_{12}\,p^2_1.
\eq
Before presenting the list of results we would like to discuss one specific
example. Consider a rank two tensor, e.g. from \sect{bcafam}, where all
indices are saturated with external momenta:
\bqa
\cV^{\bca}(0\,|\,p_1, p_1) &=& \frac{1}{2}\,\Bigl\{
       - l_{145}\,\Bigl[ \cV^{\bca}_{\sba}\,p^2_1 
       + \cV^{\bca}_{\sbb}\,p_{12}\Bigr]
       - \cV^{\bba}_{\sba}(P,P,p_1,\lstm{12365})\,p^2_1
\nl 
{}&+& \spro{p_1}{P}\,\Bigl[ \cV^{\bba}_{\sba}(P,P,0,\lstm{12364})
       - \cV^{\bba}_{\sbb}(P,P,0,\lstm{12364})\Bigr]\Bigr\}.
\eqa
After the first step in the reduction there is no Gram determinant but the 
latter may still be hidden in form factors corresponding to tensors of lower 
rank. As a matter of fact, we may iterate the procedure and consider
\bqa
\cV^{\bca}_{\sba}\,p^2_1 + \cV^{\bca}_{\sbb}\,p_{12} &=&
\cV^{\bca}(0\,|\,p_1),
\nl
\spro{p_1}{P}\,\Bigl[ \cV^{\bba}_{\sba}(P,P,0,\lstm{12364}) - 
\cV^{\bba}_{\sbb}(P,P,0,\lstm{12364}) \Bigr] &=&
\cV^{\bba}(0\,|\,p_1\,;\,P,P,0,\lstm{12364}),
\nl
\cV^{\bba}_{\sba}(P,P,p_1,\lstm{12365}).
\label{threeterms}
\eqa
A reduction, which is again free from Gram determinants, can be applied to 
the first term in \eqn{threeterms}; however, further scalarization for the 
last two can only be performed if Gram determinants do not pose a problem, 
as in proving WST identities; otherwise the reduction chain for these terms 
should stop and their evaluation will follow according to the corresponding 
defining representation (note that 
$\cV^{\bba}(0\,|\,P\,;\,P,P,0,\lstm{12364})$ is instead fully reducible). 
Alternatively, the term can be further reduced with generalized recurrence 
relations which, however, introduce additional kinematic coefficients, with 
the appearance of (physical) singularities etc, etc. 

In summarizing the whole set of results we adopt the following convention: the 
list of arguments of tensor integrals in a given class is suppressed when we 
present their reduction; therefore
\[
\ba{ll}
\cV^{\aba}_{\ssJ} \equiv \cV^{\aba}_{\ssJ}(p_2,P,\lstm{1234}),
\qquad & \qquad
\cV^{\aca}_{\ssJ} \equiv \cV^{\aca}_{\ssJ}(p_1,P,\lstm{12345}),
\\
\cV^{\ada}_{\ssJ} \equiv \cV^{\ada}_{\ssJ}(p_1,P,\lstm{12345}),
\qquad & \qquad
\cV^{\bba}_{\ssJ} \equiv \cV^{\bba}_{\ssJ}(p_1,p_1,P,\lstm{12345}),
\\
\cV^{\bca}_{\ssJ} \equiv \cV^{\bca}_{\ssJ}(P,p_1,P,\lstm{123456}),
\qquad & \qquad
\cV^{\bbb}_{\ssJ} \equiv \cV^{\bbb}_{\ssJ}(-p_2,p_1,-p_2,-p_1,\lstm{123456}),
\ea
\]
where $J$ denotes a generic form factor in the family and where, for the $M$
family we always assume $m_6 = m_3$. We also report, for completeness the 
definitions of all form factors occurring in our paper:
\bqa
\cV^{\ssJ}(\mu\,|\,0\,;\,\cdots) &=& 
\sum_{i=1,2}\,\cV^{\ssJ}_{\sagen}\,p_{i\mu},
\quad
\cV^{\ssJ}(0\,|\,\mu\,;\,\cdots) = 
\sum_{i=1,2}\,\cV^{\ssJ}_{\sbgen}\,p_{i\mu},
\qquad J = E,I,M,G,K,H,
\nl
\cV^{\aba}(\mu,\nu\,|\,0\,;\,\cdots) &=& 
\cV^{\aba}_{\saaa}\,p_{1\mu}\,p_{1\nu} + 
\cV^{\aba}_{\saab}\,p_{2\mu}\,p_{2\nu} +
\cV^{\aba}_{\saac}\,\{p_1\,p_2\}_{\mu\nu} + 
\cV^{\aba}_{\saad}\,\delta_{\mu\nu}, 
\quad \mbox{etc,} 
\nl
\cV^{\aca}(0\,|\,\mu,\nu\,;\,\cdots) &=& 
\cV^{\aca}_{\sbba}\,p_{1\mu}\,p_{1\nu} + 
\cV^{\aca}_{\sbbb}\,p_{2\mu}\,p_{2\nu} + 
\cV^{\aca}_{\sbbc}\,\{p_1 p_2\}_{\mu\nu} + 
\cV^{\aca}_{\sbbd}\,\delta_{\mu\nu}, 
\quad \mbox{etc,} 
\nl
\cV^{\ada}(0 \,|\,\mu,\nu\,;\,\cdots) &=& 
\cV^{\ada}_{\sbba}\,p_{1\mu}\,p_{1\nu} + 
\cV^{\ada}_{\sbbb}\,p_{2\mu}\,p_{2\nu} + 
\cV^{\ada}_{\sbbc}\,\{p_1 p_2\}_{\mu\nu} + 
\cV^{\ada}_{\sbbd}\,\delta_{\mu\nu}, 
\quad \mbox{etc,} 
\nl
\cV^{\bba}(0 \,|\,\mu,\nu\,;\,\cdots) &=& 
\cV^{\bba}_{\sbba}\,p_{1\mu}\,p_{1\nu} + 
\cV^{\bba}_{\sbbb}\,p_{2\mu}\,p_{2\nu} + 
\cV^{\bba}_{\sbbc}\,\{p_1 p_2\}_{\mu\nu} + 
\cV^{\bba}_{\sbbd}\,\delta_{\mu\nu} 
\nl
\cV^{\bba}(\mu\,|\,\nu\,;\,\cdots) &=& 
\cV^{\bba}_{\saba}\,p_{1\mu}\,p_{1\nu} + 
\cV^{\bba}_{\sabb}\,p_{2\mu}\,p_{2\nu} + 
\cV^{\bba}_{\sabc}\,p_{1\mu}\,p_{2\nu} +
\cV^{\bba}_{\sabe}\,p_{1\nu}\,p_{2\mu} + 
\cV^{\bba}_{\sabd}\,\delta_{\mu\nu},
\nl
\cV^{\bba}(\mu, \nu\,|\,0\,;\,\cdots) &=& 
\cV^{\bba}_{\saaa}\,p_{1\mu}\,p_{1\nu} + 
\cV^{\bba}_{\saab}\,p_{2\mu}\,p_{2\nu} + 
\cV^{\bba}_{\saac}\,\{p_1 p_2\}_{\mu\nu} + 
\cV^{\bba}_{\saad}\,\delta_{\mu\nu},
\nl
\cV^{\bca}(0\,|\,\mu, \nu\,;\,\cdots) &=& 
\cV^{\bca}_{\sbba}\,p_{1\mu}\,p_{1\nu} + 
\cV^{\bca}_{\sbbb}\,p_{2\mu}\,p_{2\nu} +
\cV^{\bca}_{\sbbc}\,\{p_1\,p_2\}_{\mu\nu} + 
\cV^{\bca}_{\sbbd}\,\delta_{\mu\nu},
\nl
\cV^{\bca}(\mu \,|\,\nu\,;\,\cdots) &=&
\cV^{\bca}_{\saba}\,p_{1\mu}\,p_{1\nu} +
\cV^{\bca}_{\sabb}\,p_{2\mu}\,p_{2\nu} +
\cV^{\bca}_{\sabc}\,p_{1\mu}\,p_{2\nu} +
\cV^{\bca}_{\sabe}\,p_{1\nu}\,p_{2\mu} +
\cV^{\bca}_{\sabd}\,\delta_{\mu\nu},
\nl
\cV^{\bca}(\mu, \nu\,|\,0\,;\,\cdots) &=& 
\cV^{\bca}_{\saaa}\,p_{1\mu}\,p_{1\nu} + 
\cV^{\bca}_{\saab}\,p_{2\mu}\,p_{2\nu} + 
\cV^{\bca}_{\saac}\,\{p_1 p_2\}_{\mu\nu} + 
\cV^{\bca}_{\saad}\,\delta_{\mu\nu},
\nl
\cV^{\bbb}(0\,|\,\mu, \nu\,;\,\cdots) &=& 
\cV^{\bbb}_{\sbba}\,p_{1\mu}\,p_{1\nu} + 
\cV^{\bbb}_{\sbbb}\,p_{2\mu}\,p_{2\nu} + 
\cV^{\bbb}_{\sbbc}\,\{ p_1 p_2\}_{\mu\nu} + 
\cV^{\bbb}_{\sbbd}\,\delta_{\mu\nu},
\nl
\cV^{\bbb}(\mu, \nu\,|\,0;\,\cdots) &=&
\cV^{\bbb}_{\saaa}\,p_{1\mu}\,p_{1\nu} +
\cV^{\bbb}_{\saab}\,p_{2\mu}\,p_{2\nu} +
\cV^{\bbb}_{\saac}\,\{ p_1 p_2\}_{\mu\nu} +
\cV^{\bbb}_{\saad}\,\delta_{\mu\nu},
\nl
\cV^{\bbb}(\mu \,|\, \nu ;\,\cdots) &=& 
\cV^{\bbb}_{\aba}\,p_{1\mu}\,p_{1\nu} + 
\cV^{\bbb}_{\sabb}\,p_{2\mu}\,p_{2\nu} + 
\cV^{\bbb}_{\sabc}\,\{p_1\,p_2\}_{\mu\nu} + 
\cV^{\bbb}_{\sabd}\,\delta_{\mu\nu} ,
\nl
\cV^{\ada}(\mu, \nu, \alpha\,|\,0 ; \cdots) &=& 
\cV^{\ada}_{\saaaa}\,\{\delta\,p_1\}_{\mu\nu\alpha} + 
\cV^{\ada}_{\saaab}\, \{\delta\,p_2\}_{\mu\nu\alpha} + 
\cV^{\ada}_{\saaac}\,\{p_1 p_1 p_2\}_{\mu\nu\alpha} + 
\cV^{\ada}_{\saaad}\,\{p_2 p_2 p_1\}_{\mu\nu\alpha}
\nl
{}&+&
\cV^{\ada}_{\saaae}\, p_{1 \mu}\,p_{1 \nu}\,p_{1 \alpha} +
\cV^{\ada}_{\saaaf}\, p_{2 \mu}\,p_{2 \nu}\,p_{2 \alpha},
\quad \mbox{etc,}
\nl
\cV^{\ada}(0 \,|\, \mu, \nu, \alpha ; \cdots) &=& 
\cV^{\ada}_{\sbbba}\,\{\delta\,p_1\}_{\mu\nu\alpha} + 
\cV^{\ada}_{\sbbbb}\, \{\delta\,p_2\}_{\mu\nu\alpha} + 
\cV^{\ada}_{\sbbbc}\,\{p_1 p_1 p_2\}_{\mu\nu\alpha} + 
\cV^{\ada}_{\sbbbd}\,\{p_2 p_2 p_1\}_{\mu\nu\alpha}
\nl
{}&+&
\cV^{\ada}_{\sbbbe}\, p_{1 \mu}\,p_{1 \nu}\,p_{1 \alpha} +
\cV^{\ada}_{\sbbbf}\, p_{2 \mu}\,p_{2 \nu}\,p_{2 \alpha},
\quad \mbox{etc,}
\nl
\cV^{\ada}(\mu\,|\, \nu ,\alpha ; \cdots) &=& 
\cV^{\ada}_{\sabba}\,\{\delta\,p_1\}_{\mu\nu\alpha} + 
\cV^{\ada}_{\sabbb}\, \{\delta\,p_2\}_{\mu\nu\alpha} + 
\cV^{\ada}_{\sabbc}\,\{p_1 p_1 p_2\}_{\mu\nu\alpha} + 
\cV^{\ada}_{\sabbd}\,\{p_2 p_2 p_1\}_{\mu\nu\alpha}
\nl
{}&+&
\cV^{\ada}_{\sabbe}\, p_{1 \mu}\,p_{1 \nu}\,p_{1 \alpha} +
\cV^{\ada}_{\sabbf}\, p_{2 \mu}\,p_{2 \nu}\,p_{2 \alpha},
\nl
\cV^{\ada}(\mu,\nu\,|\,\alpha; \cdots) &=& 
\cV^{\ada}_{\saaba}\,\{\delta\,p_1\}_{\mu\nu\alpha} +
\cV^{\ada}_{\saabb}\,\{\delta\,p_2\}_{\mu\nu\alpha} +
\cV^{\ada}_{\saabc}\,\{p_1 p_1 p_2\}_{\mu\nu\alpha} +
\cV^{\ada}_{\saabd}\,\{p_2 p_2 p_1\}_{\mu\nu\alpha} 
\nl 
{}&+& 
\cV^{\ada}_{\saabe}\, p_{1 \mu}\,p_{1 \nu}\,p_{1 \alpha} +
\cV^{\ada}_{\saabf}\, p_{2 \mu}\,p_{2 \nu}\,p_{2 \alpha} +
\cV^{\ada}_{\saabg}\,\{\delta\,p_1\}_{\mu\nu\,|\,\alpha} +
\cV^{\ada}_{\saabh}\,\{\delta\,p_2\}_{\mu\nu\,|\,\alpha},
\nl
\cV^{\bca}(\mu \,|\,  \nu , \alpha ; \cdots) &=& 
\cV^{\bca}_{\sabba}\,\{\delta\,p_1\}_{\nu\alpha\,|\,\mu} +
\cV^{\bca}_{\sabbb}\,\{\delta\,p_2\}_{\nu\alpha\,|\,\mu} +
\cV^{\bca}_{\sabbc}\,\delta_{\nu \alpha} p_{1 \mu} + 
\cV^{\bca}_{\sabbd}\,\delta_{\nu \alpha} p_{2 \mu}
\nl 
&+& 
\cV^{\bca}_{\sabbe}\,p_{1 \alpha}\,p_{1 \mu}\,p_{1 \nu} +
\cV^{\bca}_{\sabbf}\,p_{2 \alpha}\,p_{2 \mu}\,p_{2 \nu} +
\cV^{\bca}_{\sabbg}\,\{p_1\,p_1\,p_2\}_{\alpha\nu\,|\,\mu} 
\nl 
&+& 
\cV^{\bca}_{\sabbh}\, 
\{p_2\,p_2\,p_1\}_{\nu\alpha\,|\,\mu} +
\cV^{\bca}_{\sabbi}\, p_{1 \alpha}\,p_{1 \nu}\,p_{2 \mu} +
\cV^{\bca}_{\sabbj}\, p_{1 \mu}\,p_{2 \nu}\,p_{2 \alpha},
\nl
\cV^{\bca}(\mu,  \nu \,|\, \alpha ; \cdots) &=& 
\cV^{\bca}_{\saaba}\,\{\delta\,p_1\}_{\mu\nu\,|\,\alpha} +
\cV^{\bca}_{\saabb}\,\{\delta\,p_2\}_{\mu\nu\,|\,\alpha} +
\cV^{\bca}_{\saabc}\,\delta_{\mu \nu} p_{1 \alpha} + 
\cV^{\bca}_{\saabd}\,\delta_{\mu \nu} p_{2 \alpha}
\nl 
&+& 
\cV^{\bca}_{\saabe}\,p_{1 \alpha}\,p_{1 \mu}\,p_{1 \nu} +
\cV^{\bca}_{\saabf}\,p_{2 \alpha}\,p_{2 \mu}\,p_{2 \nu} +
\cV^{\bca}_{\saabg}\,\{p_1 p_1 p_2\}_{\mu\nu\,|\,\alpha} 
\nl 
&+& \cV^{\bca}_{\saabh}\,\{p_2 p_2 p_1\}_{\nu\mu\,|\,\alpha} +
\cV^{\bca}_{\saabi}\, p_{1 \mu}\,p_{1 \nu}\,p_{2 \alpha} +
\cV^{\bca}_{\saabj}\, p_{1 \alpha}\,p_{2 \mu}\,p_{2 \nu},
\nl
\cV^{\bbb}(\mu \,|\,  \nu,  \alpha ; \cdots) &=& 
\cV^{\bbb}_{\sabba}\,\{\delta\,p_1\}_{\nu\alpha\,|\,\mu} +
\cV^{\bbb}_{\sabbb}\,\{\delta\,p_2\}_{\nu\alpha\,|\,\mu} +
\cV^{\bbb}_{\sabbc}\,\delta_{\nu \alpha} p_{1 \mu} + 
\cV^{\bbb}_{\sabbd}\,\delta_{\nu \alpha} p_{2 \mu} + 
\cV^{\bbb}_{\sabbe}\,p_{1 \alpha}\,p_{1 \mu}\,p_{1 \nu}
\nl
{}&+&
\cV^{\bbb}_{\sabbf}\,p_{2 \alpha}\,p_{2 \mu}\,p_{2 \nu} +
\cV^{\bbb}_{\sabbg}\,\{p_{1}\,p_{1}\,p_{2}\}_{\mu \nu \alpha} + 
\cV^{\bbb}_{\sabbh}\,\{p_{1}\,p_{2}\,p_{2}\}_{\mu \nu \alpha},
\nl
\cV^{\bbb}(\mu,\nu \,|\, \alpha ; \cdots) &=& 
\cV^{\bbb}_{\saaba}\,\{\delta\,p_1\}_{\mu\nu\,|\,\alpha} +
\cV^{\bbb}_{\saabb}\,\{\delta\,p_2\}_{\mu\nu\,|\,\alpha} +
\cV^{\bbb}_{\saabc}\,\delta_{\mu \nu} p_{1 \alpha} + 
\cV^{\bbb}_{\saabd}\,\delta_{\mu \nu} p_{2 \alpha} +
\cV^{\bbb}_{\saabe}\,p_{1 \alpha}\,p_{1 \mu}\,p_{1 \nu}
\nl
{}&+&
\cV^{\bbb}_{\saabf}\,p_{2 \alpha}\,p_{2 \mu}\,p_{2 \nu} +
\cV^{\bbb}_{\saabg}\,\{p_{1}\,p_{1}\,p_{2}\}_{\mu \nu \alpha} + 
\cV^{\bbb}_{\saabh}\,\{p_{1}\,p_{2}\,p_{2}\}_{\mu \nu \alpha}.
\eqa
We are now ready to summarize the results; tags were introduced to
facilitate the search of the various items, for instance results related
to rank two tensor integrals of the $kl$ group ($kl = \{11, 12, 22\}$) in 
the $J$ family ($J = E,I,M,G,K$ and $H$) are to be searched under the tag 
$\bf V^{\ssJ}_{kli}$.
\subsection{\boldmath $V^{\aba}(p_2,P,\lstm{1234})$ family \label{summaaba}}
\noindent\fbox{$\bf V^{\aba}_{\sijgen}$}
Results were derived in \sect{rankoneaba}. Referring to \eqn{genaba}, for 
vector integrals we have
\bqa 
\cV^{\aba}_{\sagen} &=& \frac{1}{2}\,\Bigl[ 
 \cV^{\aba}_{\sbgen} 
+ m^2_{21}\,\cV^{\aca}_{\sbgen}(p_2,P,\lstm{12},0,\lstm{34}) 
- \,C_{\sagen}(p_2,p_1,0,\lstm{34})\,A_{\Z}([m_2,m_1])\Bigr]\, ,
\nl
\cV^{\aba}_{\sba} &=& - \omega^2\,\Bigl[ \cV^{2|1,2|1}_{\aba} + 
\cV^{1|1,2|2}_{\aba}\Bigr]\,\bmid_{n=6-\ep}, \quad 
\cV^{\aba}_{\sbb}  =  - \omega^2\,\Bigl[ \cV^{2|2,1|1}_{\aba} +
\cV^{2|1,2|1}_{\aba} + \cV^{1|2,1|2}_{\aba} + \cV^{1|1,2|2}_{\aba}
\Bigr]\,\bmid_{n=6-\ep}.
\eqa
\fbox{$\bf V^{\aba}_{\sbbgen}$}
For tensor integrals (see \sect{ranktwoaba}) we introduce a vector 
$U^{\aba}_{\sbb}$ with components
\bqa
U^{\aba}_{\sbbsca} &=& \frac{1}{2}\,\Bigl[
- \cV^{\aba}_{\sbb}\,( 2\,p_{12} + l_{134} )
+ \cS^{\ssA}_{\Z}(P,\lstm{124}) 
- \cS^{\ssA}_{\Z}(p_2,\lstm{123}) 
\nl
{}&+& \cS^{\ssA}_{\msb}(P,\lstm{124}) 
- \cS^{\ssA}_{\msb}(p_2,\lstm{123})\Bigr],
\nl
U^{\aba}_{\sbbscb} &=&
- \cV^{\aba}_{\Z}\,( p^2_2 + m_3^2 )
- \cV^{\aba}_{\sba}\,( p_{12} - \frac{1}{2}\,l_{134} )
- 2\,\cV^{\aba}_{\sbb}\,p^2_2
- (n-1)\,\cV^{\aba}_{\sbbd}
\nl
{}&+& \frac{1}{2}\,\cS^{\ssA}_{\Z}(P,\lstm{124}) 
- \frac{1}{2}\,\cS^{\aba}_{\msb}(P,\lstm{124}),
\eqa
then one obtains 
\bq
\left( \begin{array}{c} \cV^{\aba}_{\sbbc}  \\ \cV^{\aba}_{\sbbb}  
\end{array}\right) = G^{-1}\,U^{\aba}_{\sbb},
\qquad
p^2_1\,\cV^{\aba}_{\sbba} = -\,p_{12}\,\cV^{\aba}_{\sbbc} -
\cV^{\aba}_{\sbbd} + U^{\aba}_{\sbbscb},
\eq 
with a generalized function
\bq
\cV^{\aba}_{\sbbd} = \frac{1}{2}\,\omega^2\,\Bigl[
\cV^{2|1,1|1}_{\aba} + \cV^{1|1,1|2}_{\aba}\Bigr]\,
\bmid_{n = 6-\ep}.
\eq
For tensors with saturated indices we obtain
\bqa
\cV^{\aba}(0\,|\,\mu, \mu) &=&
       - \cV^{\aba}_{\Z}\,( p^2_2 + m_3^2 )
       -2\,\cV^{\aba}_{\sbb}\,p^2_2
       -2\,\cV^{\aba}_{\sba}\,p^2_1
       + \cS^{\ssA}_{\Z}(P,\lstm{124}) \,
\nl
\cV^{\aba}(0\,|\,p_1, p_1) &=&
       - (p_{12}\,\cV^{\aba}_{\sbb} + p^2_1\,\cV^{\aba}_{\sba})\,
       ( p_{12} + \frac{1}{2}\,l_{134} )
+  \frac{1}{2}\,\spro{p_1}{P}\,\Bigl[
        \cS^{\ssA}_{\Z}(P,\lstm{124})
       + \cS^{\ssA}_{\msb}(P,\lstm{124}) \Bigr]
\nl
{}&-& \frac{1}{2}\,p_{12}\,\Bigl[
         \cS^{\ssA}_{\Z}(p_2,\lstm{123})
       + \cS^{\ssA}_{\msb}(p_2,\lstm{123}) \Bigr],
\nl
p^2_1\,\cV^{\aba}(0\,|\, p_2, p_2) &=&
       - \cV^{\aba}_{\sbbd}\,( n - 2 )\,D
       - \cV^{\aba}_{\sbb}\,p^2_2\,
 ( 2\,D + \frac{1}{2}\,p_{12}\,l_{134} + p_{12}^2 )
\nl
{}&-& \cV^{\aba}_{\Z}\,( p^2_2 + m_3^2 )\,D
       - \cV^{\aba}_{\sba}\,
\Bigl[ D\,( p_{12} - \frac{1}{2}\,l_{134} ) + 
p^2_{12}\,( p_{12} + \frac{1}{2}\,l_{134} ) \Bigr]
\nl
{}&+& \frac{1}{2}\,\cS^{\ssA}_{\Z}(P,\lstm{124})\,(  D + D_2 + 
p_{12}^2 )
       -\frac{1}{2}\,\cS^{\ssA}_{\Z}(p_2,\lstm{123})\,D_2 
\nl
{}&-& \frac{1}{2}\,\cS^{\ssA}_{\msb}(P,\lstm{124})\,
( D - D_2 - p_{12}^2 )
       -\frac{1}{2}\,\cS^{\ssA}_{\msb}(p_2,\lstm{123})\, D_2,
\nl
\cV^{\aba}(0\,|\,p_1, p_2) &=&
       - (\cV^{\aba}_{\sbb}\,p^2_2 + \cV^{\aba}_{\sba}\,p_{12})
      \,( \frac{1}{2}\,l_{134} + p_{12} )
+ \frac{1}{2}\,\spro{p_2}{P}\,\Bigl[ 
       \cS^{\ssA}_{\Z}(P,\lstm{124})
     + \cS^{\ssA}_{\msb}(P,\lstm{124}) \Bigr]
\nl
{}&-& \frac{1}{2}\,p^2_2\,\Bigl[
         \cS^{\ssA}_{\Z}(p_2,\lstm{123}) 
       + \cS^{\ssA}_{\msb}(p_2,\lstm{123}) \Bigr].
\eqa
\fbox{$\bf V^{\aba}_{\sabgen}$} Furthermore we obtain
\bqa 
\cV^{\aba}_{\sabgen} &=&
\frac{1}{2}\,\Bigl[ 
\cV^{\aba}_{\sbbgen} 
+ m^2_{21}\,\cV^{\aca}_{\sbbgen}(p_2,P,\lstm{12},0,\lstm{34}) 
+\,C_{\sbgen}(p_2,p_1,0,\lstm{34})\,A_{\Z}([m_2,m_1])\Bigr].
\eqa
\fbox{$\bf V^{\aba}_{\saagen}$} Finally, for $i < 4$ we have
\bqa
4\,(n-1)\,\cV^{\aba}_{\saagen} &=&
n\,\cV^{\aba}_{\sbbgen}
+ n\,m^4_{12}\,\cV^{\ada}_{\sbbgen}(p_1,P,\lstm{12},0,\lstm{34},0)
\nl
{}&+& 2\,( n\,m^2_{21} + 2\,m^2_1)\,
\cV^{\aca}_{\sbbgen}(p_2,P,\lstm{12},0,\lstm{34}) 
\nl
&-& n\,A_{\Z}(m_1)\,\Bigl[ m^2_{12}\,
C_{\sbgen}(2,1,1\,;\,p_2,p_1,0,\lstm{34}) -
C_{\sbgen}(p_2,p_1,0,\lstm{34})\Bigr]
\nl
{}&-& A_{\Z}(m_2)\,\Bigl[ (3\,n - 4)\,C_{\sbgen}(p_2,p_1,0,\lstm{34}) +
n\,m^2_{21}\,C_{\sbgen}(2,1,1\,;\,p_2,p_1,0,\lstm{34})\Bigr],
\eqa
while, for $i = 4$ it follows that
\bqa
4\,(n-1)\,\cV^{\aba}_{\saad} &=&
 n\,\cV^{\aba}_{\sbbd}
- 2\,(m^2_1 + m^2_2)\,\cV^{\aba}_{\Z}
- \cV^{\aba}(0|\mu,\mu)
+ n\,m^4_{12}\,\cV^{\ada}_{\sbbd}(p_1,P,\lstm{12},0,\lstm{34},0)
\nl
{}&+& 2\,( n\,m^2_{21} + 2\,m^2_1)\,
\cV^{\aca}_{\sbbd}(p_2,P,\lstm{12},0,\lstm{34}) 
- m^4_{12}\,\cV^{\aca}_{\Z}(p_2,P,\lstm{12},0,\lstm{34}) 
\nl
{}&-& n\,A_{\Z}(m_1)\,\Bigl[ m^2_{12}\,
C_{\sbd}(2,1,1\,;\,p_2,p_1,0,\lstm{34}) -
C_{\sbd}(p_2,p_1,0,\lstm{34})\Bigr]
\nl
{}&-& A_{\Z}(m_2)\,\Bigl[ (3\,n - 4)\,C_{\sbd}(p_2,p_1,0,\lstm{34}) +
n\,m^2_{\sab}\,C_{\sbd}(2,1,1\,;\,p_2,p_1,0,\lstm{34})\Bigr]
\nl
{}&-& A_{\Z}(m_1)\,\Bigl[ m^2_{21}\,C_{\Z}(p_2,p_1,0,\lstm{34}) +
B_{\Z}(\,p_1,\lstm{34})\Bigr]
\nl
{}&-& A_{\Z}(m_2)\,\Bigl[ m^2_{12}\,C_{\Z}(p_2,p_1,0,\lstm{34}) +
B_{\Z}(\,p_1,\lstm{34})\Bigr].
\label{oneofprev}
\eqa
\eqn{oneofprev} requires some of the results corresponding to the 
$V^{\aca}$ family: they are collected in \sect{summaaca}.
\subsection{\boldmath $V^{\aca}(p_1,P,\lstm{12345})$ family \label{summaaca}}
\noindent\fbox{$\bf V^{\aca}_{\sbgen}$}
Results were derived in \sect{rankoneaca}. Introduce a vector $U^{\aca}_{2}$ 
with components
\bqa
U^{\aca}_{\sbsca} &=&
\frac{1}{2}\Bigl[ -l_{134}\cV^{\aca}_{\Z}
-\cV^{\aba}_{\Z}(p_1,P,\lstm{1245}) + \cV^{\aba}_{\Z}(0,P,\lstm{1235})\Bigr],
\nl
U^{\aca}_{\sbscb} &=&
\frac{1}{2}\Bigl[ (l_{154} \!\! -\!\! P^2)V^{\aca}_{\Z}
\!+\!\cV^{\aba}_{\Z}(0,p_1,\lstm{1234}) -
\!\cV^{\aba}_{\Z}(0,P,\lstm{1235})\Bigr];
\eqa
we obtain the following result:
\bq
\left( \begin{array}{c} \cV^{\aca}_{\sba}  \\ \cV^{\aca}_{\sbb}  
\end{array}\right) = G^{-1}\,U^{\aca}_{2}, \qquad
\cV^{\aca}(0\,|\,p_1) = U^{\aca}_{\sbsca}, \qquad
\cV^{\aca}(0\,|\,p_2) = U^{\aca}_{\sbscb}.
\eq
\fbox{$\bf V^{\aca}_{\sagen}$} Furthermore we have
\bqa
\cV^{\aca}_{\sagen}(p_1,P,\lstm{12345}) &=&
\frac{1}{2}\,\frac{m^2_{123}}{m^2_3}\,
\cV^{\aca}_{\sbgen}(p_1,P,\lstm{12345}) - 
\frac{1}{2}\,\frac{m^2_{12}}{m^2_3}\,
\cV^{\aca}_{\sbgen}(p_1,P,\lstm{12},0,\lstm{45}) \nl
{}&-& \frac{1}{2 m_3^2}\,A_{\Z}([m_1,m_2])\,
\Bigl\{ \delta_{\sgena}\,\sum_{j=1,2}\,\Bigl[ C_{\sajgen}(p_1,p_2,\lstm{345}) 
- C_{\sajgen}(p_1,p_2,0,\lstm{45})\Bigr] 
\nl
{}&+& \delta_{\sgenb}\,\Bigl[
C_{\sab}(p_1,p_2,\lstm{345}) - C_{\sab}(p_1,p_2,0,\lstm{45})\Bigr]\Bigr\}.
\eqa
\fbox{$\bf V^{\aca}_{\sbbgen}$}
For rank two tensors (see \sect{ranktwoaca}) we introduce a vector 
$U^{\aca}_{\sbb}$ with components
\bqa
U^{\aca}_{\sbbsca} &=& \frac{1}{2}\,\Bigl[
- l_{134}\,\cV^{\aca}_{\sba}
- \cV^{\aba}_{\sba}(p_1,P,\lstm{1245})
+ \cV^{\aba}_{\sba}(0,P,\lstm{1235})\Bigr],
\nl
U^{\aca}_{\sbbscb} &=& 
-\,\cV^{\aca}_{\sbbd}
+ \frac{1}{2}\,(p^2_1 - l_{\ssP 45})\,\cV^{\aca}_{\sbb}
- \frac{1}{2}\,\cV^{\aba}_{\sba}(0,P,\lstm{1235}) \,.
\eqa
We obtain the following result:
\bq
\left( \begin{array}{c} \cV^{\aca}_{\sbbc}  \\ \cV^{\aca}_{\sbbb}  
\end{array}\right) = G^{-1}\,U^{\aca}_{\sbb}.
\eq
Introduce a vector $W^{\aca}_{\sbb}$ with components
\bqa
W^{\aca}_{\sbbsca} &=& 
-\,\cV^{\aca}_{\sbbd}
- \frac{1}{2}\,\,l_{134}\,\cV^{\aca}_{\sbb}
- \frac{1}{2}\,\cV^{\aba}_{\sba}(p_1,P,\lstm{1245})
+ \frac{1}{2}\,\cV^{\aba}_{\sbb}(0,P,\lstm{1235}),
\nl
W^{\aca}_{\sbbscb} &=& \frac{1}{2}\,\Bigl[
(p^2_1 - l_{\ssP 45} )\,\cV^{\aca}_{\sba}
- \cV^{\aba}_{\sba}(0,P,\lstm{1235})
+ \cV^{\aba}_{\sba}(0,p_1,\lstm{1234}\Bigr].
\eqa
We obtain the following result:
\bq
\left( \begin{array}{c} \cV^{\aca}_{\sbba}  \\ \cV^{\aca}_{\sbbc}  
\end{array}\right) = G^{-1}\,W^{\aca}_{\sbb}.
\eq
Furthermore, we get
\bqa
(2-n)\,\cV^{\aca}_{\sbbd} &=& 
 \cV^{\aca}_{\Z}\,m^2_3 
- \cV^{\aba}_{\Z}(p_1,P,\lstm{1245}) 
- \frac{1}{2}\,\Bigl[ \cV^{\aca}_{\sbb}\,l_{\ssP 35} 
+ \cV^{\aba}_{\sba}(p_1,P,\lstm{1245}) 
\nl
{}&+& \cV^{\aba}_{\sba}(0,P,\lstm{1235}) 
- \cV^{\aba}_{\sbb}(0,P,\lstm{1235})\Bigr].
\eqa
For the corresponding tensors with saturated indices we obtain
\bqa
\cV^{\aca}(0\,|\,\mu,\mu) &=&
       - \cV^{\aca}_{\Z}\,m^2_3 
       + \cV^{\aba}_{\Z}(p_1,P,\lstm{1245}) ,
\nl
\cV^{\aca}(0\,|\,p_1, p_1) &=& \frac{1}{2}\,\Bigl[
       - l_{134}\,( \cV^{\aca}_{\sba}\,p_{12} 
       + \cV^{\aca}_{\sbb}\,p^2_1) 
\nl
{}&-& \cV^{\aba}_{\sba}(p_1,P,\lstm{1245})\,\spro{p_1}{P}
       + \cV^{\aba}_{\sba}(0,P,\lstm{1235})\,p_{12} 
+ \cV^{\aba}_{\sbb}(0,P,\lstm{1235})\,p^2_1\Bigr] ,
\nl
p^2_1\,\cV^{\aca}(0\,|\,p_2, p_2) &=& \frac{1}{2}\,\Bigl[
       - \cV^{\aca}_{\sba}\,D_2\,l_{134} 
+ \cV^{\aca}_{\sbb}\,( D\,p^2_1 + D\,l_{134} - 
D\,l_{\ssP 45} - D_1\,l_{134} )
       - \cV^{\aba}_{\sba}(p_1,P,\lstm{1245})\,( p^2_{12} + D_2 )
\nl
{}&+& \cV^{\aba}_{\sba}(0,P,\lstm{1235})\,(  - D + D_2 )
       + \cV^{\aba}_{\sbb}(0,P,\lstm{1235})\,p^2_{12} ) \Bigr],
\nl
\cV^{\aca}(0\,|\,p_1, p_2) &=& \frac{1}{2}\,\Bigl[
       - l_{134}\,( \cV^{\aca}_{\sba}\,p^2_2 
       + \cV^{\aca}_{\sbb}\,p_{12} ) 
\nl
{}&-& \cV^{\aba}_{\sba}(p_1,P,\lstm{1245})\,\spro{p_2}{P}
       + \cV^{\aba}_{\sba}(0,P,\lstm{1235})\,p^2_2 
+ \cV^{\aba}_{\sbb}(0,P,\lstm{1235})\,p_{12} \Bigr].
\eqa

\noindent\fbox{$\bf V^{\aca}_{\sabgen}$}
The form factors corresponding to the $12$ and $22$ groups are related by
\bqa
\cV^{\aca}_{\sabgen}(p_1,P,\lstm{12345}) &=& \frac{m^2_{123}}{2 m_3^2}\,
\cV^{\aca}_{\sbbgen}(p_1,P,\lstm{12345}) + \frac{m^2_{21}}{2 m_3^2}\,
\cV^{\aca}_{\sbbgen}(p_1,P,\lstm{12},0,\lstm{45})
\nl
{}&+& \Delta\cV^{\aca}_{\sabgen}(p_1,P,\lstm{12345}),
\eqa
\bqa
\Delta\cV^{\aca}_{\sabgen} &=& 
-\,\frac{1}{2 m_3^2}\,A_{\Z}([m_1,m_2])
\Bigl[ C_{\sbgen}(p_1,p_2,\lstm{345}) - 
C_{\sbgen}(p_1,p_2,0,\lstm{45}) \Bigr],
\quad i=  1\cdots 4.
\eqa
\fbox{$\bf V^{\aca}_{\saagen}$}
The form factors corresponding to the $11$ and $22$ groups are related by
\bqa
4\,(n - 1)\,m^2_3\,\cV^{\aca}_{\saagen} &=& 
m_3^2\,(2\,n\,m_{12}^2 + n\,m_3^2 - 4\,m_1^2)\,
\cV^{\aca}_{\sbbgen}(p_1,P,\lstm{12345})
\nl
{}&-& \,m_3^2\,\Bigl[(n-4)\,m_1^2 -2\,n\,m_2^2\Bigr]\,
\cV^{\aca}_{\sbbgen}(p_1,P,\lstm{12},0,\lstm{45})  
\nl
{}&+& n\,m_3^2\,m^4_{12}\,\Bigl[
\cV^{\ada}_{\sbbgen}(p_1,P,\lstm{12},0,\lstm{45},0)  
-\cV^{\ada}_{\sbbgen}(p_1,P,\lstm{12345},0)\Bigr]
\nl
{}&-& n\,m^2_{3}\,A_{\Z}(m_1)\,\Bigl[
C_{\sbgen}\,(p_1,p_2,\lstm{345})-
C_{\sbgen}\,(p_1,p_2,0,\lstm{45}) \Bigr]
\nl
{}&-& n\,m^2_{12}\,A_{\Z}([m_1,m_2])\,\Bigl[
m^2_3\,C_{\sbgen}(2,1,1\,;\,p_1,p_2,0,\lstm{45})
\nl
{}&-& C_{\sbgen}\,(p_1,p_2,0,\lstm{45}) 
+ C_{\sbgen}\,(p_1,p_2,\lstm{345}) \Bigr] 
\nl
{}&+& (3\,n-4)\,A_{\Z}(m_2)\,\Bigl[ C_{\sbgen}\,(p_1,p_2,\lstm{345}) - 
C_{\sbgen}\,(p_1,p_2,0,\lstm{45})\Bigr] \, ,
\eqa
for $i < 4$ and
\bqa
4\,(n - 1)\,m^2_3\,\cV^{\aca}_{\saad} &=& 
- m_3^2\,\cV^{\aca}(0|\mu ,\mu\,;\,p_1,P,\lstm{12345}) 
- m^4_{12}\,\cV^{\aca}_{\Z}(p_1,P,\lstm{12},0,\lstm{45}) 
\nl
{}&+& m_3^2\,(2\,n\,m_{12}^2 + n\,m_3^2 - 4\,m_1^2)\,
\cV^{\aca}_{\sbbd}(p_1,P,\lstm{12345})
\nl
{}&-& \,m_3^2\,\Bigl[(n-4)\,m_1^2 -2\,n\,m_2^2\Bigr]\,
\cV^{\aca}_{\sbbd}(p_1,P,\lstm{12},0,\lstm{45}) 
\nl
{}&+& n\,m_3^2\,m^4_{12}\,\Bigl[
\cV^{\ada}_{\sbbd}(p_1,P,\lstm{12},0,\lstm{45},0)  
-\cV^{\ada}_{\sbbd}(p_1,P,\lstm{12345},0)\Bigr]
\nl
{}&-& n\,m^2_{3}\,A_{\Z}(m_1)\,\Bigl[
C_{\sbd}\,(p_1,p_2,\lstm{345})-
C_{\sbd}\,(p_1,p_2,0,\lstm{45}) \Bigr]
\nl
{}&-& n\,m^2_{12}\,A_{\Z}([m_1,m_2])\,\Bigl[
m^2_3\,C_{\sbd}(2,1,1\,;\,p_1,p_2,0,\lstm{45})
\nl
{}&-& C_{\sbd}\,(p_1,p_2,0,\lstm{45}) 
+ C_{\sbd}\,(p_1,p_2,\lstm{345}) \Bigr] 
\nl
{}&+& (3\,n-4)\,A_{\Z}(m_2)\,\Bigl[ C_{\sbd}\,(p_1,p_2,\lstm{345}) - 
C_{\sbd}\,(p_1,p_2,0,\lstm{45})\Bigr] 
\nl
{}&+& \frac{1}{m_3^2}\,\Bigl\{ -A_{\Z}(m_1)\,\Bigl[ 
m^2_{123}\,C_{\Z}(p_1,p_2,\lstm{345}) + 
m^2_{21}\,C_{\Z}(p_1,p_2,0,\lstm{45})\Bigr] 
\nl
{}&-&  A_{\Z}(m_2)\,\Bigl[ 
m^2_{213}\,C_{\Z}(p_1,p_2,\lstm{345}) + 
m^2_{12}\,C_{\Z}(p_1,p_2,0,\lstm{45}) \Bigr] 
\nl
{}&+& \Bigl[ m^4_{12} - 2\,m_3^2\,(m_1^2 + m_2^2)\Bigr]\,
\cV^{\aca}_{\Z}(p_1,P,\lstm{12345}) \, .
\eqa
This expression requires results from the $V^{\ada}$ family, presented in 
\sect{summaada}.
\subsection{\boldmath $V^{\ada}(p_1,P,\lstm{12345})$ family \label{summaada}}
\noindent\fbox{$\bf V^{\ada}_{\sbgen}$}
Results were derived in \sect{rankoneada}, in particular the generalized 
scalar in \eqn{genada}. Introduce a vector $U^{\ada}_{2}$ of components
\bqa
U^{\ada}_{\sbsca} &=& \frac{1}{2}\,\Bigl[
        -\,\cV^{\ada}_{\Z} l_{134} 
- \cV^{\aca}_{\Z}\,(p_1,P,\lstm{12345})\,
       + \cV^{\aca}_{\Z}\,(0,P,\lstm{12335}) \Bigr]\, , 
\nl
U^{\ada}_{\sbscb}  &=& \frac{1}{2}\,\Bigl[
       \cV^{\ada}_{\Z}\, ( l_{154} - P^2 ) 
+ \cV^{\aca}_{\Z}\,(0,p_1,\lstm{12334})\,
- \cV^{\aca}_{\Z}\,(0,P,\lstm{12335})\, \Bigr]\, ; 
\eqa
we obtain the following result:
\bq
\left( \begin{array}{c} \cV^{\ada}_{\sba}  \\ \cV^{\ada}_{\sbb}  
\end{array}\right) = G^{-1}\,U^{\ada}_{2}, \qquad
\cV^{\ada}(0\,|\,p_1) = U^{\ada}_{\sbsca}, \qquad
\cV^{\ada}(0\,|\,p_2) = U^{\ada}_{\sbscb}.
\eq
\fbox{$\bf V^{\ada}_{\sagen}$}
We obtain
\bqa
\cV^{\ada}_{\sagen}(p_1,P,\lstm{12345}) &=&
\frac{m^2_{123}}{2 m^2_3}\,
\cV^{\ada}_{\sbgen}(p_1,P,\lstm{12345}) +
\,\frac{m^2_{12}}{2 m^4_3}\,\Bigl[
\cV^{\aca}_{\sbgen}(p_1,P,\lstm{12345})
\nl 
{}&-&
\cV^{\aca}_{\sbgen}(p_1,P,\lstm{12},0,\lstm{45})\Bigr] -
\frac{A_{\Z}([m_1,m_2])}{2 m_3^4}\,\Bigl[
C_{\sagen}(2,1,1,p_1,p_2,m_3,\lstm{345})\,m_3^2 
\nl 
{}&+& C_{\sagen}(p_1,p_2,\lstm{345}) - 
C_{\sagen}(p_1,p_2,0,\lstm{45}) \Bigr].
\eqa
\fbox{$\bf V^{\ada}_{\sbbgen}$}
For rank two tensors (see \sect{ranktwoada}) we introduce a vector 
$U^{\ada}_{\sbb}$ with components
\bqa
U^{\ada}_{\sbbsca} &=& -\,\frac{1}{2}\,\Bigl[
l_{134}\,\cV^{\ada}_{\sbb}
+ \cV^{\aca}_{\sbb}(p_1,P,\lstm{12345}) - 
\cV^{\aca}_{\sbb}(0,P,\lstm{12335}) 
\Bigr],
\nl
U^{\ada}_{\sbbscb} &=&
-\cV^{\ada}_{\sbbd}
+ \frac{1}{2}\,( p^2_1 - l_{\ssP 45})\,\cV^{\ada}_{\sbb}
- \frac{1}{2}\,\cV^{\aca}_{\sbb}(0,P,\lstm{12335}) \, ;
\eqa
we obtain the following result:
\bq
\left( \begin{array}{c} \cV^{\ada}_{\sbbc}  \\ \cV^{\ada}_{\sbbb}  
\end{array}\right) = G^{-1}\,U^{\ada}_{\sbb}.
\eq
Introduce a vector $W^{\ada}_{\sbb}$ with components
\bqa
W^{\ada}_{\sbbsca} &=&
- \cV^{\ada}_{\sbbd} - \frac{1}{2}\,\Bigl[
\cV^{\aca}_{\sba}(p_1,P,\lstm{12345}) - 
\cV^{\aca}_{\sbb}(0,P,\lstm{12335}) 
+ l_{134}\,\cV^{\ada}_{\sba}\Bigr],
\nl
W^{\ada}_{\sbbscb} &=& - \frac{1}{2}\,\Bigl[
- ( p^2_1 - l_{\ssP 45})\,\cV^{\ada}_{\sba}
+ \cV^{\aca}_{\sbb}(0,P,\lstm{12335}) - 
\cV^{\aca}_{\sbb}(0,p_1,\lstm{12334}) \Bigr]\,;
\eqa
we obtain the following result:
\bq
\left( \begin{array}{c} \cV^{\ada}_{\sbba}  \\ \cV^{\ada}_{\sbbc}  
\end{array}\right) = G^{-1}\,W^{\ada}_{\sbb}.
\eq
Furthermore we derive
\bqa
(2-n)\,\cV^{\ada}_{\sbbd} &=&
 \cV^{\ada}_{\Z}\,m^2_3 
- \frac{1}{2}\,\cV^{\ada}_{\sba}\,l_{134}
+ \frac{1}{2}\,\cV^{\ada}_{\sbb}\,(p^2_1 - l_{\ssP 45})
- \cV^{\aca}_{\Z}(p_1,P,\lstm{12345}) 
- \frac{1}{2}\,\cV^{\aca}_{\sba}(p_1,P,\lstm{12345}).
\eqa
For tensor integrals with saturated indices we obtain
\bqa
\cV^{\ada}(0\,|\,\mu,\mu) &=& 
       - \cV^{\ada}_{\Z}\,m^2_3 
       + \cV^{\aca}_{\Z}(p_1,P,\lstm{12345}),
\nl
p^2_2\,\cV^{\ada}(0\,|\,p_1, p_1) &=& \frac{1}{2}\,\Bigl[
       - l_{134}\,( \cV^{\ada}_{\sba}\,D_1 
       + \cV^{\ada}_{\sbb}\,D_2 ) 
       - \cV^{\aca}_{\sba}(p_1,P,\lstm{12345})\,D_1 
\nl
{}&-& \cV^{\aca}_{\sbb}(p_1,P,\lstm{12345})\,D_2 
   + \cV^{\aca}_{\sbb}(0,P,\lstm{12335})\,( D_1 + D_2 )\Bigr],
\nl
p^2_1\,\cV^{\ada}(0\,|\,p_2, p_2) &=& \frac{1}{2}\,\Bigl[
       - \cV^{\ada}_{\sba}\,l_{134}\,p^2_{12}
 + \cV^{\ada}_{\sbb}\,( D\,p^2_1 - D\,l_{\ssP 45} - D_2\,l_{134} )
       - \cV^{\aca}_{\sba}(p_1,P,\lstm{12345})\,p^2_{12}
\nl
{}&-& \cV^{\aca}_{\sbb}(p_1,P,\lstm{12345})\,D_2 
       - \cV^{\aca}_{\sbb}(0,P,\lstm{12335})\,(  2\,D - D_1 - D_2 )\Bigr],
\nl
\cV^{\ada}(0\,|\,p_1, p_2) &=& \frac{1}{2}\,\Bigl[
       - l_{134}\,( \cV^{\ada}_{\sba}\,p_{12} 
       + \cV^{\ada}_{\sbb}\,p^2_2 ) 
 - \cV^{\aca}_{\sba}(p_1,P,\lstm{12345})\,p_{12} 
       - \cV^{\aca}_{\sbb}(p_1,P,\lstm{12345})\,p^2_2 
\nl
{}&+& \cV^{\aca}_{\sbb}(0,P,\lstm{12335})\,\spro{p_2}{P}\Bigr].
\eqa
\fbox{$\bf V^{\ada}_{\sabgen}$}
For tensor integrals in the $12$ group we obtain
\bqa
\cV^{\ada}_{\sabgen} &=& 
\cV^{\ada}_{\sbbgen}(p_1,P,\lstm{12345}) \,\frac{m^2_{312}}{2 m_3^2} + 
\frac{m_{12}^2}{2\,m_3^4}\,\Bigl[
\cV^{\aca}_{\sbbgen}(p_1,P,\lstm{12345})
\nl
&-& \cV^{\aca}_{\sbbgen}(p_1,P,\lstm{12},0,\lstm{45})\Bigr]
+\Delta\,\cV^{\ada}_{\sabgen}(p_1,P,\lstm{12345}),
\eqa   
\bqa
\Delta\,\cV^{\ada}_{\sabgen} &=& 
-\,\frac{A_{\Z}([m_1,m_2])}{2 m_3^4}\,\Bigl[
C_{\sbgen}(p_1,p_2,\lstm{345}) 
- C_{\sbgen}(p_1,p_2,0,\lstm{45}) 
\nl
{}&+& m_3^2\,C_{\sbgen}(2,1,1\,;\,p_1,p_2,\lstm{345})\Bigr],
\qquad i = 1\cdots 4.
\eqa
\noindent\fbox{$\bf V^{\ada}_{\saagen}$}
For tensor integrals in the $11$ group we obtain
\bqa
\cV^{\ada}_{\saagen} &=& \frac{1}{4\,(n-1)}\,\frac{1}{m^4_3}\,\Bigl\{
( n\,m_{123}^4 - 4\,m_1^2\,m_3^2 ) 
\,\cV^{\ada}_{\sbbgen}(p_1,P,\lstm{12345}) +
\cV^{\ada}_{\sbbgen}(p_1,P,\lstm{12},0,\lstm{45}) \, n\,m^4_{12} 
\nl 
{}&+&
\frac{2}{m_3^2}\,( 2\,m_1^2\,m_3^2 - n\,m_{12}^2\,m_{123}^2 ) \,
\Bigl[ \cV^{\aca}_{\sbbgen}(p_1,P,\lstm{12},0,\lstm{45}) - 
\cV^{\aca}_{\sbbgen}(p_1,P,\lstm{12345})\Bigr] +
\Delta\cV^{\ada}_{\saagen}\Bigr\}.
\eqa
For $i < 4$ we have 
\bqa
\Delta\cV^{\ada}_{\saagen} &=& 
n\,\Bigl(2\,\frac{m^2_{123}}{m_3^2}-1\Bigr)\,A_{\Z}([m_1,m_2])\,\Bigl[ 
C_{\sbgen}(p_1,p_2,\lstm{345}) -
C_{\sbgen}(p_1,p_2,0,\lstm{45})\Bigr] 
\nl 
&-& n\,A_{\Z}([m_1,m_2])\,\Bigl[m_{123}^2  
\,C_{\sbgen}(2,1,1\,;\,p_1,p_2,\lstm{345}) + 
m_{12}^2\,C_{\sbgen}(2,1,1\,;\,p_1,p_2,0,\lstm{45})\Bigr] 
\nl 
{}&+& 2\,(n-2)\,A_{\Z}(m_2)\,\Bigl[ 
C_{\sbgen}(p_1,p_2,\lstm{345}) -
C_{\sbgen}(p_1,p_2,0,\lstm{45})
+ m_3^2\,C_{\sbgen}(2,1,1\,;\,p_1,p_2,\lstm{345})\Bigr].
\eqa
\bqa
\Delta\cV^{\ada}_{\saad} &=& 
n\,\Bigl(2\,\frac{m^2_{123}}{m_3^2}-1\Bigr)\,A_{\Z}([m_1,m_2])\,\Bigl[ 
C_{\sbd}(p_1,p_2,\lstm{345}) -
C_{\sbd}(p_1,p_2,0,\lstm{45})\Bigr] 
\nl 
&-& n\,A_{\Z}([m_1,m_2])\,\Bigl[m_{123}^2  
\,C_{\sbd}(2,1,1\,;\,p_1,p_2,\lstm{345}) + 
m_{12}^2\,C_{\sbd}(2,1,1\,;\,p_1,p_2,0,\lstm{45})\Bigr] 
\nl 
{}&+& 2\,(n-2)\,A_{\Z}(m_2)\,\Bigl[ 
C_{\sbd}(p_1,p_2,\lstm{345}) -
C_{\sbd}(p_1,p_2,0,\lstm{45})
+ m_3^2\, C_{\sbd}(2,1,1\,;\,p_1,p_2,\lstm{345})\Bigr]
\nl
{}&-& m^2_{12}\,A_{\Z}([m_1,m_2])\, 
\Bigl[ C_{\Z}(p_1,p_2,\lstm{345}) - 
C_{\Z}(p_1,p_2,0,\lstm{45})\Bigr] 
\nl 
{}&-&
m^2_3\,C_{\Z}(2,1,1\,;\,p_1,p_2,\lstm{345})\,\Bigl[
m^2_{123}\,A_{\Z}(m_1) +
m^2_{213}\,A_{\Z}(m_2)\Bigr].
\eqa
\subsection{\boldmath $V^{\bba}(p_1,p_1,P,\lstm{12345})$ family 
\label{summabba}}
\noindent\fbox{$\bf V^{\bba}_{\sgena}$}
Results were derived in \sect{rankonebba}, in particular the generalized 
scalar in \eqn{genbba}. Introduce a vector $U^{\bba}_1$ with components
\bqa
U^{\bba}_{\sasca} &=&
 \frac{1}{2}\Bigl[ -l_{112}\,
\cV^{\bba}_{\Z} + 
\cV^{\aba}_{\Z}(p_1,P,\lstm{1345})-
\cV^{\aba}_{\Z}(0,p_2,\lstm{2345})\Bigr],
\nl
U^{\bba}_{\sascb} &=&
\frac{1}{2}\Bigl[(- l_{245} - 2p_{12})
\cV^{\bba}_{\Z}
-\cV^{\aba}_{\Z}(-p_2,-P,\lstm{5321})
+\cV^{\aba}_{\Z}(0,-p_1,\lstm{4321})\Bigr]. 
\eqa
Referring to \eqn{genbba} we obtain
\bqa
\cV^{\bba}_{\saa} &=& \frac{1}{p^2_1}\,\Bigl[ U^{\bba}_{\sasca} +
\omega^2\,p_{12}\,\cV^{1,1|1,2|2}_{\bba}\,\bmid_{n = 6 - \ep}\Bigr],
\nl
\cV^{\bba}_{\sba} &=& \frac{1}{p_{12}}\,\Bigl\{ U^{\bba}_{\sascb} +
\,\omega^2\,p^2_2\,\Bigl[ \cV^{1,2|1,2|1}_{\bba} +
\cV^{2,1|1,2|1}_{\bba} + \,\cV^{1,1|1,2|2}_{\bba}\Bigr]\,\bmid_{n = 6-\ep}
\Bigr\}.
\eqa
\fbox{$\bf V^{\bba}_{\sgenb}$} Furthermore we find
\bqa
\cV^{\bba}_{\sab} &=& - \omega^2\,\cV^{1,1|1,2|2}_{\bba}\,\bmid_{n = 6 - \ep},
\quad
\cV^{\bba}_{\sbb} = -\,\omega^2\,\Bigl[ \cV^{1,2|1,2|1}_{\bba} +
\cV^{2,1|1,2|1}_{\bba} + \,\cV^{1,1|1,2|2}_{\bba}\Bigr]\,\bmid_{n = 6-\ep}.
\eqa
\fbox{$\bf V^{\bba}_{\sbbgen}$}
For rank two tensors (see \sect{ranktwobba}) we introduce a vector 
$U^{\bba}_{\sbb}$ with components
\bqa
U^{\bba}_{\sbbsca} &=&
       \cV^{\bba}_{\sbbd}\,( 1 - n )
       - \cV^{\bba}_{\Z}\,( p^2_1 + m^2_4 )
       - 2\,\cV^{\bba}_{\sba}\,p^2_1 
       + \frac{1}{2}\,\Bigl[
        \cV^{\bba}_{\sbb}\,(  P^2 - 4\,p_{12} - l_{154} )
\nl
{}&+& 2\,\cV^{\aba}_{\Z}(-p_2,-P,\lstm{5321}) 
       - \cV^{\aba}_{\Z}(0,-p_1,\lstm{4321}) 
       - \cV^{\aba}_{\sab}(-p_2,-P,\lstm{5321}) \Bigr],
\nl
U^{\bba}_{\sbbscb} &=& \frac{1}{2}\,\Bigl[
       - \cV^{\bba}_{\sba}\,( P^2 - l_{154} )
       + \cV^{\aba}_{\Z}(-p_2,-P,\lstm{5321}) 
\nl
{}&-& \cV^{\aba}_{\Z}(0,-p_1,\lstm{4321}) 
       + \cV^{\aba}_{\saa}(-p_2,-P,\lstm{5321}) 
- \cV^{\aba}_{\saa}(0,-p_1,\lstm{4321}) \Bigr]\,;
\eqa
we obtain the following result: 
\bqa
\left( \begin{array}{c} \cV^{\bba}_{\sbba}  \\ \cV^{\bba}_{\sbbc}  
\end{array}\right) &=& G^{-1}\,U^{\bba}_{\sbb},
\eqa
\bqa
p^2_2\,\cV^{\bba}_{\sbbb} &=& 
       -\cV^{\bba}_{\sbbd} 
       - \cV^{\bba}_{\sbbc}\,p_{12}
       - \frac{1}{2}\,\Bigl[
 \cV^{\bba}_{\sbb}\,( P^2 - l_{154} )
       - \cV^{\aba}_{\Z}(0,-p_1,\lstm{4321}) 
- \cV^{\aba}_{\sab}(-p_2,-P,\lstm{5321}) \Bigr],
\nl
\cV^{\bba}_{\sbbd} &=& \frac{1}{2}\,\omega^2\,\cV^{1,1|1,1|2}_{\bba}\,
\bmid_{n = 6 -\ep} \, .
\eqa 
When indices are saturated we obtain
\bqa
\cV^{\bba}(0\,|\,\mu,\mu) &=&
       - \cV^{\bba}_{\Z}\,(p^2_1+m^2_4) 
       -2\,\cV^{\bba}_{\sba}\,p^2_1 
       - 2\,\cV^{\bba}_{\sbb}\,p_{12} 
       + \cV^{\aba}_{\Z}(-p_2,-P,\lstm{5321}),
\nl
\cV^{\bba}(0\,|\,p_1, p_1) &=&
       - \cV^{\bba}_{\Z}\,(p^2_1+m^2_4)\,p^2_1 
       - \cV^{\bba}_{\sbbd}\,(n-1)\,p^2_1 
\nl
{}&-& 2\,\cV^{\bba}_{\sba}\,p^4_1 
     -2\,\cV^{\bba}_{\sbb}\,D_3 
       + \cV^{\aba}_{\Z}(-p_2,-P,\lstm{5321})\,p^2_1,
\nl
p^2_1\,\cV^{\bba}(0\,|\,p_2, p_2) &=&
       - \cV^{\bba}_{\Z}\,(p^2_1 + m^2_4)\,p_{12}^2 
       - \cV^{\bba}_{\sbbd}\,\Bigl[  D + (n-1)\,p_{12}^2 \Bigr]
\nl
{}&-& 2\,\cV^{\bba}_{\sba}\,D_3\,p_{12} 
       - \cV^{\bba}_{\sbb}\,\Bigl[ D\,(P^2 - l_{154}) + 2\,p_{12}^3 \Bigr]
       + \cV^{\aba}_{\Z}(-p_2,-P,\lstm{5321})\,p_{12}^2 
\nl
{}&+& \cV^{\aba}_{\Z}(0,-p_1,\lstm{4321})\,D 
       + \cV^{\aba}_{\sab}(-p_2,-P,\lstm{5321})\,D,
\nl
p_{12}\,\cV^{\bba}(0\,|\,p_1, p_2) &=&
       - \cV^{\bba}_{\Z}\,(p^2_1 + m^2_4)\,p_{12}^2 
       - \cV^{\bba}_{\sbbd}\,\Bigl[  D + (n-1)\,p_{12}^2 \Bigr]
\nl
{}&-& 2\,\cV^{\bba}_{\sba}\,D_3\,p_{12} 
       -\frac{1}{2}\,\cV^{\bba}_{\sbb}\,
\Bigl[ D\,(P^2 - l_{154}) + 4\,p_{12}^3 \Bigr]
       + \cV^{\aba}_{\Z}(-p_2,-P,\lstm{5321})\,p_{12}^2 
\nl
{}&+& \frac{1}{2}\,\cV^{\aba}_{\Z}(0,-p_1,\lstm{4321})\,D 
       +\frac{1}{2}\,\cV^{\aba}_{\sab}(-p_2,-P,\lstm{5321})\,D.
\eqa
\fbox{$\bf V^{\bba}_{\sabgen}$}
Introduce a vector $U^{\bba}_{\sab}$ with components
\bqa
U^{\bba}_{\sabsca} &=& \frac{1}{2}\,\Bigl[
       - \cV^{\bba}_{\sbb}\,l_{112} 
       + \cV^{\aba}_{\sba}(p_1,P,\lstm{1345}) 
       - \cV^{\aba}_{\sba}(0,p_2,\lstm{2345}) \Bigr],
\nl
U^{\bba}_{\sabscb} &=& \frac{1}{2}\,\Bigl[
        \cV^{\bba}(0\,|\,\mu\mu) 
       + \cV^{\bba}_{\Z}\,m^2_{31} 
       + \cV^{\bba}_{\sba}\,l_{112}
       + 2\,\cV^{\bba}_{\sabd}\,(n-1)
\nl
{}&+& 2\,\cV^{\aba}_{\Z}(0,p_2,\lstm{2345}) 
       + B_{\Z}(p_1,\lstm{12})\,B_{\Z}(p_2,\lstm{45}) 
       - \cV^{\aba}_{\sbb}(p_1,P,\lstm{1345})\Bigr]\,;
\eqa
we obtain the following result:
\bq
\left( \begin{array}{c} \cV^{\bba}_{\sabc}  \\ \cV^{\bba}_{\sabb}  
\end{array}\right) = G^{-1}\,U^{\bba}_{\sab}.
\eq
Introduce a vecror $W^{\bba}_{\sab}$ with components
\bqa
W^{\bba}_{\sabsca} &=& \frac{1}{2}\,\Bigl[
         2\,\cV^{\bba}_{\sabd} \, (  n-1 )
       - \cV^{\bba}_{\sab} \, (  - P^2 + l_{145} )
       + \cV^{\bba}_{\Z} \, m^2_{31} 
       + \cV^{\bba}(0\,|\,\mu\mu) 
       - \cV^{\aba}_{\sbb}(-p_2,-P,\lstm{5321}) 
\nl
{}&-& \cV^{\aba}_{\Z}(-p_2,-P,\lstm{5321}) 
       + \cV^{\aba}_{\Z}(0,p_2,\lstm{2345}) 
       + B_{\Z}(p_1,\lstm{12})\,B_{\Z}(p_2,\lstm{45}) \Bigr],
\nl
W^{\bba}_{\sabsca} &=& \frac{1}{2}\,\Bigl[
       - \cV^{\bba}_{\saa} \, ( P^2 - l_{145} )
       + \cV^{\aba}_{\Z}(-p_2,-P,\lstm{5321}) 
       - \cV^{\aba}_{\Z}(0,-p_1,\lstm{4321}) 
\nl
{}&-& \cV^{\aba}_{\saa}(0,-p_1,\lstm{4321}) 
       + \cV^{\aba}_{\sab}(-p_2,-P,\lstm{5321}) \Bigr]\,;
\eqa
we obtain the following result;
\bq
\left( \begin{array}{c} \cV^{\bba}_{\saba}  \\ \cV^{\bba}_{\sabc}  
\end{array}\right) = G^{-1}\,W^{\bba}_{\sab}.
\eq
Furthermore we get
\bqa
\cV^{\bba}_{\sabe} &=& - \frac{1}{p_{12}}\,\Bigl\{
       \cV^{\bba}_{\sabb} \,  p^2_2 
       + \cV^{\bba}_{\sabd} 
+ \frac{1}{2}\,\Bigl[ - \cV^{\aba}_{\Z}(-p_2,-P,\lstm{5321}) 
       - \cV^{\aba}_{\sbb}(-p_2,-P,\lstm{5321}) 
       + \cV^{\bba}_{\sab} \, ( P^2 - l_{145} )\Bigr]\Bigr\},
\nl
\cV^{\bba}_{\sabd} &=& \frac{1}{2}\,\omega^2\,
\cV^{1,1|1,1|2}_{\bba}\,\bmid_{n=6-\ep}.
\eqa
For saturated indices we have
\bqa
\cV^{\bba}(\mu\,|\,\mu) &=& \frac{1}{2}\,\Bigl[
        \cV^{\bba}_{\Z}\,m^2_{31} 
       + \cV^{\bba}(0\,|\,\mu,\mu)
       +  \cV^{\aba}_{\Z}(0,p_2,\lstm{2345}) 
+ B_{\Z}(p_1,\lstm{12})\,B_{\Z}(p_2,\lstm{45})\Bigr],
\nl
\cV^{\bba}(p_1\,|\,p_1) &=& \frac{1}{2}\,\Bigl[
       - l_{112}\,( \cV^{\bba}_{\sbb}\,p_{12} 
       + \cV^{\bba}_{\sba}\,p^2_1 )
       + \cV^{\aba}_{\sba}(p_1,P,\lstm{1345})\,p_{12} 
       - \cV^{\aba}_{\sba}(0,p_2,\lstm{2345})\,p_{12} 
\nl
{}&-& \cV^{\aba}_{\Z}(0,p_2,\lstm{2345})\,p^2_1 
       + \cV^{\aba}_{\sbb}(p_1,P,\lstm{1345})\,p^2_1 
       \Bigr],
\nl
p^2_1\,\cV^{\bba}(p_2\,|\,p_2) &=& \frac{1}{2}\,\Bigl[
       - \cV^{\bba}_{\sbb}\,D_2\,l_{112} 
       + \cV^{\bba}_{\sab}\,D\,(  - P^2 + l_{145} )
       + \cV^{\bba}_{\sba}\,( D - D_1 )\,l_{112}  
\nl
{}&+& \cV^{\aba}_{\sba}(p_1,P,\lstm{1345})\,D_2 
       - \cV^{\aba}_{\sba}(0,p_2,\lstm{2345})\,D_2 
       + \cV^{\aba}_{\Z}(-p_2,-P,\lstm{5321})\,D 
\nl
{}&-& \cV^{\aba}_{\Z}(0,p_2,\lstm{2345})\,p^2_{12}
       + \cV^{\aba}_{\sbb}(p_1,P,\lstm{1345})\,p^2_{12}
       + \cV^{\aba}_{\sbb}(-p_2,-P,\lstm{5321})\,D \Bigr],
\nl
\cV^{\bba}(p_1\,|\,p_2) &=& \frac{1}{2}\,\Bigl[
       - l_{112}\,( \cV^{\bba}_{\sbb}\,p^2_2 
       + \cV^{\bba}_{\sba}\,p_{12} )
       + \cV^{\aba}_{\sba}(p_1,P,\lstm{1345})\,p^2_2 
       - \cV^{\aba}_{\sba}(0,p_2,\lstm{2345})\,p^2_2 
\nl
{}&-& \cV^{\aba}_{\Z}(0,p_2,\lstm{2345})\,p_{12} 
       + \cV^{\aba}_{\sbb}(p_1,P,\lstm{1345})\,p_{12} 
        \Bigr],
\nl
p_{12}\,\cV^{\bba}(p_2\,|\,p_1) &=& \frac{1}{2}\,\Bigl[
       - l_{112}\,( \cV^{\bba}_{\sbb}\,D_2
          + \cV^{\bba}_{\sba}\,D_1 )
       + 2\,\cV^{\bba}_{\sabd}\,D\,( n - 2 )
          - \cV^{\bba}_{\Z}\,  D\,m^2_{31} 
          - \cV^{\bba}(0\,|\,\mu\mu)\,D 
\nl
{}&+& \cV^{\bba}_{\sab}\,D\,(  - P^2 + l_{145} )
    +    \cV^{\aba}_{\sba}(p_1,P,\lstm{1345})\,D_2 
       - \cV^{\aba}_{\sba}(0,p_2,\lstm{2345})\,D_2 
\nl
{}&+& \cV^{\aba}_{\Z}(-p_2,-P,\lstm{5321})\,D 
       - \cV^{\aba}_{\Z}(0,p_2,\lstm{2345})\,(  D + D_1 )
       + \cV^{\aba}_{\sbb}(p_1,P,\lstm{1345})\,D_1 
\nl
{}&+& \cV^{\aba}_{\sbb}(-p_2,-P,\lstm{5321})\,D 
- B_{\Z}(p_1,\lstm{12})\,B_{\Z}(p_2,\lstm{45})\,D \Bigr].
\eqa
\fbox{$\bf V^{\bba}_{\saagen}$}
For form factors belonging to the $11$ group we have
\bqa
(n-1)\,p^4_1\,\cV^{\bba}_{\saaa} &=& \Bigl\{
         \cV^{\bba}_{\saab}\,\Bigl[   (n-1)\,D_1 - (n-2)\,D  \Bigr]
       + \cV^{\bba}_{\Z}\,p^2_1\,m^2_1 
       + \frac{1}{2}\,\Bigl[
\nl
{}&-& \cV^{\bba}_{\saa}\,n\,p^2_1\,l_{112} 
 +    \cV^{\aba}_{\sab}(p_1,P,\lstm{1345})\,n\,p^2_1 
    + (n-2)\,\Bigl[ \cV^{\bba}_{\sab}\,p_{12}\,l_{112}
    + \cV^{\aba}_{\Z}(0,p_2,\lstm{2345})\,p^2_1
\nl
{}&-&    \cV^{\aba}_{\saa}(p_1,P,\lstm{1345})\,p_{12}
       + \cV^{\aba}_{\saa}(0,p_2,\lstm{2345})\,p_{12}\,\Bigr]\Bigr\},
\nl
p^2_1\,\cV^{\bba}_{\saac} &=& \frac{1}{2}\,\Bigl[
     -2\,\cV^{\bba}_{\saab}\,p_{12} 
       - \cV^{\bba}_{\sab}\,l_{112} 
       + \cV^{\aba}_{\saa}(p_1,P,\lstm{1345}) 
       - \cV^{\aba}_{\saa}(0,p_2,\lstm{2345}) \Bigr],
\nl
(n-1)\,p^2_1\,\cV^{\bba}_{\saad} &=& \frac{1}{2}\,\Bigl[
        l_{112}\,( \cV^{\bba}_{\saa}\,p^2_1
       + \cV^{\bba}_{\sab}\,p_{12} )  
       -2\,\cV^{\bba}_{\saab}\,D 
       -2\,\cV^{\bba}_{\Z}\,p^2_1\,m^2_1 
       +   \cV^{\aba}_{\Z}(0,p_2,\lstm{2345})\,p^2_1 
\nl
{}&-&    \cV^{\aba}_{\sab}(p_1,P,\lstm{1345})\,p^2_1 
       - \cV^{\aba}_{\saa}(p_1,P,\lstm{1345})\,p_{12}
       + \cV^{\aba}_{\saa}(0,p_2,\lstm{2345})\,p_{12} \Bigr],
\nl
\cV^{\bba}_{\saab} &=& 4\,\omega^4\,
\cV^{1 , 1 | 1 , 3 | 3}_{\bba} (n = 8 - \ep).
\eqa
For saturated indices we have
\bqa
\cV^{\bba}(\mu,\mu\,|\,0) &=&
       - \cV^{\bba}_{\Z}\,m^2_1 
       + \cV^{\aba}_{\Z}(0,p_2,\lstm{2345}),
\nl
\cV^{\bba}(p_1, p_1\,|\,0) &=& \frac{1}{2}\,\Bigl[
       - l_{112}\,( \cV^{\bba}_{\saa}\,p^2_1 
       + \cV^{\bba}_{\sab}\,p_{12} ) 
       + \cV^{\aba}_{\Z}(0,p_2,\lstm{2345})\,p^2_1 
\nl
{}&+&    \cV^{\aba}_{\sab}(p_1,P,\lstm{1345})\,p^2_1 
       + \cV^{\aba}_{\saa}(p_1,P,\lstm{1345})\,p_{12} 
       - \cV^{\aba}_{\saa}(0,p_2,\lstm{2345})\,p_{12} \Bigr],
\nl
(n-1)\,p^4_1\,\cV^{\bba}(p_2, p_2\,|\,0) &=& \frac{1}{2}\,\Bigl\{
       2\,\cV^{\bba}_{\saab}\,(n-2)\,D^2 
     - 2\,\cV^{\bba}_{\Z}\,D\,p^2_1\,m^2_1 
      +   \cV^{\bba}_{\saa}\,p^2_1\,l_{112}\,\Bigl[ D - (n-1)\,p^2_{12}\Bigr] 
\nl
{}&+& \cV^{\bba}_{\sab}\,p_{12}\,l_{112}\,\Bigl[ D - (n-1)\,( D+D_3)\Bigr] 
+ \cV^{\aba}_{\Z}(0,p_2,\lstm{2345})\,p^2_1\,\Bigl[ D + (n-1)\,p^2_{12}\Bigr] 
\nl
{}&+& (n-1)\,p_{12}\,( D + D_1 )\,\Bigl[
         \cV^{\aba}_{\saa}(p_1,P,\lstm{1345})\,
       - \cV^{\aba}_{\saa}(0,p_2,\lstm{2345}) \Bigr] 
\nl
{}&-& \cV^{\aba}_{\sab}(p_1,P,\lstm{1345})\,p^2_1\,
\Bigl[ D - (n-1)\,p^2_{12}\Bigr] 
       - D\,p_{12}\,\Bigl[ 
         \cV^{\aba}_{\saa}(p_1,P,\lstm{1345}) 
\nl
{}&-& \cV^{\aba}_{\saa}(0,p_2,\lstm{2345}) \Bigr] \Bigr\},
\nl
\cV^{\bba}(p_1, p_2\,|\,0) &=& \frac{1}{2}\,\Bigl\{
       - l_{112}\,( \cV^{\bba}_{\saa}\,p_{12} 
       + \cV^{\bba}_{\sab}\,p^2_2 ) 
       + p_{12}\,\Bigl[ \cV^{\aba}_{\Z}(0,p_2,\lstm{2345}) 
\nl
{}&+& \cV^{\aba}_{\sab}(p_1,P,\lstm{1345}) \Bigr] 
       + p^2_2\,\Bigl[ \cV^{\aba}_{\saa}(p_1,P,\lstm{1345}) 
       - \cV^{\aba}_{\saa}(0,p_2,\lstm{2345}) \Bigr] \Bigr\}.
\eqa
\subsection{\boldmath $V^{\bca}(P,p_1,P,\lstm{123456})$ family 
\label{summabca}}
\noindent\fbox{$\bf V^{\bca}_{\sagen}$}
Results were derived in \sect{rankonebca}. Referring to \eqn{genbca} the 
vector integrals are
\bq
\cV^{\bca}_{\saa} = \frac{\spro{p_2}{P}}{P^2}\,
\omega^2\,\cV^{1,1|1,2,1|2}_{\bca}\,\bmid_{n = 6 -\ep},
\qquad
\cV^{\bca}_{\sab} = \cV^{\bca}_{\saa} + 
\omega^2\,\cV^{1,1|1,2,1|2}_{\bca}\,\bmid_{n = 6 -\ep}.
\eq
\fbox{$\bf V^{\bca}_{\sbgen}$}
Introduce a vector $U^{\bca}_{2}$ with components
\bqa
U^{\bca}_{\sbsca} &=&
\frac{1}{2}\Bigl[l_{145}\cV^{\bca}_{\Z} + 
\cV^{\bba}_{\Z}(P,P,p_1,\lstm{12365}) 
- \cV^{\bba}_{\Z}(P,P,0,\lstm{12364})\Bigr], 
\nl
U^{\bca}_{\sbscb} &=&
-\,\frac{1}{2}\Bigl[ l_{\ssP 46}\cV^{\bca}_{\Z} + 
\cV^{\bba}_{\Z}(P,P,p_1,\lstm{12365})
- \cV^{\bba}_{\Z}(-P,-P,-p_2,\lstm{21345})\Bigr]\,;
\eqa
we obtain the following result:
\bq
\left( \begin{array}{c} \cV^{\bca}_{\sba}  \\ \cV^{\bca}_{\sbb}  
\end{array}\right) = G^{-1}\,W^{\bca}_{2},
\quad 
W^{\bca}_{\sbsca} = U^{\bca}_{\sbsca},
\quad
W^{\bca}_{\sbscb} = U^{\bca}_{\sbscb} - U^{\bca}_{\sbsca}.
\eq
\fbox{$\bf V^{\bca}_{\sbbgen}$}
For rank two tensors (see \sect{ranktwobca}) we introduce a vector 
$U^{\bca}_{\sbb}$ with components
\bqa
U^{\bca}_{\sbbsca} &=& -\,\frac{1}{2}\,\Bigl[
        \cV^{\bca}_{\sbb}\,l_{145} 
       - \cV^{\bba}_{\sba}(P,P,0,\lstm{12364}) 
       + \cV^{\bba}_{\sbb}(P,P,0,\lstm{12364}) \Bigr],
\nl
U^{\bca}_{\sbbscb} &=& -\,\frac{1}{2}\,\Bigl[
       2\,\cV^{\bca}_{\sbbd} 
       +  \cV^{\bca}_{\sbb} \, ( P^2 - l_{165} ) 
       +  \cV^{\bba}_{\Z}(-P,-P,-p_2,\lstm{21345}) 
\nl
{}&+&  \cV^{\bba}_{\sba}(-P,-P,-p_2,\lstm{21345}) 
     + \cV^{\bba}_{\sba}(P,P,0,\lstm{12364}) 
     - \cV^{\bba}_{\sbb}(P,P,0,\lstm{12364}) \Bigr]\,;
\eqa
we obtain the following result:
\bq
\left( \begin{array}{c} \cV^{\bca}_{\sbbc}  \\ \cV^{\bca}_{\sbbb}  
\end{array}\right) = G^{-1}\,U^{\bca}_{\sbb}.
\eq
Introduce a vector $W^{\bca}_{\sbb}$ with components
\bqa
W^{\bca}_{\sbbsca} &=& -\,\frac{1}{2}\,\Bigl[
      2\,\cV^{\bca}_{\sbbd} 
       + \cV^{\bca}_{\sba}\,l_{145} 
\nl
{}&+& \cV^{\bba}_{\sba}(P,P,p_1,\lstm{12365}) 
    - \cV^{\bba}_{\sba}(P,P,0,\lstm{12364}) 
    + \cV^{\bba}_{\sbb}(P,P,0,\lstm{12364}) \Bigr],
\nl
W^{\bca}_{\sbbscb} &=& -\,\frac{1}{2}\,\Bigl[
         \cV^{\bca}_{\sba}\,( P^2 - l_{165} )
       + \cV^{\bba}_{\Z}(-P,-P,-p_2,\lstm{23145}) 
+        \cV^{\bba}_{\sba}(-P,-P,-p_2,\lstm{21345}) 
\nl
{}&+& \cV^{\bba}_{\sba}(P,P,0,\lstm{12364}) 
 -    \cV^{\bba}_{\sbb}(-P,-P,-p_2,\lstm{21345}) 
       - \cV^{\bba}_{\sbb}(P,P,0,\lstm{12364}) \Bigr]\,;
\eqa
we obtain
\bq
\left( \begin{array}{c} \cV^{\bca}_{\sbba}  \\ \cV^{\bca}_{\sbbc}  
\end{array}\right) = G^{-1}\,W^{\bca}_{\sbb},
\eq
and also
\bqa
\cV^{\bca}_{\sbbd} &=& \frac{1}{2\,(2-n)}\,\Bigl[
       2\,\cV^{\bca}_{\Z}\, m^2_4 
        - \cV^{\bca}_{\sba}\,l_{145} 
        + \cV^{\bca}_{\sbb}\,(  - P^2 + l_{165} ) 
        - \cV^{\bba}_{\Z}(-P,-P,-p_2,\{m\}_{21345}) 
\nl
{}&-& 2\,\cV^{\bba}_{\Z}(P,P,p_1,\{m\}_{12365}) 
        -\cV^{\bba}_{\sba}(-P,-P,-p_2,\{m\}_{21345}) 
       - \cV^{\bba}_{\sba}(P,P,p_1,\{m\}_{12365}) \Bigr]. 
\eqa
For saturated indices we get
\bqa
\cV^{\bca}(0\,|\,\mu,\mu) &=&
      -\cV^{\bca}_{\Z}\,m^2_4+
       \cV^{\bba}_{\Z}(P,P,p_1,\{m\}_{12365}),
\nl
\cV^{\bca}(0\,|\,p_1, p_1) &=& \frac{1}{2}\,\Bigl\{
       - l_{145}\,( \cV^{\bca}_{\sba}\,p^2_1 
       + \cV^{\bca}_{\sbb}\,p_{12} )
       - \cV^{\bba}_{\sba}(P,P,p_1,\{m\}_{12365})\,p^2_1 
\nl
{}&+& \spro{p_1}{P}\,\Bigl[ \cV^{\bba}_{\sba}(P,P,0,\{m\}_{12364})
      + \cV^{\bba}_{\sbb}(P,P,0,\{m\}_{12364}) \Bigr] \Bigr\},
\nl
p^2_1\,\cV^{\bca}(0\,|\,p_2, p_2) &=& \frac{1}{2}\,\Bigl[
       - \cV^{\bca}_{\sba}\,l_{145}\,p^2_{12}
       + \cV^{\bca}_{\sbb}\,
\Bigl[ ( l_{165}  - P^2 )\,D - D_2\,l_{145} \Bigr] 
       - \cV^{\bba}_{\sba}(-P,-P,-p_2,\{m\}_{21345})\,D 
\nl
{}&-& \cV^{\bba}_{\sba}(P,P,p_1,\{m\}_{12365})\,p^2_{12}
       - \cV^{\bba}_{\sba}(P,P,0,\{m\}_{12364})\,( 2\,D - D_1 - D_2 )
\nl
{}&+& \cV^{\bba}_{\sbb}(P,P,0,\{m\}_{12364})\,(  2\,D - D_1 - D_2 )
       - \cV^{\bba}_{\Z}(-P,-P,-p_2,\{m\}_{21345})\,D \Bigr],
\nl
\cV^{\bca}(0\,|\,p_1, p_2) &=& \frac{1}{2}\,\Bigl\{
       - l_{145}\,( \cV^{\bca}_{\sba}\,p_{12} 
       + \cV^{\bca}_{\sbb}\,p^2_2 )
       - \cV^{\bba}_{\sba}(P,P,p_1,\{m\}_{12365})\,p_{12} 
\nl
{}&+& \spro{p_2}{P}\,\Bigl[ \cV^{\bba}_{\sba}(P,P,0,\{m\}_{12364})
      + \cV^{\bba}_{\sbb}(P,P,0,\{m\}_{12364}) \Bigr]\Bigr\}.
\eqa
\fbox{$\bf V^{\bca}_{\sabgen}$}
Introduce a vector $U^{\bca}_{\sab}$ with components
\bqa
U^{\bca}_{\sabsca} &=& -\,\frac{1}{2}\,\Bigl[
        \cV^{\bca}_{\sab}\,l_{145} 
       - \cV^{\bba}_{\sab}(P,P,p_1,\lstm{12365}) 
\nl
{}&+&    \cV^{\bba}_{\sab}(P,P,0,\lstm{12364}) 
       + \cV^{\bba}_{\saa}(P,P,p_1,\lstm{12365}) 
       - \cV^{\bba}_{\saa}(P,P,0,\lstm{12364}) \Bigr],
\nl
U^{\bca}_{\sabscb} &=& -\,\frac{1}{2}\,\Bigl[
       2\,\cV^{\bca}_{\sabd} 
       + \cV^{\bca}_{\sab}\,( P^2 - l_{165} ) 
       + \cV^{\bba}_{\Z}(-P,-P,-p_2,\lstm{21345}) 
\nl
{}&-& \cV^{\bba}_{\sab}(P,P,0,\lstm{12364}) 
    + \cV^{\bba}_{\saa}(-P,-P,-p_2,\lstm{21345}) 
    + \cV^{\bba}_{\saa}(P,P,0,\lstm{12364}) \Bigr]\,;
\eqa
we obtain the following result:
\bq
\left( \begin{array}{c} \cV^{\bca}_{\sabe}  \\ \cV^{\bca}_{\sabb}  
\end{array}\right) = G^{-1}\,U^{\bca}_{\sab}.
\eq
Introduce a vector $W^{\bca}_{\sab}$ with components
\bqa
W^{\bca}_{\sabsca} &=& -\,\frac{1}{2}\,\Bigl[
       2\,\cV^{\bca}_{\sabd} 
       + \cV^{\bca}_{\saa}\,l_{145} 
       + \cV^{\bba}_{\saa}(P,P,p_1,\lstm{12365}) \Bigr],
\nl
W^{\bca}_{\sabscb} &=& -\,\frac{1}{2}\,\Bigl[
         \cV^{\bca}_{\saa}\,( P^2 - l_{165} )
       + \cV^{\bba}_{\Z}(-P,-P,-p_2,\lstm{21345}) 
       -    \cV^{\bba}_{\sab}(-P,-P,-p_2,\lstm{21345}) 
\nl
{}&-& \cV^{\bba}_{\sab}(P,P,0,\lstm{12364}) 
       + \cV^{\bba}_{\saa}(-P,-P,-p_2,\lstm{21345}) 
       + \cV^{\bba}_{\saa}(P,P,0,\lstm{12364}) \Bigr]\,;
\eqa
we obtain
\bq
\left( \begin{array}{c} \cV^{\bca}_{\saba}  \\ \cV^{\bca}_{\sabc}  
\end{array}\right) = G^{-1}\,W^{\bca}_{\sab},
\eq
and also
\bqa
(2-n)\,\cV^{\bca}_{\sabd} &=& \frac{1}{2}\,\Bigl[
         \cV^{\bca}_{\Z}\,m^2_{134} 
       - \cV^{\bca}_{\sab}\,(  P^2 - l_{165} )
       - \cV^{\bca}_{\saa}\,l_{145} 
       - \cV^{\bba}_{\Z}(P,P,p_1,\lstm{12365}) 
\nl
{}&-&    \cV^{\bba}_{\Z}(-P,-P,-p_2,\lstm{21345}) 
       - \cV^{\aca}_{\Z}(-p_2,-P,\lstm{23654}) 
       - \cV^{\bba}_{\saa}(P,P,p_1,\lstm{12365}) 
\nl
{}&-& \cV^{\bba}_{\saa}(-P,-P,-p_2,\lstm{21345}) 
- B_{\Z}(P,\lstm{12})\,C_{\Z}(p_1,p_2,\lstm{456}) \Bigr]. 
\eqa
For saturated indices we obtain
\bqa
\cV^{\bca}(\mu\,|\,\mu) &=& \frac{1}{2}\,\Bigl[
       - \cV^{\bca}_{\Z}\,m^2_{134} 
       + \cV^{\bba}_{\Z}(P,P,p_1,\lstm{12365}) 
\nl
{}&+& \cV^{\aca}_{\Z}(-p_2,-P,\lstm{23654}) 
       + B_{\Z}(P,\lstm{12})\,C_{\Z}(p_1,p_2,\lstm{456}) \Bigr],
\nl
\cV^{\bca}(p_1\,|\,p_1) &=& \frac{1}{2}\,\Bigl\{
       - l_{145}\,( \cV^{\bca}_{\saa}\,p^2_1 
       + \cV^{\bca}_{\sab}\,p_{12} ) 
       + p_{12}\,\cV^{\bba}_{\sab}(P,P,p_1,\lstm{12365}) 
\nl
{}&-& \spro{p_1}{P}\,\Bigl[
     \cV^{\bba}_{\sab}(P,P,0,\lstm{12364}) 
   - \cV^{\bba}_{\saa}(P,P,0,\lstm{12364}) 
   + \cV^{\bba}_{\saa}(P,P,p_1,\lstm{12365})\,\Bigr]
       \Bigr\},
\nl
\cV^{\bca}(p_2\,|\,p_2) &=& \frac{1}{2}\,\Bigl\{
       - ( P^2 - l_{165} )\,( \cV^{\bca}_{\saa}\,p_{12}\,
       + \cV^{\bca}_{\sab}\,p^2_2 )
\nl
{}&-& \spro{p_2}{P}\,\Bigl[ \cV^{\bba}_{\Z}(-P,-P,-p_2,\lstm{21345})
       - \cV^{\bba}_{\sab}(P,P,0,\lstm{12364})
       + \cV^{\bba}_{\saa}(-P,-P,-p_2,\lstm{21345}) 
\nl
{}&+& \cV^{\bba}_{\saa}(P,P,0,\lstm{12364})\Bigr]
    + \cV^{\bba}_{\sab}(-P,-P,-p_2,\lstm{21345})\,p_{12} \Bigr\},
\nl
\cV^{\bca}(p_1\,|\,p_2) &=& \frac{1}{2}\,\Bigl\{
       - (  P^2 - l_{165} )\,( \cV^{\bca}_{\saa}\,p^2_1
       + \cV^{\bca}_{\sab}\,p_{12} )
\nl
{}&-& \spro{p_1}{P}\,\Bigl[ \cV^{\bba}_{\Z}(-P,-P,-p_2,\lstm{21345})
       - \cV^{\bba}_{\sab}(P,P,0,\lstm{12364})
       + \cV^{\bba}_{\saa}(-P,-P,-p_2,\lstm{21345}) 
\nl
{}&+& \cV^{\bba}_{\saa}(P,P,0,\lstm{12364})\Bigr]
    + \cV^{\bba}_{\sab}(-P,-P,-p_2,\lstm{21345})\,p^2_1 \Bigr\},
\nl
\cV^{\bca}(p_2\,|\,p_1) &=& \frac{1}{2}\,\Bigl\{
       - l_{145}\,( \cV^{\bca}_{\saa}\,p_{12} 
       + \cV^{\bca}_{\sab}\,p^2_2 ) 
  - \spro{p_2}{P} \,\Bigl[ 
  \cV^{\bba}_{\saa}(P,P,p_1,\lstm{12365})
\nl
{}&+& \cV^{\bba}_{\sab}(P,P,0,\lstm{12364})
    - \cV^{\bba}_{\saa}(p,p,0,\lstm{12364}) \Bigr] 
  + p^2_2\,\Bigl[ \cV^{\bba}_{\sab}(P,P,p_1,\lstm{12365}) \Bigr\} 
\eqa
\fbox{$\bf V^{\bca}_{\saagen}$} for the $11$ group we introduce
auxiliary quantities (they only appear in the present subsection)
\bqa
\cV^{\bca}_{\saa\ssA} &=& 4\,\omega^4\,\cV^{1,1|1,3,1|3}_{\bca}\,
\bmid_{n=8-\ep},
\nl
v^{\bca}_{1} &=& -m^2_1\,\cV^{\bca}_{\Z} +
\cV^{\aca}_{\Z}(-P,-p_2,\lstm{23645}),
\nl
v^{\bca}_{2} &=& \frac{1}{2}\,\Bigl[ -l_{\ssP 12}\,\cV^{\bca}_{\saa} +
\cV^{\aca}_{\saa}(p_1,P,\lstm{13456})
+ \cV^{\aca}_{\sab}(-P,-p_2,\lstm{23645}) +
\cV^{\aca}_{\Z}(-P,-p_2,\lstm{23645})\Bigr].
\nl
v^{\bca}_{3} &=& \frac{1}{2}\,\Bigl[ -l_{\ssP 12}\,
\cV^{\bca}_{\sab} + \cV^{\aca}_{\sab}(p_1,P,\lstm{13456})
+ \cV^{\aca}_{\saa}(-P,-p_2,\lstm{23645}) + 
\cV^{\aca}_{\Z}(-P,-p_2,\lstm{23645})\Bigr],
\eqa
to obtain
\bqa
\cV^{\bca}_{\saaa} &=& 2\,\cV^{\bca}_{\saac}-\cV^{\bca}_{\saab} +
\cV^{\bca}_{\saa\ssA},
\quad
P^2\,\cV^{\bca}_{\saab} =  \cV^{\bca}_{\saac}\,P^2 
       - v^{\bca}_{2} + v^{\bca}_{3} + \frac{1}{2}\,
\cV^{\bca}_{\saa\ssA}\,(P^2 + p^2_1 - p^2_2),
\nl
(n-1)\,P^2\,\cV^{\bca}_{\saac} &=& \frac{1}{4}\,\Bigl\{ 
       2\,\frac{v^{\bca}_{3}-v^{\bca}_{2}}{P^2}\,(n-2)\,(p^2_1-p^2_2) 
-  4\,v^{\bca}_{1} + 2\,v^{\bca}_{2}\,n + 2\,v^{\bca}_{3}\,n 
\nl
{}&+&\,\cV^{\bca}_{\saa\ssA}\,\Bigl[\frac{n-2}{P^2}\,(p^2_1-p^2_2)^2 
       -\,P^2\,n 
       +2\,(p^2_1+p^2_2)\Bigr]\Bigr\},
\nl
\cV^{\bca}_{\saad} &=& - \cV^{\bca}_{\saac}\,\spro{p_1}{P}
       - \cV^{\bca}_{\saab}\,\spro{p_2}{P}
       + v^{\bca}_{3}.
\eqa
\subsection{\boldmath $V^{\bbb}(-p_2,p_1,-p_2,-p_1,\lstm{123456})$ family 
\label{summabbb}}
\noindent\fbox{$\bf V^{\bbb}_{\sijgen}$}
Results were derived in \sect{rankonebbb}. Referring to \eqn{genbbb} we have
\bqa
\cV^{\bbb}_{\sbb} &=&
\omega^2\,\Bigl[ \cV^{1 , 2 | 1 , 1 | 2 , 1 }_{\bbb} -
\cV^{2 , 1 | 1 , 1 | 1 , 2 }_{\bbb}\Bigr]\,\bmid_{n = 6-\ep},  \quad
\cV^{\bbb}_{\saa} =
\omega^2\,\Bigl[ \cV^{1 , 1 | 1 , 2 | 1 , 2 }_{\bbb} -
\cV^{1 , 1 | 2 , 1 | 2 , 1 }_{\bbb}\Bigr]\,\bmid_{n = 6-\ep} \, .
\eqa
Introduce a vector $U^{\bbb}$ with components
\bqa
U^{\bbb}_{1} &=&
\!\frac{1}{2}\!\Bigl[ l_{212}\cV^{\bbb}_{\Z}
\!-\! \cV^{\bba}_{\Z}(p_1,p_1,-p_2,\lstm{56134})
\!+\! \cV^{\bba}_{\Z}(-P,-P,-p_2,\lstm{34256})\!\Bigr],
\nl
U^{\bbb}_{2} &=&
\!\frac{1}{2}\!\Bigl[ l_{156}\cV^{\bbb}_{\Z}
\!+\! \cV^{\bba}_{\Z}(p_2,p_2,-p_1, \lstm{21634})
\!-\! \cV^{\bba}_{\Z}(p_2,p_2,-p_1, \lstm{12543})\!\Bigr],
\eqa
\bq
\cV^{\bbb}_{\sab} = \frac{1}{p^2_2}\,( U^{\bbb}_{1} - p_{12}\,
\cV^{\bbb}_{\saa}),
\quad
\cV^{\bbb}_{\sba} = \frac{1}{p^2_1}\,( U^{\bbb}_{2} - p_{12}\,
\cV^{\bbb}_{\sbb}).
\eq
\fbox{$\bf V^{\bbb}_{\sbbgen}$}
For rank two tensors (see \sect{ranktwobbb}) we introduce a vector 
$U^{\bbb}_{\sbb}$ with components
\bqa
U^{\bbb}_{\sbbsca} &=& -\,\frac{1}{2}\,\Bigl[
       - \cV^{\bbb}_{\sbb} 
    - \cV^{\bba}_{\saa}(-p_2,-p_2,p_1,\lstm{12543}) 
    + \cV^{\bba}_{\sab}(-p_2,-p_2,p_1,\lstm{12543}) 
\nl
{}&+& \cV^{\bba}_{\sba}(-p_2,-p_2,p_1,\lstm{12543})
    - \cV^{\bba}_{\sbb}(-p_2,-p_2,p_1,\lstm{12543}) 
    - \cV^{\bba}_{\saa}(p_2,p_2,-p_1,\lstm{21634}) 
\nl
{}&+& \cV^{\bba}_{\sab}(p_2,p_2,-p_1,\lstm{21634}) 
    + \cV^{\bba}_{\sba}(p_2,p_2,-p_1,\lstm{21634}) 
    - \cV^{\bba}_{\sbb}(p_2,p_2,-p_1,\lstm{21634})  \Bigr],
\nl
U^{\bbb}_{\sbbscb} &=& -\,\frac{1}{2}\,\Bigl[
        2\,\cV^{\bbb}_{\Z} 
       + \cV^{\bbb}_{\sba}\,l_{156}
       + 2\,\cV^{\bbb}_{\sbbd}\,(n-1)
\nl
{}&-& \cV^{\bba}_{\Z}(p_2,p_2,-p_1,\lstm{21634}) 
  - \cV^{\bba}_{\sab}(p_2,p_2,-p_1,\lstm{21634}) 
  + \cV^{\bba}_{\sbb}(p_2,p_2,-p_1,\lstm{21634}) 
\nl
{}&-& \cV^{\bba}_{\sab}(-p_2,-p_2,p_1,\lstm{12543}) 
       + \cV^{\bba}_{\sbb}(-p_2,-p_2,p_1,\lstm{12543}) \Bigr]\,;
\eqa
we obtain
\bq
\left( \begin{array}{c} \cV^{\bbb}_{\sbbc}  \\ \cV^{\bbb}_{\sbbb}  
\end{array}\right) = G^{-1}\,U^{\bbb}_{\sbb},
\eq
and also
\bqa
p^2_1\,\cV^{\bbb}_{\sbba} &=&
       - \cV^{\bbb}_{\sbbc} 
       - \cV^{\bbb}_{\sbbd}
       +  \frac{1}{2}\,\Bigl[
 \cV^{\bba}_{\Z}(p_2,p_2,-p_1,\lstm{21634}) 
       + \cV^{\bbb}_{\sba}\,l_{156} 
       - \cV^{\bba}_{\sab}(p_2,p_2,-p_1,\lstm{21634}) 
\nl
{}&-& \cV^{\bba}_{\sab}(-p_2,-p_2,p_1,\lstm{12543}) 
       + \cV^{\bba}_{\sbb}(p_2,p_2,-p_1,\lstm{21634}) 
       + \cV^{\bba}_{\sbb}(-p_2,-p_2,p_1,\lstm{12543}) \Bigr],
\nl
\cV^{\bbb}_{\sbbd} &=& \frac{\omega^2}{2}\,\Bigl[
\cV^{1 , 1 | 1 , 1 | 1 , 2 }_{\bbb}  +
\cV^{1 , 1 | 1 , 1 | 2  , 1  }_{\bbb} +
\cV^{2 , 1 | 1, 1|1 , 1 }_{\bbb}  +
\cV^{1 , 2 | 1 ,  1 | 1 , 1 }_{\bbb}\, \Bigr]\,\bmid_{n = 6-\ep}.
\eqa
When indices are saturated we obtain
\bqa
\cV^{\bbb}(0\,|\,\mu,\mu) &=&
       - \cV^{\bbb}_{\Z} 
       + \cV^{\bba}_{\Z}(p_2,p_2,-p_1,\lstm{21634}), 
\nl\nl
\cV^{\bbb}(0\,|\,p_1, p_1) &=& \frac{1}{2}\,\Bigl\{
       l_{156}\,( \cV^{\bbb}_{\sba}\,p^2_1 
       + \cV^{\bbb}_{\sbb}\,p_{12} ) 
       + \cV^{\bba}_{\Z}(p_2,p_2,-p_1,\lstm{21634})\,p^2_1 
\nl
{}&-& \spro{p_1}{P}\,\Bigl[ \cV^{\bba}_{\sab}(p_2,p_2,-p_1,\lstm{21634})
       - \cV^{\bba}_{\sbb}(p_2,p_2,-p_1,\lstm{21634})
\nl
{}&+& \cV^{\bba}_{\sab}(-p_2,-p_2,p_1,\lstm{12543})
       - \cV^{\bba}_{\sbb}(-p_2,-p_2,p_1,\lstm{12543}) \Bigr]
\nl
{}&+& p_{12}\,\Bigl[ \cV^{\bba}_{\saa}(p_2,p_2,-p_1,\lstm{21634}) 
       - \cV^{\bba}_{\sba}(p_2,p_2,-p_1,\lstm{21634}) 
\nl
{}&+& \cV^{\bba}_{\saa}(-p_2,-p_2,p_1,\lstm{12543}) 
       - \cV^{\bba}_{\sba}(-p_2,-p_2,p_1,\lstm{12543}) \Bigr] \Bigr\},
\nl\nl
p_1^2\,\cV^{\bbb}(0\,|\,p_2, p_2) &=& \frac{1}{2}\,\Bigl\{
       -2\,\cV^{\bbb}_{\Z}\,D\,m^2_5 
       + l_{156}\,\Bigl[ \cV^{\bbb}_{\sba}\,(  - 2\,D + D_1 )
       + \cV^{\bbb}_{\sbb}\,D_2 \Bigr] 
\nl
{}&-& 2\,(n-2)\,\cV^{\bbb}_{\sbbd}\,D 
       + \cV^{\bba}_{\Z}(p_2,p_2,-p_1,\lstm{21634})\,D_1 
\nl
{}&+& ( 2\,D - D_1 - D_2 )\,\Bigl[ 
         \cV^{\bba}_{\sab}(p_2,p_2,-p_1,\lstm{21634})
       - \cV^{\bba}_{\sbb}(p_2,p_2,-p_1,\lstm{21634})
\nl
{}&+& \cV^{\bba}_{\sab}(-p_2,-p_2,p_1,\lstm{12543})
       - \cV^{\bba}_{\sbb}(-p_2,-p_2,p_1,\lstm{12543}) \Bigr]
\nl
{}&+& D_2\,\Bigl[ 
         \cV^{\bba}_{\saa}(p_2,p_2,-p_1,\lstm{21634}) 
       - \cV^{\bba}_{\sba}(p_2,p_2,-p_1,\lstm{21634}) 
\nl
{}&+& \cV^{\bba}_{\saa}(-p_2,-p_2,p_1,\lstm{12543}) 
       - \cV^{\bba}_{\sba}(-p_2,-p_2,p_1,\lstm{12543}) \Bigr] \Bigr\},
\nl\nl
\cV^{\bbb}(0\,|\,p_1, p_2) &=& \frac{1}{2}\,\Bigl\{
       l_{156}\,( \cV^{\bbb}_{\sba}\,p_{12} 
       + \cV^{\bbb}_{\sbb}\,p^2_2 ) 
       + \cV^{\bba}_{\Z}(p_2,p_2,-p_1,\lstm{21634})\,p_{12} 
\nl
{}&+& \spro{p_2}{P}\,\Bigl[
         \cV^{\bba}_{\sbb}(p_2,p_2,-p_1,\lstm{21634})
       - \cV^{\bba}_{\sab}(p_2,p_2,-p_1,\lstm{21634})
\nl
{}&-& \cV^{\bba}_{\sab}(-p_2,-p_2,p_1,\lstm{12543})
       + \cV^{\bba}_{\sbb}(-p_2,-p_2,p_1,\lstm{12543})\Bigr]
\nl
{}&+& p^2_2\,\Bigl[ 
         \cV^{\bba}_{\saa}(p_2,p_2,-p_1,\lstm{21634}) 
       - \cV^{\bba}_{\sba}(p_2,p_2,-p_1,\lstm{21634}) 
\nl
{}&+& \cV^{\bba}_{\saa}(-p_2,-p_2,p_1,\lstm{12543}) 
       - \cV^{\bba}_{\sba}(-p_2,-p_2,p_1,\lstm{12543}) \Bigr]\Bigr\}.
\eqa 
\fbox{$\bf V^{\bbb}_{\saagen}$}
Introduce a vector $U^{\bbb}_{\saa}$ with components
\bqa
U^{\bbb}_{\saasca} &=& - \frac{1}{2}\,\Bigl[
         2\,\cV^{\bbb}_{\saad}\,( n - 1 )
       + 2\,\cV^{\bbb}_{\Z}\, m_1^2 
       + \cV^{\bbb}_{\sab}\,l_{212}
\nl
{}&-& \cV^{\bba}_{\sab}(p_1,p_1,-p_2,\lstm{56234}) 
       - \cV^{\bba}_{\Z}(-P,-P,-p_2,\lstm{34256}) 
       - \cV^{\bba}_{\saa}(-P,-P,-p_2,\lstm{34256}) 
\nl
{}&+& \cV^{\bba}_{\sba}(-P,-P,-p_2,\lstm{34256}) 
       + \cV^{\bba}_{\sbb}(p_1,p_1,-p_2,\lstm{56234})  \Bigr],
\nl
U^{\bbb}_{\saascb} &=& - \frac{1}{2}\,\Bigl[
       - \cV^{\bbb}_{\saa}\,l_{212}
       - \cV^{\bba}_{\saa}(p_1,p_1,-p_2,\lstm{56234}) 
       + \cV^{\bba}_{\sba}(p_1,p_1,-p_2,\lstm{56234}) 
\nl
{}&-& \cV^{\bba}_{\sbb}(p_1,p_1,-p_2,\lstm{56234}) 
       + \cV^{\bba}_{\sab}(p_1,p_1,-p_2,\lstm{56234}) \Bigr]\,;
\eqa
we obtain
\bq
\left( \begin{array}{c} \cV^{\bbb}_{\saaa}  \\ \cV^{\bbb}_{\saac}  
\end{array}\right) = G^{-1}\,U^{\bbb}_{\saa},
\eq
and also
\bqa
p^2_2\,\cV^{\bbb}_{\saab} &=& \frac{1}{2}\,\Bigl[
        \cV^{\bbb}_{\sab}\,l_{212}
       -2\,\cV^{\bbb}_{\saac}\,p_{12} 
       -2\,\cV^{\bbb}_{\saad} 
       + \cV^{\bba}_{\Z}(-P,-P,-p_2,\lstm{34256}) 
       - \cV^{\bba}_{\saa}(-P,-P,-p_2,\lstm{34256}) 
\nl
{}&+& \cV^{\bba}_{\sba}(-P,-P,-p_2,\lstm{34256}) 
       + \cV^{\bba}_{\sbb}(p_1,p_1,-p_2,\lstm{56234}) 
       - \cV^{\bba}_{\sab}(p_1,p_1,-p_2,\lstm{56234}) \Bigr].
\eqa
For saturated indices we obtain
\bqa
\cV^{\bbb}(\mu,\mu\,|\,0) &=&
       - \cV^{\bbb}_{\Z}\,m^2_1
       + \cV^{\bba}_{\Z}(-P,-P,-p_2,\lstm{34256}), 
\nl\nl 
p^2_2\,\cV^{\bbb}(p_1, p_1\,|\,0) &=& \frac{1}{2}\,\Bigl\{
        2\,(2-n)\,D\,\cV^{\bbb}_{\saad} 
       - 2\,\cV^{\bbb}_{\Z}\,D\,m^2_1 
\nl
{}&+& l_{212}\,\Bigl[ \cV^{\bbb}_{\sab}\,(  - 2\,D + D_1 )
       + \cV^{\bbb}_{\saa}\,D_3 \Bigr] 
       + \cV^{\bba}_{\Z}(-P,-P,-p_2,\lstm{34256})\,D_1 
\nl
{}&+& ( 2\,D - D_1 )\,\Bigl[ \cV^{\bba}_{\saa}(-P,-P,-p_2,\lstm{34256})
       - \cV^{\bba}_{\sba}(-P,-P,-p_2,\lstm{34256}) \Bigr]
\nl
{}&+& D_3\,\Bigl[ \cV^{\bba}_{\saa}(p_1,p_1,-p_2,\lstm{56234}) 
       - \cV^{\bba}_{\sba}(p_1,p_1,-p_2,\lstm{56234}) \Bigr] 
\nl       
{}&+& ( 2\,D - D_1 - D_3 )\,\Bigl[
         \cV^{\bba}_{\sab}(p_1,p_1,-p_2,\lstm{56234})\,
       - \cV^{\bba}_{\sbb}(p_1,p_1,-p_2,\lstm{56234}) \Bigr] \Bigr\},
\nl\nl
\cV^{\bbb}(p_2, p_2\,|\,0) &=& \frac{1}{2}\,\Bigl\{
        l_{212}\,( \cV^{\bbb}_{\sab}\,p^2_2 
       + \cV^{\bbb}_{\saa}\,p_{12} ) 
       + p^2_2\,\Bigl[ \cV^{\bba}_{\Z}(-P,-P,-p_2,\lstm{34256})  
\nl
{}&-& \cV^{\bba}_{\saa}(-P,-P,-p_2,\lstm{34256}) 
       + \cV^{\bba}_{\sba}(-P,-P,-p_2,\lstm{34256}) \Bigr] 
\nl
{}&+& p_{12}\,\Bigl[
         \cV^{\bba}_{\saa}(p_1,p_1,-p_2,\lstm{56234}) 
       - \cV^{\bba}_{\sba}(p_1,p_1,-p_2,\lstm{56234}) \Bigr] 
\nl
{}&+& \spro{p_2}{P}\,\Bigl[ 
         \cV^{\bba}_{\sbb}(p_1,p_1,-p_2,\lstm{56234})
       - \cV^{\bba}_{\sab}(p_1,p_1,-p_2,\lstm{56234}) \Bigr] \Bigr\},
\nl\nl
\cV^{\bbb}(p_1, p_2,|\,0) &=& \frac{1}{2}\,\Bigl\{
       l_{212}\,(  \cV^{\bbb}_{\sab}\,p_{12} 
       + \cV^{\bbb}_{\saa}\,p^2_1 )
       + p_{12}\,\Bigl[ \cV^{\bba}_{\Z}(-P,-P,-p_2,\lstm{34256}) 
\nl
{}&-& \cV^{\bba}_{\saa}(-P,-P,-p_2,\lstm{34256}) 
       + \cV^{\bba}_{\sba}(-P,-P,-p_2,\lstm{34256}) \Bigr] 
\nl
{}&+& p^2_1\,\Bigl[ \cV^{\bba}_{\saa}(p_1,p_1,-p_2,\lstm{56234}) 
       - \cV^{\bba}_{\sba}(p_1,p_1,-p_2,\lstm{56234}) \Bigr] 
\nl       
{}&+& \spro{p_1}{P}\,\Bigl[ \cV^{\bba}_{\sbb}(p_1,p_1,-p_2,\lstm{56234})
       - \cV^{\bba}_{\sab}(p_1,p_1,-p_2,\lstm{56234}) \Bigr] \Bigr\}.
\eqa
\fbox{$\bf V^{\bbb}_{\sabgen}$}
Introduce a vector $U^{\bbb}_{\sab}$ with components
\bqa
U^{\bbb}_{\sabsca} &=& - \frac{1}{2}\,\Bigl[
      2\,\cV^{\bbb}_{\sabd} 
       - \cV^{\bbb}_{\saa}\,l_{156}
       + \cV^{\bba}_{\sab}(p_2,p_2,-p_1,\lstm{21634}) 
       + \cV^{\bba}_{\sab}(-p_2,-p_2,p_1,\lstm{12543}) \Bigr],
\nl
U^{\bbb}_{\sabscb} &=& - \frac{1}{2}\,\Bigl[
     - \cV^{\bbb}_{\sba}\,l_{212}
     + \cV^{\bba}_{\sab}(p_1,p_1,-p_2,\lstm{56234}) 
     - \cV^{\bba}_{\saa}(p_1,p_1,-p_2,\lstm{56234}) 
\nl
{}&-& \cV^{\bba}_{\Z}(-P,-P,-p_2,\lstm{34256}) \Bigr]\,;
\eqa
we obtain
\bq
\left( \begin{array}{c} \cV^{\bbb}_{\saba}  \\ \cV^{\bbb}_{\sabc}  
\end{array}\right) = G^{-1}\,U^{\bbb}_{\sab},
\eq
and also
\bqa
D\,\cV^{\bbb}_{\sabb} &=& \frac{1}{2}\,\Bigl\{
       -2\,\cV^{\bbb}_{\sabd}\,p^2_1 
       - \cV^{\bbb}_{\sab}\,p_{12}\,l_{156} 
       + \cV^{\bbb}_{\sbb}\,p^2_1\,l_{212} 
\nl
{}&+& p^2_1\,\Bigl[ \cV^{\bba}_{\Z}(-P,-P,-p_2,\lstm{34256}) 
       + \cV^{\bba}_{\sba}(-P,-P,-p_2,\lstm{34256}) 
       - \cV^{\bba}_{\sab}(p_1,p_1,-p_2,\lstm{56234}) \Bigr] 
\nl
{}&+& p_{12}\,\Bigl[
         \cV^{\bba}_{\sab}(p_2,p_2,-p_1,\lstm{21634}) 
       - \cV^{\bba}_{\saa}(p_2,p_2,-p_1,\lstm{21634}) 
       - \cV^{\bba}_{\Z}(p_2,p_2,-p_1,\lstm{21634}) 
\nl
{}&+& \cV^{\bba}_{\sab}(-p_2,-p_2,p_1,\lstm{12543}) 
       - \cV^{\bba}_{\saa}(-p_2,-p_2,p_1,\lstm{12543}) \Bigr] \Bigr\},
\nl
\cV^{\bbb}_{\sabd} &=& \frac{\omega^2}{2}\,
\Bigl[ \cV^{1 , 1| 1,1 | 1 , 2}_{\bbb} +
\cV^{1 , 1 | 1 , 1 | 2 , 1 }_{\bbb}\Bigr]\,\bmid_{n = 6 - \ep}. 
\eqa
For tensors with saturated indices we have
\bqa
\cV^{\bbb}(\mu\,|\,\mu) &=& \frac{1}{2}\,\Bigl[
       2\,\cV^{\bbb}_{\sabd}\,(  n - 2  )
       + \cV^{\bbb}_{\saa}\,l_{156} 
       + \cV^{\bbb}_{\sbb}\,l_{212} 
\nl
{}&-& \cV^{\bba}_{\sab}(p_1,p_1,-p_2,\lstm{56234}) 
       - \cV^{\bba}_{\sab}(p_2,p_2,-p_1,\lstm{21634}) 
       - \cV^{\bba}_{\sab}(-p_2,-p_2,p_1,\lstm{12543}) 
\nl
{}&+& \cV^{\bba}_{\Z}(-P,-P,-p_2,\lstm{34256}) 
       + \cV^{\bba}_{\sba}(-P,-P,-p_2,\lstm{34256}) \Bigr],
\nl
\nl
\cV^{\bbb}(p_1\,|\,p_1) &=& \frac{1}{2}\,\Bigl[
         l_{156}\,( \cV^{\bbb}_{\saa}\,p^2_1 
       + \cV^{\bbb}_{\sab}\,p_{12} ) 
       + p_{12}\,\Bigl[ 
         \cV^{\bba}_{\saa}(p_2,p_2,-p_1,\lstm{21634}) 
\nl
{}&+& \cV^{\bba}_{\Z}(p_2,p_2,-p_1,\lstm{21634}) 
       + \cV^{\bba}_{\saa}(-p_2,-p_2,p_1,\lstm{12543}) \Bigr]
\nl
{}&-& \spro{p_1}{P}\,\Bigl[
         \cV^{\bba}_{\sab}(p_2,p_2,-p_1,\lstm{21634})
       + \cV^{\bba}_{\sab}(-p_2,-p_2,p_1,\lstm{12543}) \Bigr]\Bigr\},
\nl
\nl
p^2_1\,\cV^{\bbb}(p_2\,|\,p_2) &=& \frac{1}{2}\,\Bigl[
        l_{156}\,( \cV^{\bbb}_{\saa}\,p^2_{12}
       + \cV^{\bbb}_{\sab}\,D_2 ) 
       + \cV^{\bbb}_{\sbb}\,D\,l_{212} 
  + D\,\Bigl[ \cV^{\bba}_{\Z}(-P,-P,-p_2,\lstm{34256}) 
\nl
{}&+& \cV^{\bba}_{\sba}(-P,-P,-p_2,\lstm{34256}) 
       - \cV^{\bba}_{\sab}(p_1,p_1,-p_2,\lstm{56234}) \Bigr] 
\nl
{}&-& ( p^2_{12} + D_2 )\,\Bigl[
         \cV^{\bba}_{\sab}(p_2,p_2,-p_1,\lstm{21634})
       + \cV^{\bba}_{\sab}(-p_2,-p_2,p_1,\lstm{12543}) \Bigr]
\nl
{}&+& D_2\,\Bigl[ \cV^{\bba}_{\saa}(p_2,p_2,-p_1,\lstm{21634}) 
       + \cV^{\bba}_{\saa}(-p_2,-p_2,p_1,\lstm{12543}) 
\nl
{}&+& \cV^{\bba}_{\Z}(p_2,p_2,-p_1,\lstm{21634}) \Bigr] \Bigr\},
\nl
\nl
\cV^{\bbb}(p_1\,|\,p_2) &=&
\cV^{\bbb}(p_2\,|\,p_1) = \frac{1}{2}\,\Bigl[
       l_{156}\,( \cV^{\bbb}_{\saa}\,p_{12} 
       + \cV^{\bbb}_{\sab}\,p^2_2 ) 
       + p^2_2\,\Bigl[
         \cV^{\bba}_{\saa}(p_2,p_2,-p_1,\lstm{21634}) 
\nl
{}&+& \cV^{\bba}_{\Z}(p_2,p_2,-p_1,\lstm{21634}) 
       + \cV^{\bba}_{\saa}(-p_2,-p_2,p_1,\lstm{12543}) \Bigr] 
\nl
{}&-& \spro{p_2}{P}\,\Bigl[
         \cV^{\bba}_{\sab}(p_2,p_2,-p_1,\lstm{21634})
       + \cV^{\bba}_{\sab}(-p_2,-p_2,p_1,\lstm{12543}) \Bigr]\Bigr\}.
\eqa
\subsection{Further reduction of rank two integrals}
In this Section we collect all combinations of vector form factors that can
be further reduced without occurrence of inverse powers of Gram determinants.
For each combination we list the equation where the r.h.s. can be found.
\[
\ba{llll}
p^2_1\,\cV^{\aba}_{\sba} + p_{12}\,\cV^{\aba}_{\sbb}, 
\qquad & \qquad 
\qquad & \qquad 
\qquad & \qquad  
\mbox{\eqn{unknownone}}, \\
p^2_1\,\cV^{\aca}_{\sba} + p_{12}\,\cV^{\aca}_{\sbb}, 
\qquad & \qquad
p_{12}\,\cV^{\aca}_{\sba} + p^2_2\,\cV^{\aca}_{\sbb}, 
\qquad & \qquad
\qquad & \qquad
\mbox{\eqn{redI2}},  \\
p^2_1\,\cV^{\ada}_{\sba} + p_{12}\,\cV^{\ada}_{\sbb}, 
\qquad & \qquad
p_{12}\,\cV^{\ada}_{\sba} + p^2_2\,\cV^{\ada}_{\sbb}, 
\qquad & \qquad
\qquad & \qquad
\mbox{\eqn{omada}}, \\
p^2_1\,\cV^{\bba}_{\saa} + p_{12}\,\cV^{\bba}_{\sab}, 
\qquad & \qquad
p_{12}\,\cV^{\bba}_{\sba} + p^2_2\,\cV^{\bba}_{\sbb}, 
\qquad & \qquad
\qquad & \qquad
\mbox{\eqn{ombba}}, \\
\spro{p_1}{P}\,\cV^{\bca}_{\saa} + \spro{p_2}{P}\,\cV^{\bca}_{\sab},
\qquad & \qquad
\spro{p_1}{P}\,\cV^{\bca}_{\sba} + \spro{p_2}{P}\,\cV^{\bca}_{\sbb},
\qquad & \qquad
p^2_1\,\cV^{\bca}_{\sba} + p_{12}\,\cV^{\bca}_{\sbb}, 
\qquad & \qquad
\mbox{\eqn{af30}}, \\
p_{12}\,\cV^{\bbb}_{\saa} + p^2_2\,\cV^{\bbb}_{\sab}, 
\qquad & \qquad
p^2_1\,\cV^{\bbb}_{\sba} + p_{12}\,\cV^{\bbb}_{\sbb}, 
\qquad & \qquad
\qquad & \qquad
\mbox{\eqn{afy30}}.
\ea
\]
Once again we stress that form factors are introduced with respect to a certain
basis of vectors which is fully specified by the corresponding list of
arguments; therefore, form factors appearing in the reduction of other form
factors should be interpreted in the appropriate way. We have collected in 
\tabn{bases} the bases for expansion of tensor integrals occurring in the 
reduction procedure. A typical example is
\bqa
\cV^{\bba}(\mu\,|\,0; p_1,p_1,-p_2,\cdots) &=& 
\cV^{\bba}_{\saa}(p_1,p_1,-p_2,\cdots)\,p_{1\mu} -
\cV^{\bba}_{\sab}(p_1,p_1,-p_2,\cdots)\,P_{\mu},
\nl
\cV^{\bba}(\mu\,|\,0; -P,-P,-p_2,\cdots) &=& 
- \cV^{\bba}_{\saa}(-P,-P,-p_2,\cdots)\,P_{\mu} +
\cV^{\bba}_{\sab}(-P,-P,-p_2,\cdots)\,p_{1\mu}.
\label{examplebas}
\eqa
\tiny
\begin{table}[th]\centering
\renewcommand\arraystretch{0.9}
\begin{tabular}{|c|c|c|c|}
\hline
& & & \\
Family & argument & $p_{\rm first}$ & $p_{\rm second}$ \\
& & & \\
\hline
& & & \\
E & $p_2\,,\,P$ & $p_1$ & $p_2$ \\
E & $p_1\,,\,P$ & $p_2$ & $p_1$ \\
E & $0\,,\,P$ & $P$ & $0$ \\
E & $0\,,\,p_1$ & $p_1$ & $0$ \\
E & $0\,,\,p_2$ & $p_2$ & $0$ \\
E & $-p_2\,,\,-P$ & $-p_1$ & $-p_2$ \\
E & $0\,,\,-p_1$ & $-p_1$ & $0$\\
& & & \\
\hline
& & & \\
I & $p_1\,,\,P$ & $p_1$ & $p_2$ \\
I & $p_2\,,\,P$ & $p_2$ & $p_1$ \\
I & $0\,,\,p_1$ & $0$ & $p_1$ \\
I & $0\,,\,P$ & $0$ & $P$ \\
I & $-p_2\,,\,-P$ & $-p_2$ & $-p_1$ \\
I & $-P\,,\,-p_2$ & $-p_2$ & $-p_1$ \\
I & $p_1\,,\,0$ & $p_1$ & $-p_1$ \\
& & & \\
\hline
& & & \\
G & $p_1\,,\,p_1\,,\,P$ & $p_1$ & $p_2$ \\
G & $P\,,\,P\,,\,p_1$ & $P$ & $-p_2$ \\
G & $-P\,,\,-P\,,-p_2$ & $-P$ & $p_1$ \\
G & $P\,,\,P\,,\,0$ & $P$ & $-P$ \\
G & $p_1\,,\,p_1\,,\,-p_2$ & $p_1$ & $-P$ \\
G & $p_2\,,\,p_2\,,\,-p_1$ & $p_2$ & $-P$ \\
G & $-p_2\,,\,-p_2\,,\,p_1$ & $-p_2$ & $P$ \\ 
& & & \\
\hline
\hline
\end{tabular}
\vspace*{3mm}
\caption[]{The basis $p_{\rm first}\,,\,p_{\rm second}$ for expanding form
factors occurring in the reduction of tensor integrals corresponding to 
diagrams with a larger number of propagators. First entry is always the 
defyining representation. An example is given in \eqn{examplebas}.}
\label{bases}
\end{table}
\normalsize
When reducing recursively all symbols must be interpreted as referred to the
appropriated basis, i.e. 
\bq
D = p^2_{\rm first}\,p^2_{\rm second} - 
(\spro{p_{\rm first}}{p_{\rm second}})^2, \qquad \mbox{etc,}
\eq
(see \eqn{defmisc}) where, at the first level of reduction, we always 
have $p_{\rm first} = p_1$ and $p_{\rm second} = p_2$.
\subsection{Reduction for rank three tensors}
For rank three tensors the number of form factors and of contractions
(tensors with saturated indices) increases considerably and it is not
convenient to write down all cases explicitly; we prefer to adopt a different 
way of collecting the results.
The reduction technique is based on two algorithms which we illustrate in the 
case of the $V^{\ada}$ family. 
\begin{description}
\item{A1.} Contraction of tensor integrals with $\delta^{\mu\nu}$ and 
decomposition of tensors of lower rank
\end{description}
\bqa 
{}&{}&
\delta^{\mu\nu}\,\intmomii{n}{q_1}{q_2}\,
\frac{q_{2\mu}\,q_{2\nu}\,q_{2\alpha}}
{[1][2]_{\ada}[3]_{\ada}[4]_{\ada}[5]_{\ada}[6]_{\ada}}
\nl
{}&=& 
\intmomii{n}{q_1}{q_2}\,
\frac{q_{2\alpha}}
{[1][2]_{\ada}[4]_{\ada}[5]_{\ada}[6]_{\ada}} -
m^2_3\,
\intmomii{n}{q_1}{q_2}\,
\frac{\,q_{2\alpha}}
{[1][2]_{\ada}[3]_{\ada}[4]_{\ada}[5]_{\ada}[6]_{\ada}}
\nl
{}&=&
\frac{\pi^4}{\mu^{2\ep}}\,
\sum_{i=1,2}\,\Bigl[ V^{\aca}_{\sbgen} - 
m^2_3\,V^{\ada}_{\sbgen}\Bigr]\,p_{i\alpha} = 
\sum_{i=1,2}\,v^{\ada}_{\sbbbgen}\,p_{i\alpha}.
\eqa
\begin{description}
\item{A2.} Contraction of tensor integrals with with $p^{\mu}_1$ or 
$p^{\mu}_2$ (or $P^{\mu}$) and decomposition of tensors of lower rank: for 
instance we obtain
\end{description}
\bqa 
\cV^{\ada}(0\,|\,p_1, \nu,\alpha)= 
v^{\ada}_{\sbbbc}\,p_{1\nu}\,p_{1\alpha}+
v^{\ada}_{\sbbbd}\,p_{2\nu}\,p_{2\alpha}+
v^{\ada}_{\sbbbe}\,\{p_1 p_2\}_{\nu\alpha}+
v^{\ada}_{\sbbbf}\,\delta_{\nu\alpha}.
\eqa
The $v^{\ada}_i$ originate from the decomposition of a rank two tensor
after using \eqn{enummada} and writing 
\bq
2\,\spro{q_2}{p_1} = [4]_{\ada} - [3]_{\ada} - p^2_1 + m^2_4 - m^2_3,
\eq
and after shifting the loop momenta in order to recover the standard form
of diagrams with fewer propagators. Therefore we have 
\bqa
 \cV^{\ada}_{\sbbbd}\,p^2_1
+ \cV^{\ada}_{\sbbbc}\,p_{12}
+ 2\,\cV^{\ada}_{\sbbba} &=& v^{\ada}_{\sbbbc},
\qquad
\cV^{\ada}_{\sbbbd}\,p^2_1
+ \cV^{\ada}_{\sbbbf}\,p_{12} = v^{\ada}_{\sbbbd},
\nl
 \cV^{\ada}_{\sbbbc}\,p^2_1
+ \cV^{\ada}_{\sbbbd}\,p_{12} 
+ \cV^{\ada}_{\sbbbb} &=& v^{\ada}_{\sbbbe},
\qquad
 \cV^{\ada}_{\sbbba}\,p^2_1
+ \cV^{\ada}_{\sbbbb}\,p_{12} = v^{\ada}_{\sbbbf}.
\eqa
The choice of contractions is limited by the request that the resulting
scalar products be reducible. In each case one obtains a system of
equations for the rank three form factors to be solved in terms of lower
rank form factors and of generalized scalars. Decompositions of vector
integrals are defined in \eqn{ffaba}, \eqn{ffaca}, \eqn{ffada}, \eqn{ffbba},
\eqn{ffbca} and \eqn{ffbbb}. Decompositions for rank two tensor integrals
are defined in \eqn{tensoraba}, for $V^{\aba}$, in \eqn{tensoraca} for
$V^{\aca}$, in \eqn{tensorada} for $V^{\ada}$, in \eqn{tensorbba},
\eqn{tensorbba2}, \eqn{tensorbba3} for $V^{\bba}$, in \eqn{tensorbca3},
\eqn{tensorbca2}, \eqn{tensorbca1} for $V^{\bca}$ and in \eqn{tensorbbb3},
\eqn{tensorbbb2}, \eqn{tensorbbb1} for $V^{\bbb}$.
\subsubsection{Contractions of rank three tensor integrals}
In this Section we collect the results for all contractions of rank three
tensors (with a Kronecker delta functions or with an external momentum) that
give rise to reducible scalar products. These definitions will be used in
\sectm{satuone}{satuthree} to construct tensors with three saturated indices
and to build systems of equations that can be solved for the corresponding form
factors. First we define the relevant contractions, once again those that
are leading to reducible scalar products in the numerators:
\begin{itemize}
\item{\bf \boldmath $M$ family}
\end{itemize}
\bqa
\cV^{\ada}(0\,|\,\mu,\mu,\nu) &=& v^{\ada}_{\sbbba}\,p_{1\nu} +
v^{\ada}_{\sbbbb}\,p_{2\nu},
\nl
\cV^{\ada}(0\,|\,p_1, \mu,\nu) &=&
v^{\ada}_{\sbbbc}\,p_{1\mu}\,p_{1\nu} +
v^{\ada}_{\sbbbd}\,p_{2\mu}\,p_{2\nu} +
v^{\ada}_{\sbbbe}\,\{p_1 p_2\}_{\mu\nu} +
v^{\ada}_{\sbbbf}\,\delta_{\mu\nu},
\nl
\cV^{\ada}(0\,|\,p_2, \mu,\nu) &=&
v^{\ada}_{\sbbbg}\,p_{1\mu}\,p_{1\nu} +
v^{\ada}_{\sbbbh}\,p_{2\mu}\,p_{2\nu} +
v^{\ada}_{\sbbbi}\,\{p_1 p_2\}_{\mu\nu} +
v^{\ada}_{\sbbbj}\,\delta_{\mu\nu},
\nl\nl
\cV^{\ada}(\nu\,|\,\mu,\mu) &=& 
v^{\ada}_{\sabba}\,p_{1\nu} +
v^{\ada}_{\sabbb}\,p_{2\nu},
\nl
\cV^{\ada}(\mu\,|\,p_1 ,\nu) &=&
v^{\ada}_{\sabbc}\,p_{1\mu}\,p_{1\nu} +
v^{\ada}_{\sabbd}\,p_{2\mu}\,p_{2\nu} +
v^{\ada}_{\sabbe}\,\{p_1 p_2\}_{\mu\nu} +
v^{\ada}_{\sabbf}\,\delta_{\mu\nu},
\nl
\cV^{\ada}(\mu\,|\,p_2 ,\nu) &=&
v^{\ada}_{\sabbg}\,p_{1\mu}\,p_{1\nu} +
v^{\ada}_{\sabbh}\,p_{2\mu}\,p_{2\nu} +
v^{\ada}_{\sabbi}\,\{p_1 p_2\}_{\mu\nu} +
v^{\ada}_{\sabbj}\,\delta_{\mu\nu},
\nl
\cV^{\ada}(\mu\,|\,\mu,\nu) &=& 
v^{\ada}_{\sabbl}\,p_{1\nu} +
v^{\ada}_{\sabbm}\,p_{2\nu},
\nl\nl
\cV^{\ada}(\mu,\mu\,|\,\nu) &=& v^{\ada}_{\saaba}\,p_{1\nu} +
v^{\ada}_{\saabb}\,p_{2\nu},
\nl
\cV^{\ada}(\mu,\nu\,|\,p_1) &=&
v^{\ada}_{\saabc}\,p_{1\mu}\,p_{1\nu} +
v^{\ada}_{\saabd}\,p_{2\mu}\,p_{2\nu} +
v^{\ada}_{\saabe}\,\{p_1 p_2\}_{\mu\nu} +
v^{\ada}_{\saabf}\,\delta_{\mu\nu},
\nl
\cV^{\ada}(\mu,\nu\,|\,p_2) &=& 
v^{\ada}_{\saabg}\,p_{1\mu}\,p_{1\nu} +
v^{\ada}_{\saabh}\,p_{2\mu}\,p_{2\nu} +
v^{\ada}_{\saabi}\,\{p_1 p_2\}_{\mu\nu} +
v^{\ada}_{\saabj}\,\delta_{\mu\nu},
\nl
\cV^{\ada}(\mu,\nu\,|\,\mu) &=& v^{\ada}_{\saabl}\,p_{1\nu} +
v^{\ada}_{\saabm}\,p_{2\nu}.
\nl\nl
\cV^{\ada}(\mu,\mu,\nu\,|\,0) &=& v^{\ada}_{\saaaa}\,p_{1\nu} +
v^{\ada}_{\saaab}\,p_{2\nu},
\label{contraM}
\eqa
\begin{itemize}
\item{\bf \boldmath $K$ family}
\end{itemize}
\bqa
\cV^{\bca}(0\,|\,\nu,\nu,\mu) &=&
v^{\bca}_{\sbbba}\,p_{1\mu} +
v^{\bca}_{\sbbbb}\,p_{2\mu},
\nl
\cV^{\bca}(0\,|\,p_1, \mu,\nu) &=&
v^{\bca}_{\sbbbc}\,p_{1\mu}\,p_{1\nu} +
v^{\bca}_{\sbbbd}\,p_{2\mu}\,p_{2\nu} +
v^{\bca}_{\sbbbe}\,\{ p_1 p_2\}_{\mu\nu} +
v^{\bca}_{\sbbbf}\,\delta_{\mu\nu},
\nl
\cV^{\bca}(0\,|\,p_2, \mu,\nu) &=&
v^{\bca}_{\sbbbg}\,p_{1\mu}\,p_{1\nu} +
v^{\bca}_{\sbbbh}\,p_{2\mu}\,p_{2\nu} +
v^{\bca}_{\sbbbi}\,\{ p_1 p_2\}_{\mu\nu} +
v^{\bca}_{\sbbbj}\,\delta_{\mu\nu},
\nl
\nl
\cV^{\bca}(\nu,\nu,\mu\,|\,0) &=&
v^{\bca}_{\saaaa}\,p_{1\mu} +
v^{\bca}_{\saaab}\,p_{2\mu}, 
\nl
\cV^{\bca}(P, \mu,\nu\,|\,0) &=&
v^{\bca}_{\saaac}\,p_{1\mu}\,p_{1\nu} +
v^{\bca}_{\saaad}\,p_{2\mu}\,p_{2\nu} +
v^{\bca}_{\saaae}\,\{ p_1 p_2\}_{\mu\nu} +
v^{\bca}_{\saaaf}\,\delta_{\mu\nu},
\nl
\nl
\cV^{\bca}(\nu,\nu\,|\,\mu) &=&
v^{\bca}_{\saaba}\,p_{1\mu} +
v^{\bca}_{\saabb}\,p_{2\mu}, 
\nl
\cV^{\bca}(P ,\mu\,|\,\nu) &=&
v^{\bca}_{\saabc}\,p_{1\mu}\,p_{1\nu} +
v^{\bca}_{\saabd}\,p_{2\mu}\,p_{2\nu} +
v^{\bca}_{\saabe}\,p_{1\mu}\,p_{2\nu} +
v^{\bca}_{\saabf}\,p_{1\nu}\,p_{2\mu} +
v^{\bca}_{\saabg}\,\delta_{\mu\nu},
\nl
\cV^{\bca}(\mu,\nu\,|\,p_1) &=&
v^{\bca}_{\saabh}\,p_{1\mu}\,p_{1\nu} +
v^{\bca}_{\saabi}\,p_{2\mu}\,p_{2\nu} +
v^{\bca}_{\saabj}\,\{ p_1 p_2\}_{\mu\nu} +
v^{\bca}_{\saabl}\,\delta_{\mu\nu},
\nl
\cV^{\bca}(\mu,\nu\,|\,p_2) &=&
v^{\bca}_{\saabm}\,p_{1\mu}\,p_{1\nu} +
v^{\bca}_{\saabn}\,p_{2\mu}\,p_{2\nu} +
v^{\bca}_{\saabo}\,\{ p_1 p_2\}_{\mu\nu} +
v^{\bca}_{\saabp}\,\delta_{\mu\nu},
\nl
\cV^{\bca}(\nu,\mu\,|\,\nu) &=&
v^{\bca}_{\saabq}\,p_{1\mu} +
v^{\bca}_{\saabr}\,p_{2\mu}, 
\nl
\nl
\cV^{\bca}(\mu\,|\,\nu,\nu) &=&
v^{\bca}_{\sabba}\,p_{1\mu} +
v^{\bca}_{\sabbb}\,p_{2\mu}, 
\nl
\cV^{\bca}(P\,|\,\mu,\nu) &=&
v^{\bca}_{\sabbc}\,p_{1\mu}\,p_{1\nu} +
v^{\bca}_{\sabbd}\,p_{2\mu}\,p_{2\nu} +
v^{\bca}_{\sabbe}\,\{ p_1 p_2\}_{\mu\nu} +
v^{\bca}_{\sabbf}\,\delta_{\mu\nu},
\nl
\cV^{\bca}(\mu\,|\,\nu, p_1) &=&
v^{\bca}_{\sabbg}\,p_{1\mu}\,p_{1\nu} +
v^{\bca}_{\sabbh}\,p_{2\mu}\,p_{2\nu} +
v^{\bca}_{\sabbi}\,p_{1\mu}\,p_{2\nu} +
v^{\bca}_{\sabbj}\,p_{1\nu}\,p_{2\mu} +
v^{\bca}_{\sabbl}\,\delta_{\mu\nu},
\nl
\cV^{\bca}(\mu\,|\,\nu, p_2) &=&
v^{\bca}_{\sabbm}\,p_{1\mu}\,p_{1\nu} +
v^{\bca}_{\sabbn}\,p_{2\mu}\,p_{2\nu} +
v^{\bca}_{\sabbo}\,p_{1\mu}\,p_{2\nu} +
v^{\bca}_{\sabbp}\,p_{1\nu}\,p_{2\mu} +
v^{\bca}_{\sabbq}\,\delta_{\mu\nu},
\nl
\cV^{\bca}(\nu\,|\,\nu, \mu) &=&
v^{\bca}_{\sabbr}\,p_{1\mu} +
v^{\bca}_{\sabbs}\,p_{2\mu}.
\label{contraK}
\eqa
\begin{itemize}
\item{\bf {\boldmath $H$} family}
\end{itemize}
\bqa
\cV^{\bbb}(0\,|\,\nu,\nu,\mu) &=&
v^{\bbb}_{\sbbba}\,p_{1\mu} +
v^{\bbb}_{\sbbbb}\,p_{2\mu} ,
\nl
\cV^{\bbb}(0\,|\,p_1, \mu,\nu) &=&
v^{\bbb}_{\sbbbc}\,p_{1\mu}\,p_{1\nu} +
v^{\bbb}_{\sbbbd}\,p_{2\mu}\,p_{2\nu} +
v^{\bbb}_{\sbbbe}\,\{p_1 p_2\}_{\mu\nu} +
v^{\bbb}_{\sbbbf}\,\delta_{\mu\nu},
\nl
\nl
\cV^{\bbb}(\nu,\nu,\mu\,|\,0) &=&
v^{\bbb}_{\saaaa}\,p_{1\mu} +
v^{\bbb}_{\saaab}\,p_{2\mu} ,
\nl
\cV^{\bbb}(p_2 ,\mu,\nu\,|\,0) &=&
v^{\bbb}_{\saaac}\,p_{1\mu}\,p_{1\nu} +
v^{\bbb}_{\saaad}\,p_{2\mu}\,p_{2\nu} +
v^{\bbb}_{\saaae}\,\{p_1 p_2\}_{\mu\nu} +
v^{\bbb}_{\saaaf}\,\delta_{\mu\nu},
\nl
\nl
\cV^{\bbb}(\nu,\nu\,|\,\mu) &=&
v^{\bbb}_{\saaba}\,p_{1\mu} +
v^{\bbb}_{\saabb}\,p_{2\mu} ,
\nl
\cV^{\bbb}(p_2, \mu\,|\,\nu) &=&
v^{\bbb}_{\saabc}\,p_{1\mu}\,p_{1\nu} +
v^{\bbb}_{\saabd}\,p_{2\mu}\,p_{2\nu} +
v^{\bbb}_{\saabe}\,p_{1\mu}\,p_{2\nu} +
v^{\bbb}_{\saabf}\,p_{1\nu}\,p_{2\mu} +
v^{\bbb}_{\saabg}\,\delta_{\mu\nu},
\nl
\cV^{\bbb}(\mu,\nu\,|\,p_1) &=&
v^{\bbb}_{\saabh}\,p_{1\mu}\,p_{1\nu} +
v^{\bbb}_{\saabi}\,p_{2\mu}\,p_{2\nu} +
v^{\bbb}_{\saabj}\,\{p_1 p_2\}_{\mu\nu} +
v^{\bbb}_{\saabl}\,\delta_{\mu\nu},
\nl
\nl
\cV^{\bbb}(\mu\,|\,\nu,\nu) &=&
v^{\bbb}_{\sabba}\,p_{1\mu} +
v^{\bbb}_{\sabbb}\,p_{2\mu} ,
\nl
\cV^{\bbb}(p_2\,|\,\mu,\nu) &=&
v^{\bbb}_{\sabbc}\,p_{1\mu}\,p_{1\nu} +
v^{\bbb}_{\sabbd}\,p_{2\mu}\,p_{2\nu} +
v^{\bbb}_{\sabbe}\,\{ p_1 p_2\}_{\mu\nu} +
v^{\bbb}_{\sabbf}\,\delta_{\mu\nu},
\nl
\cV^{\bbb}(\mu\,|\,\nu, p_1) &=&
v^{\bbb}_{\sabbg}\,p_{1\mu}\,p_{1\nu} +
v^{\bbb}_{\sabbh}\,p_{2\mu}\,p_{2\nu} +
v^{\bbb}_{\sabbi}\,p_{1\mu}\,p_{2\nu} +
v^{\bbb}_{\sabbj}\,p_{1\nu}\,p_{2\mu} +
v^{\bbb}_{\sabbl}\,\delta_{\mu\nu}.
\label{contraH}
\eqa
\subsubsection{Evaluation of contracted rank three tensor integrals}
Successively all contractions of \eqns{contraM}{contraH} are expressed as 
linear combinations of form factors of lower rank. These relations can be used
as they stand or we can insert, recursively, results for rank two and rank
one form factors (listed in \sectm{summaaba}{summabbb}) until one 
reaches a result which is written in terms of scalar integrals only.
\begin{itemize}
\item{\bf {\boldmath $M$} family} \eqn{contraM}
\end{itemize}
\bqa
v^{\ada}_{\sbbbgen} &=& -m^2_3\,\cV^{\ada}_{\sbgen} +
                         \cV^{\aca}_{\sbgen}(p_1,P,\lstm{12345}),
\qquad i = 1,2
\nl
v^{\ada}_{\sbbb\,i+2} &=& \frac{1}{2}\,\Bigl[
- l_{134}\,\cV^{\ada}_{\sbbgen}
- \cV^{\aca}_{\sbbgen}(p_1,P,\lstm{12345})
+ \cV^{\aca}_{\sbbb}(0,P,\lstm{12335}) \Bigr],
\qquad i = 1 \cdots 3,
\nl
v^{\ada}_{\sbbbf} &=& \frac{1}{2}\,\Bigl[
- l_{134}\,\cV^{\ada}_{\sbbd}
- \cV^{\aca}_{\sbbd}(p_1,P,\lstm{12345})
+ \cV^{\aca}_{\sbbd}(0,P,\lstm{12335}) \Bigr],
\nl
v^{\ada}_{\sbbbg} &=& \frac{1}{2}\,\Bigl[
( l_{134} - l_{\ssP 35} )\,\cV^{\ada}_{\sbba}
- \cV^{\aca}_{\sbbb}(0,P,\lstm{12335})
+ \cV^{\aca}_{\sbba}(p_1,0,\lstm{12343})
+ \cV^{\aca}_{\sbbb}(p_1,0,\lstm{12343})
\nl
{}&-& 2\,\cV^{\aca}_{\sbbc}(p_1,P,\lstm{12343}) \Bigr],
\nl
v^{\ada}_{\sbbb\,i+6} &=& \frac{1}{2}\,\Bigl[
( l_{134} - l_{\ssP 35} )\,\cV^{\ada}_{\sbbgen}
- \cV^{\aca}_{\sbbb}(0,P,\lstm{12335}) \Bigr],
\quad i = 2,3,
\nl
v^{\ada}_{\sbbbj} &=& \frac{1}{2}\,\Bigl[
( l_{134} - l_{\ssP 35} )\,\cV^{\ada}_{\sbbd}
+ \cV^{\aca}_{\sbbd}(p_1,0,\lstm{12343})
- \cV^{\aca}_{\sbbd}(0,P,\lstm{12335}) \Bigr],
\label{smallvMa}
\eqa
\bqa
v^{\ada}_{\saaagen} &=& - m^2_1\,\cV^{\ada}_{\sagen} -
A_{\Z}(m_2)\,C_{\sagen}(2,1,1\,;\,p_1,p_2,\lstm{345}),
\qquad i = 1,2,
\label{smallvMb}
\eqa
\bqa
v^{\ada}_{\saabgen} &=& - m^2_1\,\cV^{\ada}_{\sbgen} -
A_{\Z}(m_2)\,C_{\sagen}(2,1,1\,;\,p_1,p_2,\lstm{345}),
\qquad i = 1,2,
\nl
v^{\ada}_{\saab\,i+2} &=& \frac{1}{2}\,\Bigl[
- l_{134}\,\cV^{\ada}_{\saagen}
- \cV^{\aca}_{\saagen}(p_1,P,\lstm{12345})
+ \cV^{\aca}_{\saab}(0,P,\lstm{12335}) \Bigr],
\quad i = 1 \cdots 3,
\nl
v^{\ada}_{\saabf} &=& \frac{1}{2}\,\Bigl[
- l_{134}\,\cV^{\ada}_{\saad}
- \cV^{\aca}_{\saad}(p_1,P,\lstm{12345})
+ \cV^{\aca}_{\saad}(0,P,\lstm{12335}) \Bigr],
\nl
v^{\ada}_{\saabg} &=& \frac{1}{2}\,\Bigl[
( l_{134} - l_{\ssP 35} )\,\cV^{\ada}_{\saaa}
- \cV^{\aca}_{\saab}(0,P,\lstm{12335})
+ \cV^{\aca}_{\saaa}(p_1,0,\lstm{12343})
+ \cV^{\aca}_{\saab}(p_1,0,\lstm{12343})
\nl
{}&-& 2\,\cV^{\aca}_{\saaa}(p_1,0,\lstm{12343}) \Bigr],
\nl
v^{\ada}_{\saab\,i+6} &=& \frac{1}{2}\,\Bigl[
( l_{134} - l_{\ssP 35} )\,\cV^{\ada}_{\saagen}
- \cV^{\aca}_{\saab}(0,P,\lstm{12335}) \Bigr],
\quad i = 2,3,
\nl
v^{\ada}_{\saabj} &=& \frac{1}{2}\,\Bigl[
( l_{134} - l_{\ssP 35} )\,\cV^{\ada}_{\saad}
- \cV^{\aca}_{\saad}(0,P,\lstm{12335})
+ \cV^{\aca}_{\saad}(p_1,0,\lstm{12343})\Bigr],
\nl 
v^{\ada}_{\saab\,i+10} &=& \frac{1}{2}\,\Bigl[
- m^2_{123}\,\cV^{\ada}_{\sagen}
+ \cV^{\aca}_{\sagen}(p_1,P,\lstm{12345}) -
A_{\Z}(m_2)\,C_{\sagen}(2,1,1\,;\,p_1,p_2,\lstm{345}) \Bigr],
\quad i = 1,2,
\label{smallvMc}
\eqa
\bqa
v^{\ada}_{\sabbgen} &=& - m^2_3\,\cV^{\ada}_{\sagen}
+ \cV^{\aca}_{\sagen}(p_1,P,\lstm{12345}),
\qquad i = 1,2,
\nl
v^{\ada}_{\sabb\,i+2} &=& \frac{1}{2}\,\Bigl[
- l_{134}\,\cV^{\ada}_{\sabgen}
- \cV^{\aca}_{\sabgen}(p_1,P,\lstm{12345})
+ \cV^{\aca}_{\sabb}(0,P,\lstm{12335}) \Bigr],
\quad i = 1 \cdots 3,
\nl
v^{\ada}_{\sabbf} &=& \frac{1}{2}\,\Bigl[
- l_{134}\,\cV^{\ada}_{\sabd}
- \cV^{\aca}_{\sabd}(p_1,P,\lstm{12345})
+ \cV^{\aca}_{\sabd}(0,P,\lstm{12335}) \Bigr],
\nl
v^{\ada}_{\sabbg} &=& \frac{1}{2}\,\Bigl[
( l_{134} - l_{\ssP 35} )\,\cV^{\ada}_{\saba}
- \cV^{\aca}_{\sabb}(0,P,\lstm{12335})
+ \cV^{\aca}_{\saba}(p_1,0,\lstm{12343})
+ \cV^{\aca}_{\sabb}(p_1,0,\lstm{12343}) 
\nl
{}&-& 2\,\cV^{\aca}_{\sabc}(p_1,0,\lstm{12343}) \Bigr],
\nl
v^{\ada}_{\sabb\,i+6} &=& \frac{1}{2}\,\Bigl[
( l_{134} - l_{\ssP 35} )\,\cV^{\ada}_{\sabgen}
- \cV^{\aca}_{\sabb}(0,P,\lstm{12335}) \Bigr],
\quad i = 2,3,
\nl
v^{\ada}_{\sabbj} &=& \frac{1}{2}\,\Bigl[
( l_{134} - l_{\ssP 35} )\,\cV^{\ada}_{\sabd}
+ \cV^{\aca}_{\sabd}(p_1,0,\lstm{12343})
- \cV^{\aca}_{\sabd}(0,P,\lstm{12335}) \Bigr],
\nl
v^{\ada}_{\sabb\,i+10} &=& \frac{1}{2}\,\Bigl[
m^2_{123}\,\cV^{\ada}_{\sbgen}
+ \cV^{\aca}_{\sbgen}(p_1,P,\lstm{12345})
+ A_{\Z}([m_1,m_2])\,C_{\sagen}(2,1,1\,;\,p_1,p_2,\lstm{345})\Bigr],
\quad i = 1,2.
\label{smallvMd}
\eqa
Furthermore we define
\bqa
v^{\ada}_{\saaac} &=& 36\,\omega^6\,
\cV^{1 |2,4,1| 4 }_{\ada}\,\bmid_{n  = 10 -\ep}, 
\qquad
v^{\ada}_{\saaad} = 12\,\omega^6\,
\cV^{1 |2,3,2| 4 }_{\ada}\,\bmid_{n  = 10 -\ep}, 
\nl
v^{\ada}_{\saaae} &=& 12\,\omega^6\,
\cV^{1 |2,2,3|4 }_{\ada}\,\bmid_{n  = 10 -\ep}, 
\qquad
v^{\ada}_{\saaaf} = 36\,\omega^6\,
\cV^{1 |2,1,4|4 }_{\ada}\,\bmid_{n  = 10 -\ep}.
\label{genaaa}
\eqa
\begin{itemize}
\item{\bf {\boldmath $K$} family}
\eqn{contraK}
\end{itemize}
\bqa
v^{\bca}_{\sbbba} &=&
          - \cV^{\bca}_{\sba}\,m^2_4
          + \cV^{\bba}_{\sba}(P,P,p_1,\lstm{12365}),
          \nl
v^{\bca}_{\sbbbb} &=& 
          - \cV^{\bca}_{\sbb}\,m^2_4
          + \cV^{\bba}_{\sba}(P,P,p_1,\lstm{12365})
          - \cV^{\bba}_{\sbb}(P,P,p_1,\lstm{12365}),
\nl          
v^{\bca}_{\sbbbc} &=& \frac{1}{2}\,\Bigl[
          - \cV^{\bca}_{\sbba}\,l_{145}
          - \cV^{\bba}_{\sbba}(P,P,p_1,\lstm{12365})
          + \cV^{\bba}_{\sbba}(P,P,0,\lstm{12364})
\nl
{}&+& \cV^{\bba}_{\sbbb}(P,P,0,\lstm{12364})
          - 2\,\cV^{\bba}_{\sbbc}(P,P,0,\lstm{12364})\Bigr],          
\nl
v^{\bca}_{\sbbbd} &=& \frac{1}{2}\,\Bigl[
          - \cV^{\bca}_{\sbbb}\,l_{145}
          - \cV^{\bba}_{\sbba}(P,P,p_1,\lstm{12365})
          + \cV^{\bba}_{\sbba}(P,P,0,\lstm{12364})
\nl
{}&-& \cV^{\bba}_{\sbbb}(P,P,p_1,\lstm{12365})
          + \cV^{\bba}_{\sbbb}(P,P,0,\lstm{12364})
          + 2\,\cV^{\bba}_{\sbbc}(P,P,p_1,\lstm{12365})
\nl
{}&-& 2\,\cV^{\bba}_{\sbbc}(P,P,0,\lstm{12364})\Bigr],          
\nl
v^{\bca}_{\sbbbe} &=& \frac{1}{2}\,\Bigl[
          - \cV^{\bca}_{\sbbc}\,l_{145}
          - \cV^{\bba}_{\sbba}(P,P,p_1,\lstm{12365})
          + \cV^{\bba}_{\sbba}(P,P,0,\lstm{12364})
\nl
{}&+& \cV^{\bba}_{\sbbb}(P,P,0,\lstm{12364})
          + \cV^{\bba}_{\sbbc}(P,P,p_1,\lstm{12365})
          - 2\,\cV^{\bba}_{\sbbc}(P,P,0,\lstm{12364})\Bigr],
\nl          
v^{\bca}_{\sbbbf} &=& \frac{1}{2}\,\Bigl[
          - \cV^{\bca}_{\sbb}4\,l_{145}
          - \cV^{\bba}_{\sbbd}(P,P,p_1,\lstm{12365})
          + \cV^{\bba}_{\sbbd}(P,P,0,\lstm{12364})\Bigr],
\nl          
v^{\bca}_{\sbbbg} &=& \frac{1}{2}\,\Bigl[
           \cV^{\bca}_{\sbba}\,(l_{165}-P^2)
          - \cV^{\bba}_{\sbba}(P,P,0,\lstm{12364})
\nl
{}&+& \cV^{\bba}_{\sbba}(-P,-P,-p_2,\lstm{21345})
          - \cV^{\bba}_{\sbbb}(P,P,0,\lstm{12364})
          + \cV^{\bba}_{\sbbb}(-P,-P,-p_2,\lstm{21345})
\nl
{}&+& 2\,\cV^{\bba}_{\sbbc}(P,P,0,\lstm{12364})
          - 2\,\cV^{\bba}_{\sbbc}(-P,-P,-p_2,\lstm{21345})
          + 2\,\cV^{\bba}_{\sba}(-P,-P,-p_2,\lstm{21345})
\nl
{}&-& 2\,\cV^{\bba}_{\sbb}(-P,-P,-p_2,\lstm{21345})
          + \cV^{\bba}_{\Z}(-P,-P,-p_2,\lstm{21345})\Bigr],
\nl          
v^{\bca}_{\sbbbh} &=& \frac{1}{2}\,\Bigl[
           \cV^{\bca}_{\sbbb}\,(l_{165}-P^2)
          - \cV^{\bba}_{\sbba}(P,P,0,\lstm{12364})
\nl
{}&+& \cV^{\bba}_{\sbba}(-P,-P,-p_2,\lstm{21345})
          - \cV^{\bba}_{\sbbb}(P,P,0,\lstm{12364})
          + 2\,\cV^{\bba}_{\sbbc}(P,P,0,\lstm{12364})
\nl
{}&+& 2\,\cV^{\bba}_{\sba}(-P,-P,-p_2,\lstm{21345})
          + \cV^{\bba}_{\Z}(-P,-P,-p_2,\lstm{21345})\Bigr],
\nl         
v^{\bca}_{\sbbbi} &=& \frac{1}{2}\,\Bigl[
           \cV^{\bca}_{\sbbc}\,(l_{165}-P^2)
          - \cV^{\bba}_{\sbba}(P,P,0,\lstm{12364})
          + \cV^{\bba}_{\sbba}(-P,-P,-p_2,\lstm{21345})
\nl
{}&-& \cV^{\bba}_{\sbbb}(P,P,0,\lstm{12364})
          + 2\,\cV^{\bba}_{\sbbc}(P,P,0,\lstm{12364})
          - \cV^{\bba}_{\sbbc}(-P,-P,-p_2,\lstm{21345})
\nl
{}&+& 2\,\cV^{\bba}_{\sba}(-P,-P,-p_2,\lstm{21345})
          - \cV^{\bba}_{\sbb}(-P,-P,-p_2,\lstm{21345})
          + \cV^{\bba}_{\Z}(-P,-P,-p_2,\lstm{21345})\Bigr],
\nl          
v^{\bca}_{\sbbbj} &=& \frac{1}{2}\,\Bigl[
           \cV^{\bca}_{\sbbd}\,(l_{165}-P^2)
          - \cV^{\bba}_{\sbbd}(P,P,0,\lstm{12364})
          + \cV^{\bba}_{\sbbd}(-P,-P,-p_2,\lstm{21345})\Bigr],
\label{smallvKa}
\eqa         
\bqa
v^{\bca}_{\saaaa} &=&
          - \cV^{\bca}_{\saa}\,m^2_1
          - \cV^{\aca}_{\sab}(-p_2,-P,\lstm{23654})
          - \cV^{\aca}_{\Z}(-p_2,-P,\lstm{23654}),
\nl          
v^{\bca}_{\saaab} &=&
          - \cV^{\bca}_{\sab}\,m^2_1
          - \cV^{\aca}_{\saa}(-p_2,-P,\lstm{23654})
          - \cV^{\aca}_{\Z}(-p_2,-P,\lstm{23654}),
\nl         
v^{\bca}_{\saaac} &=& \frac{1}{2}\,\Bigl[
          - \cV^{\bca}_{\saaa}\,l_{\ssP 12}
          + \cV^{\aca}_{\saaa}(p_1,P,\lstm{13456})
          - \cV^{\aca}_{\saab}(-p_2,-P,\lstm{23654})
\nl
{}&-& 2\,\cV^{\aca}_{\sab}(-p_2,-P,\lstm{23654})
          - \cV^{\aca}_{\Z}(-p_2,-P,\lstm{23654})\Bigr],
\nl          
v^{\bca}_{\saaad} &=& \frac{1}{2}\,\Bigl[
          - \cV^{\bca}_{\saab}\,l_{\ssP 12}
          - \cV^{\aca}_{\saaa}(-p_2,-P,\lstm{23654})
          + \cV^{\aca}_{\saab}(p_1,P,\lstm{13456})
\nl
{}&-& 2\,\cV^{\aca}_{\saa}(-p_2,-P,\lstm{23654})
          - \cV^{\aca}_{\Z}(-p_2,-P,\lstm{23654})\Bigr],
\nl          
v^{\bca}_{\saaae} &=& \frac{1}{2}\,\Bigl[
          - \cV^{\bca}_{\saac}\,l_{\ssP 12}
          + \cV^{\aca}_{\saac}(p_1,P,\lstm{13456})
          - \cV^{\aca}_{\saac}(-p_2,-P,\lstm{23654})
\nl
{}&-& \cV^{\aca}_{\saa}(-p_2,-P,\lstm{23654})
          - \cV^{\aca}_{\sab}(-p_2,-P,\lstm{23654})
          - \cV^{\aca}_{\Z}(-p_2,-P,\lstm{23654})\Bigr],
\nl          
v^{\bca}_{\saaaf} &=& \frac{1}{2}\,\Bigl[
          - \cV^{\bca}_{\saad}\,l_{\ssP 12}
          + \cV^{\aca}_{\saad}(p_1,P,\lstm{13456})
          - \cV^{\aca}_{\saad}(-p_2,-P,\lstm{23654})\Bigr],
\label{smallvKbb}
\eqa         
\bqa
v^{\bca}_{\saaba} &=&
          - \cV^{\bca}_{\sba}\,m^2_1
          - \cV^{\aca}_{\sbb}(-p_2,-P,\lstm{23654})
          - \cV^{\aca}_{\Z}(-p_2,-P,\lstm{23654}),
\nl          
v^{\bca}_{\saabb} &=&
          - \cV^{\bca}_{\sbb}\,m^2_1
          - \cV^{\aca}_{\sba}(-p_2,-P,\lstm{23654})
          - \cV^{\aca}_{\Z}(-p_2,-P,\lstm{23654}),
\nl          
v^{\bca}_{\saabc} &=& \frac{1}{2}\,\Bigl[
          - \cV^{\bca}_{\saba}\,l_{\ssP 12}
          + \cV^{\aca}_{\saba}(p_1,P,\lstm{13456})
          - \cV^{\aca}_{\sabb}(-p_2,-P,\lstm{23654})
\nl
{}&-& \cV^{\aca}_{\sbb}(-p_2,-P,\lstm{23654})
          - \cV^{\aca}_{\sab}(-p_2,-P,\lstm{23654})
          - \cV^{\aca}_{\Z}(-p_2,-P,\lstm{23654})\Bigr],
\nl          
v^{\bca}_{\saabd} &=& \frac{1}{2}\,\Bigl[
          - \cV^{\bca}_{\sabb}\,l_{\ssP 12}
          - \cV^{\aca}_{\saba}(-p_2,-P,\lstm{23654})
          + \cV^{\aca}_{\sabb}(p_1,P,\lstm{13456})
\nl
{}&-& \cV^{\aca}_{\sba}(-p_2,-P,\lstm{23654})
          - \cV^{\aca}_{\saa}(-p_2,-P,\lstm{23654})
          - \cV^{\aca}_{\Z}(-p_2,-P,\lstm{23654})\Bigr],
\nl          
v^{\bca}_{\saabe} &=& \frac{1}{2}\,\Bigl[
          - \cV^{\bca}_{\sabc}\,l_{\ssP 12}
          + \cV^{\aca}_{\sabc}(p_1,P,\lstm{13456})
          - \cV^{\aca}_{\sabe}(-p_2,-P,\lstm{23654})
\nl
{}&-& \cV^{\aca}_{\sba}(-p_2,-P,\lstm{23654})
          - \cV^{\aca}_{\sab}(-p_2,-P,\lstm{23654})
          - \cV^{\aca}_{\Z}(-p_2,-P,\lstm{23654})\Bigr],
\nl          
v^{\bca}_{\saabf} &=& \frac{1}{2}\,\Bigl[
          - \cV^{\bca}_{\sabe}\,l_{\ssP 12}
          - \cV^{\aca}_{\sabc}(-p_2,-P,\lstm{23654})
          + \cV^{\aca}_{\sabe}(p_1,P,\lstm{13456})
\nl
{}&-& \cV^{\aca}_{\sbb}(-p_2,-P,\lstm{23654})
          - \cV^{\aca}_{\saa}(-p_2,-P,\lstm{23654})
          - \cV^{\aca}_{\Z}(-p_2,-P,\lstm{23654})\Bigr],
\nl          
v^{\bca}_{\saabg} &=& \frac{1}{2}\,\Bigl[
          - \cV^{\bca}_{\sabd}\,l_{\ssP 12}
          + \cV^{\aca}_{\sabd}(p_1,P,\lstm{13456})
          - \cV^{\aca}_{\sabd}(-p_2,-P,\lstm{23654})\Bigr],
\nl          
v^{\bca}_{\saabh} &=& \frac{1}{2}\,\Bigl[
          - \cV^{\bca}_{\saaa}\,l_{145}
          - \cV^{\bba}_{\saaa}(P,P,p_1,\lstm{12365})
          + \cV^{\bba}_{\saaa}(P,P,0,\lstm{12364})
\nl
{}&+& \cV^{\bba}_{\saab}(P,P,0,\lstm{12364})
          - 2\,\cV^{\bba}_{\saac}(P,P,0,\lstm{12364})\Bigr],
\nl          
v^{\bca}_{\saabi} &=& \frac{1}{2}\,\Bigl[
          - \cV^{\bca}_{\saab}\,l_{145}
          - \cV^{\bba}_{\saaa}(P,P,p_1,\lstm{12365})
          + \cV^{\bba}_{\saaa}(P,P,0,\lstm{12364})
\nl
{}&-& \cV^{\bba}_{\saab}(P,P,p_1,\lstm{12365})
          + \cV^{\bba}_{\saab}(P,P,0,\lstm{12364})
          + 2\,\cV^{\bba}_{\saac}(P,P,p_1,\lstm{12365})
\nl
{}&-& 2\,\cV^{\bba}_{\saac}(P,P,0,\lstm{12364})\Bigr],
\nl          
v^{\bca}_{\saabj} &=& \frac{1}{2}\,\Bigl[
          - \cV^{\bca}_{\saac}\,l_{145}
          - \cV^{\bba}_{\saaa}(P,P,p_1,\lstm{12365})
          + \cV^{\bba}_{\saaa}(P,P,0,\lstm{12364})
\nl
{}&+& \cV^{\bba}_{\saab}(P,P,0,\lstm{12364})
          + \cV^{\bba}_{\saac}(P,P,p_1,\lstm{12365})
          - 2\,\cV^{\bba}_{\saac}(P,P,0,\lstm{12364})\Bigr],
\nl          
v^{\bca}_{\saabl} &=& \frac{1}{2}\,\Bigl[
          - \cV^{\bca}_{\saad}\,l_{145}
          - \cV^{\bba}_{\saad}(P,P,p_1,\lstm{12365})
          + \cV^{\bba}_{\saad}(P,P,0,\lstm{12364})\Bigr],
\nl          
v^{\bca}_{\saabm} &=& \frac{1}{2}\,\Bigl[
           \cV^{\bca}_{\saaa}\,(l_{165}-P^2)
          - \cV^{\bba}_{\saaa}(P,P,0,\lstm{12364})
          + \cV^{\bba}_{\saaa}(-P,-P,-p_2,\lstm{21345})
\nl
{}&-& \cV^{\bba}_{\saab}(P,P,0,\lstm{12364})
          + \cV^{\bba}_{\saab}(-P,-P,-p_2,\lstm{21345})
          + 2\,\cV^{\bba}_{\saac}(P,P,0,\lstm{12364})
\nl
{}&-& 2\,\cV^{\bba}_{\saac}(-P,-P,-p_2,\lstm{21345})
          + 2\,\cV^{\bba}_{\saa}(-P,-P,-p_2,\lstm{21345})
          - 2\,\cV^{\bba}_{\sab}(-P,-P,-p_2,\lstm{21345})
\nl
{}&+& \cV^{\bba}_{\Z}(-P,-P,-p_2,\lstm{21345})\Bigr],
\nl          
v^{\bca}_{\saabn} &=& \frac{1}{2}\,\Bigl[
           \cV^{\bca}_{\saab}\,(l_{165}-P^2)
          - \cV^{\bba}_{\saaa}(P,P,0,\lstm{12364})
          + \cV^{\bba}_{\saaa}(-P,-P,-p_2,\lstm{21345})
\nl
{}&-& \cV^{\bba}_{\saab}(P,P,0,\lstm{12364})
          + 2\,\cV^{\bba}_{\saac}(P,P,0,\lstm{12364})
          + 2\,\cV^{\bba}_{\saa}(-P,-P,-p_2,\lstm{21345})
\nl
{}&+& \cV^{\bba}_{\Z}(-P,-P,-p_2,\lstm{21345})\Bigr],
\nl          
v^{\bca}_{\saabo} &=& \frac{1}{2}\,\Bigl[
           \cV^{\bca}_{\saac}\,(l_{165}-P^2)
          - \cV^{\bba}_{\saaa}(P,P,0,\lstm{12364})
          + \cV^{\bba}_{\saaa}(-P,-P,-p_2,\lstm{21345})
\nl
{}&-& \cV^{\bba}_{\saab}(P,P,0,\lstm{12364})
          + 2\,\cV^{\bba}_{\saac}(P,P,0,\lstm{12364})
          - \cV^{\bba}_{\saac}(-P,-P,-p_2,\lstm{21345})
\nl
{}&+& 2\,\cV^{\bba}_{\saa}(-P,-P,-p_2,\lstm{21345})
          - \cV^{\bba}_{\sab}(-P,-P,-p_2,\lstm{21345})
          + \cV^{\bba}_{\Z}(-P,-P,-p_2,\lstm{21345})\Bigr],
\nl          
v^{\bca}_{\saabp} &=& \frac{1}{2}\,\Bigl[
           \cV^{\bca}_{\saad}\,(l_{165}-P^2)
          - \cV^{\bba}_{\saad}(P,P,0,\lstm{12364})
          + \cV^{\bba}_{\saad}(-P,-P,-p_2,\lstm{21345})\Bigr],
\nl          
v^{\bca}_{\saabq} &=& \frac{1}{2}\,\Bigl[
          - \cV^{\bca}_{\saa}\,m^2_{134}
          + \cV^{\bba}_{\saa}(P,P,p_1,\lstm{12365})
          - \cV^{\aca}_{\sab}(-p_2,-P,\lstm{23654})
\nl
{}&+& B_{\sa}(P,\lstm{12})\,C_{\Z}(p_1,p_2,\lstm{456})
          - \cV^{\aca}_{\Z}(-p_2,-P,\lstm{23654})\Bigr],
\nl          
v^{\bca}_{\saabr} &=& \frac{1}{2}\,\Bigl[
          - \cV^{\bca}_{\sab}\,m^2_{134}
          + \cV^{\bba}_{\saa}(P,P,p_1,\lstm{12365})
          - \cV^{\bba}_{\sab}(P,P,p_1,\lstm{12365})
\nl
{}&-& \cV^{\aca}_{\saa}(-p_2,-P,\lstm{23654})
          + B_{\sa}(P,\lstm{12})\,C_{\Z}(p_1,p_2,\lstm{456})
          - \cV^{\aca}_{\Z}(-p_2,-P,\lstm{23654})\Bigr],
\label{smallvKc}
\eqa          
\bqa
v^{\bca}_{\sabba} &=&
          - \cV^{\bca}_{\saa}\,m^2_4
          + \cV^{\bba}_{\saa}(P,P,p_1,\lstm{12365}),
\nl          
v^{\bca}_{\sabbb} &=&
          - \cV^{\bca}_{\sab}\,m^2_4
          + \cV^{\bba}_{\saa}(P,P,p_1,\lstm{12365})
          - \cV^{\bba}_{\sab}(P,P,p_1,\lstm{12365}),
\nl          
v^{\bca}_{\sabbc} &=& \frac{1}{2}\,\Bigl[
          - \cV^{\bca}_{\sbba}\,l_{\ssP 12}
          + \cV^{\aca}_{\sbba}(p_1,P,\lstm{13456})
          - \cV^{\aca}_{\sbbb}(-p_2,-P,\lstm{23654})
\nl
{}&-& 2\,\cV^{\aca}_{\sbb}(-p_2,-P,\lstm{23654})
          - \cV^{\aca}_{\Z}(-p_2,-P,\lstm{23654})\Bigr],
\nl          
v^{\bca}_{\sabbd} &=& \frac{1}{2}\,\Bigl[
          - \cV^{\bca}_{\sbbb}\,l_{\ssP 12}
          - \cV^{\aca}_{\sbba}(-p_2,-P,\lstm{23654})
          + \cV^{\aca}_{\sbbb}(p_1,P,\lstm{13456})
\nl
{}&-& 2\,\cV^{\aca}_{\sba}(-p_2,-P,\lstm{23654})
          - \cV^{\aca}_{\Z}(-p_2,-P,\lstm{23654})\Bigr],
\nl          
v^{\bca}_{\sabbe} &=& \frac{1}{2}\,\Bigl[
          - \cV^{\bca}_{\sbbc}\,l_{\ssP 12}
          + \cV^{\aca}_{\sbbc}(p_1,P,\lstm{13456})
          - \cV^{\aca}_{\sbbc}(-p_2,-P,\lstm{23654})
\nl
{}&-& \cV^{\aca}_{\sba}(-p_2,-P,\lstm{23654})
          - \cV^{\aca}_{\sbb}(-p_2,-P,\lstm{23654})
          - \cV^{\aca}_{\Z}(-p_2,-P,\lstm{23654})\Bigr],
\nl          
v^{\bca}_{\sabbf} &=& \frac{1}{2}\,\Bigl[
          - \cV^{\bca}_{\sbbd}\,l_{\ssP 12}
          + \cV^{\aca}_{\sbbd}(p_1,P,\lstm{13456})
          - \cV^{\aca}_{\sbbd}(-p_2,-P,\lstm{23654})\Bigr],
\nl         
v^{\bca}_{\sabbg} &=& \frac{1}{2}\,\Bigl[
          - \cV^{\bca}_{\saba}\,l_{145}
          - \cV^{\bba}_{\saba}(P,P,p_1,\lstm{12365})
          + \cV^{\bba}_{\saba}(P,P,0,\lstm{12364})
\nl
{}&+& \cV^{\bba}_{\sabb}(P,P,0,\lstm{12364})
          - \cV^{\bba}_{\sabc}(P,P,0,\lstm{12364})
          - \cV^{\bba}_{\sabe}(P,P,0,\lstm{12364})\Bigr],
\nl          
v^{\bca}_{\sabbh} &=& \frac{1}{2}\,\Bigl[
          - \cV^{\bca}_{\sabb}\,l_{145}
          - \cV^{\bba}_{\saba}(P,P,p_1,\lstm{12365})
          + \cV^{\bba}_{\saba}(P,P,0,\lstm{12364})
\nl
{}&-& \cV^{\bba}_{\sabb}(P,P,p_1,\lstm{12365})
          + \cV^{\bba}_{\sabb}(P,P,0,\lstm{12364})
          + \cV^{\bba}_{\sabc}(P,P,p_1,\lstm{12365})
\nl
{}&-& \cV^{\bba}_{\sabc}(P,P,0,\lstm{12364})
          + \cV^{\bba}_{\sabe}(P,P,p_1,\lstm{12365})
          - \cV^{\bba}_{\sabe}(P,P,0,\lstm{12364})\Bigr],
\nl          
v^{\bca}_{\sabbi} &=& \frac{1}{2}\,\Bigl[
          - \cV^{\bca}_{\sabc}\,l_{145}
          - \cV^{\bba}_{\saba}(P,P,p_1,\lstm{12365})
          + \cV^{\bba}_{\saba}(P,P,0,\lstm{12364})
\nl
{}&+& \cV^{\bba}_{\sabb}(P,P,0,\lstm{12364})
          + \cV^{\bba}_{\sabc}(P,P,p_1,\lstm{12365})
          - \cV^{\bba}_{\sabc}(P,P,0,\lstm{12364})
\nl
{}&-& \cV^{\bba}_{\sabe}(P,P,0,\lstm{12364})\Bigr],
\nl          
v^{\bca}_{\sabbj} &=& \frac{1}{2}\,\Bigl[
          - \cV^{\bca}_{\sabe}\,l_{145}
          - \cV^{\bba}_{\saba}(P,P,p_1,\lstm{12365})
          + \cV^{\bba}_{\saba}(P,P,0,\lstm{12364})
          + \cV^{\bba}_{\sabb}(P,P,0,\lstm{12364})
\nl
{}&-&       \cV^{\bba}_{\sabc}(P,P,0,\lstm{12364})
          + \cV^{\bba}_{\sabe}(P,P,p_1,\lstm{12365})
          - \cV^{\bba}_{\sabe}(P,P,0,\lstm{12364})\Bigr],
\nl          
v^{\bca}_{\sabbl} &=& \frac{1}{2}\,\Bigl[
          - \cV^{\bca}_{\sabd}\,l_{145}
          - \cV^{\bba}_{\sabd}(P,P,p_1,\lstm{12365})
          + \cV^{\bba}_{\sabd}(P,P,0,\lstm{12364})\Bigr],
\nl
v^{\bca}_{\sabbm} &=& \frac{1}{2}\,\Bigl[
           \cV^{\bca}_{\saba}\,(l_{165}-P^2)
          - \cV^{\bba}_{\saba}(P,P,0,\lstm{12364})
          + \cV^{\bba}_{\saba}(-P,-P,-p_2,\lstm{21345})
\nl
{}&-& \cV^{\bba}_{\sabb}(P,P,0,\lstm{12364})
          + \cV^{\bba}_{\sabb}(-P,-P,-p_2,\lstm{21345})
          + \cV^{\bba}_{\sabc}(P,P,0,\lstm{12364})
\nl
{}&-& \cV^{\bba}_{\sabc}(-P,-P,-p_2,\lstm{21345})
          + \cV^{\bba}_{\sabe}(P,P,0,\lstm{12364})
          - \cV^{\bba}_{\sabe}(-P,-P,-p_2,\lstm{21345})
\nl
{}&+& \cV^{\bba}_{\sba}(-P,-P,-p_2,\lstm{21345})
          - \cV^{\bba}_{\sbb}(-P,-P,-p_2,\lstm{21345})
          + \cV^{\bba}_{\saa}(-P,-P,-p_2,\lstm{21345})
\nl
{}&-& \cV^{\bba}_{\sab}(-P,-P,-p_2,\lstm{21345})
          + \cV^{\bba}_{\Z}(-P,-P,-p_2,\lstm{21345})\Bigr],
\nl          
v^{\bca}_{\sabbn} &=& \frac{1}{2}\,\Bigl[
           \cV^{\bca}_{\sabb}\,(l_{165}-P^2)
          - \cV^{\bba}_{\saba}(P,P,0,\lstm{12364})
          + \cV^{\bba}_{\saba}(-P,-P,-p_2,\lstm{21345})
\nl
{}&-& \cV^{\bba}_{\sabb}(P,P,0,\lstm{12364})
          + \cV^{\bba}_{\sabc}(P,P,0,\lstm{12364})
          + \cV^{\bba}_{\sabe}(P,P,0,\lstm{12364})
\nl
{}&+& \cV^{\bba}_{\sba}(-P,-P,-p_2,\lstm{21345})
          + \cV^{\bba}_{\saa}(-P,-P,-p_2,\lstm{21345})
          + \cV^{\bba}_{\Z}(-P,-P,-p_2,\lstm{21345})\Bigr],
\nl          
v^{\bca}_{\sabbo} &=& \frac{1}{2}\,\Bigl[
           \cV^{\bca}_{\sabc}\,(l_{165}-P^2)
          - \cV^{\bba}_{\saba}(P,P,0,\lstm{12364})
          + \cV^{\bba}_{\saba}(-P,-P,-p_2,\lstm{21345})
\nl
{}&-& \cV^{\bba}_{\sabb}(P,P,0,\lstm{12364})
          + \cV^{\bba}_{\sabc}(P,P,0,\lstm{12364})
          + \cV^{\bba}_{\sabe}(P,P,0,\lstm{12364})
\nl
{}&-& \cV^{\bba}_{\sabe}(-P,-P,-p_2,\lstm{21345})
          + \cV^{\bba}_{\sba}(-P,-P,-p_2,\lstm{21345})
          + \cV^{\bba}_{\saa}(-P,-P,-p_2,\lstm{21345})
\nl
{}&-& \cV^{\bba}_{\sab}(-P,-P,-p_2,\lstm{21345})
          + \cV^{\bba}_{\Z}(-P,-P,-p_2,\lstm{21345})\Bigr],
\nl          
v^{\bca}_{\sabbp} &=& \frac{1}{2}\,\Bigl[
            \cV^{\bca}_{\sabe}\,(l_{165}-P^2)
          - \cV^{\bba}_{\saba}(P,P,0,\lstm{12364})
          + \cV^{\bba}_{\saba}(-P,-P,-p_2,\lstm{21345})
\nl
{}&-& \cV^{\bba}_{\sabb}(P,P,0,\lstm{12364})
          + \cV^{\bba}_{\sabc}(P,P,0,\lstm{12364})
          - \cV^{\bba}_{\sabc}(-P,-P,-p_2,\lstm{21345})
\nl 
{}&+& \cV^{\bba}_{\sabe}(P,P,0,\lstm{12364})
          + \cV^{\bba}_{\sba}(-P,-P,-p_2,\lstm{21345})
          - \cV^{\bba}_{\sbb}(-P,-P,-p_2,\lstm{21345})
\nl
{}&+& \cV^{\bba}_{\saa}(-P,-P,-p_2,\lstm{21345})
          + \cV^{\bba}_{\Z}(-P,-P,-p_2,\lstm{21345})\Bigr],
\nl          
v^{\bca}_{\sabbq} &=& \frac{1}{2}\,\Bigl[
           \cV^{\bca}_{\sabd}\,(l_{165}-P^2)
          - \cV^{\bba}_{\sabd}(P,P,0,\lstm{12364})
          + \cV^{\bba}_{\sabd}(-P,-P,-p_2,\lstm{21345})\Bigr],
\nl          
v^{\bca}_{\sabbr} &=& \frac{1}{2}\,\Bigl[
          - \cV^{\bca}_{\sba}\,m^2_{134}
          + \cV^{\bba}_{\sba}(P,P,p_1,\lstm{12365})
          - \cV^{\aca}_{\sbb}(-p_2,-P,\lstm{23654})
\nl
{}&+& B_{\Z}(P,\lstm{12})\,C_{\saa}(p_1,p_2,\lstm{456})
          - \cV^{\aca}_{\Z}(-p_2,-P,\lstm{23654})\Bigr],
\nl          
v^{\bca}_{\sabbs} &=& \frac{1}{2}\,\Bigl[
          - \cV^{\bca}_{\sbb}\,m^2_{134}
          + \cV^{\bba}_{\sba}(P,P,p_1,\lstm{12365})
          - \cV^{\bba}_{\sbb}(P,P,p_1,\lstm{12365})
\nl
{}&-& \cV^{\aca}_{\sba}(-p_2,-P,\lstm{23654})
          + B_{\Z}(P,\lstm{12})\,C_{\sab}(p_1,p_2,\lstm{456})
          - \cV^{\aca}_{\Z}(-p_2,-P,\lstm{23654})\Bigr].
\label{smallvKd}
\eqa
\begin{itemize}
\item{\bf {\boldmath $H$} family}
\eqn{contraK}
\end{itemize}
It is convenient to define certain combinations of form factors to be used
in this family (they only appear in the present subsection):
\bqa
\cV^{\aca}_{\cagen} &=& \cV^{\aca}_{\saagen} + \cV^{\aca}_{\sbbgen} -
2\,\cV^{\aca}_{\sabgen},
\quad
\cV^{\aca}_{\cbgen} = \cV^{\aca}_{\sagen} + \cV^{\aca}_{\sbgen},
\quad
\cV^{\bba}_{\cagen} = \cV^{\bba}_{\saagen} - \cV^{\bba}_{\sabgen},
\quad
\cV^{\bba}_{\cbgen} = \cV^{\bba}_{\sbbgen} - \cV^{\bba}_{\sabgen},
\nl
\cV^{\bba}_{\ccgen} &=& \cV^{\bba}_{\sagen} - \cV^{\bba}_{\sbgen},
\quad
\cV^{\bba}_{\sab\ssA} = \cV^{\bba}_{\sabe} - \cV^{\bba}_{\sabc}.
\eqa
We obtain
\bqa
v^{\bbb}_{\sbbba} &=&
          - \cV^{\bbb}_{\sba}\,m^2_5
          + \cV^{\bba}_{\Z}(p_2,p_2,-p_1,\lstm{21634})
          - \cV^{\bba}_{\ccb}(p_2,p_2,-p_1,\lstm{21634}),
\nl
v^{\bbb}_{\sbbbb} &=& 
          - \cV^{\bbb}_{\sbb}\,m^2_5
          + \cV^{\bba}_{\cca}(p_2,p_2,-p_1,\lstm{21634})
          - \cV^{\bba}_{\ccb}(p_2,p_2,-p_1,\lstm{21634}),
\nl         
v^{\bbb}_{\sbbbc} &=& \frac{1}{2}\,\Bigl[ 
           \cV^{\bbb}_{\sbba}\,(2\,p^2_1-l_{165})
          + \cV^{\bba}_{\Z}(p_2,p_2,-p_1,\lstm{21634})
          + \cV^{\bba}_{\cab}(p_2,p_2,-p_1,\lstm{21634})
\nl
{}&-& \cV^{\bba}_{\cab}(-p_2,-p_2,p_1,\lstm{12543})
          + \cV^{\bba}_{\cbb}(p_2,p_2,-p_1,\lstm{21634})
          - \cV^{\bba}_{\cbb}(-p_2,-p_2,p_1,\lstm{12543})
\nl
{}&-& 2\,\cV^{\bba}_{\ccb}(p_2,p_2,-p_1,\lstm{21634})\Bigr],
\nl          
v^{\bbb}_{\sbbbd} &=& \frac{1}{2}\,\Bigl[ 
           \cV^{\bbb}_{\sbbb}\,(2\,p^2_1-l_{165})
          + \cV^{\bba}_{\caa}(p_2,p_2,-p_1,\lstm{21634})
          - \cV^{\bba}_{\caa}(-p_2,-p_2,p_1,\lstm{12543})
\nl
{}&+& \cV^{\bba}_{\cba}(p_2,p_2,-p_1,\lstm{21634})
          - \cV^{\bba}_{\cba}(-p_2,-p_2,p_1,\lstm{12543})
          + \cV^{\bba}_{\cab}(p_2,p_2,-p_1,\lstm{21634})
\nl
{}&-& \cV^{\bba}_{\cab}(-p_2,-p_2,p_1,\lstm{12543})
          + \cV^{\bba}_{\cbb}(p_2,p_2,-p_1,\lstm{21634})
          - \cV^{\bba}_{\cbb}(-p_2,-p_2,p_1,\lstm{12543})
\nl
{}&-& 2\,\cV^{\bba}_{\cac}(p_2,p_2,-p_1,\lstm{21634})
          + 2\,\cV^{\bba}_{\cac}(-p_2,-p_2,p_1,\lstm{12543})
          - 2\,\cV^{\bba}_{\cbc}(p_2,p_2,-p_1,\lstm{21634})
\nl
{}&+& 2\,\cV^{\bba}_{\cbc}(-p_2,-p_2,p_1,\lstm{12543})
          + 2\,\cV^{\bba}_{\sab\ssA}(p_2,p_2,-p_1,\lstm{21634})
          - 2\,\cV^{\bba}_{\sab\ssA}(-p_2,-p_2,p_1,\lstm{12543})\Bigr],
\nl          
v^{\bbb}_{\sbbbe} &=& \frac{1}{2}\,\Bigl[ 
           \cV^{\bbb}_{\sbbc}\,(2\,p^2_1-l_{165})
          + \cV^{\bba}_{\cca}(p_2,p_2,-p_1,\lstm{21634})
          + \cV^{\bba}_{\cab}(p_2,p_2,-p_1,\lstm{21634})
\nl
{}&-& \cV^{\bba}_{\cab}(-p_2,-p_2,p_1,\lstm{12543})
          + \cV^{\bba}_{\cbb}(p_2,p_2,-p_1,\lstm{21634})
          - \cV^{\bba}_{\cbb}(-p_2,-p_2,p_1,\lstm{12543})
\nl
{}&-& \cV^{\bba}_{\ccb}(p_2,p_2,-p_1,\lstm{21634})
          - \cV^{\bba}_{\cac}(p_2,p_2,-p_1,\lstm{21634})
          + \cV^{\bba}_{\cac}(-p_2,-p_2,p_1,\lstm{12543})
\nl
{}&-& \cV^{\bba}_{\cbc}(p_2,p_2,-p_1,\lstm{21634})
          + \cV^{\bba}_{\cbc}(-p_2,-p_2,p_1,\lstm{12543})
          + \cV^{\bba}_{\sab\ssA}(p_2,p_2,-p_1,\lstm{21634})
\nl
{}&-& \cV^{\bba}_{\sab\ssA}(-p_2,-p_2,p_1,\lstm{12543})\Bigr],
\nl          
v^{\bbb}_{\sbbbf} &=& \frac{1}{2}\,\Bigl[ 
           \cV^{\bbb}_{\sbbd}\,(2\,p^2_1-l_{165})
          + \cV^{\bba}_{\cad}(p_2,p_2,-p_1,\lstm{21634})
          - \cV^{\bba}_{\cad}(-p_2,-p_2,p_1,\lstm{12543})
\nl
{}&+& \cV^{\bba}_{\cbd}(p_2,p_2,-p_1,\lstm{21634})
          - \cV^{\bba}_{\cbd}(-p_2,-p_2,p_1,\lstm{12543})\Bigr],
\label{smallvHa}
\eqa          
\bqa
v^{\bbb}_{\saaaa} &=&
          - \cV^{\bbb}_{\saa}\,m^2_1
          - \cV^{\bba}_{\cca}(P,P,p_2,\lstm{34256})
          + \cV^{\bba}_{\ccb}(P,P,p_2,\lstm{34256}),
\nl          
v^{\bbb}_{\saaab} &=&
          - \cV^{\bbb}_{\sab}\,m^2_1
          + \cV^{\bba}_{\Z}(P,P,p_2,\lstm{34256})
          - \cV^{\bba}_{\cca}(P,P,p_2,\lstm{34256}),
\nl          
v^{\bbb}_{\saaac} &=& \frac{1}{2}\,\Bigl[ 
           \cV^{\bbb}_{\saaa}\,l_{212}
          + \cV^{\bba}_{\caa}(P,P,p_2,\lstm{34256})
          - \cV^{\bba}_{\caa}(-P,-P,-p_2,\lstm{21345})
\nl
{}&+& \cV^{\bba}_{\cba}(P,P,p_2,\lstm{34256})
          - \cV^{\bba}_{\cba}(-P,-P,-p_2,\lstm{21345})
          + \cV^{\bba}_{\cab}(P,P,p_2,\lstm{34256})
\nl
{}&-& \cV^{\bba}_{\cab}(-P,-P,-p_2,\lstm{21345})
          + \cV^{\bba}_{\cbb}(P,P,p_2,\lstm{34256})
          - \cV^{\bba}_{\cbb}(-P,-P,-p_2,\lstm{21345})
\nl
{}&-& 2\,\cV^{\bba}_{\cac}(P,P,p_2,\lstm{34256})
          + 2\,\cV^{\bba}_{\cac}(-P,-P,-p_2,\lstm{21345})
          - 2\,\cV^{\bba}_{\cbc}(P,P,p_2,\lstm{34256})
\nl
{}&+& 2\,\cV^{\bba}_{\cbc}(-P,-P,-p_2,\lstm{21345})
          + 2\,\cV^{\bba}_{\sab\ssA}(P,P,p_2,\lstm{34256})
          - 2\,\cV^{\bba}_{\sab\ssA}(-P,-P,-p_2,\lstm{21345})\Bigr],
\nl          
v^{\bbb}_{\saaad} &=& \frac{1}{2}\,\Bigl[ 
           \cV^{\bbb}_{\saab}\,l_{212}
          + \cV^{\bba}_{\Z}(P,P,p_2,\lstm{34256})
          + \cV^{\bba}_{\caa}(P,P,p_2,\lstm{34256})
\nl
{}&-& \cV^{\bba}_{\caa}(-P,-P,-p_2,\lstm{21345})
          + \cV^{\bba}_{\cba}(P,P,p_2,\lstm{34256})
          - \cV^{\bba}_{\cba}(-P,-P,-p_2,\lstm{21345})
\nl
{}&-& 2\,\cV^{\bba}_{\cca}(P,P,p_2,\lstm{34256})\Bigr],
\nl          
v^{\bbb}_{\saaae} &=& \frac{1}{2}\,\Bigl[ 
           \cV^{\bbb}_{\saac}\,l_{212}
          + \cV^{\bba}_{\caa}(P,P,p_2,\lstm{34256})
          - \cV^{\bba}_{\caa}(-P,-P,-p_2,\lstm{21345})
\nl
{}&+& \cV^{\bba}_{\cba}(P,P,p_2,\lstm{34256})
          - \cV^{\bba}_{\cba}(-P,-P,-p_2,\lstm{21345})
          - \cV^{\bba}_{\cca}(P,P,p_2,\lstm{34256})
\nl
{}&+& \cV^{\bba}_{\ccb}(P,P,p_2,\lstm{34256})
          - \cV^{\bba}_{\cac}(P,P,p_2,\lstm{34256})
          + \cV^{\bba}_{\cac}(-P,-P,-p_2,\lstm{21345})
\nl
{}&-& \cV^{\bba}_{\cbc}(P,P,p_2,\lstm{34256})
          + \cV^{\bba}_{\cbc}(-P,-P,-p_2,\lstm{21345})
          + \cV^{\bba}_{\sab\ssA}(P,P,p_2,\lstm{34256})
\nl
{}&-& \cV^{\bba}_{\sab\ssA}(-P,-P,-p_2,\lstm{21345})\Bigr],
\nl          
v^{\bbb}_{\saaaf} &=&  \frac{1}{2}\,\Bigl[ 
           \cV^{\bbb}_{\saad}\,l_{212}
          + \cV^{\bba}_{\cad}(P,P,p_2,\lstm{34256})
          - \cV^{\bba}_{\cad}(-P,-P,-p_2,\lstm{21345})
\nl
{}&+& \cV^{\bba}_{\cbd}(P,P,p_2,\lstm{34256})
          - \cV^{\bba}_{\cbd}(-P,-P,-p_2,\lstm{21345})\Bigr],
\label{smallvHb}          
\eqa
\bqa
v^{\bbb}_{\saaba} &=&
          - \cV^{\bbb}_{\sba}\,m^2_1
          + \cV^{\bba}_{\sba}(P,P,p_2,\lstm{34256})
          - \cV^{\bba}_{\sbb}(P,P,p_2,\lstm{34256})
          + \cV^{\bba}_{\Z}(P,P,p_2,\lstm{34256}),
\nl          
v^{\bbb}_{\saabb} &=&
          - \cV^{\bbb}_{\sbb}\,m^2_1
          + \cV^{\bba}_{\sba}(P,P,p_2,\lstm{34256})
          + \cV^{\bba}_{\Z}(P,P,p_2,\lstm{34256}),
\nl          
v^{\bbb}_{\saabc} &=& \frac{1}{2}\,\Bigl[ 
           \cV^{\bbb}_{\saba}\,l_{212}
          + \cV^{\bba}_{\cba}(P,P,p_2,\lstm{34256})
          - \cV^{\bba}_{\cba}(-P,-P,-p_2,\lstm{21345})
\nl
{}&-& \cV^{\bba}_{\cca}(P,P,p_2,\lstm{34256})
          + \cV^{\bba}_{\cbb}(P,P,p_2,\lstm{34256})
          - \cV^{\bba}_{\cbb}(-P,-P,-p_2,\lstm{21345})
\nl
{}&+& \cV^{\bba}_{\ccb}(P,P,p_2,\lstm{34256})
          - 2\,\cV^{\bba}_{\cbc}(P,P,p_2,\lstm{34256})
          + 2\,\cV^{\bba}_{\cbc}(-P,-P,-p_2,\lstm{21345})
\nl
{}&+& \cV^{\bba}_{\sab\ssA}(P,P,p_2,\lstm{34256})
          - \cV^{\bba}_{\sab\ssA}(-P,-P,-p_2,\lstm{21345})\Bigr],
\nl
v^{\bbb}_{\saabd} &=& \frac{1}{2}\,\Bigl[ 
           \cV^{\bbb}_{\sabb}\,l_{212}
          + \cV^{\bba}_{\sba}(P,P,p_2,\lstm{34256})
          + \cV^{\bba}_{\Z}(P,P,p_2,\lstm{34256})
\nl
{}&+& \cV^{\bba}_{\cba}(P,P,p_2,\lstm{34256})
          - \cV^{\bba}_{\cba}(-P,-P,-p_2,\lstm{21345})
          - \cV^{\bba}_{\cca}(P,P,p_2,\lstm{34256})
\nl
{}&+& \cV^{\bba}_{\cca}(-P,-P,-p_2,\lstm{21345})\Bigr],
\nl          
v^{\bbb}_{\saabe} &=& \frac{1}{2}\,\Bigl[ 
           \cV^{\bbb}_{\sabc}\,l_{212}
          + \cV^{\bba}_{\cba}(P,P,p_2,\lstm{34256})
          - \cV^{\bba}_{\cba}(-P,-P,-p_2,\lstm{21345})
\nl
{}&-& \cV^{\bba}_{\cca}(P,P,p_2,\lstm{34256})
          + \cV^{\bba}_{\cca}(-P,-P,-p_2,\lstm{21345})
          + \cV^{\bba}_{\ccb}(P,P,p_2,\lstm{34256})
\nl
{}&-& \cV^{\bba}_{\ccb}(-P,-P,-p_2,\lstm{21345})
          - \cV^{\bba}_{\cbc}(P,P,p_2,\lstm{34256})
          + \cV^{\bba}_{\cbc}(-P,-P,-p_2,\lstm{21345})
\nl
{}&+& \cV^{\bba}_{\sab\ssA}(P,P,p_2,\lstm{34256})
          - \cV^{\bba}_{\sab\ssA}(-P,-P,-p_2,\lstm{21345})\Bigr],
\nl          
v^{\bbb}_{\saabf} &=& \frac{1}{2}\,\Bigl[ 
           \cV^{\bbb}_{\sabe}\,l_{212}
          + \cV^{\bba}_{\sba}(P,P,p_2,\lstm{34256})
          - \cV^{\bba}_{\sbb}(P,P,p_2,\lstm{34256})
\nl
{}&+& \cV^{\bba}_{\Z}(P,P,p_2,\lstm{34256})
          + \cV^{\bba}_{\cba}(P,P,p_2,\lstm{34256})
          - \cV^{\bba}_{\cba}(-P,-P,-p_2,\lstm{21345})
\nl
{}&-& \cV^{\bba}_{\cca}(P,P,p_2,\lstm{34256})
          - \cV^{\bba}_{\cbc}(P,P,p_2,\lstm{34256})
          + \cV^{\bba}_{\cbc}(-P,-P,-p_2,\lstm{21345})\Bigr],
\nl          
v^{\bbb}_{\saabg} &=& \frac{1}{2}\,\Bigl[ 
           \cV^{\bbb}_{\sabd}\,l_{212}
          + \cV^{\bba}_{\cbd}(P,P,p_2,\lstm{34256})
          - \cV^{\bba}_{\cbd}(-P,-P,-p_2,\lstm{21345})\Bigr],
\nl          
v^{\bbb}_{\saabh} &=& \frac{1}{2}\,\Bigl[ 
           \cV^{\bbb}_{\saaa}\,(2\,p^2_1-l_{165})
          + \cV^{\bba}_{\sabb}(p_2,p_2,-p_1,\lstm{21634})
          - \cV^{\bba}_{\sabb}(-p_2,-p_2,p_1,\lstm{12543})
\nl
{}&+& \cV^{\bba}_{\cab}(p_2,p_2,-p_1,\lstm{21634})
          - \cV^{\bba}_{\cab}(-p_2,-p_2,p_1,\lstm{12543})\Bigr],
\nl          
v^{\bbb}_{\saabi} &=& \frac{1}{2}\,\Bigl[ 
           \cV^{\bbb}_{\saab}\,(2\,p^2_1-l_{165})
          + \cV^{\bba}_{\saba}(p_2,p_2,-p_1,\lstm{21634})
          - \cV^{\bba}_{\saba}(-p_2,-p_2,p_1,\lstm{12543})
\nl
{}&+& \cV^{\bba}_{\sabb}(p_2,p_2,-p_1,\lstm{21634})
          - \cV^{\bba}_{\sabb}(-p_2,-p_2,p_1,\lstm{12543})
          - 2\,\cV^{\bba}_{\sabc}(p_2,p_2,-p_1,\lstm{21634})
\nl
{}&+& 2\,\cV^{\bba}_{\sabc}(-p_2,-p_2,p_1,\lstm{12543})
          + 2\,\cV^{\bba}_{\sba}(p_2,p_2,-p_1,\lstm{21634})
          - 2\,\cV^{\bba}_{\sbb}(p_2,p_2,-p_1,\lstm{21634})
\nl
{}&+& \cV^{\bba}_{\Z}(p_2,p_2,-p_1,\lstm{21634})
          + \cV^{\bba}_{\caa}(p_2,p_2,-p_1,\lstm{21634})
          - \cV^{\bba}_{\caa}(-p_2,-p_2,p_1,\lstm{12543})
\nl
{}&+& 2\,\cV^{\bba}_{\cca}(p_2,p_2,-p_1,\lstm{21634})
          + \cV^{\bba}_{\cab}(p_2,p_2,-p_1,\lstm{21634})
          - \cV^{\bba}_{\cab}(-p_2,-p_2,p_1,\lstm{12543})
\nl
{}&-& 2\,\cV^{\bba}_{\ccb}(p_2,p_2,-p_1,\lstm{21634})
          - 2\,\cV^{\bba}_{\cac}(p_2,p_2,-p_1,\lstm{21634})
          + 2\,\cV^{\bba}_{\cac}(-p_2,-p_2,p_1,\lstm{12543})\Bigr],
\nl          
v^{\bbb}_{\saabj} &=& \frac{1}{2}\,\Bigl[ 
           \cV^{\bbb}_{\saac}\,(2\,p^2_1-l_{165})
          + \cV^{\bba}_{\sabb}(p_2,p_2,-p_1,\lstm{21634})
          - \cV^{\bba}_{\sabb}(-p_2,-p_2,p_1,\lstm{12543})
\nl
{}&-& \cV^{\bba}_{\sabc}(p_2,p_2,-p_1,\lstm{21634})
          + \cV^{\bba}_{\sabc}(-p_2,-p_2,p_1,\lstm{12543})
          - \cV^{\bba}_{\sbb}(p_2,p_2,-p_1,\lstm{21634})
\nl
{}&+& \cV^{\bba}_{\cab}(p_2,p_2,-p_1,\lstm{21634})
          - \cV^{\bba}_{\cab}(-p_2,-p_2,p_1,\lstm{12543})
          - \cV^{\bba}_{\ccb}(p_2,p_2,-p_1,\lstm{21634})
\nl
{}&-& \cV^{\bba}_{\cac}(p_2,p_2,-p_1,\lstm{21634})
          + \cV^{\bba}_{\cac}(-p_2,-p_2,p_1,\lstm{12543})\Bigr],
\nl          
v^{\bbb}_{\saabl} &=& \frac{1}{2}\,\Bigl[ 
           \cV^{\bbb}_{\saad}\,(2\,p^2_1-l_{165})
          + \cV^{\bba}_{\sabd}(p_2,p_2,-p_1,\lstm{21634})
          - \cV^{\bba}_{\sabd}(-p_2,-p_2,p_1,\lstm{12543})
\nl
{}&+& \cV^{\bba}_{\cad}(p_2,p_2,-p_1,\lstm{21634})
          - \cV^{\bba}_{\cad}(-p_2,-p_2,p_1,\lstm{12543})\Bigr],
\label{smallvHc}          
\eqa
\bqa
v^{\bbb}_{\sabba} &=&  
          - \cV^{\bbb}_{\saa}\,m^2_5
          - \cV^{\bba}_{\sbb}(p_2,p_2,-p_1,\lstm{21634})
          - \cV^{\bba}_{\ccb}(p_2,p_2,-p_1,\lstm{21634}),
\nl          
v^{\bbb}_{\sabbb} &=&  
          - \cV^{\bbb}_{\sab}\,m^2_5
          + \cV^{\bba}_{\sba}(p_2,p_2,-p_1,\lstm{21634})
          - \cV^{\bba}_{\sbb}(p_2,p_2,-p_1,\lstm{21634})
\nl
{}&+& \cV^{\bba}_{\Z}(p_2,p_2,-p_1,\lstm{21634})
          + \cV^{\bba}_{\cca}(p_2,p_2,-p_1,\lstm{21634})
          - \cV^{\bba}_{\ccb}(p_2,p_2,-p_1,\lstm{21634}),
\nl          
v^{\bbb}_{\sabbc} &=& \frac{1}{2}\,\Bigl[ 
           \cV^{\bbb}_{\sbba}\,l_{212}
          + \cV^{\bba}_{\saba}(P,P,p_2,\lstm{34256})
          - \cV^{\bba}_{\saba}(-P,-P,-p_2,\lstm{21345})
\nl
{}&+& \cV^{\bba}_{\sabb}(P,P,p_2,\lstm{34256})
          - \cV^{\bba}_{\sabb}(-P,-P,-p_2,\lstm{21345})
          - 2\,\cV^{\bba}_{\sabc}(P,P,p_2,\lstm{34256})
\nl
{}&+& 2\,\cV^{\bba}_{\sabc}(-P,-P,-p_2,\lstm{21345})
          + 2\,\cV^{\bba}_{\sba}(P,P,p_2,\lstm{34256})
          - 2\,\cV^{\bba}_{\sbb}(P,P,p_2,\lstm{34256})
\nl
{}&+& \cV^{\bba}_{\Z}(P,P,p_2,\lstm{34256})
          + \cV^{\bba}_{\cba}(P,P,p_2,\lstm{34256})
          - \cV^{\bba}_{\cba}(-P,-P,-p_2,\lstm{21345})
\nl
{}&+& \cV^{\bba}_{\cbb}(P,P,p_2,\lstm{34256})
          - \cV^{\bba}_{\cbb}(-P,-P,-p_2,\lstm{21345})
          - 2\,\cV^{\bba}_{\cbc}(P,P,p_2,\lstm{34256})
\nl
{}&+& 2\,\cV^{\bba}_{\cbc}(-P,-P,-p_2,\lstm{21345})\Bigr],
\nl          
v^{\bbb}_{\sabbd} &=& \frac{1}{2}\,\Bigl[ 
           \cV^{\bbb}_{\sbbb}\,l_{212}
          + \cV^{\bba}_{\saba}(P,P,p_2,\lstm{34256})
          - \cV^{\bba}_{\saba}(-P,-P,-p_2,\lstm{21345})
\nl
{}&+& 2\,\cV^{\bba}_{\sba}(P,P,p_2,\lstm{34256})
          - 2\,\cV^{\bba}_{\sba}(-P,-P,-p_2,\lstm{21345})
          + \cV^{\bba}_{\Z}(P,P,p_2,\lstm{34256})
\nl
{}&-& \cV^{\bba}_{\Z}(-P,-P,-p_2,\lstm{21345})
          + \cV^{\bba}_{\cba}(P,P,p_2,\lstm{34256})
          - \cV^{\bba}_{\cba}(-P,-P,-p_2,\lstm{21345})\Bigr],
\nl          
v^{\bbb}_{\sabbe} &=& \frac{1}{2}\,\Bigl[ 
           \cV^{\bbb}_{\sbbc}\,l_{212}
          + \cV^{\bba}_{\saba}(P,P,p_2,\lstm{34256})
          - \cV^{\bba}_{\saba}(-P,-P,-p_2,\lstm{21345})
\nl
{}&-& \cV^{\bba}_{\sabc}(P,P,p_2,\lstm{34256})
          + \cV^{\bba}_{\sabc}(-P,-P,-p_2,\lstm{21345})
          + 2\,\cV^{\bba}_{\sba}(P,P,p_2,\lstm{34256})
\nl
{}&-& \cV^{\bba}_{\sba}(-P,-P,-p_2,\lstm{21345})
          - \cV^{\bba}_{\sbb}(P,P,p_2,\lstm{34256})
          + \cV^{\bba}_{\sbb}(-P,-P,-p_2,\lstm{21345})
\nl
{}&+& \cV^{\bba}_{\Z}(P,P,p_2,\lstm{34256})
          + \cV^{\bba}_{\cba}(P,P,p_2,\lstm{34256})
          - \cV^{\bba}_{\cba}(-P,-P,-p_2,\lstm{21345})
\nl
{}&-& \cV^{\bba}_{\cbc}(P,P,p_2,\lstm{34256})
          + \cV^{\bba}_{\cbc}(-P,-P,-p_2,\lstm{21345})\Bigr],
\nl          
v^{\bbb}_{\sabbf} &=& \frac{1}{2}\,\Bigl[ 
           \cV^{\bbb}_{\sbbd}\,l_{212}
          + \cV^{\bba}_{\sabd}(P,P,p_2,\lstm{34256})
          - \cV^{\bba}_{\sabd}(-P,-P,-p_2,\lstm{21345})
\nl
{}&+& \cV^{\bba}_{\cbd}(P,P,p_2,\lstm{34256})
          - \cV^{\bba}_{\cbd}(-P,-P,-p_2,\lstm{21345})\Bigr],
\nl         
v^{\bbb}_{\sabbg} &=& \frac{1}{2}\,\Bigl[ 
           \cV^{\bbb}_{\saba}\,(2\,p^2_1-l_{165})
          - \cV^{\bba}_{\sbb}(p_2,p_2,-p_1,\lstm{21634})
          + \cV^{\bba}_{\cab}(p_2,p_2,-p_1,\lstm{21634})
\nl
{}&-& \cV^{\bba}_{\cab}(-p_2,-p_2,p_1,\lstm{12543})
          - \cV^{\bba}_{\ccb}(p_2,p_2,-p_1,\lstm{21634})\Bigr],
\nl          
v^{\bbb}_{\sabbh} &=& \frac{1}{2}\,\Bigl[ 
           \cV^{\bbb}_{\sabb}\,(2\,p^2_1-l_{165})
          + \cV^{\bba}_{\caa}(p_2,p_2,-p_1,\lstm{21634})
          - \cV^{\bba}_{\caa}(-p_2,-p_2,p_1,\lstm{12543})
\nl
{}&+& \cV^{\bba}_{\cca}(p_2,p_2,-p_1,\lstm{21634})
          + \cV^{\bba}_{\cab}(p_2,p_2,-p_1,\lstm{21634})
          - \cV^{\bba}_{\cab}(-p_2,-p_2,p_1,\lstm{12543})
\nl
{}&-& \cV^{\bba}_{\ccb}(p_2,p_2,-p_1,\lstm{21634})
          - 2\,\cV^{\bba}_{\cac}(p_2,p_2,-p_1,\lstm{21634})
          + 2\,\cV^{\bba}_{\cac}(-p_2,-p_2,p_1,\lstm{12543})
\nl
{}&+& \cV^{\bba}_{\sab\ssA}(p_2,p_2,-p_1,\lstm{21634})
          - \cV^{\bba}_{\sab\ssA}(-p_2,-p_2,p_1,\lstm{12543})\Bigr],
\nl          
v^{\bbb}_{\sabbi} &=& \frac{1}{2}\,\Bigl[ 
           \cV^{\bbb}_{\sabc}\,(2\,p^2_1-l_{165})
          + \cV^{\bba}_{\cab}(p_2,p_2,-p_1,\lstm{21634})
          - \cV^{\bba}_{\cab}(-p_2,-p_2,p_1,\lstm{12543})
\nl
{}&-& \cV^{\bba}_{\cac}(p_2,p_2,-p_1,\lstm{21634})
          + \cV^{\bba}_{\cac}(-p_2,-p_2,p_1,\lstm{12543})
          + \cV^{\bba}_{\sab\ssA}(p_2,p_2,-p_1,\lstm{21634})
\nl
{}&-& \cV^{\bba}_{\sab\ssA}(-p_2,-p_2,p_1,\lstm{12543})\Bigr],
\nl          
v^{\bbb}_{\sabbj} &=& \frac{1}{2}\,\Bigl[ 
           \cV^{\bbb}_{\sabe}\,(2\,p^2_1-l_{165})
          + \cV^{\bba}_{\sba}(p_2,p_2,-p_1,\lstm{21634})
          - \cV^{\bba}_{\sbb}(p_2,p_2,-p_1,\lstm{21634})
\nl
{}&+& \cV^{\bba}_{\Z}(p_2,p_2,-p_1,\lstm{21634})
          + \cV^{\bba}_{\cca}(p_2,p_2,-p_1,\lstm{21634})
          + \cV^{\bba}_{\cab}(p_2,p_2,-p_1,\lstm{21634})
\nl
{}&-& \cV^{\bba}_{\cab}(-p_2,-p_2,p_1,\lstm{12543})
          - 2\,\cV^{\bba}_{\ccb}(p_2,p_2,-p_1,\lstm{21634})
          - \cV^{\bba}_{\cac}(p_2,p_2,-p_1,\lstm{21634})
\nl
{}&+& \cV^{\bba}_{\cac}(-p_2,-p_2,p_1,\lstm{12543})\Bigr],
\nl          
v^{\bbb}_{\sabbl} &=& \frac{1}{2}\,\Bigl[ 
           \cV^{\bbb}_{\sabd}\,(2\,p^2_1-l_{165})
          + \cV^{\bba}_{\cad}(p_2,p_2,-p_1,\lstm{21634})
          - \cV^{\bba}_{\cad}(-p_2,-p_2,p_1,\lstm{12543})\Bigr].
\label{smallvHd}
\eqa
In the following Appendices results for rank three tensors with completely
saturated indices are presented. At the same time we give an explicit
solution for the form factors which are needed, implicitly, for some of the
contracted expressions and, explicitly, for testing WST identities.
We recall that tensors with saturated indices are of the upmost importance for
applications related to projection techniques (see \sect{prered}).

\subsubsection{The $V^{\ada}$ family \label{satuone}}
Almost all contracted tensors in this family can be trivially 
obtained using \eqn{contraM} and the definitions of 
\eqns{smallvMa}{smallvMd}.
Only for the $111$ group we have to solve first for the form factors and 
then to replace the resulting expressions into the decomposition of the 
saturated tensors; in this way we are able to obtain
\bqa
\cV^{\ada}(\mu,\mu, p_1\,|\,0) &=&
       v^{\ada}_{\saaaa}\,p^2_1 
       + v^{\ada}_{\saaab}\,p_{12},
\quad
\cV^{\ada}(\mu,\mu ,p_2\,|\,0) =
       v^{\ada}_{\saaaa}\,p_{12} 
       + v^{\ada}_{\saaab}\,p^2_2,
\eqa
\bqa
{}&{}& (n+2)\,\cV^{\ada}(p_1, p_1, p_2\,|\,0) =
       3\,v^{\ada}_{\saaaa}\,D_3 
       - v^{\ada}_{\saaab}\,( 2\,D - 3\,D_1 )
       - v^{\ada}_{\saaac}\,p^2_1\,(n-1)\,D_3 
\nl
{}&+&
 v^{\ada}_{\saaaf}\,p^2_2\,\Bigl[ n\,D - (n-1)\,D_1 \Bigr]
 - v^{\ada}_{\saaaf}\,p^2_1\,
\Bigl[  6\,(n-2)\,D - (n-1)\,(9\,D_1 - 11\,D_3) \Bigr]
\nl
{}&+& \Bigl[ v^{\ada}_{\saaaf} + v^{\ada}_{\saaae} \Bigr]\,p_{12}\,
\Bigl[ (n-2)\,D - 3\,(n-1)\,D_1 \Bigr]
\nl
{}&+& v^{\ada}_{\saaad}\,p^2_1\,
\Bigl[ 2\,(n-2)\,D - 3\,(n-1)\,( D_1 - D_3 ) \Bigr],
\nl
\nl
{}&{}& (n+2)\,\cV^{\ada}(p_2, p_2, p_1\,|\,0) =
       - v^{\ada}_{\saaaa}\,( 2\,D - 3\,D_1 )
       +3\,v^{\ada}_{\saaab}\,D_2 
\nl
{}&+& v^{\ada}_{\saaaf}\,p^2_2\,
\Bigl[ 2\,(n-2)\,D - (n-1)\,(3\,D_1 + D_2) \Bigr]
\nl
{}&+& 11\,v^{\ada}_{\saaaf}\,p^2_1\,
\Bigl[ n\,D - (n-1)\,D_1 \Bigr]
    - 3\,v^{\ada}_{\saaaf}\,p_{12}\,
\Bigl[  (n-2)\,D - 3\,(n-1)\,D_1 \Bigr]
\nl
{}&+& v^{\ada}_{\saaae}\,p^2_2\,
\Bigl[ 2\,(n-2)\,D - 3\,(n-1)\,D_1 \Bigr]
       +9\,v^{\ada}_{\saaae}\,p^2_1\,
\Bigl[ n\,D - (n-1)\,D_1 \Bigr]
\nl
{}&-& 2\,v^{\ada}_{\saaae}\,p_{12}\,
\Bigl[  (n-2)\,D - 3\,(n-1)\,D_1 \Bigr]
       + v^{\ada}_{\saaac}\,p^2_1\,
\Bigl[ n\,D - (n-1)\,D_1 \Bigr]
\nl
{}&-& 3\,v^{\ada}_{\saaad}\,p^2_1\,
\Bigl[  n\,D - (n-1)\,D_1 \Bigr]
  + v^{\ada}_{\saaad}\,p_{12}\,
\Bigl[ (n-2)\,D - 3\,(n-1)\,D_1 \Bigr],
\nl
\nl
{}&{}& (n+2)\,p^2_2\,\cV^{\ada}(p_1, p_1, p_1\,|\,0) =
       3\,v^{\ada}_{\saaaa}\,p^2_1\,D_1 
       +3\,v^{\ada}_{\saaab}\,p_{12}\,D_1 
- 11\,v^{\ada}_{\saaaf}\,p^4_1\,D_1\,( n-1 )
\nl
{}&+& v^{\ada}_{\saaaf}\,
\Bigl[ 3\,n\,D\,D_1 + (n+2)\,D\,D_2 - (n-1)\,D_1\,(D_2 - 9\,D_3 
+ 3\,D_1)\Bigr] 
\nl
{}&-& 9\,v^{\ada}_{\saaae}\,p^4_1\,D_1\,( n-1 )
    +3\,v^{\ada}_{\saaae}\,D_1\,
\Bigl[ n\,D + (n-1)\,(2\,D_3 - D_1) \Bigr]
\nl
{}&-& (n-1)\,\Bigl[ v^{\ada}_{\saaac}\,p^4_1
               - 3\,v^{\ada}_{\saaad}\,p^4_1
               + 3\,v^{\ada}_{\saaad}\,D_3 \Bigr]\,D_1
\nl
\nl
{}&{}& (n+2)\,p^4_1\,\cV^{\ada}(p_2, p_2, p_2\,|\,0) =
       \Bigl[ 3\,v^{\ada}_{\saaaa}\,D_3 
       +3\,v^{\ada}_{\saaab}\,\,D_1 
       - v^{\ada}_{\saaaf}\,p^2_2\,D_1\,( n-1 )
       - 3\,v^{\ada}_{\saaaf}\,p_{12}\,D_1\,( n-1 ) \Bigr]\,D_1
\nl
{}&-& v^{\ada}_{\saaaf}\,p^2_1\,
\Bigl[ 9\,n\,D\,D_1 - 11\,(n+2)\,D\,D_3 + (n-1)\,D_1\,(11\,D_3 
- 9\,D_1) \Bigr]
\nl
{}&-& 3\,v^{\ada}_{\saaae}\,p^2_1\,
\Bigl[ 2\,n\,D\,D_1 - 3\,(n+2)\,D\,D_3 + (n-1)\,D_1\,(3\,D_3 
- 2\,D_1) \Bigr]
\nl
{}&-& 3\,v^{\ada}_{\saaae}\,p_{12}\,D_1^2\,( n-1 )
       + v^{\ada}_{\saaac}\,p^2_1\,D_3\,
\Bigl[ (n+2)\,D - (n-1)\,D_1 \Bigr]
\nl
{}&+& 3\,v^{\ada}_{\saaad}\,p^2_1\,
\Bigl[ n\,D\,D_1 - (n+2)\,D\,D_3 + (n-1)\,D_1\,(D_3 - D_1) \Bigr].
\eqa
We still need the explicit form of the form factors which requires generalized
scalars of \eqn{genaaa}:
\[ \left( \begin{array}{c}
\cV^{\ada}_{\sbbba} \\ 
\cV^{\ada}_{\sbbbb} \\ 
\cV^{\ada}_{\sbbbc} \\ 
\cV^{\ada}_{\sbbbd} \\ 
\cV^{\ada}_{\sbbbe} \\ 
\cV^{\ada}_{\sbbbf}
\end{array} \right) \;=\;
\left( \begin{array}{cccccc}
n+2 \,&\, 0   \,&\, 2\,p_{12} \,&\, p^2_2  \,&\, p^2_1 \,&\, 0\\
2   \,&\, 0   \,&\, p_{12}   \,&\, 0    \,&\, p^2_1  \,&\, 0\\
0   \,&\, 1   \,&\, p^2_1   \,&\, p_{12}  \,&\, 0    \,&\, 0\\
0   \,&\, 0   \,&\, 0     \,&\, p^2_1  \,&\, 0    \,&\, p_{12}\\
p^2_1 \,&\, p_{12} \,&\, 0     \,&\, 0    \,&\, 0  \,&\,  0\\
0   \,&\, 2   \,&\, 0     \,&\, p_{12}  \,&\, 0   \,&\, p^2_2
\end{array} \right)^{-1}\;\;
\left( \begin{array}{c}
v^{\ada}_{\sbbba} \\ 
v^{\ada}_{\sbbbc} \\ 
v^{\ada}_{\sbbbe} \\ 
v^{\ada}_{\sbbbd} \\ 
v^{\ada}_{\sbbbf} \\
v^{\ada}_{\sbbbh}
\end{array} \right)
\]
\[ \left( \begin{array}{c}
\cV^{\ada}_{\sabba} \\ 
\cV^{\ada}_{\sabbb} \\ 
\cV^{\ada}_{\sabbc} \\ 
\cV^{\ada}_{\sabbd} \\ 
\cV^{\ada}_{\sabbe} \\ 
\cV^{\ada}_{\sabbf}
\end{array} \right) \;=\;
\left( \begin{array}{cccccc}
2     \,&\,  0    \,&\,   p_{12} \,&\,    0   \,&\,    p^2_1 \,&\,    0\\
0     \,&\,  1    \,&\,   p^2_1  \,&\,   p_{12} \,&\,    0   \,&\,    0\\
0     \,&\,  0    \,&\,  0    \,&\,   p^2_1  \,&\,   0   \,&\,    p_{12}\\
p^2_1 \,&\,    p_{12}  \,&\,   0  \,&\,     0  \,&\,     0  \,&\,     0\\
0     \,&\,  0   \,&\,    p_{12}  \,&\,   0    \,&\,   p_{12}  \,&\,   0\\
p_{12}\,&\,     p^2_2  \,&\,   0  \,&\,     0  \,&\,     0   \,&\,    0
\end{array} \right)^{-1}\;\;
\left( \begin{array}{c}
v^{\ada}_{\sabbc} \\ 
v^{\ada}_{\sabbe} \\ 
v^{\ada}_{\sabbd} \\ 
v^{\ada}_{\sabbf} \\ 
v^{\ada}_{\sabbg} \\
v^{\ada}_{\sabbj}
\end{array} \right)
\]
\[ \left( \begin{array}{c}
\cV^{\ada}_{\saaba} \\ 
\cV^{\ada}_{\saabb} \\ 
\cV^{\ada}_{\saabc} \\ 
\cV^{\ada}_{\saabd} \\ 
\cV^{\ada}_{\saabe} \\ 
\cV^{\ada}_{\saabf} \\
\cV^{\ada}_{\saabg} \\
\cV^{\ada}_{\saabh}
\end{array} \right) \;=\;
\left( \begin{array}{cccccccc}
0    \,&\,   2+n   \,&\,  p^2_1   \,&\,  2\,p_{12} \,&\,  0    \,&\,   p^2_2  \,&\,   0  \,&\,     2 \\
2    \,&\,   0     \,&\,  p_{12}   \,&\,  0     \,&\,  p^2_1  \,&\,   0    \,&\,   2  \,&\,     0\\
0    \,&\,   0     \,&\,  0     \,&\,  p^2_1   \,&\,  0    \,&\,   p_{12}  \,&\,   0  \,&\,     0\\
0    \,&\,   0     \,&\,  p^2_2   \,&\,  0     \,&\,  p_{12}  \,&\,   0    \,&\,   0  \,&\,     0\\
p^2_1  \,&\,   p_{12}   \,&\,  0     \,&\,  0     \,&\,  0    \,&\,   0    \,&\,   0  \,&\,     0\\
1    \,&\,   0     \,&\,  p_{12}   \,&\,  p^2_2   \,&\,  0    \,&\,   0    \,&\,   1  \,&\,     0\\
0    \,&\,   2     \,&\,  0     \,&\,  p_{12}   \,&\,  0    \,&\,   p^2_2  \,&\,   0  \,&\,     2\\
p_{12}  \,&\,   p^2_2   \,&\,  0     \,&\,  0     \,&\,  0    \,&\,   0    \,&\,   0  \,&\,     0
\end{array} \right)^{-1}\;\;
\left( \begin{array}{c}
v^{\ada}_{\saabb} \\ 
v^{\ada}_{\saabc} \\ 
v^{\ada}_{\saabd} \\ 
v^{\ada}_{\saabg} \\ 
v^{\ada}_{\saabf} \\
v^{\ada}_{\saabi} \\
v^{\ada}_{\saabh} \\
v^{\ada}_{\saabj} \\
\end{array} \right)
\]
For the $111$ group we have
\bqa
(n+2)\,\cV^{\ada}_{\saaaa} &=&
       v^{\ada}_{\saaac}\, p^2_1 
       + v^{\ada}_{\saaad}\,(  - 3\,p^2_1 + 2\,p_{12} )
       + v^{\ada}_{\saaae}\,( 9\,p^2_1 - 4\,p_{12} + p^2_2 )
\nl
{}&+& v^{\ada}_{\saaaf}\,( 11\,p^2_1 - 6\,p_{12} + p^2_2 )
       + v^{\ada}_{\saaaa},
\nl
(n+2)\,\cV^{\ada}_{\saaab} &=&
        v^{\ada}_{\saaad}\,p^2_1 
       +2\,v^{\ada}_{\saaae}\,(  - p^2_1 + p_{12} )
+ v^{\ada}_{\saaaf}\,(  - 3\,p^2_1 + 2\,p_{12} + p^2_2 )
       + v^{\ada}_{\saaab},
\eqa
\bqa
\cV^{\ada}_{\saaac} &=& 
- v^{\ada}_{\saaad} + 2\,v^{\ada}_{\saaae} + 3\,v^{\ada}_{\saaaf},
\qquad
\cV^{\ada}_{\saaad} = -v^{\ada}_{\saaae}-v^{\ada}_{\saaaf},
\nl
\cV^{\ada}_{\saaae} &=& 
- v^{\ada}_{\saaac} + 3\,v^{\ada}_{\saaad} - 9\,v^{\ada}_{\saaae} - 
11\,v^{\ada}_{\saaaf},
\qquad
\cV^{\ada}_{\saaaf} = -v^{\ada}_{\saaaf}.
\eqa
\subsubsection{The $V^{\bca}$ family \label{satutwo}}
Most of the fully saturated rank three tensors in this family can be
trivially obtained from the partial contractions of \eqn{contraK} and
evaluated with the help of \eqns{smallvKa}{smallvKd}. Some of them, however,
correspond to contractions leading to irreducible scalar products and,
therefore, they require an explicit solution for the form factors. The
latter are given in the following list where we use shorthand notation,
$P_i = \spro{p_i}{P}$:
\bqa
{}&{}& P^2\,P_2\,\cV^{\bca}(p_1, p_1, p_2\,|\,0) =
      v^{\bca}_{\saaaa}\,p_{12}\,
\Bigl[ D\,(P^2 + p_{12}) + p_{12}\,(P^2\,p_{12} +  P_1^2)\Bigr]
\nl
{}&+& v^{\bca}_{\saaab}\,
\Bigl[ D\,(P_2^2 + p_{12}^2) + p_{12}^2\,(P^2\,p_{12} + P_2^2)\Bigr]
- v^{\bca}_{\saaad}\,D\,P_2\,p^2_2 
       + v^{\bca}_{\saaae}\,D\,( P_1\,p_{12} + D )
\nl
{}&-& v^{\bca}_{\saaaf}\,
\Bigl[  (n-1)\,(D\,P_2 + P^2\,p_{12}^2) + D\,p_{12} \Bigr]
+ V^{\bca}_{\saaaa}\,D^2\,(n-2),
\nl
\nl
{}&{}& P^2\,P_2\,\cV^{\bca}(p_2, p_2, p_1\,|\,0) =
       v^{\bca}_{\saaaa}\,p_{12}^2\,( D + P^2\,p_{12} + P_2^2  )
+ v^{\bca}_{\saaab}\,P_2\,
\Bigl[ D\,(p_{12} - P_2) + P_2^2\,p_{12} \Bigr]
- V^{\bca}_{\saaaa}\,D^2\,(n-2), 
\nl
{}&-& v^{\bca}_{\saaad}\,D\,P_2\,( p_{12} - P_2 )
       + v^{\bca}_{\saaae}\,D\,P_2\,( p_{12} + P_2 )
+ v^{\bca}_{\saaaf}\,
\Bigl[ (n-1)\,P_2\,( D - P_2\,p_{12}) - D\,(n-2)\,p_{12} \Bigr],
\nl
\nl
{}&{}& P^2\,P_2\,\cV^{\bca}(p_1, p_1, p_1\,|\,0) =
  v^{\bca}_{\saaaa}\,( D + P_1\,p_{12})\,( p_1^2 + D ) 
- V^{\bca}_{\saaaa}\,D^2\,(n-2) 
\nl
{}&+& v^{\bca}_{\saaab}\,
\Bigl[  D\,( D + 2\,P^2\,p_{12} + p_{12}^2) + p_{12}^2\,(P^2\,p_{12} 
+ P_1^2) \Bigr] 
- v^{\bca}_{\saaad}\,D\,(  2\,P_2\,p_{12} + P_1\,p_{12} + D )
\nl
{}&-& v^{\bca}_{\saaae}\,D\,(  P_1\,p_{12} + D )
- v^{\bca}_{\saaaf}\,
\Bigl[ (n-1)\,(D\,P_1 - P_1^2\,p_{12}) + D\,(n-2)\,p_{12} \Bigr],
\nl
\nl
{}&{}& P^2\,P_2\,p^2_1\,\cV^{\bca}(p_2, p_2, p_2\,|\,0) =
       v^{\bca}_{\saaaa}\,p_{12}^3\,
( D + P^2\,p_{12} + P_2^2 )
+ v^{\bca}_{\saaab}\,P_2\,
\Bigl[ p^2_{12}\,( D + P_2^2 ) + D\,( D - P_2\,p_{12} )\Bigr] 
\nl
{}&-& v^{\bca}_{\saaad}\,D\,P_2\,( p_{12}^2 - P_2^2 )
       + v^{\bca}_{\saaae}\,D\,P_2\,( p_{12}^2 + 2\,P_2\,p_{12} - D )
\nl
{}&+& v^{\bca}_{\saaaf}\,
\Bigl[  D\,n\,P_2\,p_{12} - (n-2)\,D\,(p_{12}^2 + D) 
+ D\,P_2^2 - (n-1)\,P_2^2\,p_{12}^2 \Bigr] 
+ V^{\bca}_{\saaaa}\,D^2\,(n-2)\,p^2_1.
\eqa
For some special purpose, one may need to have direct access to the
explicit expressions of the form factors, or of the uncontracted tensors,
which is the same. Here is their solution:
\[
\left( \begin{array}{c} \cV^{\bca}_{\sbbba}  \\ \cV^{\bca}_{\sbbbb}  
\end{array}\right) = G^{-1}\,
\left( \begin{array}{c} v^{\bca}_{\sbbbf}  \\ v^{\bca}_{\sbbbj}  
\end{array}\right),
\qquad
\left( \begin{array}{c}
\cV^{\bca}_{\sbbbc} \\ 
\cV^{\bca}_{\sbbbd} \\ 
\cV^{\bca}_{\sbbbe} \\ 
\cV^{\bca}_{\sbbbf}
\end{array} \right) \;=\;
\left( \begin{array}{cccc}
2\,p_{12} &  p^2_2  &   p^2_1  &   0  \\
p_{12}   &  0      &   p^2_1  &   0  \\
p^2_1    &  p_{12} &   0      &   0  \\
0        &  p^2_1  &   0      &   p_{12}
\end{array} \right)^{-1}\;\;
\left( \begin{array}{c}
v^{\bca}_{\sbbba} - (n+2)\,V^{\bca}_{\sbbba} \\ 
v^{\bca}_{\sbbbc} - 2\,V^{\bca}_{\sbbba} \\ 
v^{\bca}_{\sbbbe} - V^{\bca}_{\sbbbb} \\ 
v^{\bca}_{\sbbbd} 
\end{array} \right)
\]
\bqa
\cV^{\bca}_{\saaaa} &=& -\omega^4 \Bigl[
	    \cV_{\bca}^{1, 1 | 1, 1, 3 | 2 }
          + \cV_{\bca}^{1, 1 | 1, 2, 2 | 2 }
          + \cV_{\bca}^{1, 1 | 1, 3, 1 | 2 }
     + \frac{1}{2} \cV_{\bca}^{1, 1 | 2, 1, 2 | 2 }
     + \frac{1}{2} \cV_{\bca}^{1, 1 | 2, 2, 1 | 2 } 
\nn \\  &&
          + \cV_{\bca}^{1, 1 | 1, 1, 2 | 3 }
          + \cV_{\bca}^{1, 1 | 1, 2, 1 | 3 }
          + \cV_{\bca}^{1, 2 | 1, 1, 3 | 1 }
          + \cV_{\bca}^{1, 2 | 1, 2, 2 | 1 }
          + \cV_{\bca}^{1, 2 | 1, 3, 1 | 1 }
          + \cV_{\bca}^{1, 2 | 2, 1, 2 | 1 } 
\nn \\  &&
          + \cV_{\bca}^{1, 2 | 2, 2, 1 | 1 }
          + \cV_{\bca}^{1, 2 | 3, 1, 1 | 1 }
          + \cV_{\bca}^{1, 2 | 1, 1, 2 | 2 }
          + \cV_{\bca}^{1, 2 | 1, 2, 1 | 2 }
          + \cV_{\bca}^{1, 2 | 2, 1, 1 | 2 }
          + \cV_{\bca}^{1, 2 | 1, 1, 1 | 3 }\Bigr]\,,
\nn \\
        \cV^{\bca}_{\saaab} &=& \frac{1}{{p_2}{P}}\,\Bigl[ 
\spro{p_1}{P}\,V^{\bca}_{\saaaa} + v^{\bca}_{\saaaf}\Bigr] 
\nn
\eqa
\[
\left( \begin{array}{c}
\cV^{\bca}_{\saaac} \\ 
\cV^{\bca}_{\saaad} \\ 
\cV^{\bca}_{\saaae} \\ 
\cV^{\bca}_{\saaaf}
\end{array} \right) \;=\;
\left( \begin{array}{cccc}
2\,p_{12}      &  p^2_2         &  p^2_1  &   0             \\
p^2_1          &  2\,p_{12}     &  0      &   p^2_2         \\
\spro{p_1}{P}  &  \spro{p_2}{P} &  0      &   0             \\
0              &  \spro{p_1}{P} &  0      &   \spro{p_2}{P} \\
\end{array} \right)^{-1}\;\;
\left( \begin{array}{c}
v^{\bca}_{\saaaa} - (n+2)\,V^{\bca}_{\saaaa} \\ 
v^{\bca}_{\saaac} - (n+2)\,V^{\bca}_{\saaab} \\ 
v^{\bca}_{\saaae} - V^{\bca}_{\saaaa} \\ 
v^{\bca}_{\saaad} - 2\,V^{\bca}_{\saaab}
\end{array} \right)
\]
Note that for the $111$ group one form factor must be written in terms of
generalized functions. This is a typical aspect of the procedure where, 
sometimes, the equations that one obtains are not all linearly independent.
\bq
\left( \begin{array}{c} \cV^{\bca}_{\saabc}  \\ \cV^{\bca}_{\saabd}  
\end{array}\right) = G^{-1}\,
\left( \begin{array}{c} v^{\bca}_{\saabg}  \\ v^{\bca}_{\saabp}  
\end{array}\right),
\eq
\[
\left( \begin{array}{c}
\cV^{\bca}_{\saaba} \\ 
\cV^{\bca}_{\saabb} \\ 
\cV^{\bca}_{\saabe} \\ 
\cV^{\bca}_{\saabf} \\ 
\cV^{\bca}_{\saabg} \\ 
\cV^{\bca}_{\saabh} \\ 
\cV^{\bca}_{\saabi} \\ 
\cV^{\bca}_{\saabj}  
\end{array} \right) \;=\;
\left( \begin{array}{cccccccc}
2 \,&\, 0 \,&\, p^2_1 \,&\, 0 \,&\, 2\,p_{12} \,&\, 0 \,&\, 0 \,&\, p^2_2  \\
1 \,&\, 0 \,&\, 0   \,&\, 0 \,&\, 0      \,&\, \spro{p_1}{P} \,&\,  
\spro{p_1}{P} \,&\, 0  \\
0 \,&\, 1 \,&\, 0   \,&\, 0 \,&\, \spro{p_1}{P} \,&\, 0 \,&\, 0 \,&\, 
\spro{p_2}{P}  \\
\spro{p_1}{P} \,&\, \spro{p_2}{P} \,&\, 0 \,&\, 0 
\,&\, 0 \,&\, 0 \,&\, 0 \,&\, 0  \\
2 \,&\, 0 \,&\, p^2_1 \,&\, 0 \,&\, 0 \,&\, 0 \,&\, p_{12} \,&\, 0  \\     
0 \,&\, 0 \,&\, 0   \,&\, p_{12} \,&\, 0 \,&\, 0 \,&\, 0 \,&\, p^2_1  \\
1 \,&\, 0 \,&\, 0 \,&\, 0 \,&\, p^2_1 \,&\, p_{12} \,&\, 0 \,&\, 0  \\
0 \,&\, 0 \,&\, p_{12} \,&\, 0 \,&\, 0 \,&\, 0 \,&\, p^2_2 \,&\, 0
\end{array} \right)^{-1}\;\;
\left( \begin{array}{c}
v^{\bca}_{\saaba} - n\,V^{\bca}_{\saabc} \\ 
v^{\bca}_{\saabj} - \,V^{\bca}_{\saabd} \\ 
v^{\bca}_{\saabl} - \,V^{\bca}_{\saabc} \\ 
v^{\bca}_{\saabc} \\ 
v^{\bca}_{\saabd} \\ 
v^{\bca}_{\saabe} \\ 
v^{\bca}_{\saabf} \\ 
v^{\bca}_{\saabm}  
\end{array} \right)
\]
\bqa
\left( \begin{array}{c} \cV^{\bca}_{\sabba}  \\ \cV^{\bca}_{\sabbb}  
\end{array}\right) &=& G^{-1}\,
\left( \begin{array}{c} v^{\bca}_{\sabbg}  \\ v^{\bca}_{\sabbq}  
\end{array}\right),
\nn
\eqa
\[
\left( \begin{array}{c}
\cV^{\bca}_{\sabbc} \\ 
\cV^{\bca}_{\sabbd} \\ 
\cV^{\bca}_{\sabbe} \\ 
\cV^{\bca}_{\sabbf} \\ 
\cV^{\bca}_{\sabbg} \\ 
\cV^{\bca}_{\sabbh} \\ 
\cV^{\bca}_{\sabbi} \\ 
\cV^{\bca}_{\sabbj}  
\end{array} \right) \;=\;
\left( \begin{array}{cccccccc}
n  \,&\,   0  \,&\,   p^2_1  \,&\, 0  \,&\,   2\,p_{12} \,&\,0  \,&\,   
0  \,&\,   p^2_2 \\
0  \,&\,   n  \,&\,   0   \,&\,  p^2_2 \,&\,  0  \,&\,   
2\,p_{12} \,&\, p^2_1  \,&\, 0  \\
1  \,&\,   0  \,&\,   p^2_1  \,&\, 0   \,&\,  p_{12} \,&\,  0   \,&\,  
0  \,&\,   0 \\
0  \,&\,   0  \,&\,   0  \,&\,   0  \,&\,   p^2_1 \,&\,  0  \,&\,   
0  \,&\,   p_{12} \\
0  \,&\,   0  \,&\,   0  \,&\,   p_{12}  \,&\, 0  \,&\,   p^2_1 \,&\,  
0   \,&\,  0 \\
0  \,&\,   0  \,&\,   p_{12} \,&\,  0   \,&\,  p^2_2 \,&\,  0   \,&\,  
0   \,&\,  0 \\
0  \,&\,   0  \,&\,   0  \,&\,   0 \,&\,    
0  \,&\,   p^2_2 \,&\,  p_{12} \,&\,  0 \\
0  \,&\,   1  \,&\,   0  \,&\,   
p^2_2  \,&\, 0   \,&\,  p_{12} \,&\,  0    \,&\, 0
\end{array} \right)^{-1}\;\;
\left( \begin{array}{c}
v^{\bca}_{\sabba} - 2\,\cV^{\bca}_{\sabba} \\ 
v^{\bca}_{\sabbb} - 2\,\cV^{\bca}_{\sabbb} \\ 
v^{\bca}_{\sabbc} - \,\cV^{\bca}_{\sabba} \\ 
v^{\bca}_{\sabbe} - \,\cV^{\bca}_{\sabbb} \\ 
v^{\bca}_{\sabbd} \\
v^{\bca}_{\sabbm}  \\ 
v^{\bca}_{\sabbp} - \,\cV^{\bca}_{\sabba} \\ 
v^{\bca}_{\sabbn} - \,\cV^{\bca}_{\sabbb}  
\end{array} \right)
\]
\subsubsection{The $V^{\bbb}$ family \label{satuthree}}
We obtain tensors with saturated indices in the $111$ and $222$ groups
from the corresponding results in the $V^{\ada}$ family by replacing 
$v^{\ada}_{\saaagen}$ and $v^{\ada}_{\sbbbgen}$ with $v^{\bbb}_{\saaagen}$ 
and $v^{\bbb}_{\sbbbgen}$. Once again there are saturated tensors
leading to contractions with irreducible scalar products that require an
explicit solution for the form factors. The latter are given in the following 
list:
\bqa
p_{12}\,p^2_1\,\cV^{\bbb}(\mu\,|\,\mu, p_2) &=& 
        v^{\bbb}_{\sabbb}\,D\,p_{12} 
       - v^{\bbb}_{\sabbh}\,D\,p^2_2\,(n-1)
\nl
{}&-& v^{\bbb}_{\sabbf}\, n\,D\,p_{12} 
       - v^{\bbb}_{\sabbj}\,(n-1)\,(  D - p_{12}^2 )
       + v^{\bbb}_{\sabbc}\,D_3\,p^2_1 
\nl
{}&+& v^{\bbb}_{\sabbd}\,p_{12}\,\Bigl[ (n-1)\,D  + p_{12}^2 \Bigr]
       + v^{\bbb}_{\sabbe}\,
     \Bigl[ (n-1)\,D\,p^2_1 + 2\,D_3\,p_{12} \Bigr]
       + v^{\bbb}_{\sabbf}\,D_3 ,
\nl
p_{12}\,p^2_1\,\cV^{\bbb}(p_2\,|\,p_2, p_2) &=& 
            v^{\bbb}_{\sabbc}\,D_3\,p_{12} 
          + v^{\bbb}_{\sabbd}\,( D\,p^2_2 + D_2\,p_{12} )
       + 2\,v^{\bbb}_{\sabbe}\,p_{12}\,D_1
          + v^{\bbb}_{\sabbf}\,D_1 ,
\nl
\cV^{\bbb}(p_2\,|\,p_2, p_1) &=& 
         v^{\bbb}_{\sabbc}\,D_3 
       + v^{\bbb}_{\sabbd}\,D_2 
       + v^{\bbb}_{\sabbe}\,( D + 2\,p_{12}^2 )
       + v^{\bbb}_{\sabbf}\,p_{12} ,
\nl
\cV^{\bbb}(p_2\,|\,p_1, p_1) &=& 
         v^{\bbb}_{\sabbc}\,p^4_1 
       + v^{\bbb}_{\sabbd}\,p_{12}^2 
    + 2\,v^{\bbb}_{\sabbe}\,D_3 
       + v^{\bbb}_{\sabbf}\,p^2_1,
\nl
p^2_2\,\cV^{\bbb}(\mu, p_1\,|\,\mu) &=&
         v^{\bbb}_{\saaba}\,D 
       + v^{\bbb}_{\saabl}\,D_3 
       + v^{\bbb}_{\saabc}\,D_2 
\nl
{}&+& v^{\bbb}_{\saabd}\,\Bigl[ D\,(n-1) + p_{12}^2 \Bigr]
       - v^{\bbb}_{\saabe}\,\Bigl[  D\,(n-1) - p_{12}^2 \Bigr]
       + v^{\bbb}_{\saabf}\,p_{12}^2, 
\nl
\cV^{\bbb}(\mu, p_2\,|\,\mu) &=&
         v^{\bbb}_{\saabl}\,p^2_1 
       + v^{\bbb}_{\saabc}\,p^2_2 
       + v^{\bbb}_{\saabd}\,p_{12} 
       + v^{\bbb}_{\saabe}\,p_{12} 
       + v^{\bbb}_{\saabf}\,n,
\nl
\cV^{\bbb}(p_2, p_2\,|\,p_1) &=&
         v^{\bbb}_{\saabl}\,D_3 
       + v^{\bbb}_{\saabc}\,D_2 
       + v^{\bbb}_{\saabd}\,p_{12}^2 
+ v^{\bbb}_{\saabe}\,D_1
       + v^{\bbb}_{\saabf}\,p_{12}.
\eqa 
The form factors are obtained as follows:
\[
\cV^{\bbb}_{\sbbba} = \frac{1}{p_1^2}\,\Bigl[ 
v^{\bbb}_{\sbbbe} - p_{12}\,\cV^{\bbb}_{\sbbbb}\Bigr],
\qquad 
\left( \begin{array}{c}
\cV^{\bbb}_{\sbbbc} \\ 
\cV^{\bbb}_{\sbbbd} \\ 
\cV^{\bbb}_{\sbbbe} \\ 
\cV^{\bbb}_{\sbbbf}
\end{array} \right) \;=\;
\left( \begin{array}{cccc}
2\,p_{12} \,&\,  p^2_2 \,&\,   p^2_1 \,&\,  0 \\
p_{12}   \,&\, 0    \,&\,  p^2_1  \,&\, 0 \\
p^2_1   \,&\, p_{12}  \,&\,  0    \,&\, 0 \\
0     \,&\, p^2_1  \,&\,  0    \,&\, p_{12}
\end{array} \right)^{-1}\;\;
\left( \begin{array}{c}
v^{\bbb}_{\sbbba} - (n+2)\,V^{\bbb}_{\sbbba} \\ 
v^{\bbb}_{\sbbbc} - 2\,V^{\bbb}_{\sbbba} \\ 
v^{\bbb}_{\sbbbe} - V^{\bbb}_{\sbbbb} \\ 
v^{\bbb}_{\sbbbd} 
\end{array} \right)
\]
\bqa
\cV^{\bbb}_{\sbbbb} &=& \omega^4 \Bigl[
	\frac{1}{2} \cV_{\bbb}^{1, 2 | 1, 1| 2, 2 }
          + \cV_{\bbb}^{1, 2 | 1, 1| 3, 1 }
          + \cV_{\bbb}^{1, 3 | 1, 1| 2, 1 }
          - \cV_{\bbb}^{2, 1 | 1, 1| 1, 3 }
      -  \frac{1}{2} \cV_{\bbb}^{2, 1 | 1, 1| 2, 2 }
      -  \frac{1}{2} \cV_{\bbb}^{2, 2 | 1, 1| 1, 2 } \nl
      &+&  \frac{1}{2} \cV_{\bbb}^{2, 2 | 1, 1| 2, 1 }
          - \cV_{\bbb}^{3, 1 | 1, 1| 1, 2 }\Bigr]\,,
\nn
\eqa
\[
\cV^{\bbb}_{\saaab} = \frac{1}{p_{2}^2}\,\Bigl[
v^{\bbb}_{\saaaf} - p_{12}\,\cV^{\bbb}_{\saaaa}\Bigr],
\qquad 
\left( \begin{array}{c}
\cV^{\bbb}_{\saaac} \\ 
\cV^{\bbb}_{\saaad} \\ 
\cV^{\bbb}_{\saaae} \\ 
\cV^{\bbb}_{\saaaf}
\end{array} \right) \;=\;
\left( \begin{array}{cccc}
p^2_1  \,&\,   2\,p_{12} \,&\,  0  \,&\,     p^2_2 \\
p^2_2  \,&\,   0     \,&\,  p_{12} \,&\,    0 \\
p_{12}  \,&\,   p^2_2   \,&\,  0   \,&\,    0 \\
0    \,&\,   p_{12}   \,&\,  0   \,&\,    p^2_2
\end{array} \right)^{-1}\;\;
\left( \begin{array}{c}
v^{\bbb}_{\saaab} - (n+2)\,V^{\bbb}_{\saaab} \\ 
v^{\bbb}_{\saaac}  \\ 
v^{\bbb}_{\saaae} - V^{\bbb}_{\saaaa} \\ 
v^{\bbb}_{\saaad} - 2\,V^{\bbb}_{\saaab}
\end{array} \right)
\]
\bqa
\cV^{\bbb}_{\saaaa} &=& \omega^4 \Bigl[
            \cV_{\bbb}^{1 , 1 | 1 , 3| 1 , 2 }
      + \frac{1}{2}\cV_{\bbb}^{1 , 1 | 2 , 2| 1 , 2 }
          + \cV_{\bbb}^{1 , 1 | 1 , 2| 1 , 3 }
      - \frac{1}{2}\cV_{\bbb}^{1 , 1 | 2 , 2| 2 , 1 } 
          - \cV_{\bbb}^{1 , 1 | 3 , 1| 2 , 1 }
      + \frac{1}{2}\cV_{\bbb}^{1 , 1 | 1 , 2| 2 , 2 } \nl
      &-& \frac{1}{2}\cV_{\bbb}^{1 , 1 | 2 , 1| 2 , 2 }
          - \cV_{\bbb}^{1 , 1 | 2 , 1| 3 , 1 }\Bigr]\,,
\nn
\eqa
\[ 
\left( \begin{array}{c}
\cV^{\bbb}_{\saaba} \\ 
\cV^{\bbb}_{\saabb} \\ 
\cV^{\bbb}_{\saabc} \\ 
\cV^{\bbb}_{\saabd} \\ 
\cV^{\bbb}_{\saabe} \\ 
\cV^{\bbb}_{\saabf} \\
\cV^{\bbb}_{\saabg} \\
\cV^{\bbb}_{\saabh}
\end{array} \right) \;=\;
\left( \begin{array}{cccccccc}
2 \,&\,  0\,&\,   n \,&\,  0 \,&\,  p^2_1\,&\,   0\,&\,   
2\,p_{12}\,&\, p^2_2\\
2 \,&\,  0\,&\,   0 \,&\,  0 \,&\,  p^2_1\,&\,   0\,&\,   p_{12} \,&\,0\\
0 \,&\,  1\,&\,   0 \,&\,  0 \,&\,  0 \,&\,  0\,&\,   p^2_1 \,&\,p_{12}\\
0 \,&\,  0\,&\,   0 \,&\,  0 \,&\,  0 \,&\,  p_{12}\,&\, 0   \,&\,p^2_1\\
0 \,&\,  0\,&\,   p^2_1\,&\,   p_{12}\,&\,   0 \,&\,  0  \,&\, 0   \,&\,0\\
0 \,&\,  0\,&\,   0 \,&\,  0 \,&\,  p_{12}\,&\,   0  \,&\, p^2_2 \,&\,0\\
0 \,&\,  0\,&\,   1 \,&\,  0 \,&\,  0 \,&\,  0  \,&\, p_{12} \,&\,p^2_2\\
p_{12}\,&\,   p^2_2\,&\,   0 \,&\,  0 \,&\,  0 \,&\,  0  \,&\, 0   \,&\,0\\
\end{array} \right)^{-1}\;\;
\left( \begin{array}{c}
v^{\bbb}_{\saaba} \\ 
v^{\bbb}_{\saabc} \\ 
v^{\bbb}_{\saabe} \\ 
v^{\bbb}_{\saabd} \\ 
v^{\bbb}_{\saabf} \\
v^{\bbb}_{\saabg} \\
v^{\bbb}_{\saabj} \\
v^{\bbb}_{\saabl} \\
\end{array} \right)
\]
\[ 
\left( \begin{array}{c}
\cV^{\bbb}_{\sabba} \\ 
\cV^{\bbb}_{\sabbb} \\ 
\cV^{\bbb}_{\sabbc} \\ 
\cV^{\bbb}_{\sabbd} \\ 
\cV^{\bbb}_{\sabbe} \\ 
\cV^{\bbb}_{\sabbf} \\
\cV^{\bbb}_{\sabbg} \\
\cV^{\bbb}_{\sabbh}
\end{array} \right) \;=\;
\left( \begin{array}{cccccccc}
0 \,&\,    2  \,&\,   0  \,&\,   n  \,&\,   0  \,&\,   
p^2_2 \,&\,  p^2_1 \,&\,  2\,p_{12}\\
1 \,&\,    0  \,&\,   1  \,&\,   0  \,&\,   p^2_1\,&\,   
0   \,&\,  p_{12} \,&\,  0\\
0 \,&\,    1  \,&\,   0  \,&\,   0  \,&\,   0  \,&\,   
0   \,&\,  p^2_1 \,&\,  p_{12}\\
0 \,&\,    0  \,&\,   0  \,&\,   1  \,&\,   0  \,&\,   
0   \,&\,  p^2_1 \,&\,  p_{12}\\
0 \,&\,    0  \,&\,   0  \,&\,   0  \,&\,   0  \,&\,   
p_{12} \,&\,  0   \,&\,  p^2_1\\
p^2_1\,&\,   p_{12}\,&\,   0  \,&\,   0  \,&\,   0  \,&\,   
0   \,&\,  0   \,&\,  0\\
0  \,&\,   0  \,&\,   0  \,&\,   0  \,&\,   p_{12}\,&\,   
0   \,&\,  p^2_2 \,&\,  0\\
0  \,&\,   0  \,&\,   p_{12}\,&\,   p^2_2\,&\,   
0  \,&\,   0   \,&\,  0   \,&\,  0
\end{array} \right)^{-1}\;\;
\left( \begin{array}{c}
v^{\bbb}_{\sabbb} \\ 
v^{\bbb}_{\sabbl} \\ 
v^{\bbb}_{\sabbd} \\ 
v^{\bbb}_{\sabbc} \\ 
v^{\bbb}_{\sabbf} \\
v^{\bbb}_{\sabbg} \\
v^{\bbb}_{\sabbi} \\
v^{\bbb}_{\sabbj} \\
\end{array} \right)
\]
%
\section{Symmetry properties \label{symme}}
Before discussing the symmetry properties of two-loop functions we briefly
describe our strategy to generate Feynman diagrams. We use the $\GS$ 
code~\cite{GS}, written in
FORM~\cite{Vermaseren:2000nd}, which uses the same logic introduced
in~\cite{vanderBij:1983bw} and has also been applied to one-loop diagrams
in~\cite{Andonov:2002jg}. Well-known packages for diagram handling are
listed in~\cite{Tentyukov:1999is}.

The basic algorithm is inspired by an efficient way of accounting for 
combinatorial factors in diagrams and uses the results of~\cite{Goldberg:hg}.

\begin{figure}[th]
\vspace{0.5cm}
$$
S^{\ssE}_{\ssQ\ssE\ssD} 
\quad= \quad
C_{131}\quad\Bigl[ \quad
  \vcenter{\hbox{
  \SetScale{0.8}
  \SetWidth{2.}
  \begin{picture}(150,0)(0,0)
  \Photon(0,0)(50,0){2}{7}
  \ArrowArc(75,0)(25,0,90)
  \ArrowArc(75,0)(25,90,180)
  \ArrowArc(75,0)(25,-180,0)
  \Photon(100,0)(150,0){2}{7}
  \GCirc(75,25){4.}{0.5}
  \end{picture}}}
\!\!\!+
\quad
  \vcenter{\hbox{
  \SetScale{0.8}
  \SetWidth{2.}
  \begin{picture}(150,0)(0,0)
  \Photon(0,0)(50,0){2}{7}
  \ArrowArcn(75,0)(25,180,90)
  \ArrowArcn(75,0)(25,90,0)
  \ArrowArcn(75,0)(25,360,180)
  \Photon(100,0)(150,0){2}{7}
  \GCirc(75,25){4.}{0.5}
  \end{picture}}}
\!\! \Bigr]
$$
\vspace{0.5cm}
$$
  \vcenter{\hbox{
  \SetScale{0.8}
  \SetWidth{2.}
  \begin{picture}(100,0)(0,0)
  \ArrowLine(0,0)(45,0)
  \ArrowLine(55,0)(100,0)
  \GCirc(50,0){10.}{0.5}
  \end{picture}}}
=
\quad
  \vcenter{\hbox{
  \SetScale{0.8}
  \SetWidth{2.}
  \begin{picture}(120,0)(0,0)
  \ArrowLine(0,0)(50,0)
  \ArrowLine(100,0)(150,0)
  \ArrowArcn(75,0)(25,180,0)
  \PhotonArc(75,0)(25,180,360){2}{10}
  \end{picture}}}
\quad +
  \vcenter{\hbox{
  \SetScale{0.8}
  \SetWidth{2.}
  \begin{picture}(120,0)(0,0)
  \ArrowLine(0,0)(50,0)
  \ArrowLine(100,0)(150,0)
  \ArrowArc(75,0)(25,180,360)
  \PhotonArc(75,0)(25,0,180){2}{10}
  \end{picture}}}
$$
\vspace{0.5cm}
\caption[]{The QED $S^{\ssE}$ graph as generated by $\GS$~\cite{GS}. The
combinatoric factor $C_{131}$ corresponds to the scalar $S^{\ssE}$ graph with
identical lines and is equal to  $1/2$.}
\label{generation}
\end{figure}
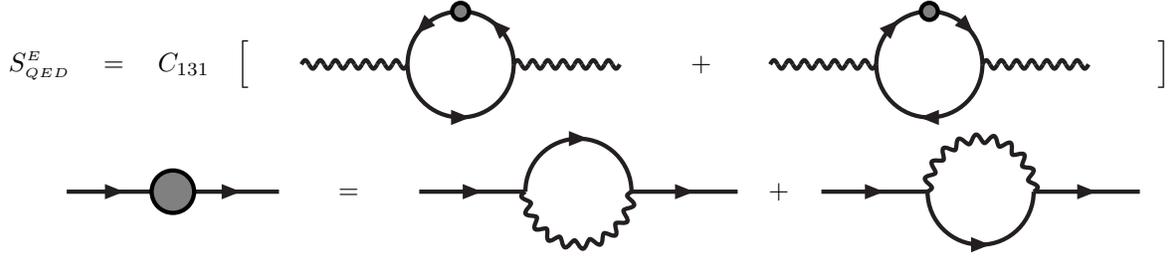

$\GS$ has a table of all the vertices occurring in the Standard Model, which
involves the particles $\gamma, \,\cdots f, \,\cdots$. Starting from the
class of diagrams one wants to evaluate, $\GS$ generates all the graphs by
inserting every possible combination of propagators $\gamma\gamma, \barf f,
f\barf,\,\cdots$ in the topology, discarding those containing non existing
vertices. Combinatorial factors are included for each graph corresponding to
the situation where all propagators are scalar and identical. An example is
given in \fig{generation}.
After their generation, the diagrams are ready for evaluation or for a check
of the corresponding WST identities of the theory. For the latter case we 
always produce a full scalarization of the result and look for symmetries 
among the various terms in the result. A typical intermediate output of $\GS$ 
is given in \eqn{onesexa}.

Let us give an example of the symmetries: for the $S^{\ssA}$ family we have
the inverse propagators $[1] = q^2_1 + m^2_1, [2]_{\ssA} = (q_1 - q_2 + p)^2
+ m^2_2$, and $[3]_{\ssA} = q^2_2 + m^2_3$.  Consider a set of
transformations (where $\oplus$ stands for `followed by'):
\bqa
{\rm a}) \quad &:& \quad q_1 \to q_1 + q_2 - p \,\oplus\, q_2 \to - q_2,
\nl
{\rm b}) \quad &:& \quad q_1 \leftrightarrow q_2,
\nl
{\rm c}) \quad &:& \quad q_2 \to -q_2 + q_1 + p;
\eqa
they correspond to symmetries specified by
\bqa
{\rm a}) \,&\to&\, S^{\ssA}_{\Z}(p,\lstm{123}) = S^{\ssA}_{\Z}(-p,\lstm{213}),
\nl
{}&{}& S^{\ssA}_{\sa}(p,\lstm{123}) =  -S^{\ssA}_{\sa}(-p,\lstm{213}) +
S^{\ssA}_{\msb}(-p,\lstm{213}) - S^{\ssA}_{\Z}(-p,\lstm{213}),
\nl
{}&{}& S^{\ssA}_{\msb}(p,\lstm{123}) =  S^{\ssA}_{\msb}(-p,\lstm{213}),
\nl
{\rm b}) \,&\to&\, S^{\ssA}_{\Z}(p,\lstm{123}) = S^{\ssA}_{\Z}(-p,\lstm{321}),
\nl
{}&{}& S^{\ssA}_{\sa}(p,\lstm{123}) = - S^{\ssA}_{\msb}(-p,\lstm{321}),
\nl
{\rm c}) \,&\to&\, S^{\ssA}_{\Z}(p,\lstm{123}) = S^{\ssA}_{\Z}(p,\lstm{132}),
\nl
{}&{}& S^{\ssA}_{\sa}(p,\lstm{123}) = S^{\ssA}_{\sa}(p,\lstm{132}),
\nl
{}&{}& S^{\ssA}_{\msb}(p,\lstm{123}) = - S^{\ssA}_{\msb}(p,\lstm{132}) +
S^{\ssA}_{\sa}(p,\lstm{132}) + S^{\ssA}_{\Z}(p,\lstm{132}).
\eqa
For the other two-point functions we recall the conventions:
each propagator will be denoted by $[i] = k^2_i + m^2_1$ and
\[
\ba{llllll}
C, \qquad & k_1= q_1,\, & k_2 = q_1-q_2, \, & k_3 = q_2, \, & k_4 = q_2+p ,&
\\
D, \qquad & k_1= q_1,\, & k_2 = q_1+p,\, & k_3 = q_1-q_2,\, & k_4 = q_2,\, 
& k_5 = q_2+p,
\\
E, \qquad & k_1= q_1,\, & k_2 = q_1-q_2,\, & k_3 = q_2,\, 
& k_4 = q_2+p,\, & k_5 = q_2.
\ea
\]
We simply indicate the symmetry property of the scalar configurations;
for instance, a change of variables $q_1 \to q_1 + q_2$ followed by 
$q_2 \to - q_2$ corresponds to a symmetry of the $S^{\ssC}$ family 
with respect to the exchange $p \to - p$ and $m_1 \leftrightarrow m_2$:
\bqa
S^{\ssC} \quad &:& \quad q_1 \to q_1 + q_2 \,\oplus\, q_2 \to - q_2, \quad
\Rightarrow \quad p \to - p, \; m_1 \leftrightarrow m_2,
\nl
S^{\ssE} \quad &:& \quad m_3 \leftrightarrow m_5,
\nl
{}&{}&
\quad q_1 \to q_1 + q_2 \,\oplus\, q_2 \to - q_2, \quad
\Rightarrow \quad p \to - p, \; m_1 \leftrightarrow m_2,
\nl
S^{\ssD} \quad &:& \quad  q_1 \leftrightarrow q_2, \quad
\Rightarrow \quad m_1 \leftrightarrow m_4, \quad m_2 \leftrightarrow m_5,
\nl
{}&{}&  \quad q_1 \to q_1 - p \ ,  \quad q_2 \to q_2 -p, \quad
\Rightarrow m_1 \leftrightarrow m_2 , \quad m_4 \leftrightarrow m_5.
\eqa
Finally, let us consider the symmetries of the three-point functions; 
for the general class $V^{\ono}$ we obtain
\bqa
V^{\ono} \quad &:& \quad q_1 \to q_1 + q_2 \,\oplus\, q_2 \to - q_2, \quad
\Rightarrow \quad p_i \to - p_i, \; m_1 \leftrightarrow m_2,
\nl
V^{\ada} \quad &:& \quad \quad m_3 \leftrightarrow m_6,
\nl
{}&{}& \quad \quad m_4 \leftrightarrow m_5 \,\ , \  p_1 \leftrightarrow P   
\,\ ,  \  p_2 \to - p_2.
\label{sym1}
\eqa
For symmetries in the $V^{\bba}, V^{\bca}$ and $V^{\bbb}$ families we have
\bqa
V^{\bba} \quad &:& \quad q_1 \leftrightarrow q_2, \quad
\Rightarrow m_2 \leftrightarrow m_4, \quad m_1 \leftrightarrow m_5, \quad
p_1 \leftrightarrow - p_2.
\nl
V^{\bca} \quad &:& \quad q_i \to q_i - P, \quad
\Rightarrow m_1 \leftrightarrow m_2, \quad m_4 \leftrightarrow m_6, \quad
p_1 \leftrightarrow - p_2.
\nl
V^{\bbb} \quad &:& \quad q_1 \leftrightarrow q_2, \quad
\Rightarrow m_1 \leftrightarrow m_5, \quad
m_2 \leftrightarrow m_6, \quad
m_3 \leftrightarrow m_4, \quad
p_1 \leftrightarrow p_2.
\eqa
All symmetry properties refer to the scalar configurations.

\clearpage

\end{document}